\documentclass{report}

\usepackage{amsfonts}
\usepackage{amsmath}
\usepackage{amssymb}
\usepackage{tcolorbox}
\usepackage{color,colortbl}
\definecolor{cinza}{rgb}{0.8,0.8,0.8}

\usepackage{hyperref}
\hypersetup{
    colorlinks=true,
    linkcolor=blue,
    filecolor=magenta,      
    urlcolor=cyan,
      citecolor=blue,
    }

\usepackage{makeidx}         
\usepackage{graphicx}        
\usepackage[bottom]{footmisc}

\newcommand{\V}[0]{\mathbb{V}}
\newcommand{\N}[0]{\mathcal{N}}
\newcommand{\LN}[0]{\mathcal{LN}}

\makeindex             

\graphicspath{{./}{fig/}}

\begin{document}


\thispagestyle{empty}
\begin{center}

\LARGE{\sc Statistical Laws in Complex Systems}
\bigskip

\Large{Monograph\footnote{To be submitted for publication to the Springer series "Understanding Complex Systems".}}

\vspace{5cm}

\Large{Eduardo G. Altmann\footnote{School of Mathematics and Statistics \& Centre for Complex Systems, The University of Sydney, Sydney, Australia.}\footnote{Max Planck Institute for the Physics of Complex Systems, Dresden, Germany.}\footnote{E-mail: eduardo.altmann@sydney.edu.au}}

\vspace{8cm}

\today

\end{center}


\newpage 
$\;$
\newpage

\tableofcontents
\newpage

\addcontentsline{toc}{chapter}{Abstract}
\begin{center}
  {\sc \bf Abstract}
  \end{center}
Statistical laws describe regular patterns observed in diverse scientific domains, ranging from the magnitude of earthquakes (Gutenberg-Richter law) and metabolic rates in organisms (Kleiber's law), to the frequency distribution of words in texts (Zipf's and Herdan-Heaps' laws), and productivity metrics of cities (urban scaling laws). The origins of these laws, their empirical validity, and the insights they provide into underlying systems have been subjects of scientific inquiry for centuries. This monograph provides an unifying approach to the study of statistical laws, critically evaluating their role in the theoretical understanding of complex systems and the different data-analysis methods used to evaluate them.
Through a historical review and a unified analysis, we uncover that the persistent controversies on the validity of statistical laws are predominantly rooted not in novel empirical findings but in the discordance among data-analysis techniques, mechanistic models, and the interpretations of statistical laws.
Starting with simple examples and progressing to more advanced time-series and statistical methods, this monograph and its accompanying repository provide comprehensive material for researchers interested in analyzing data, testing and comparing different laws, and interpreting results in both existing and new datasets.

\newpage

\addcontentsline{toc}{chapter}{Preface}
\begin{center}
  {\sc \bf Preface}
  \end{center}

In an era where information inundates every aspect of our lives and underpins economic activities, the identification of regular patterns has become vitally important. This challenge is not exclusive to our daily lives but is also prevalent in the scientific quantification of physical, biological, and social phenomena. The advent of "big data" has propelled this issue to the forefront of scientific discourse.
The aim of this monograph is to provide a critical examination of {\it statistical laws}, a methodology extensively employed across various disciplines to summarize regularities in observational data and to incorporate them into theory.

 Prominent exemplars of statistical laws go back to \index{Pareto's law} Pareto's law of income distribution (from the late 19th century), include Zipf's law\index{Zipf's law} of word frequencies and \index{Gutenberg-Richter's law} Gutenberg-Richter law of earthquake magnitudes (from the 20th century), and extend to contemporary claims of universality in the observation of \index{scale-free networks} scale-free networks, the fat-tailed distribution \index{fat-tailed distribution} of attention to online items, the stretched-exponential distribution of intervals between extreme events, the bursty temporal patterns in digital communication, and urban scaling laws (all in the 21st century). These instances, among others reviewed in this monograph, illustrate that statistical laws are not merely curiosities or summaries of empirical observations (stylized facts), they play a crucial role in the validation of mechanistic models and theories of the underlying system.  \index{stretched exponential distribution}
 
From a complex-systems perspective, statistical laws are emergent properties with inherent statistical characteristics: while they can be violated in controlled settings, they are universally observed across different scenarios. The explanation of these laws is obtained by considering {\it microscopic} models that lead to the observation of the statistical law at {\it macroscopic} scales. Numerous scientific disciplines have adopted this paradigm to gain a theoretical understanding of the predominant processes within a system, as discerned through the identification of statistical patterns. However, the influx of data inundating science and technology in the 21st century has brought not only opportunities for applications of statistical laws but also provoked the reevaluation of their relevance and validity. At a time in which these laws are under intensified scrutiny, this monograph intends to provide a much needed critical review of the potential and limitations of the complex-systems approach to statistical laws. 

In traditional "big-data" analysis, regularities are "learned" directly from the data without the need of parametric functions, the formulation of empirical laws, or the theoretization about their origin or significance. 
This stands in contrast to the traditions underpinning statistical laws, highlighting the striking differences between the \index{machine learning} machine-learning and the natural-science approaches to "Data Science". \index{Data Science}
In the natural-science approach, progress is achieved through the meticulous confrontation of theoretical predictions to empirical observations.  Conversely, in the machine-learning approach, progress is driven by the creation of generic, scalable, algorithms that exploit patterns in the data. Success is quantified by their performance in improving scores (in test datasets), in reproducing human outputs (e.g., in retrieving labels or human annotations), or in obtaining useful predictions (on particular cases). Advancement in image recognition, such as accurately identifying images of cats, are not contingent upon, nor do they influence, our theoretical understanding of feline nature or the neural processing of visual stimuli in our brains. Similarly, the deployment of large language models -- representing some of the latest mass applications of artificial intelligence --  does not derive from, nor does it alter,  our scientific comprehension of natural language.
The prevailing notion suggests that progress will not stem from the  understanding, manipulation, and application of (universally valid) scientific principles or theories. Rather, it is posited that progress will be driven by autonomous learning machines, achieving general-purpose intelligence through the training of generic algorithms with parameters and datasets of a scale beyond individual human grasp.

By reviewing and reflecting on the role of statistical laws in complex systems, a data-driven approach rooted in the natural sciences, this monograph aims to contribute to one of the crucial scientific debates of our time: the place of theory in data-driven science. The role of statistical models, data size, and assumptions of independent observations are recurrent issues in debates around the validity of statistical laws and are prevalent also in different scientific fields which are increasingly driven by data. More generally, we hope that by showing the intricate relationships between data and models in the analysis of statistical laws in Complex Systems we will show how theory is not only inseparable from the data-driven approaches, but that it can be beneficial to and benefit from the increasing availability of data. In this wider context, our aim is to contribute towards a more scientifically grounded alternative to the illusory "theory-free" approach to Data Science. \index{Data Science}

We start this monograph with a definition and the historical context in which statistical laws appear (Chap.~\ref{chap.introduction}). Subsequently, we provide an exposition of various laws (Chap.~\ref{chap.examples}), illustrating their analogous function in the development of theoretical models, from urban systems to tectonic plates. This parallelism justifies our unified approach to statistical laws and informs our more abstract theory around their interpretation and role in complex-systems research. We then examine (Chap.~\ref{chap.data}) statistical methods employed to study and test the validity of statistical laws. The need for an improved interpretation of these laws becomes apparent from the recent challenges to the validity of laws that had been long considered as well established. These questionings are a consequence not only of the modern availability of large databases, which invariably make deviations of statistical laws to be statistically significant, but also of the employment of different data-analysis methodologies. We discuss in detail the applicability of the different statistical methods and some of the pitfalls on making na\"ive interpretations of their results. We conclude (in Chap.~\ref{chap.synthesis}) with a discussion of different interpretations of statistical laws, their consequences to theoretical models, and we make recommendations for practitioners. The data and codes used in all our figures and statistical analyses are part of our coding repository (as described in Appendix~\ref{chap.appendices}),  an invitation for readers to replicate, expand, and apply the ideas developed here.

\newpage
\begin{center}
  {\sc \bf Acknowledgements}
  \end{center}

This monograph would not be possible without the support of many colleagues and institutions. The project and idea of writing a longer text in this subject goes back to 2018, but little progress was made before my 2023/2024 sabbatical, supported by The University of Sydney (Australia) and the Max Planck Institute for the Physics of Complex Systems (Germany). 
The ideas included in this monograph reflect research I have performed in the last 20 years, across three continents and in collaborations with numerous colleagues. While most of the projects involved specific questions and applications, and citations to the published works is included in context, the influence and contributions of my co-authors to the ideas I expose here greatly extrapolate the content of our joint publications.  This applies most strongly in the case of my collaboration with {\bf Martin Gerlach}, with whom ideas of statistical laws in linguistics~\cite{gerlach_stochastic_2013,altmann_statistical_2016} were expanded~\cite{gerlach_testing_2019} to the more general context presented here, but also to all my co-authors in this subject: E.C. da Silva, I.L. Caldas, H. Kantz, J. Pierrehumbert, A. E. Motter, D. R. Amancio, O. N. Oliveira Jr, G. Cristadoro, M. Degli Esposti, R. Dickman, N. R. Moloney, {\bf D. Rybski}, J. M. Miotto, F. Ghanbarnejad, J. C. Leitao, F. Font-Clos, T. P. Peixoto, H.H. Chen, D.F. M. Oliveria, T. Alexander, and {\bf J.M. Moore} (the highlighted names indicate those that kindly provided helpful feedback on a previous version of this text). I have also benefited from extensive discussions on statistical laws with E. Arcaute, R. Ferrer-i-Cancho, A. Corral, M. Prokopenko, S. Sarkar,  K. Tanaka-Ishi, and many other colleagues from the complex-systems community.

\chapter{Introduction}\label{chap.introduction}

\section{Paradigmatic examples}\label{sec.1example}

In the early 20th century, physicist Felix \index{Auerbach, Felix} Auerbach (1856-1933)~\cite{auerbach_felix_gesetz_1913,rybski_commentary_2013}   noticed a striking regularity in the population of German cities: when {\it r}anking cities from largest ($r=1$) to  smallest ($r=R$), their population $N_r$ followed the simple relationship
\begin{equation}\label{eq.zipfcities}
 N_r = A/r,
\end{equation}
where $A \in \mathbb{R}$ is a constant approximately equal to the population of the largest city ($A\approx N_1$). This ratio predicted, for example, that Dresden (the $r=6$-th largest city in Germany at the time) would have a population about one sixth ($N_6/N_1 = 1/6$) that of Berlin ($r=1$), a good approximation of the observed ratio $\approx 1/6.22$. Auerbach noticed that his observation extended to other countries and, through later generalizations by Lotka and Zipf, led to one of the most celebrated statistical laws: the Auerbach-Lotka-Zipf\index{Auerbach-Lotka-Zipf's law} law ~\cite{rybski_auerbach_2023} of city sizes. It is a particular case of a discrete power-law distribution, a functional form underlying famous statistical laws, that goes back to the work of \index{Pareto's law} Pareto on the distribution of income in the late 19th century and includes many modern applications such as \index{scale-free networks} scale-free networks and Internet data (to be reviewed in Sec. \ref{sec:power-law-distr}).  \index{power-law distribution}

\begin{figure*}[!tb]
\begin{center}
  \includegraphics[width=0.5\linewidth]{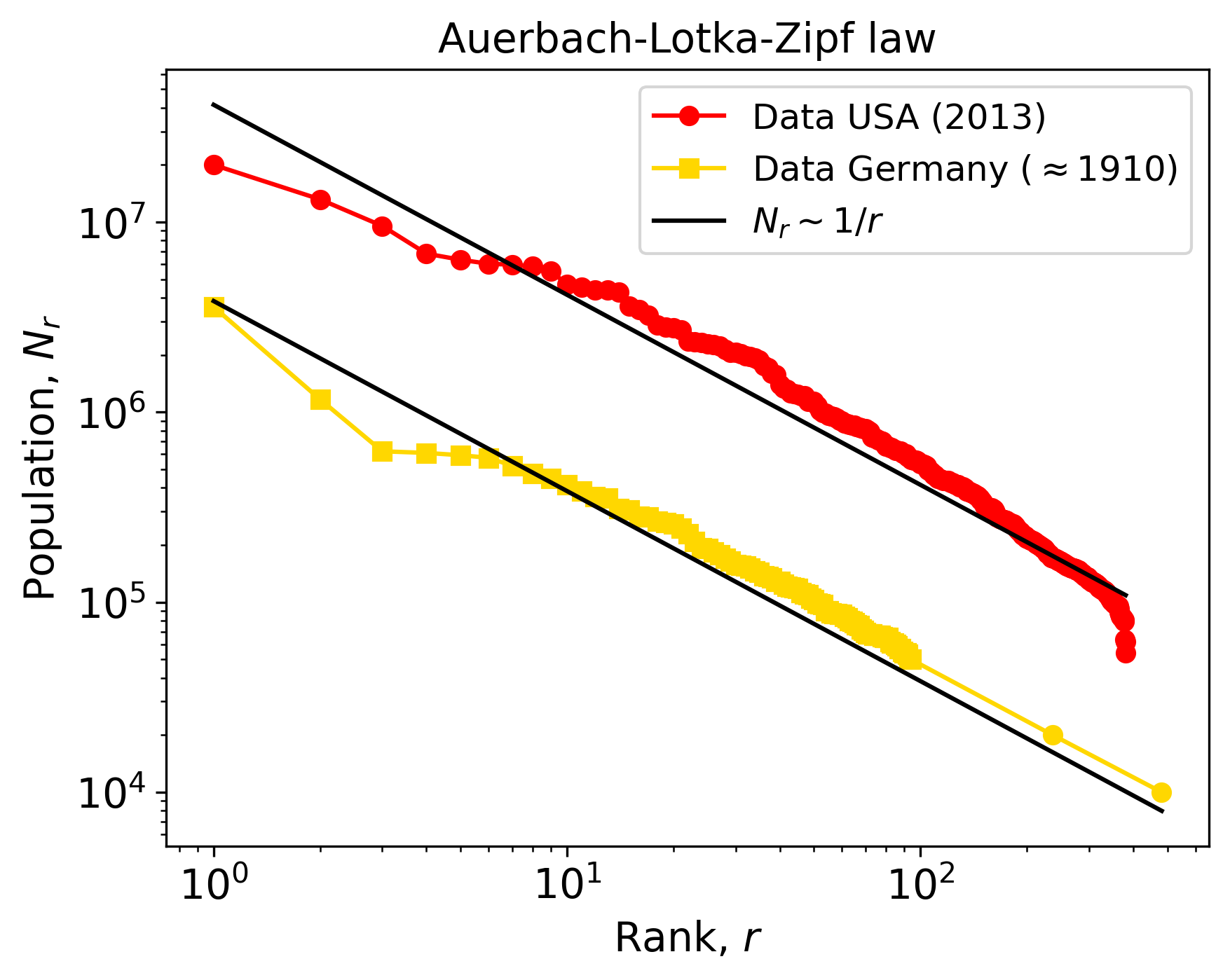}\includegraphics[width=0.5\linewidth]{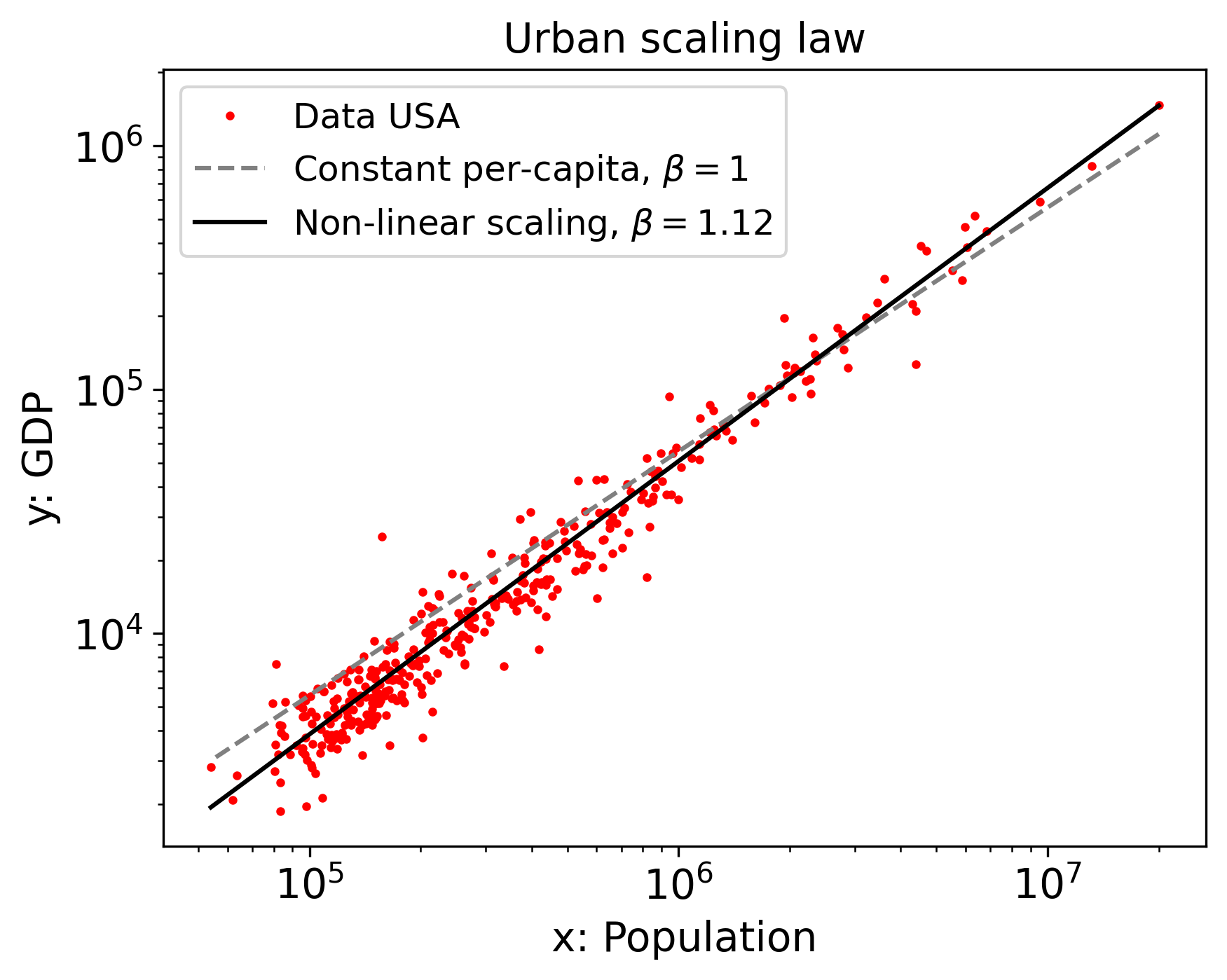}
  \caption{Statistical laws in urban system. Left: the population $N_r$ of the $r$-th largest city of two countries (symbols) is compared to the \index{Auerbach, Felix} Auerbach-Lotka-Zipf\index{Auerbach-Lotka-Zipf's law} law (straight line) in Eq.~(\ref{eq.zipfcities}). The German data is from Auerbach's historical paper from 1913~\cite{auerbach_felix_gesetz_1913} while the USA data corresponds to metropolitan urban areas from 2013. Right: the gross domestic product (GDP\index{Gross Domestic Product, GDP}) of different cities in the USA (symbols) is compared to the constant per-capita expectation $\beta=1$ (dashed line) and to the \index{Urban scaling laws} urban scaling law~(\ref{eq.urbanscaling}) with $\beta=1.12$ (solid line). See Appendix~\ref{chap.appendices} for information on code and data.}
\end{center}
\label{fig:urban}%
\end{figure*}

A century later, in the early 21st century, another statistical law in urban data sparked the interest of physicists and researchers from various disciplines. \index{Urban scaling laws} Urban scaling laws~\cite{bettencourt_growth_2007,rybski_urban_2019} posit that various city attributes $y$ (e.g., the length of all roads, the number of patents filed, the economic output) scale nonlinearly with city population~$x=N_r$ as 
\begin{equation}\label{eq.urbanscaling}
y = B x^\beta,
\end{equation}
where $B,\beta \in \mathbb{R}$ are constants with a non-trivial $\beta \neq 1$ exponent being typical. This law draw parallels with biological \index{allometric laws} allometric scalings which describe how properties of different species scale with their size (to be reviewed in Sec.~\ref{ssec.allometry}). The efficacy of these two urban statistical laws to describe historical and contemporary data are illustrated in Figure~\ref{fig:urban}. 

The power of statistical laws is their combination of simplicity and generality: they are stated as functional forms which have only a few fitting parameters but yet they are proposed to describe a large amount of data-points (cities) in many different settings (countries). This provides not only a summary of the data, it allows for analytical calculations and is thus appealing for theoretical analysis. Numerous such Statistical laws have been proposed across various disciplines,  as a probability distribution -- like the \index{Auerbach, Felix} Auerbach-Lotka-Zipf\index{Auerbach-Lotka-Zipf's law}'s law~(\ref{eq.zipfcities}) \footnote{As we explain in Sec.~\ref{ssec.rankrepresenation} of this monograph, the rank-frequency laws discussed above can be interpreted in this sense (i.e., as a probability of a city to have a given population or the probability of a person to live in a given city).} -- or as simple relationship between variables -- like the scaling law~(\ref{eq.urbanscaling}). The subsequent chapters will list several other examples of statistical laws (Chap.~\ref{chap.examples}), introduce data-analysis methods used to assess the validity of these laws (Chap.~\ref{chap.data}), and discuss their interpretation (Chap.~\ref{chap.synthesis}). Before that, the remaining of this chapter will discuss general aspects of statistical laws, including how they are defined, the scientific contexts in which they appear, and the similar role they play in complex-systems research.

\section{Historical context}\label{sec.history}

\subsection{Statistical laws in \index{social physics} Social Physics} \label{ssec.socialphysics}

The study of statistical laws dates back to the birth of many scientific disciplines in the 17th century. The origins of the idea that different datasets and phenomena can be described by the same universal function or distribution is intimately related with the attempt to expand the success of quantitative methods in the natural sciences (classical Physics) \index{Physics} to biological and social sciences via statistical methodologies. This idea plays a central role in the works of the French scientist Pierre-Simon Laplace (1749-1827) and the Belgian polymath Adolphe Quetelet (1796-1874) in the first half of the 19th  century~\cite{stewart_suggested_1947,ball_statistics_2002,ball_critical_2006,west_scale_2018}. For example, Quetelet used Binomial distributions \index{Binomial distribution} to describe measurements of the human body and proposed that the square of the weight is proportional to the height to the power five. Patterns were identified also in data related to birth, age at marriage, criminal activities, and mortality rates. 

The term "\index{social physics}social physics" became associated to this nascent field of quantitative social studies, a term also adopted by the French positivist Auguste Comte (1798-1857). Although Comte later transitioned to the terms "Sociology" and "Social Sciences", which became more prevalent, "socio-physics" or "social physics" persisted into the 20th century~\cite{stewart_suggested_1947} and is still in use, often associated with models inspired by (condensed-matter) physics~\cite{schweitzer_sociophysics_2018}.

These early statistical laws were instrumental in the birth of Statistics as a discipline~\cite{sheynin__1986} as they conveyed the potential of statistical and probabilistic thinking across the sciences.  The development of these ideas successfully explained many of the observed regularities as the consequence of random (Gaussian) \index{Gaussian distribution} fluctuations and contributed to the development of Physics through the work of Maxwell and Boltzmann in Statistical Mechanics~\cite{ball_statistics_2002}, an ironical turn of events\footnote{The irony is that a program that started with Physics as the role-model \index{Physics} science, against which the others sciences should be measured, developed ideas that turned out to be essential for Physics to overcome its own mechanistic and deterministic limitations.}. Observations of the regularities were considered characteristic of the individuals or societies and statistical laws used in the analysis of empirical observations (e.g., to detect signal among random fluctuations) or to detect fraud (e.g., under-reporting of height by soldiers)~\cite{ball_statistics_2002}. A contemporary example of this approach is Benford's law (see Sec.~\ref{sec.otherlaws} below).

Such early studies of statistical laws can create the misconception that they are mere curiosities or manifestations of the law of large numbers. This perspective overlooks the fact that not all regularities are described by distributions arising naturally as a result of random processes (e.g., Gaussian, Poisson, Binomial) \index{Poisson distribution} \index{Binomial distribution} and that these laws intended to reveal more fundamental properties of the underlying systems. For instance, a heated debate revolved around the possibilities of reconciling collective statistical laws and individual free will~\cite{ball_critical_2006}. In \index{social physics} social physics tradition, statistical laws were seen akin to empirical laws in Physics. As pointed by Ball~\cite{ball_statistics_2002}, the term social physics was associated to {\it "the search of law-like behaviour in society"} and {\it "the idea that there were laws that stood in relation to society as Newton's mechanics stood to the motion of the planets was shared by many"} (in the late 17th century).  The expectation in social physics was -- and in some extent still is -- that it will evolve as a discipline similarly to the historically-reconstructed development of classical mechanics:

\begin{tcolorbox}
\begin{center}
Observations \\
$\;$\\
$\downarrow$ \\
$\;$\\
Empirical Laws (Galileo, Kepler) \\
$\;$\\
$\downarrow$ \\
$\;$\\
Universal Laws (Newton)
\end{center}
\end{tcolorbox}
\noindent In this perspective, statistical laws play the role of empirical laws, the crucial intermediate step between empirical observations and the development of theories of general validity. This simplistic analogy overlooks the crucial statistical nature of statistical laws, which are fundamentally different from Kepler's law \index{Kepler's law} (a point further elaborated in this monograph). To date, the expectations for the development of \index{social physics} social physics have not been vindicated, as there are no indications that an unified theory akin to Classical Mechanics will appear. Nevertheless, an aspect of this na\"ive view retained in contemporary applications of statistical laws is the expectation that they connect observations and theoretical models, even if the theory is not of Newtonian generality. 

\subsection{Statistical laws in complex systems} \label{ssec.complexsystems}

The use of statistical laws in the field of complex systems builds on the "socio-physics" tradition but \index{social physics} goes beyond it in important aspects. Firstly, it does not view statistical laws simply as the effect of independent random influences that can be expected to act in individual parts and are explained naturally (e.g., using the central limit theorem or law of large numbers). Instead, they are considered an emergent property, a non-trivial consequence of underlying interactions among the system's constituents.
Secondly, the interest in these laws extend beyond practical applications or philosophical discussions, as it is used as a motivation or justification for the proposal of mechanistic models of the underlying system. Thirdly, these theoretical explanations of the laws do not follow the classical mechanics paradigm of determinism and instead are based typically on probabilistic (Statistical Mechanics) models. 

Historically, the complex-systems approach to statistical laws developed starting from the mid-20th century. Seminal work includes the debates between Herbert \index{Simon, Herbert} Simon~\cite{simon_class_1955,simon_size_1958} and Benoit \index{Mandelbrot, Benoit} Mandelbrot~\cite{mandelbrot_informational_1953,mandelbrot_note_1959} on the rich-get-richer mechanisms underlying the origin of power-law distributions (e.g., of city sizes). More generally, complex systems are composed of multiple (microscopic) components that interact with each other giving rise to non-trivial phenomena at larger (macroscopic) scales. These non-trivial phenomena are said to be {\it emergent} because they are neither designed nor an obvious consequence of the properties or interactions between the components. In complex-systems research, statistical laws are interpreted as an emergent phenomenon. As such, the universality attributed to statistical laws in their complex-systems interpretation is akin to other sources of  universality (in Mathematics and Statistical Physics):  bifurcations (normal forms), phase transitions, critical phenomena, etc. It is understood that the statistical laws are capturing only part of the system and that fluctuations and small deviations are expected, in line with the mathematical-modeling tradition of using simple models that capture essential properties of the system.    \index{power-law distribution}

\subsection{Statistical laws in the age of big data.}\label{ssec.bigdata}

The increasing availability of data for scientific investigation sparked a renewed interest on statistical laws in the 21st century. Prominent examples include the claim of ubiquity of networks with scale-free degree distribution~\cite{barabasi_emergence_1999} and the renewed interest in \index{Urban scaling laws} urban scaling laws~\cite{bettencourt_growth_2007}. While these and numerous other publications report the widespread occurrence of these statistical laws, in line with their claim of universality, their validity is far from consensual and has been consistently questioned (see, e.g., Ref.~\cite{stumpf_critical_2012} for the case of power laws, Ref.~\cite{broido_scale-free_2019} for the case of  scale-free networks, Refs.~\cite{arcaute_constructing_2015,louf_scaling_2014} for urban-scaling laws~\cite{arcaute_constructing_2015,louf_scaling_2014},  Ref.~\cite{dodds_re-examination_2001} for \index{Kleiber's law} Kleiber's law of metabolism, and Ref.~\cite{eeckhout_gibrats_2004} for city-size distributions). One of the goals of this monograph is to explain the persistence of controversies on the validity of statistical laws, many of which persist over many decades or re-emerge after being seemingly resolved.

\begin{figure*}[!tb]
\begin{center}
  \includegraphics[width=0.8\linewidth]{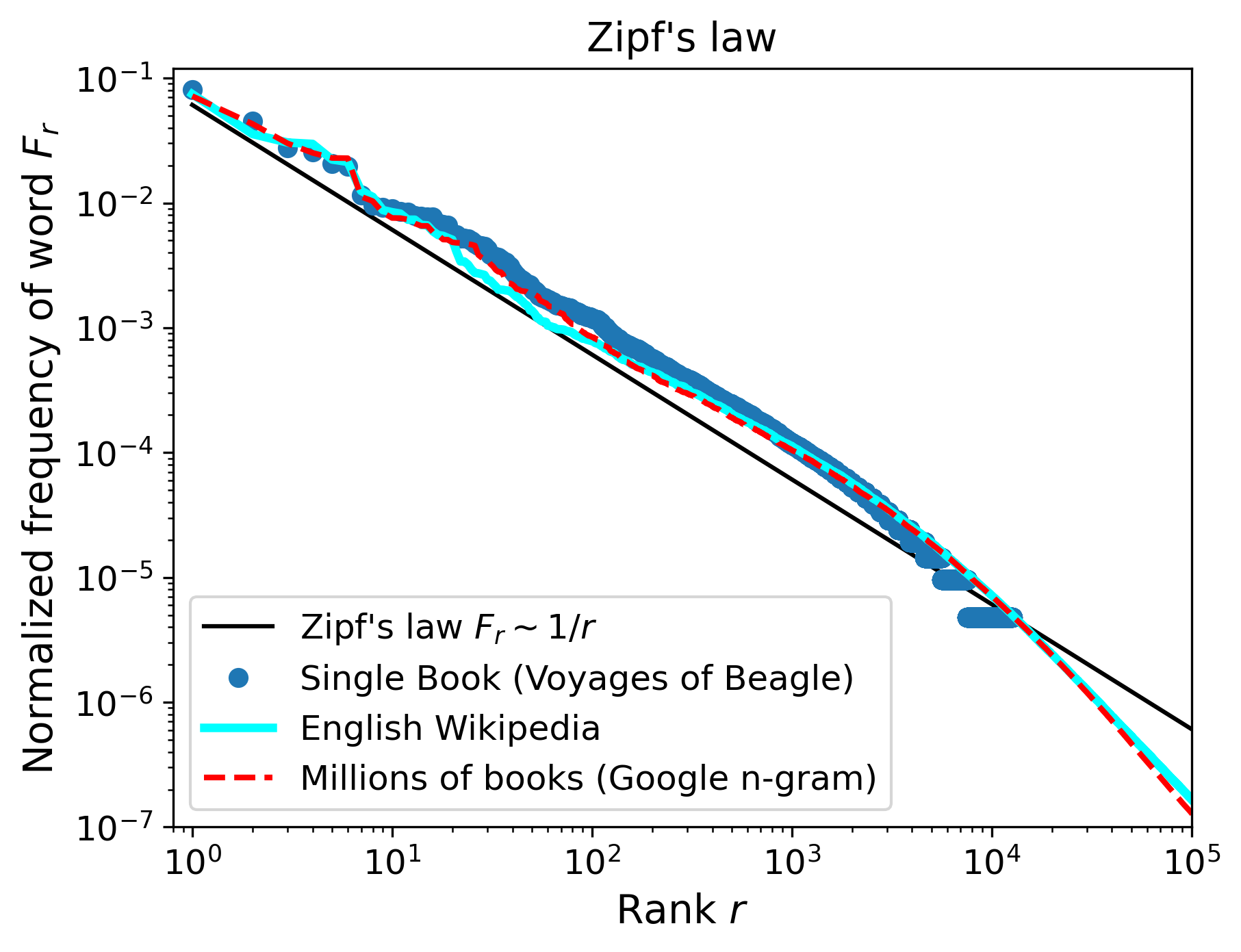}
  \caption{Large dataset confirm the statistical patterns that motivated the proposal of Statistical laws. The rank-frequency representation of Zipf's law for word frequencies is shown for three different dataset: a single book ("The voyages of the Beagle", by Charles Darwin, published in 1839), the complete English Wikipedia, and millions of books (Google n-gram). See Appendix~\ref{chap.appendices} for information on code and data.}
\end{center}
\label{fig.zipfdata}
\end{figure*}

There is a long tradition of the application of statistical laws to new datasets, which often lead also to a re-examination of previous proposals. The same functional form of the ALZ's law of city sizes discussed above was proposed to describe the frequencies of words in texts. In this context, the different words (word types) of a language assume the role of the different cities in a country, while each specific word in a text (word tokens) assumes the role of an inhabitant of the country, which can be ``attributed'' to each word type. Zipf, a linguist working in Chicago in the first half of the 20th century, investigated this regularity using the frequency of words in various books. Today, we can investigate this law using large textual datasets as typical in the 21st century. Figure~\ref{fig.zipfdata} shows the results for a single book, as analyzed in the 1930's by Zipf, the complete English Wikipedia, and the combined result over millions of English books (Google n-gram corpus). The new datasets confirm that the most frequent words (smaller ranks $r$) follow the same pattern observed by Zipf. The difference is that today we are able to evaluate the distribution of less frequent words (larger ranks $r$). We see that a faster decay of the distribution $P_r$, which was already seen in some large books, is in fact the origin of a new regime. In Ref.~\cite{gerlach_stochastic_2013} -- to be reviewed in Sec.~\ref{ssec.zipf} and~\ref{ssec.ml-freq} -- we tested different proposals to generalize Zipf's law\index{Zipf's law} to two free parameters and found out that a double power-law provided the best description. Maybe the most impressive observation is that large datasets of various origins (as the two shown in the figure) show a remarkably similar behaviour that is well described by the same generalized Zipf's law (with the same parameters). The same parametric form (with different parameters) describes datasets in different languages. In this case, datasets of extremely large magnitudes seem to corroborate Zipf's law -- not only by showing the same small $r$ behaviour but specially by suggesting that they can be described by simple parametric forms -- thus  keeping their core message of universality across datasets and languages. 

The modern availability of large datasets and computers allows not only the reproduction of previously-proposed statistical laws and their application to new cases. It opens the possibility to look beyond average values and expected behaviour, as typically described by statistical laws, and instead to consider fluctuations around the statistical laws~\cite{gerlach_scaling_2014}. It also made clear the need for improved statistical methods~\cite{clauset_power-law_2009} and for more careful evaluations of the claims of universal validity of statistical laws~\cite{stumpf_critical_2012,louf_scaling_2014,arcaute_constructing_2015,altmann_statistical_2016,leitao_is_2016,broido_scale-free_2019}.

The developments described above indicate a contradictory picture of the recent developments in the study of statistical laws: on the one hand, large datasets seem to reproduce previous claims of statistical laws in an even larger amount of cases, pointing thus towards their increased applicability. On the other hand, they pose a challenge due to new conclusions derived from the application of more rigorous statistical methods (e.g., statistical tests refuting the validity of statistical laws that otherwise were considered as well established~\cite{broido_scale-free_2019,stumpf_critical_2012}). The goal of this monograph is to shed some light on this crises, trying to re-concile how statistical laws are treated in the field of complex systems with an improved statistical interpretation of these laws (as required in view of the increasingly large datasets). This type of crisis contains many elements of scientific developments happening more generally: the chance of refuting a hypothesis \index{hypothesis testing} increases with the size of the database (assuming the null hypothesis is false) and the applicability of statistical tests based on independence of datasets is of limited applicability.

Another important contemporary development that changes the perspective on statistical laws comes from \index{machine learning} machine learning, the dominating paradigm employed in the study of big data. Statistical laws are typically formulated in form of simple parametric distributions or functions. Fitting such functions to given datasets is a traditional method of statistical analysis, which aims to, e.g., summarize the data, estimate the probability of (unobserved) events (risk estimation), and facilitate analytical reasoning. In contrast, \index{machine learning} machine learning methods typically do not use simple parametric fittings and tend to favour flexible functional forms (algorithms) that have the ability to detect arbitrary statistical correlations. \index{machine learning} On the one hand, the success of machine-learning approaches in applications can be viewed as a challenge to statistical laws, as it raises questions not only about its usefulness in practice but also about the relevance of its goal of revealing general-applicable laws. On the other hand, the lack of interpretability of machine learning methods is increasingly recognized as a limitation, highlighting the positive aspects of the scientific tradition which statistical laws build upon~\cite{mainzer_berechnung_2014} and suggesting that there are opportunities for statistical laws to complement or be incorporated in machine-learning methodology.

\section{Formalization}\label{sec.formalization}

\subsection{Definition}\label{ssec.definition}

In the next chapter, a variety of statistical laws will be reviewed and discussed. While the list is not exhaustive, it is intended to include the most prominent cases and enough variety of examples to allow for comparative studies and generalizations. It is thus worthwhile to start with an explicit statement about the type of statistical laws that we intend to review in this monograph, sharpening the focus of our analysis: 

\bigskip

\begin{tcolorbox}

\noindent {\bf Definition:} a statistical law (in Complex Systems) is a function that:

\begin{enumerate}
   \item[(i)]  has been proposed to describe a large number of observations in different settings (universality); 
     \item[]
     \item[(ii)]  is either elementary or a composition of elementary functions with a small number of parameters and dimensions (simplicity); 
    \item[]
     \item[(iii)] plays an important role in a theory or model (theoretical connection). 
\end{enumerate}

\end{tcolorbox}

\bigskip

Typically, statistical laws apply to observational data and describe either the frequency of types of observations or the relationship between (two) properties of observed items. The two examples of urban statistical laws discussed at the start of this Introduction in Sec.~\ref{sec.1example} -- Auerbach-Lotka-Zipf\index{Auerbach-Lotka-Zipf's law}'s law of population distribution and \index{Urban scaling laws} urban scaling laws --- are paradigmatic examples of these two cases. The universality condition (i) states that the law is conjectured to be valid in all similar cases, or at least not be restricted to the (few) examples already studied. In the urban example, the "large number of observations" mentioned in point (i) refers to a large number of cities and "different settings" refers to different countries, years, and types of observables $y$. The universality and simplicity conditions (i and ii) are the key points for the use of statistical laws as summaries of observations or stylized facts. The simplicity condition (ii) can also be formulated in comparison to the number of observations, which is much larger than the function's dimensions ($\mathbb{R}^d \mapsto \mathbb{R}$ with $d\le 3$, typically $d=1$) and the number parameters (not more than 3). These parameters are typically estimated from data and interpreted by theoretical models. The central role of these models, as stated in condition (iii) and in line with their sociophysics tradition~\cite{schweitzer_sociophysics_2018}, is to provide a mechanistic explanation of the law and/or to use it in a more general theory of the underlying system. 

Considering the above definition and clarifications, a Gaussian distribution  \index{Gaussian distribution}describing the heights of humans is not consider a statistical law in this monograph because, while it satisfies conditions (i) and (ii), it fails at condition (iii) as it has a trivial statistical explanation (e.g., based on the central limit theorem). Another counterexample is the use of parametric statistical models (e.g., linear models) fitted to specific data: while this approach satisfies condition (ii), in isolation it does not imply their general validity -- violating condition (i) -- and does not provide mechanistic insights on their origin -- failing condition (iii). 

The definition above does not include the veracity or empirical validity of a statistical law. As in the case of other scientific laws, we can thus expect to be able to evaluate a proposed law based on empirical evidence, possibly concluding that a specific proposal is not valid (i.e., there is no empirical support). As we will see in Chap.~\ref{chap.data}, refuting a statistical law can be more difficult than refuting a (traditional) scientific law and assessing the validity of statistical laws is a subtle matter. In fact, one of the main goals of this work is to shed light into interpretations and limitations of statements about the truth, validity, and usefulness of statistical laws. Before discussing this crucial point in Chap.~\ref{chap.synthesis}, we will review in Chap.~\ref{chap.examples} different examples of statistical laws that satisfy the definition above, identifying common aspects across different examples. Our focus during this review is on how statistical laws appear in scientific work (i.e., how they are introduced and used), leaving a critical discussion of the data-analysis methods (Chap.~\ref{chap.data}) and interpretation (Chap.~\ref{chap.synthesis}) for the later parts of the monograph.

\subsection{Reasoning with statistical laws}\label{ssec.reasoning}

 A crucial point in our argument for an unified treatment of statistical laws -- proposed to describe various types of data in a variety of scientific disciplines -- is that they are motivated, justified, and used very similarly. All cases discussed in the next chapter not only satisfy the definition introduced above, they have been studied using similar methods, they are used similarly in applications and theories, and they received similar criticisms or were subject to similar controversies.
In particular, we identify and distinguish the following three logically connected steps:

\begin{tcolorbox}
\paragraph{1. Empirical analysis.} The initial step in the the study of statistical laws involves the analysis of observations. This typically starts with the proposal of the statistical law based on observations of a few cases. The finding is then reproduced in other datasets, often with larger sample sizes, until eventually the proposed law is, explicitly or implicitly, considered to be empirically validated. Frequently, soon after the proposal of the law, new parametric forms of the statistical laws are proposed, often as generalizations of the original law. Depending on their descriptive power, the law is either re-formulated in the new term or, more frequently, the new proposals are dismissed as having a marginal additional descriptive power. 
\end{tcolorbox}

After this foundational step, there are typically two steps that take place in parallel:

\begin{tcolorbox}
\paragraph{2. Generative model.} After the statistical law is considered to be empirically valid, an obvious question is about its origin. This is typically addressed by proposing a simple mechanistic model that gives rise to observations satisfying the statistical law. Often, models claiming that the observed law is trivial compete with more involved models claiming that the law reveals important properties of the underlying system.
\end{tcolorbox}

\begin{tcolorbox}
\paragraph{3. Consequences of the law.} Another line of research following the establishment of the empirical law is the exploration of the consequences of its validity. This could involve predictions based on it, relationship to other statistical laws, using the laws as constraints for generative processes, and using the laws to derive additional expectations, in other theories, and in data-analysis methods.  (e.g., classification tasks based on Zipf's law\index{Zipf's law} exponent, assessment of risk of extreme events). 
\end{tcolorbox}

\begin{figure*}[!h]
\begin{center}
  \includegraphics[width=0.75\linewidth]{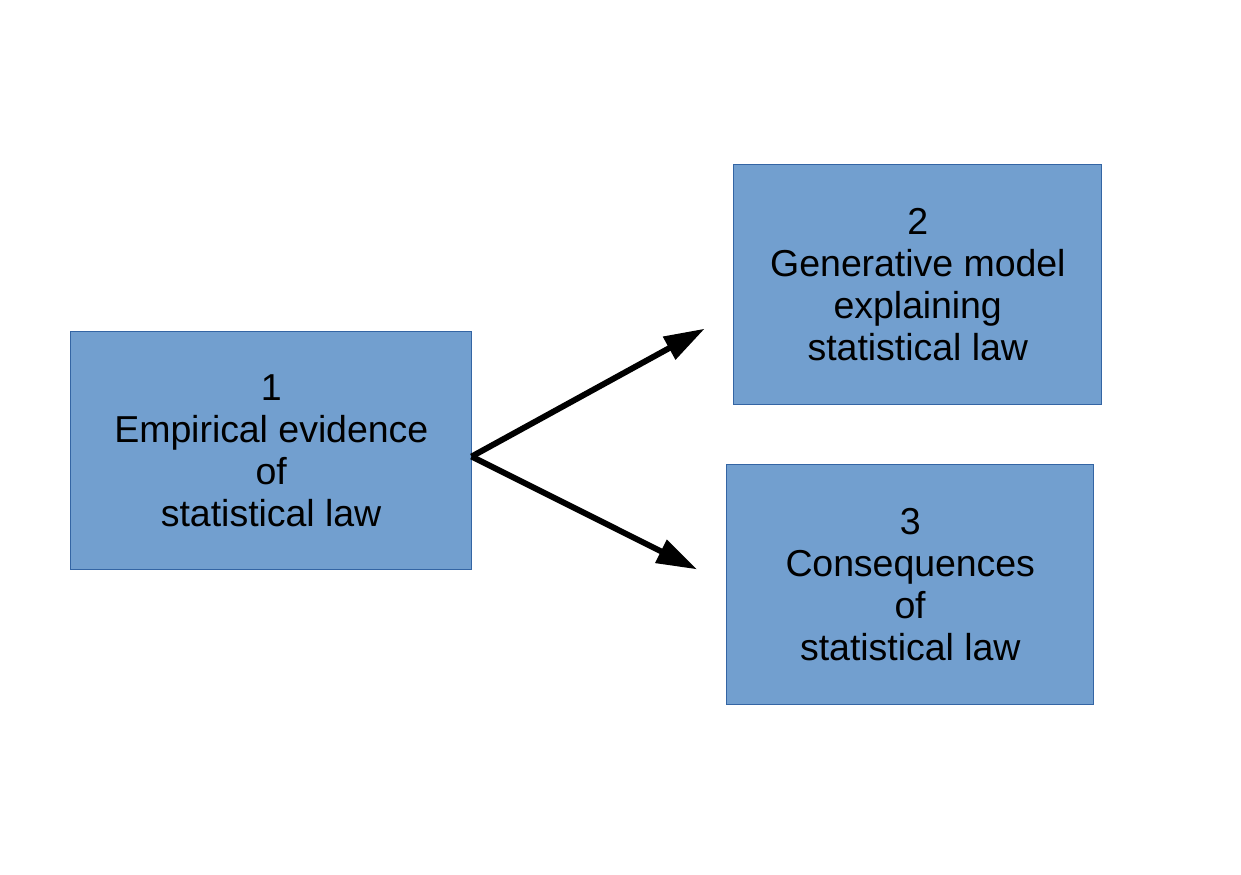}%
\end{center}
  \caption{Schematic depiction of the three-steps approach to the study of statistical laws in complex-systems research.}%
  \label{fig:fullfig}%
\end{figure*}

These three steps of the study in statistical laws in Complex Systems are depicted in Fig.~\ref{fig:fullfig} and will be referred to in the case-studies of Chap.~\ref{chap.examples}.  In Chap.~\ref{chap.data} we will critically discuss the methods used in step {\it 1. Empirical analysis}. The benefits, limitations, and potentially pitfall of this simplified approach will be discussed in further detail in Chap.~\ref{chap.synthesis}, benefiting from the concrete examples and revised methodology.  An important question in this discussion is the extent into which this first step can be separated from steps {\it 2. Generative model} and {\it 3. Consequences of the law}.

\subsection{Classification} 

The discussions above show that a classification of Statistical laws needs to go beyond a list of statements of the functional forms and settings in which statistical laws (have been proposed to) apply. It includes also the theories, models, and methods related to them, as these are essential elements to understand and, as we will argue later, evaluate them.  In particular, one of the aims of this review is to reveal how statistical laws in different fields play a very similar role in the reasoning to motivate and support the validity of mechanistic models.  

In addition to this law-models relationship, our review and classification of statistical laws is intended to be an useful overview and introduction for those interested in particular laws or in laws in particular fields. The different statistical laws can thus be usefully classified according to their type (e.g., frequency, scaling, temporal), the functional form used (e.g., power-laws, stretched exponentials), the type of data they use (e.g., urban, networks, time series), and the date in which they were proposed or started being used\footnote{Accurately tracing the first use of scientific and mathematical concepts is notoriously difficult and it is not the main focus of this work.}. Table~\ref{tab:listOfLaws} summarizes the statistical laws covered in this monograph and their classification. The three main groups -- power laws, scaling laws, and inter-event times -- are chosen to facilitated the analogy and connection between some of the most famous laws.  \index{power-law distribution}   \index{stretched exponential distribution}

\begin{table}[]
  \centering
  \small
    \begin{tabular}{|c|c|c|}
    \hline
        Statistical Law & Mechanistic Model & Section \\
        \hline
     \rowcolor{cinza}
        \multicolumn{3}{|c|}{Power-law rank-frequency distributions $F_r \sim r^{-\alpha}$}\\
        \hline
         \index{Pareto's law} Pareto's law (income) & Rich-get richer processes (Yule, \index{Simon, Herbert} Simon, \index{Mandelbrot, Benoit} Mandelbrot) & \ref{ssec.pareto} \\
         Auerbach-Lotka-Zipf's law (city sizes) & Proportional Growth (Gabaix)&  \ref{ssec.alz} \\
         Zipf's law (word frequencies) & \index{Simon, Herbert} Simon model& \ref{ssec.zipf}   \\
         \index{Gutenberg-Richter's law} Gutenberg-Ricther's law, Avalanches & Critical phenomena (SOC, Bak, Mandelbrot) &\ref{ssec.gutenberg}  \\
         Scale-free networks & Preferential attachment (Barabasi-Albert) &  \ref{ssec.scalefree} \\
         \hline
              \rowcolor{cinza}
         \multicolumn{3}{|c|}{Scaling laws $y \sim x^\beta$}\\
         \hline
         Urban scaling & Efficiency, accessible contacts & \ref{ssec.urbanscaling}  \\
                 Herdan-Heaps' law \index{Herdan-Heaps' law} (vocabulary size) & \index{Simon, Herbert} Simon Model, Urn models & \ref{ssec.heaps} \\
                         \index{Kleiber's law} Kleiber's law and \index{allometric laws} allometric scaling  & Fractal Geometry & \ref{ssec.allometry} \\
                 \hline
                      \rowcolor{cinza}
                                        \multicolumn{3}{|c|}{Burstiness and inter-event time distribution $P(\tau)  \sim \tau^\delta$ or $P(\tau) \sim e^{-a \tau^b}$} \\
                                        \hline
                                  Stretched exponential (words)   & Renewal process&  \ref{ssec.burstywords}\\ 
                                                            Stretched exponential (\index{earthquakes} earthquakes) & Epidemic-like, record breaking  &  \ref{ssec.burstyearthquakes}\\
                          Extreme events & \index{long-range correlations} Long-range correlation (Bunde et al.) &  \ref{ssec.bunde}\\
                                      Truncated power-law (human activities) & Queues (Barabasi et al.)  & \ref{ssec.burstysocial} \\
         \hline
    \end{tabular}
    \caption{List of the main statistical laws reviewed in this monograph. The name or context of the statistical law is given in the first column; the mechanistic model proposed to explain it is given in the second column; and more details can be obtained in the Section of this monograph listed in the last column.}
    \label{tab:listOfLaws}
\end{table}

\chapter{Examples of statistical laws}\label{chap.examples}
 
 This chapter contains a case by case description of paradigmatic statistical laws. While acknowledging their distinctiveness, our focus is on the common aspects across different statistical laws, in particular the similar role they have played in different research areas. The aim is to facilitate a comparative analysis that highlights the significance of this concept in complex-systems studies, supporting the unified treatment proposed in this monograph. In each case, we briefly describe how these laws were proposed, the most prominent explanations for their origin, and some of their uses. While we attempt to refer to original work, and to give credit to the original proponents of the laws, the description should be interpreted as a historical narrative that justified (and still justifies) the use of statistical laws and not as an attempt to reproduce the historical steps involved in this process. For readers interested in specific statistical laws, we hope the content of this chapter will provide a contextual introduction and point to relevant work where more specific aspects are discussed.  We leave to the next two chapters the technical discussion on statistical methods employed to study these laws (Chap.~\ref{chap.data}) and the critical debates on their interpretation (Chap.~\ref{chap.synthesis}). 
 
\section{Frequency distributions (power laws)}\label{sec:power-law-distr}\label{sec.powerlaw}

Some of the most common statistical laws specify the functional form of the frequency of events. This is typically done in one of the following two formulations:  \index{power-law distribution}

\begin{enumerate}
    \item[(Count)] When the observations are numerical quantities $x$ (e.g., $x \in \mathbb{R}$ or $x\in\mathbb{N}$), each of the $i=1, \ldots, N$ individual events correspond to a value $x_i$. The statistical law prescribes the distribution or probability density function $p(x)$ of the observations. Sometimes the complementary cumulative distribution $P(x) \equiv \int_x^\infty p(y) dy$ is used.
    \item[(Rank)] When the observations are tokens (e.g., words or objects), the events correspond to each type of token. These types can be ranked $r=1,2, \ldots, R$ according to their frequency $F_r$ of appearance (i.e., $F_1\ge F_2 \ge \ldots \ge F_R$) and the statistical law prescribes the functional form of $F_r=F(r)$.
    \end{enumerate}
\noindent The two formulations above can be related to each other by considering the frequency of a type (used in the rank formulation) to be the numerical observation $x$ (used in the count formulation) so that the $p(x)$ is estimated by the fraction of types with frequency $x$. Often, two representations of the same statistical law exist, each using one of the two formulations above, i.e., the related functional forms of $p(x)$ and $F_r$ are referred to represent the statistical law.

The most famous examples of statistical laws in form of frequency distributions use power-law functions. In the two formulations discussed above, they are written, respectively, as
\begin{equation}\label{eq.powerlaw}
   p(x) = C_\gamma x^{-\gamma} \text{ and } F_r = C_\alpha r^{-\alpha},  
\end{equation}
where typically $x \ge x_{min} > 0$, the scaling parameters are  $\gamma,\alpha \in \mathbb{R}$, and the proportionality constants $C_\gamma, C_\alpha$ are often fixed by normalization or other constraints\footnote{Some analysis focus on the counts or absolute frequency, in which case the proportionality constants $C_\gamma$ and $C_\alpha$ are considered either free parameters or fixed by properties of the observed data (e.g., the total population size or the value of the $r=1$ observation).}. Interestingly, the connection between the formulations map the two types of power-law to each other with the relationship between the exponent
$$ \gamma = \frac{1}{\alpha} +1.$$
This will be shown in Sec.~\ref{ssec.rankrepresenation} below, together with a discussion of the extent into which the two power-law formulations can be considered equivalent representations of the same statistical law. A power-law distribution in $p(x)$ as in Eq.~(\ref{eq.powerlaw}) is also equivalent to a power-law cumulative distribution 
\begin{equation}\label{eq.cumulativepowerlaw}
P(x) = \left(\frac{x}{x_{min}}\right)^{-\tilde{\gamma}} \text{, with } \tilde{\gamma} = \gamma-1.
\end{equation}

 Examples of each formulation (further discussed below) include \index{Pareto's law} Pareto's distribution of income -- the fraction of the population that has income $x$ -- or Zipf's distribution of word frequencies -- the fraction of words that are of the type $r$ in Eq.~(\ref{eq.powerlaw}). The city-size law introduced in Eq.~(\ref{eq.zipfcities}) is retrieved taking $\alpha=1$. 
The generalization for $\alpha>1$  is natural in view of the fact that, for any fixed $C_\alpha$, there exists a value $r^*$ such that $\sum_{r=1}^{r^*} C_\alpha/r > a$ for any $a$. In many statistical laws, $F_1$ does not vary with system size (e.g., the population of the largest city or the frequency of the most frequent word in texts), and thus $\alpha>1$ is required to avoid divergences (since  $C_\alpha \approx F_1$ and $\sum_r F_r$ is finite) and ensure that these laws (with unbounded domain, $R\rightarrow \infty$) can be applied to finite (but arbitrary large) populations.

Some power-law statistical laws are proposed or interpreted to be valid only in the tails, in which case the functional form in Eq.~(\ref{eq.powerlaw}) are interpreted to be valid only after cut-offs $x_c$ and/or $r_c$ (i.e., for large $x$, $x>x_c$, which correspond to small ranks, $r<r_c$), with the corresponding adjustment to the normalization constants $C$~\cite{perline_strong_2005}. Similarly, in some cases the end of the domain of validity of the law can be considered to be the maximum observed $x$, $x_{max}$, or the number of unique types $r_{max}$. In other cases, assuming the finitude of observations might not be justified and one can take $r_{max},x_{max} \rightarrow \infty$. The latter case naturally imposes that $\alpha>1,\gamma>1$ because $\sum_{1}^\infty 1/r$ diverges. In most (historical) formulations of the laws discussed below, these choices are not explicitly mentioned, leaving an ambiguity in their interpretation that results in inconsistent usage in data analysis. 

\paragraph{Mechanistic models of power-law distributions}

Different mechanisms to generate power-law distribution often build on the same mathematical background and an unified explanations for different power-laws has been the subject of investigation since at least the works of Zipf~\cite{zipf_human_2012} and \index{Simon, Herbert} Simon~\cite{simon_class_1955}. Some of the most accepted explanations of each statistical law are briefly mentioned in their disciplinary context below, but we refer to the (contemporary) reviews~\cite{simon_class_1955,mitzenmacher_brief_2004,newman_power_2005,simkin_re-inventing_2011,bak_how_2013,eliazar_power_2020} for an unified view and a comparison of the different type of explanations of power laws.  Conceptually, two broad classes of explanations can be identified:  \index{power-law distribution}

\begin{tcolorbox}
\begin{itemize}
    \item {\bf Preferential growth}: in which the probability of observing of an item is (linearly) proportional to the number of times $x$ it has been observed so far (i.e., its current frequency). The power-law distribution is interpreted as a consequence of a stochastic process, obtained computing the probability of having items with $x$ observations at a long time. Models specific to each data and problem include additional assumptions that prescribe, e.g., how new items are introduced in the system or boundary conditions for the probability of small items.  Famous examples of this type of explanation go back to Yule and \index{Simon, Herbert} Simon's model~\cite{simon_class_1955}, include the "cumulative advantage"~\cite{price_general_1976} and "preferential attachment"~\cite{barabasi_emergence_1999} mechanisms on networks, and modern extensions of Gibrat's principle~\cite{gabaix_zipfs_1999,rozenfeld_laws_2008,malevergne_gibrats_2009}.
    \item {\bf Optimization}: in which the values of $x$ are viewed as the result of the interaction between different components of an underlying (dynamical) system. The power-law distribution $p(x)$ is derived as the functional form that maximizes an utility function, minimizes a cost function, or appears when the underlying system is at a critical state. Famous examples of this explanation include Zipf's principle of least effort~\cite{zipf_human_2012}, \index{Mandelbrot, Benoit} Mandelbrot's approach based on the effectiveness of communication~\cite{mandelbrot_note_1959,mitzenmacher_brief_2004}, and more recent examples include language models~\cite{prokopenko_phase_2010} and self-organized-criticality~\cite{bak_how_2013} or other models in which the system is close to a phase transitions~\cite{nicoletti_emergence_2023}.
\end{itemize}

\end{tcolorbox}

 Mathematically, many of the mechanisms date back to the work of Yule and Willis~\cite{simkin_re-inventing_2011} and have at their core the same algebraic derivation~\cite{newman_power_2005}\footnote{We write "$A \sim B$" to denote that $A/B \rightarrow $ constant in a proper limit, usually $B\rightarrow \infty$.}:

 \begin{tcolorbox}
 {\bf Power-laws as composition of two exponential relationships:} assume the variable $x$ is related to $y$ by
$$x \sim \exp(by),$$
and $y$ is distributed exponentially as 
$$p(y) \sim \exp(ay).$$
It follows that $x$ is distributed as
\begin{equation}\label{eq.expToPower}
p(x) = p(y) \frac{dy}{dx}  \sim x^{-1+a/b},
\end{equation}
which corresponds to a power law as in Eq.~(\ref{eq.powerlaw}), with $\gamma = 1-a/b$. 
 \end{tcolorbox}

The power of this general argument is that exponential relationships appear more naturally  -- e.g., in random processes leading to Poisson or Binomial distributions, \index{Poisson distribution} \index{Binomial distribution} in multiplicative processes, and in solutions of linear differential equations -- so that the derivation is viewed as explaining a non-trivial (unexpected) observation (i.e., the power-law distribution) based on naturally appearing ones (i.e., the exponential distribution and relationship).

Below we discuss different examples of statistical laws that correspond to frequency distributions in form of power laws.

\subsection{Income (Pareto's law)}\label{ssec.pareto}

\index{Pareto's law} Pareto's law of inequality in income distribution is the most influential and possibly the earliest example of a statistical law as a power-law distribution. Its paradigmatic status is a consequence not only of its simplicity and significance, but also of the controversies it experienced and the work it motivated since its proposal in the late 19th century. It inspired similar approaches in other areas and many of the models proposed to explain it, and also the controversies about its validity and consequences, appear in very similar form also in later studies of other statistical laws. 

\paragraph{Empirical Evidence.} Pareto's empirical analysis of the income distribution in different countries led him to study the proportion $N(x)$ of the population with an income larger than $x$~\cite{pareto_cours_1897}. He proposed that for all incomes $x$ larger than a minimum income $x_m$, the following relationship holds
\begin{equation}\label{eq.pareto}
    \ln N(x) = \ln A - \tilde{\gamma} \ln(x),
\end{equation}
with the same $\tilde{\gamma} \approx 1.5$ observed in completely different settings. As he noted and emphasized, his approach resembles the complementary cumulative distribution in Eq.~(\ref{eq.cumulativepowerlaw}), with the difference that $N(x)$ is not normalized (as in $P(x)$) so that the constant $A$ plays a role similar to $C_\gamma$ in Eq.~(\ref{eq.powerlaw}). The claim that this distribution describes the income (and wealth) of different countries or regions is known as \index{Pareto's law} Pareto's law. 

\begin{figure}[!h]
    \centering
    \includegraphics[scale=0.75]{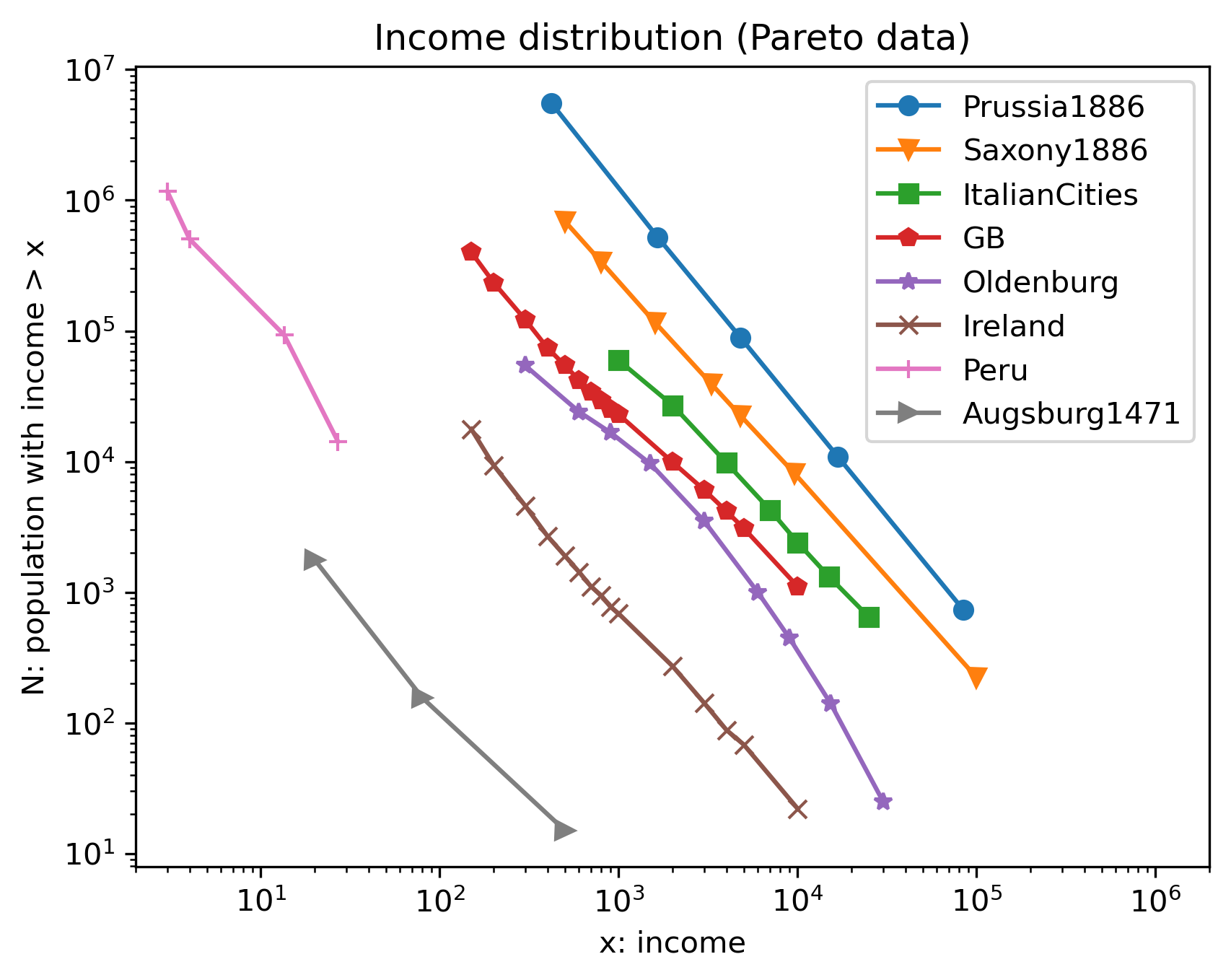}
    \caption{Pareto's law using Pareto's data. A straight line behaviour of the data corresponds to Eq.~(\ref{eq.pareto}), closely related to the cumulative distribution in Eq.~(\ref{eq.cumulativepowerlaw}). The fact that most cases are virtually parallel to each other suggests an universal exponent $\tilde{\gamma}$, which Pareto proposed to be $\tilde{\gamma}=1.5$. The data for different cities, regions, and countries was compiled by \index{Pareto's law} Pareto based on original sources. They are mostly from the late 19th century, except the case of Augsburg from 1471. The income in each case is measured on different (local) currencies. Data extracted from the tables available in Pareto's original work~\cite{pareto_cours_1897} and available in our repository, see Appendix~\ref{chap.appendices} for details.}
    \label{fig.pareto}
\end{figure}

Figure~\ref{fig.pareto} shows a reproduction of some of Pareto's original data, confirming a remarkable straight-line behaviour -- as predicted in Eq.~(\ref{eq.pareto}) -- in a variety of settings.  The straight line behaviour of $\ln(N)$ vs. $\ln(x)$ was observed by Pareto by plotting the $N$ vs. $x$ data in logarithmic paper. The use of logarithmic paper by Engineers was widespread at the time, suggesting a direct connection between Pareto's finding and his training as an Engineer before his focus on Economics~\cite{persky_retrospectives_1992}. 

Pareto noticed deviations of his simple proposal in some of the cases he analyzed (e.g., Oldenburg in Fig.~\ref{fig.pareto}). He suggested that, in more general cases, the generalized functional form holds (Ref.~\cite{pareto_cours_1897} p. 306)
\begin{equation}
    N(x) = \dfrac{A}{(x+a)^{\tilde{\gamma}}} e^{-b x},
\end{equation}
with $a,b$ constants that are close to zero in most cases so that Eq.~(\ref{eq.pareto}) is recovered (observed). 

As indicated in Persky's retrospective from 1992
\begin{quote}
{\it "The question of how well the law fits the data became a perennial one".~\cite{persky_retrospectives_1992}}
\end{quote}
Still, more than a century later, and despite numerous controversies about the interpretation, validity, and consequences of \index{Pareto's law} Pareto's law~\cite{persky_retrospectives_1992}, the usefulness of Pareto-type distributions to characterize income distributions is recognized in modern economical analysis~\cite{blanchet_generalized_2022}.

\paragraph{Mechanistic Models}

As emphasized by~\cite{persky_retrospectives_1992}, Pareto immediately set out to explain the origin of his law: {\it "possible sources of income inequality included chance, social institution, and human nature"}. The first possibility was ruled out in view of the striking difference of the law from a simple binomial distribution and the second was ruled out based on the validity of the law in societies with radically different social institutions, leaving "human nature" as the preferred option. Qualitatively, \index{Pareto's law} Pareto, and later Zipf, argue for the distribution to be the equilibrium between different forces in the society. Zipf is more explicit in this explanation in his Chapter 11 of~\cite{zipf_human_2012} -- on {\it "The distribution of economic power and social status"}, which addresses Pareto's law -- mentioning an equilibrium between exploiters and exploited or between forces of unification and diversification. 

More mathematical and quantitative explanations were obtained considering stochastic processes that capture plausible mechanisms of wealth distribution and that converge to a Pareto distribution. An early influential example is the work of Chapernowne~\cite{champernowne_model_1953}, \index{Chapernowne} who divided the income in brackets of exponentially large sizes (i.e., from 50-100, 100-200, 200-400, etc.) and considered a transition matrix between neighbouring brackets. As noticed already by \index{Simon, Herbert} Simon~\cite{simon_class_1955}, despite its different motivation, this model is compatible with the proportional growth process explanation that he introduced more generally. A simple case of Simon's model for wealth distribution considers that tokens of income are distributed in a population with a small chance of being allocated to a new individual (i.e., one that reached for the first time the minimum income) or otherwise a probability to be allocated to an existing individual with a probability proportional to its current (past) income $x$.  Stochastic processes as mechanistic models of Pareto type remain an area of investigation to these days \cite{gabaix_power_2009}.

\paragraph{Consequences}

\index{Pareto's law} Pareto's primary interest was to explore the consequences of the law to the question of wealth distribution and inequality. As the exponent~$\tilde{\gamma}$ was the main quantity that seemed to vary (slightly) from case to case, a critical debate was on how its (lack of) variation affects welfare and inequality. Economical discussions about how to best improve them (e.g., by raising minimum or average income) were addressed assuming the validity of the law and the constancy of its parameter over time. The claim of invariance suggested that attempts to change wealth distribution were purposeless or against human nature. Unavoidably, the political and economic consequence of these conclusions were not free of controversies, we refer to \cite{persky_retrospectives_1992} for an interesting historical account. The concluding part of this monograph, Chap.~\ref{chap.synthesis} below, warns about the dangers of attributing a degree of truth to statistical laws that is incompatible with its empirical support or with the large fluctuations that exist around them. In particular, statements that can be analytically computed assuming a statistical law, as performed in the case of Pareto's law, may show substantially less agreement with the data than the methods to directly evaluate the law (because, e.g., of the choice of observable, non-linear transformations, and data representations). 

Possibly the most popular consequence of Pareto's work is the 80/20 rule (sometimes called  Pareto's principle) which conveys that in many settings 80\% of the outputs are done by 20\% of the cases. It reflects the heavily skewed character of Pareto's law and provides an illustration of the consequences of fat-tailed distribution (i.e., the concentration of wealth in a few individuals).  \index{fat-tailed distribution}

\subsection{City-sizes (Auerbach-Lotka-Zipf's Law)}\label{sec:city-sizes}\label{ssec.alz}

As discussed in Sec.~\ref{sec.1example},  one of the first examples in which the power-law~(\ref{eq.powerlaw}) distribution was suggested to describe empirical data is the case of the population $x$ of different cities in a country or region.  This empirical law (with $\alpha=1$) was first proposed by the German physicist Felix \index{Auerbach, Felix} Auerbach in 1913~\cite{rybski_commentary_2013} but it is now mostly known as Zipf's law (for cities) due to the work of the American linguist George K. Zipf~\cite{zipf_human_2012}.  Refs.~\cite{rybski_commentary_2013,rybski_auerbach_2023} provide an insightful account of the (early) history of this law and we follow their suggestion to refer to this law as Auerbach-Lotka-Zipf\index{Auerbach-Lotka-Zipf's law}'s (ALZ\index{Auerbach-Lotka-Zipf's law}) law.

\paragraph{Empirical Evidence} 

\begin{figure*}[!t]
  \includegraphics[width=0.5\linewidth]{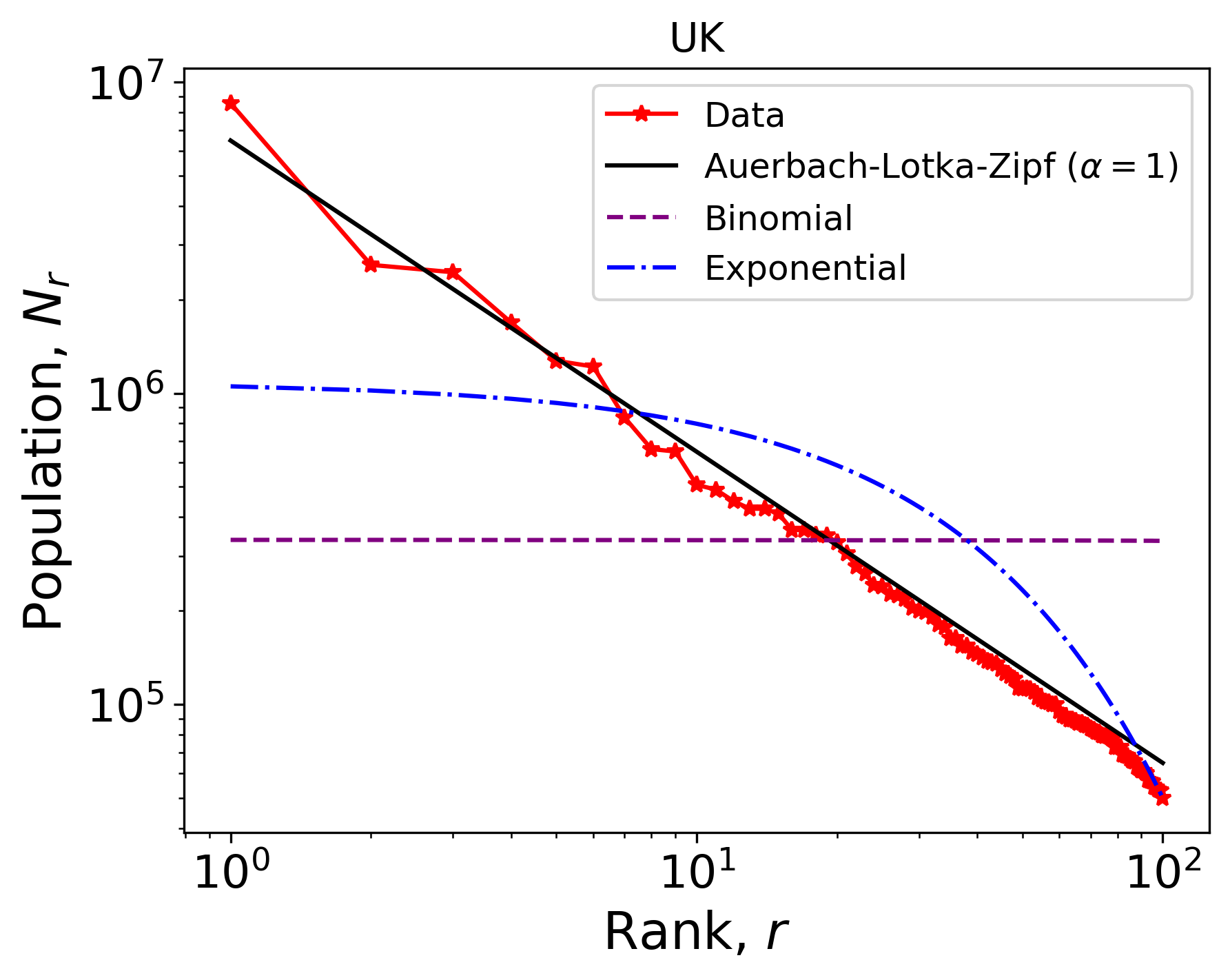}   \includegraphics[width=0.5\linewidth]{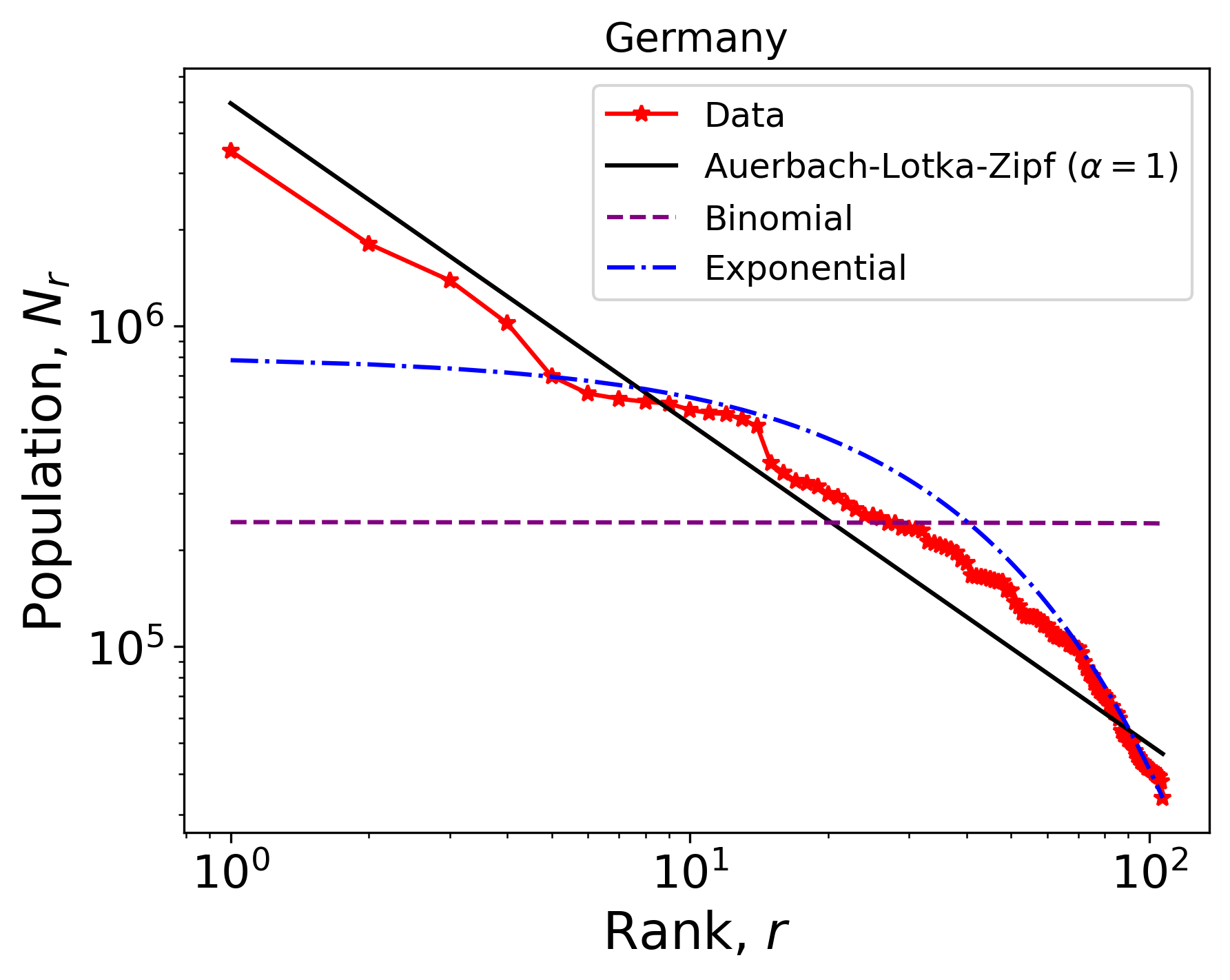}
  \includegraphics[width=0.5\linewidth]{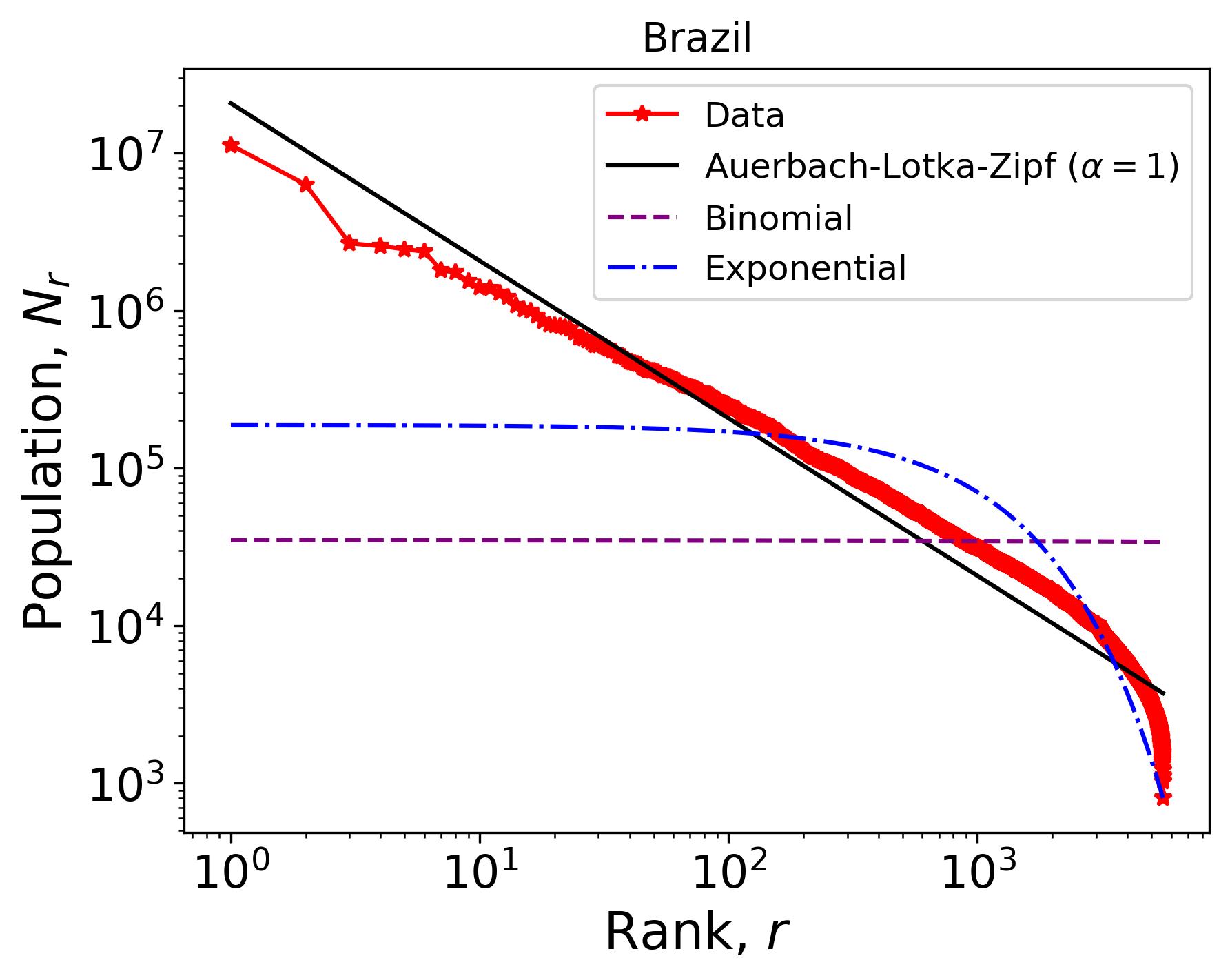}  \includegraphics[width=0.5\linewidth]{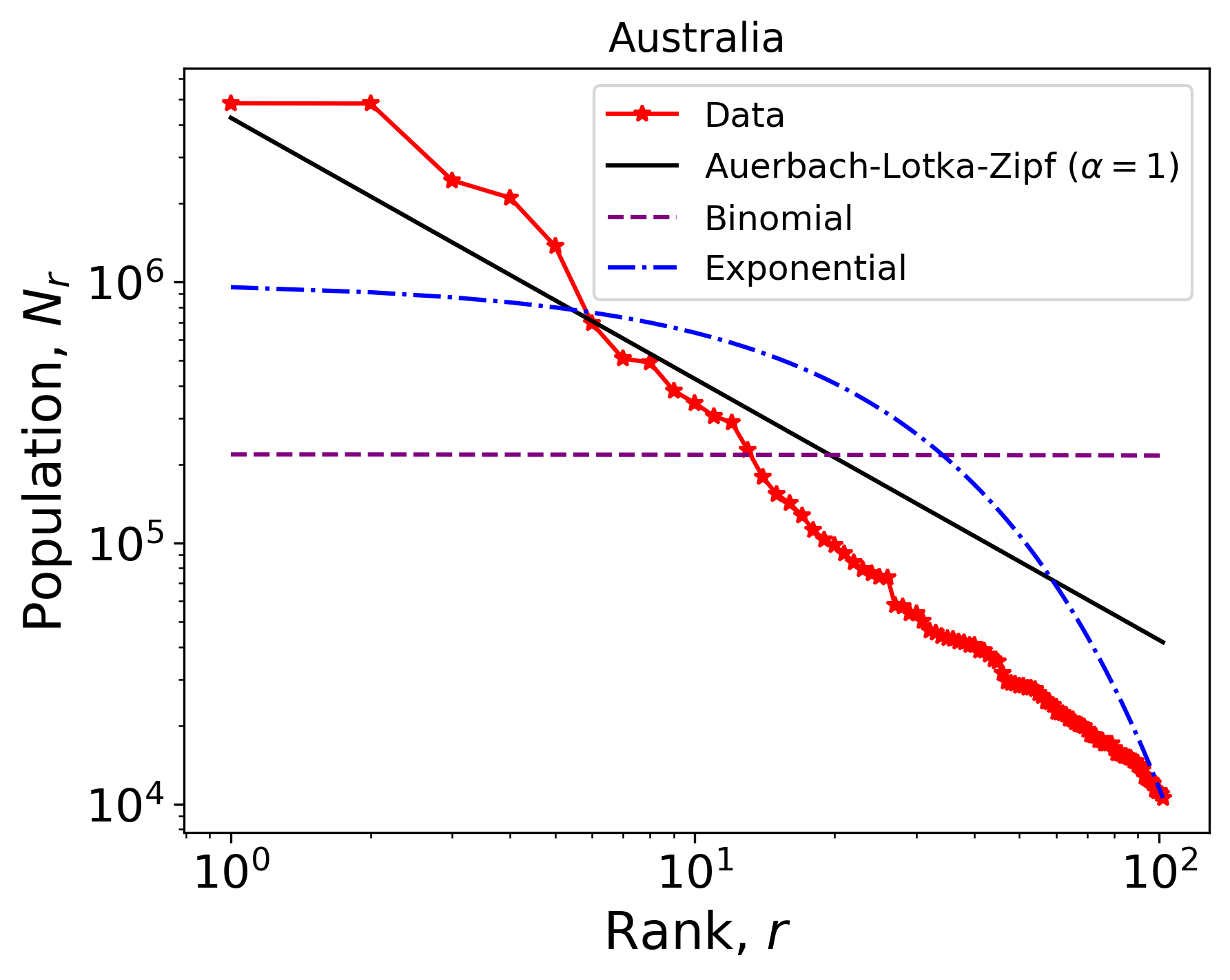} 
  \caption{\index{Auerbach, Felix} Auerbach-Lotka-Zipf's laws of city sizes. The population $N_r$ of the $r$-th largest city of a country is shown for four different countries (indicated in the title of each plot). The empirical data is shown by red symbols and the three different lines correspond to different curves. The solid line correspond to ALZ\index{Auerbach-Lotka-Zipf's law}'s law~\ref{eq.powerlaw} with $\alpha=1$ and $A = N / \sum_{r=1}^R 1/r$, where $N\equiv \sum_r^R N_r$ is the total population of the country (obtained from the data). The dashed line corresponds to a model in which each of the $N$ inhabitants choose one of the $R$ cities by chance, leading to a Binomial distribution \index{Binomial distribution} and population values that are almost identical to all cities. The dot-dashed line corresponds to an exponential distribution $N_r = C e^{-\beta r}$, where $C,\beta$ are determined imposing $\sum N_r = N$ (normalization) and equating the size of the smallest cities $N_R = X$, where $X$ is obtained from the data (this is usually an arbitrary threshold used in the definition of what a city is). In more detail, using the second constraint to fix $A$ we have that $R/X = \sum_{r=1}^R e^{-\alpha(r-R)} \approx \int_0^R e^{\alpha(R-r)} = (e^{\alpha(R-a)}-1)/\alpha$, where we use the integral from $0$ to $R$ as an approximation of the sum. This implies that $1-e^{\alpha(R-a)} +R \alpha/X=0$. We solve this equation for $\alpha$ using a bisection method, picking the $\alpha>0$ solution. See Appendix~\ref{chap.appendices} for the data and code.}
\label{fig:zipfcities}\label{fig.zipfcities}%
\end{figure*}

In Fig.~\ref{fig:zipfcities} we repeat \index{Auerbach, Felix} Auerbach's analysis for modern datasets of four different countries. If we consider the population of all cities $P = \sum_{r=1}^N P_r$ as a known quantity and the parameter $A$ a normalization parameter such that $A = P / \sum_{r=1}^N 1/r$,  Eq.~(\ref{eq.zipfcities}) has no free parameter to be adjusted to the data. Taking this into account, there is a remarkable agreement between the data (red curve with symbols) and Auerbach's prediction~(\ref{eq.zipfcities}) (straight black line) for cities in the United Kingdom (UK). In the other countries, it still provides a much better description than obvious alternative curves, with the Australian case showing a particularly poor agreement due to the exceptional case of its two largest cities (Sydney and Melbourne) having similar size\footnote{It has been suggested~\cite{cristelli_there_2012} that ALZ\index{Auerbach-Lotka-Zipf's law} law is visible only when cities within a coherent political-economic region are considered. Deviations from ALZ\index{Auerbach-Lotka-Zipf's law} law are visible when aggregating data from different countries (e.g., all cities in the European Union) or splitting cities from a country (e.g., considering cities within regions in a country independently). In this interpretation, the results for Australia could reflect the lack of integration of the country, possibly due to the independent development of its different states.}. In all cases, the straight line behaviour is clearly better described by a slope different than $\alpha=1$ (potentially, even $\alpha<1$ for a finite range of cities), in line with the generalization of Auerbach's proposal in Eq.~(\ref{eq.zipfcities}) to the more general ALZ law in Eq.~(\ref{eq.powerlaw}).

The main competitor of ALZ's law is the proposal that the data is better described by a log-normal distribution
\begin{equation}\label{eq.lognormal}
    p_{LN}(x) = \frac{1}{\sqrt{2\pi} \sigma x} e^{\dfrac{-(\ln x - \mu)^2}{(2\sigma^2 )}} = \frac{e^{-\frac{\mu^2}{2\sigma^2}}}{\sqrt{2\pi} \sigma} \cdot x^{-1+\frac{\mu}{\sigma^2}-\frac{\ln x}{2\sigma^2}},
\end{equation}
where $\mu,\sigma$ are parameters and the right hand side emphasizes that for $\sigma^2 \gg \mu,\ln(x)$ it approaches \index{Auerbach, Felix} Auerbach's proposal of Eq.~(\ref{eq.powerlaw}) with $\alpha=1$~\cite{montroll_1f_1982,perline_strong_2005,mitzenmacher_brief_2004}. There have been numerous debates about which distribution -- Eq.~(\ref{eq.powerlaw}) or Eq.~(\ref{eq.lognormal}) -- better describes the city size distribution in different countries~\cite{perline_strong_2005,eeckhout_gibrats_2004,levy_gibrats_2009,eeckhout_gibrats_2009,rozenfeld_area_2011}. Collectively, these studies suggest that these distributions provide alternative descriptions of city sizes, with the log-normal distribution \index{log-normal distribution} describing the majority of (small) cities~\cite{eeckhout_gibrats_2004} and ALZ's power-law describing \index{power-law distribution} the largest cities (small ranks) where most populations lives~\cite{levy_gibrats_2009,malevergne_testing_2011}. Methods for model comparison will be further discussed in Chap.~\ref{chap.data} and the findings related to ALZ's law will be further explained in Sec.~\ref{ssec.ml-freq} as a consequence of the difference in the statistical representation between the count and rank formulations in Eq.~(\ref{eq.powerlaw}).

\paragraph{Mechanistic models} 

In modern complex-system's research of urban systems~\cite{batty_new_2017,barthelemy_structure_2016}, there is a widespread acceptance of the significance of ALZ\index{Auerbach-Lotka-Zipf's law}'s law as one of the key characteristic of urban systems and as the starting point for theoretical work. As put by Barthelemy (\cite{barthelemy_structure_2016}, Chap. 8) 
\begin{quote}
    {\it "Such a robust, quantitative fact calls for a theoretical explanation."}
\end{quote} 
Similarly, Gabaix and Ionnides~\cite{gabaix_chapter_2004} formulate it as: 
\begin{quote}
{\it "if the empirical research establishes that the data are typical well described by a power law ... it prompts to seek theoretical explanations of why this should be true. "}    
\end{quote}

This "from law to models" reasoning has a long tradition in the complex-systems study of statistical law, as emphasized in Sec.~\ref{ssec.reasoning}. Zipf's proposed an explanation for the ALZ\index{Auerbach-Lotka-Zipf's law} law in his seminal 1948 book~\cite{zipf_human_2012}, which involved a combination of scaling relationships (e.g., between area, radius, and population of cities) and optimality (e.g., of transportation and exploration of resources).  The most popular approaches of recent works fall into the class of preferential-growth explanations as they focus on the growth of cities over time. This is often connected to Gibrat's law (also known as rule of proportionate growth or law of proportional effect) which states that the relative rate of growth is independent of city size (i.e., the absolute growth is linearly proportional to the size)~\cite{gabaix_zipfs_1999,gabaix_chapter_2004}. As already noted by \index{Simon, Herbert} Simon~\cite{simon_class_1955}, the origin of these type of explanations for power-law distributions  \index{power-law distribution} goes back to Yule's work from 1924~\cite{simkin_re-inventing_2011} obtained in mathematical studies of the evolution and distributions of species in genera. 

At the heart of preferential-growth explanation is a linear growth relationship of the population $N$ of a city with time $t$ as
\begin{equation}\label{eq.linearGrowth}
N_{t+1} = \gamma_t N_t,    
\end{equation}
where the growth rate $\gamma$ is independent of $N$. Considering $\gamma$ to be a random variable (fluctuates across $t$ and cities), the evolution of the logarithm of the population -- according to Eq.~(\ref{eq.linearGrowth}) -- can be seen as a random walk
\begin{equation}\label{eq.lnrw}
\ln N_{t+1} = \ln N_t + \ln \gamma_t,
\end{equation}
assuming $\ln \gamma_t$ is a random variable. For many choices of distributions from which $\ln \gamma$ is assumed to drawn, and after suitable re-scaling, the distribution of the log-populations $\ln N$ converges (by the central-limit theorem) to a normal distribution, i.e., a log-normal distribution~(\ref{eq.lognormal}) for the population across different cities (viewed as realization of the random walk).

As mentioned after Eq.~(\ref{eq.lognormal}), the log-normal distribution becomes very close to a ALZ\index{Auerbach-Lotka-Zipf's law}'s power-law distribution  \index{power-law distribution} for large $\sigma$ and finite $x$ (but potentially very large)~\cite{montroll_1f_1982,mitzenmacher_brief_2004,perline_strong_2005}. A power-law distribution as in ALZ's law is obtained by imposing additional modifications to the processes leading to a log-normal~\cite{eliazar_power_2020}, such as the random walk resulting from Eq.~(\ref{eq.linearGrowth}): the most famous being the addition of a small noise into Eq.~(\ref{eq.lnrw}) or reflecting boundary conditions for small $N$ which prevent small cities from becoming too small (disappear), proposed by Gabaix~\cite{gabaix_zipfs_1999,gabaix_chapter_2004,malevergne_gibrats_2009} (see also Barthelemy~\cite{barthelemy_structure_2016}-Chap. 8 for a derivation and alternative approaches). Deviations of Gibrat's law are also used to propose additional spatial correlations that affect city growth and interactions~\cite{rozenfeld_laws_2008}. Finally, models that incorporate the spatial dimension of urban growth have been able to reproduce ALZ's law and other spatial distributions of cities~\cite{schweitzer_estimation_1998,rybski_distance-weighted_2013}.


\subsection{Words (Zipf's law)}\label{ssec.zipf}
The long tradition of studying statistical properties of texts gave rise to several statistical laws, none more famous than Zipf's law\index{Zipf's law} of word frequencies: the power-law distribution~(\ref{eq.powerlaw}) of the frequency $x$ (or counts) of different words, i.e., $x \in \mathcal{N}$ is the number of repeated appearance (word tokens) of a given word (word type) in a text or corpus and $p(x)$ (or $F_r$) is the distribution over the different word types. While there is no unique definition of what a "word" is -- Should plurals be counted as different words than their singular form?--, the statistical regularities are fairly robust against different choices and counting methods, a key element of the widespread study of statistical laws in linguistics~\cite{kohler_quantitative_2005,altmann_statistical_2016,tanaka-ishii_statistical_2021}.

\paragraph{Empirical Evidence}

The origin of Zipf's law\index{Zipf's law} is typically attributed to the french stenograph Jean-Baptiste Estoup in the beginning of the 20th century (published in 1912-1916, as cited in~\cite{zipf_human_2012,mandelbrot_informational_1953}), a remarkable proximity to Auerbach's proposal discussed in Sec.~\ref{ssec.alz}. Not surprisingly, the original versions of this law considered the simple $1/r$ decay (or $\alpha=1$) as proposed by \index{Auerbach, Felix} Auerbach. Zipf's extensive studies of different books in the decades thereafter~\cite{zipf_human_2012} contributed to its dissemination and further study.

\begin{figure*}[!tb]
\begin{center}
  \includegraphics[width=0.7\linewidth]{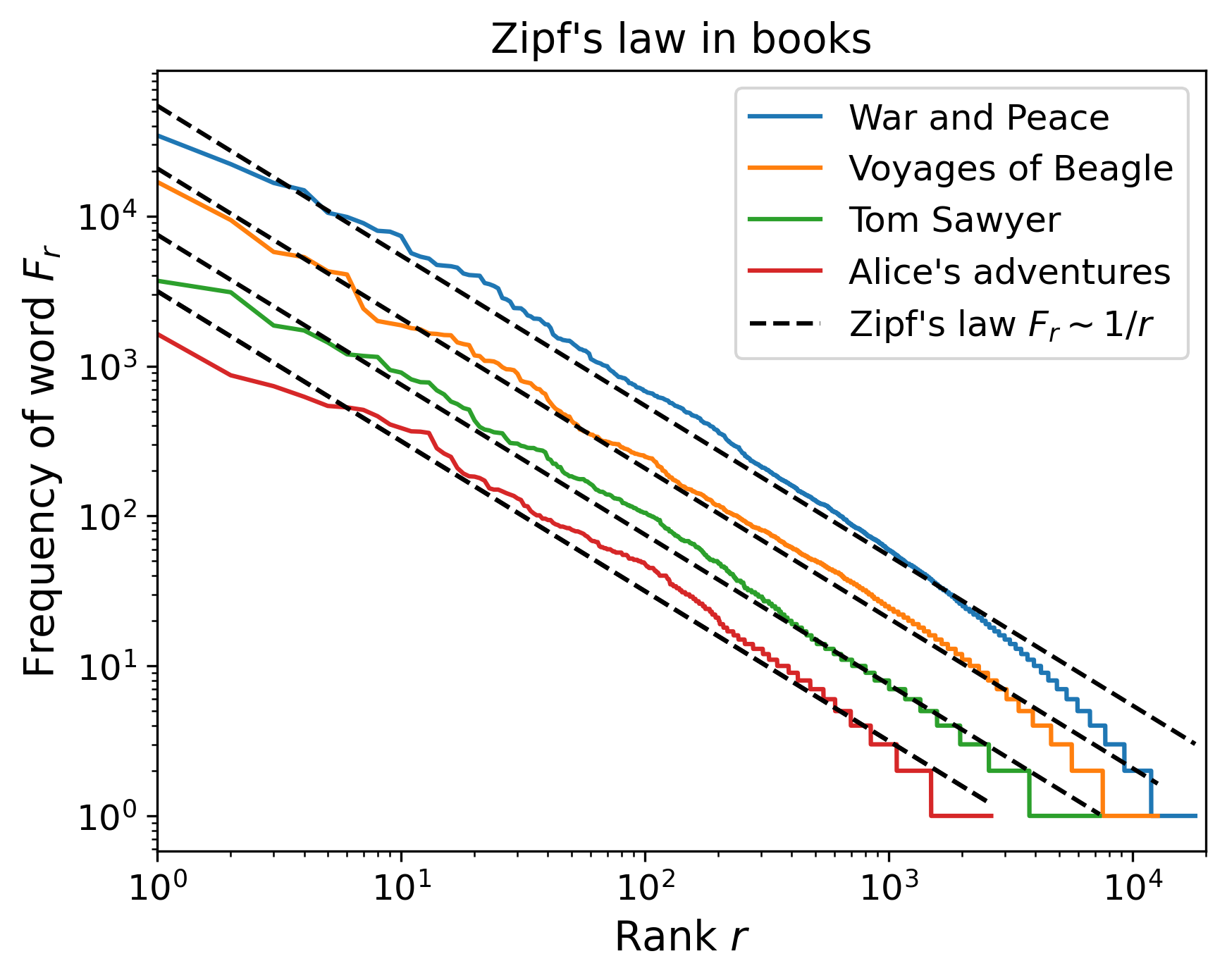}   \\\includegraphics[width=0.7\linewidth]{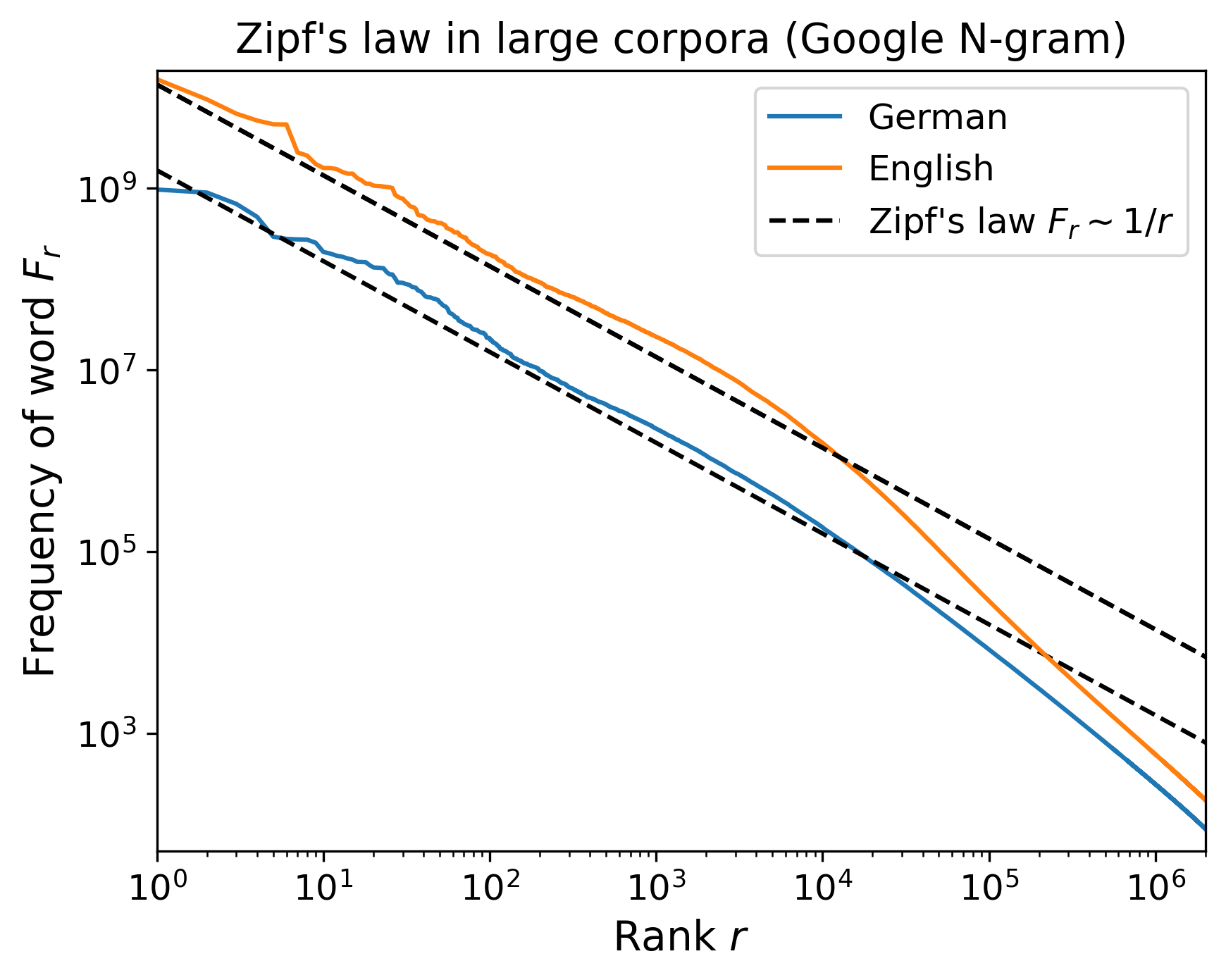}
  \end{center}
  \caption{Zipf's law\index{Zipf's law} of word frequency. The number of word tokens $F_r$ of the $r-th$ most frequent word (type) is shown in a double-logarithmic plot. Top: results for four books (an English translation of "War and Peace" by Tolstoy; "The Voyages of the Beagle" by Darwin; "The Adventures of Tom Sawyer" by Twain; and "Alice's Adventures in Wonderland" by Carrol). Bottom: Google n-gram corpus, containing millions of books published over the last centuries in English and German. The dashed lines correspond to Zipf's law~(\ref{eq.powerlaw}) with $\alpha=1$, fixing the proportionality constant $C_1$ by imposing that both the fitting and data sum to the same value $\sum_r F_r = F_{total}$. See Appendix~\ref{chap.appendices} for further information on the data and code used in this figure.}
\label{fig:zipf}%
\end{figure*}

Figure~\ref{fig:zipf} shows the rank-frequency distribution for corpora of different sizes. For small books, the Zipfian $F_r \sim 1/r$ proposal ($\alpha=1$) provides a remarkable good agreement considering that it involves no fitting parameters. For larger book sizes, a faster decay from this simple curve is observed, as expected considering that the frequency of the most frequent word $r=1$ does not change with corpus size and that $\sum 1/r$ diverges for $R \rightarrow \infty$. This has motivated the extended Zipf's law\index{Zipf's law} as in Eq.~(\ref{eq.powerlaw}), with the exponent $\alpha \gtrapprox 1$ as the single fitting parameter. Looking at even larger corpora -- containing millions of books and millons of different word types, as shown in the right side of Fig.~\ref{fig:zipf} -- we see that the large $r$ deviation clearly contains a curvature~\cite{naranan_models_1998,montemurro_beyond_2001,ferrer_i_cancho_two_2001,petersen_languages_2012,gerlach_stochastic_2013,williams_text_2015}. A detailed analysis of different corpora and $5$ different languages in Ref.~\cite{gerlach_stochastic_2013} showed that the best two-parameter generalization of Zipf's law\index{Zipf's law} is a double power-law (dp) distribution
\begin{equation}\label{eq.modeldp}
f(r) = F^{(dp)}(r;\gamma,b)=C
\begin{cases}
r^{-1}, & r < b\\
b^{\alpha-1}r^{-\alpha} & r\geq b,
\end{cases}
\end{equation}
where $b$ and $\alpha$ are the two free parameters, $C=C(\alpha,b)$ is the normalization constant (which can be approximated as $C \approx 1/(G^1_{b-1}+1/(\alpha-1))$, and $G^{a}_{b} \equiv \sum_{r=1}^{b} r^{-a}$. The double power-aw representation in Eq.~(\ref{eq.modeldp}) fixes the first power-law exponent to one $\alpha=1$, so that it corresponds to a simplified version of alternative proposals with multiple regimes~\cite{naranan_models_1998,montemurro_beyond_2001,ferrer_i_cancho_two_2001,petersen_languages_2012,williams_text_2015}. The traditional power-law distribution~(\ref{eq.powerlaw}) is recovered for $b\rightarrow 1$. Further details of this analysis will be presented together with the statistical methods used to reach this conclusion in Sec.~\ref{ssec.ml-freq} below.  \index{power-law distribution}

\paragraph{Mechanistic models}

Providing an explanation for Zipf's law\index{Zipf's law} has been an obsession in different disciplines for over a century. This led to a variety of different approaches and models, see Refs.~\cite{naranan_models_1998,piantadosi_zipfs_2014} for reviews specifically related to Zipf's law of word frequencies.

Preferential-growth explanations go back to \index{Simon, Herbert} Simon's work~\cite{simon_class_1955}. In its simplest form, it considers that a text is written token-by-token and that at each step there is a small probability $p_{\text{new}} \ll 1$ of choosing a new word type and a large probability $1-p_{\text{new}}$ of choosing a previously used one. In the latter case, the probability that the word token is of type $r$ is proportional to the frequency $F_r=x$ of each of the types (in the existing text).  For a constant $p_{\text{new}}$, the $\alpha=1$ case is recovered, suggesting a natural explanation for this case. An $\alpha>1$ is obtained considering $p_{\text{new}}$ to decay with text size, a direct connection to Herdan-Heaps' law \index{Herdan-Heaps' law} discussed in Sec.~\ref{ssec.heaps} and one of the key points in the amusing Simon-Mandelbrot \index{Mandelbrot, Benoit} \index{Simon, Herbert} exchange~\cite{mandelbrot_note_1959}. Following this tradition, in Ref.~\cite{gerlach_stochastic_2013}  a combination of constant and varying $p_{\text{new}}$ was used to obtain the double power-law distribution~(\ref{eq.modeldp}), associating the transition point $b$ to the size of a core vocabulary. 
As discussed by \index{Simon, Herbert} Simon~\cite{simon_class_1955}, the mechanistic interpretation of the preferential-growth explanation of Zipf's law is subtle: the frequency of words is highly correlated across texts (e.g., the most frequent words are the same in all texts of the same language) so that each text cannot be considered as a new realization of these processes. In particular, the initial condition of the process is unclear -- the beginning of each text cannot be associated to the time at which the stochastic process starts -- and it has a high impact on the probability of reusing a word (if the initial condition satisfies Zipf's law, is the argument circular?). Simon argues that the process of writing involves a combination of two processes: association (i.e., sampling from the past sequence of the same text) and  imitation (i.e., sampling from past sequence of other texts from the same or other authors), with Zipf's law being robust against different combinations of these processes.
 \index{power-law distribution}

The main alternative explanations are based on arguments of optimality (and criticality) of communication. They focus on the question of {\it why} Zipf's law exists, in contrast to growth models that prescribe {\it how} it emerges. This type of explanation is in line with Zipf's reasoning based on a "least effort principle"~\cite{zipf_human_2012}.  \index{Mandelbrot, Benoit} Mandelbrot~\cite{mandelbrot_informational_1953} \index{Mandelbrot, Benoit} was the first to propose an information-theoretic model that mathematically derives Zipf's law as the function that minimizes the cost of communication per transmitted information. This line of research remains active and has motivated the proposal of different models, which typically connect the onset of Zipf's law to phase transitions and criticality~\cite{ferrer_i_cancho_zipfs_2005,prokopenko_phase_2010,dickman_analysis_2012}.

Underlying the debate around the origin of Zipf's law is the question whether it reveals an important (fundamental) property of human language (cognition) or whether it is a trivial consequence of a statistical or combinatoric process. The mechanisms mentioned above, in particular the explanations based on optimality and criticality, suggest that Zipf's law provides insights on a fundamental property of the underlying system. In contrast, a trivial origin of Zipf's law of word frequency is given by the Monkey type \index{Monkey typist} writing process~\cite{li_random_1992,mitzenmacher_brief_2004,piantadosi_zipfs_2014}. In this model, a Monkey writer types a text by randomly choosing the $k$ letters on a {\bf k}eyboard with a fixed probability $p_K$, smaller than the probability $p_s$ of typing the large {\bf s}pace bar key (so that $K p_K + p_S=1$ with $p_S> p_K$). In this case we have: 

\begin{itemize}
    \item[(i)] the probability of typing a word (i.e., a sequence of letters between space bars) of length $T$ is proportional to $$p(T) \sim e^{-p_S T};$$ 
    \item[(ii)] the number of unique words of length $T$ is $K^T$ and therefore a frequency of each of them is $$x \sim \frac{1}{K^T} = e^{-\ln(K) T}.$$ 
\end{itemize}
    Zipf's power-law~(\ref{eq.powerlaw}) of word frequencies is obtained combining these two exponential distributions, as shown in Eq.~(\ref{eq.expToPower}). The value of this type of model is to act as a null model that shows how Zipf's law\index{Zipf's law} can emerge naturally as a statistical process. A linguistic argument against this explanation is that texts generated by the Monkey typist differ from real texts in important aspects: the distribution of word lengths is not exponential and the frequency of words of the same length is very far from being a constant. 

Despite the quantity and variety of explanations, there is no consensus regarding the explanation of the origin of Zipf's law or its significance. Piantadosi's recent review~\cite{piantadosi_zipfs_2014}, published a century after the first observations of Zipf's law, finishes with a sober evaluation:

\begin{quote}
{\it ``...literature on Zipf's law has mainly demonstrated there are many ways to derive Zipf's law. It has not provided any means to determine which explanation, if any, is on the right track.''~\cite{piantadosi_zipfs_2014}}
\end{quote}

\paragraph{Consequences}

Zipf's law\index{Zipf's law} plays an important role in statistical natural language processing and in methods for text analysis~\cite{baayen_word_2001}. Statistical estimation of information-theoretic measures (entropies, Jensen-Shannon divergence, etc.) are directly affected by Zipf's law, whose exponent affects the finite-size bias and fluctuations of estimators~\cite{gerlach_similarity_2016,dias_using_2018,altmann_generalized_2017}. Refs.~\cite{sato_topic_2010,lim_nonparametric_2016} considered Zipf's law as a motivation for extensions of traditional "topic modelling" methods for unsupervised classifications of collections of documents.

A direct use of Zipf's law is the association between the Zipfian exponent~$\alpha$ and characteristics of the text such as its author, language, and styles (e.g., the speech of children in different age groups~\cite{baixeries_evolution_2013}). The sub-linear growth of the vocabulary size (unique words) with document size (word tokenks) -- known as Heaps'-Herdan's law, \index{Herdan-Heaps' law} as discussed in Sec.~\ref{ssec.heaps} below -- can be connected to Zipf's law~\cite{mandelbrot_note_1959,montemurro_beyond_2001,eliazar_growth_2011,gerlach_stochastic_2013} and be seen as a consequence of it (we review this law and its consequences in Sec.~\ref{ssec.heaps} below).

\subsection{Earthquakes (Gutenberg-Richter's law) and Natural disasters}\label{ssec.gutenberg}

The \index{Gutenberg-Richter's law} Gutenberg-Richter law specifies the number $N$ of \index{earthquakes} earthquakes of a given magnitude~$M$ in a fault or region. In its original formulation~\cite{gutenberg_earthquake_1942,gutenberg_frequency_1944}, it specifies a relationship
\begin{equation}\label{eq.gutenberg}
    \ln N = a - b M,
\end{equation}
where $a,b$ are constants. The discovery of this law is often attributed~\cite{fagereng_geology_2011} to a 1939 work by Ishimoto and Iida. The identification of the \index{Gutenberg-Richter's law} Gutenberg-Richter's law in Eq.~(\ref{eq.gutenberg}) and the power-law distribution $p(x)$ in Eq.~(\ref{eq.powerlaw}) is established by noting that the magnitude $M$ is defined to be proportional to the logarithm of the energy~$x$ released by an \index{earthquakes} earthquake $M \sim \ln x$ and considering $p(x) = N(x) / \sum N$.

The Gutenberg-Richter law became a paradigmatic example for the description of many different natural disasters, including extensions to forest fires~\cite{malamud_forest_1998,nicoletti_emergence_2023} and snow avalanches~\cite{birkeland_power-laws_2002}. Power-laws are one of the three most popular distributions -- together with Gaussians  \index{Gaussian distribution}and Exponentials -- used in the analysis of natural disasters~\cite{pisarenko_distributions_2010}.  In this context, the main significance of this distribution is the heavy-tail component of power-laws, in which events with large $x$ are significantly more likely than under the alternative distributions. These extreme events, while rare, cause a disproportionally large impact so that the the behaviour of $p(x)$ in the tails of the (power-law) distribution is of foremost interest.  \index{power-law distribution}

\paragraph{Empirical Evidence}

The striking linearity of frequency distributions as in~(\ref{eq.gutenberg}) has been repeatedly observed in different faults, regions, and data of other natural disasters~\cite{malamud_forest_1998,pisarenko_distributions_2010}.

\paragraph{Mechanistic models}

While the geological origin (e.g., friction, properties of rocks) of Gutenberg-Richter is well established~\cite{fagereng_geology_2011}, the appearance of power-law distributions in different natural disasters has motivated the search for more general explanations. This has been a key motivation for the proposal that these distributions are a manifestation of critical phenomena~\cite{sornette_critical_2006}, i.e., of an underlying system that is at a critical state, and of self-organized criticality~\cite{bak_how_2013} as the key process explaining why these systems tend towards such states. 

\begin{figure}
    \centering
    \includegraphics[width=0.9\linewidth]{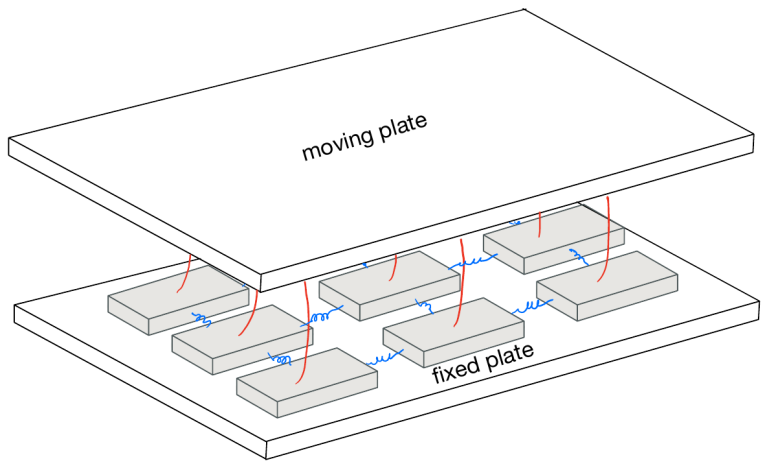}
    \caption{Mechanistic model of the \index{Gutenberg-Richter's law} Gutenberg-Richter law of \index{earthquakes} earthquakes. Earthquakes at the surface (fixed plate) happen because of its attachment (red strings) to blocks (gray parallelipids) that can move with friction on a fixed plate and that are attached to each other (blue springs). This illustrative figure is inspired by Ref.~\cite{bak_how_2013}.}
    \label{fig.blocks}
\end{figure}

Figure~\ref{fig.blocks} shows an example of a block-and-spring model to explain Gutenberg-Richter law~\cite{bak_how_2013}, similar examples exist for forest fire models~\cite{malamud_forest_1998,nicoletti_emergence_2023} and for models of the neural activity in the brain~\cite{chialvo_emergent_2010}. In all cases, the justification and use of these models follows essentially two steps:

\begin{itemize}
    \item [(i)] a model containing the main mechanisms of a system of interest is proposed and showed to be (or evolve towards) a critical state;
    \item[(ii)] a power-law distribution of event magnitudes of the model -- obtained from simulations or analytical calculations -- is considered as a successful reproduction of the statistical law and, often, as an empirical support of the model. 
\end{itemize}
 
 This data-model divide is in line with the tradition of statistical laws discussed in Sec.~\ref{ssec.reasoning}.  The advantages and limitations of this approach will be discussed in Chap.~\ref{chap.synthesis}.

\paragraph{Consequences} 

The main application of power-law distributions in natural disasters is for risk analysis and the estimation of the probability of tail events. The socio-economical impact of such extreme events is disproportionally larger than the one of typical events, highlighting the importance of fat-tailed distributions. Power-law distributions are paradigmatic examples of fat-tailed distributions and the importance underlying their validity and characterization is that its exponent -- $\alpha$ or $\gamma$ in Eq.~(\ref{eq.powerlaw}) -- is directly connected to exponents appearing in the generalized central limit theorem, extreme-value statistics, and large deviation theory~\cite{coles_introduction_2013}.  \index{fat-tailed distribution}  \index{power-law distribution}

Another major question is the predictability of \index{earthquakes} earthquakes and natural disasters. Within the self-organized-criticality paradigm, at criticality the occurrence of extreme events happens due to small perturbations at any location and are thus essentially unpredictable. This point will be further discussed in Sec.~\ref{ssec.burstyearthquakes} below, when temporal patterns in the appearance of large earthquakes will be themselves described using statistical laws.

\subsection{Scale-free networks (Price, Barabasi-Albert)}\label{ssec.scalefree}

A very powerful representation of interconnected systems or data is in form of a network (or graph) in which nodes (vertices) connect to each other via links (edges). One of the most important characteristics of a node $i$ is the number of links attached to it, denoted as its degree $k_i$. Networks that have a power-law degree distribution, Eq.~(\ref{eq.powerlaw}) with $x=k$, are denoted scale-free \index{scale-free networks} networks. The claim that scale-free networks are commonly found in empirical networks across different datasets is a statistical law that plays an important role in the field of Complex Networks or Network Theory~\cite{barabasi_network_2016}.

The name "scale free" indicates the  lack of  a characteristic scale (i.e., $P(\lambda x) = f(\lambda) P(x)$) of this distribution and suggests that a large variety of node types exist in the network, from very central hubs (large $k$) all the way to weakly connected leaves (e.g., $k=1$). The significance of this statistical law is thus to indicate a crucial property of the network that is in stark contrast to simple random graphs (e.g., Erd\"os -R\'enyi graphs)~\cite{newman_networks_2018}.

\paragraph{Empirical Evidence}

Possibly the first claim of scale free network is due to Price's analysis in the 1960s and 1970s of citation networks~\cite{price_networks_1965, price_general_1976}, i.e., networks built by scientific papers as nodes and citations between them as links. Price associated its finding to previously proposed statistical laws, including power-laws in publications (Bradford's and Lotka's law, mentioned in \index{Bradford's law} \index{Lotka's law}  Sec.~\ref{ssec.otherpowerlaws} below) and the cases discussed by Zipf and \index{Simon, Herbert} Simon (discussed at the start of Sec.~\ref{sec:power-law-distr} above). In the late 1990s, similar power-law distributions were observed~\cite{huberman_growth_1999,albert_diameter_1999} studying the connectivity of the world-wide-web data, with webpages playing the role of publications and hyperlinks the role of citations.   \index{power-law distribution}

A substantially stronger claim of the ubiquity of \index{scale-free networks} scale-free networks is due to the work of Barabasi and Albert~\cite{barabasi_emergence_1999,barabasi_network_2016}. Besides the world wide web, their original work reported on data from actor collaborations and of power grid, and included later not only numerous other social networks but also metabolic, protein, and linguistic networks (see Ref.~\cite{barabasi_network_2016} p. 128 for a historical account). The majority of reported cases have an estimated power law exponent in the range $2<\gamma < 3$. The ubiquity of scale free networks is a paradigmatic example of the more general complex-systems approach of looking for common (universal) properties in networks of radically different origins, benefiting from the recent large availability of data. 

Recent works have questioned the ubiquity of \index{scale-free networks} scale-free networks, culminating at Refs.~\cite{broido_scale-free_2019,klarreich_clara_scant_2018}. One of the main reasons for this questioning is the application of new statistical analysis techniques~\cite{clauset_power-law_2009}, a point we will discussed in Sec.~\ref{ssec.ml-freq} below.

\paragraph{Mechanistic Models}

The claim of ubiquity of scale free networks played an important role in the development and justification of mechanistic network-growth models, following the statistical law tradition we described in Sec.~\ref{ssec.reasoning}. As put by Barabasi

\begin{quote}
{\it "Given the diversity of the systems that display the scale-free property, the explanation must be simple and fundamental".}~\cite{barabasi_network_2016} 
\end{quote}

The explanation provided in the {\it preferential attachment} model proposed by Barabasi and Albert~\cite{barabasi_emergence_1999} is based on two effects:
\begin{itemize}
    \item[i)] Growth: at each time step a new node is added and connected to $m$ other nodes.
    \item[ii)] Preferential attachment: the probability $\Pi$ that one of the new links is associated to node $i$ is linearly proportional to $k_i$,i.e., $\Pi(i) = k_i/\sum_j k_j$.
\end{itemize}

This model follows Simon-Yule preferential growth processes~\cite{perline_strong_2005,simkin_re-inventing_2011}  -- as introduced at the start of Sec.~\ref{sec:power-law-distr}  -- and the key "preferential-attachment" or "cumulative advantage" part (ii) is present in previous network models~\cite{price_general_1976,huberman_growth_1999}. The timely proposal of Barabasi-Albert's paper, at a time of growth of the Internet and the resulting networks (and data) -- made their work and model extremely influential, recognized as a foundational paper of the field of Complex Networks or Network Science. Mechanism and the statistical law are so closely connected in the case of networks that observations of \index{scale-free networks} scale-free networks are often taken as evidence of the preferential attachment mechanism. Several variations and alternatives to the preferential attachment model have been introduced~\cite{small_growing_2015,barabasi_network_2016,falkenberg_identifying_2020}, for instance, to account for temporal variations in $\Pi(k)$, addition of new links between existing nodes, the inclusion of fitness in nodes, and the incorporation of other network features. One of the motivations for these models is to obtain variations in the resulting exponent $\gamma$, in view of the fact that the original model leads to $\gamma=3$.

The preferential-attachment model generates networks which have many additional features beyond the power-law degree distribution. In fact, networks generated by this mechanism differ significantly from random graphs with power-law degree distribution~\cite{judd_what_2013,small_growing_2015,zhang_exactly_2015}. Some of the additional features present in the preferential-attachment model are not found in real networks, leading to debates of the extent into which the model provides an explanation of specific cases~\cite{amaral_classes_2000,perline_strong_2005}. For instance, an early debate~\cite{adamic_power-law_2000} involved the relationship between the age of websites and their degree, comparing the strong correlation predicted by the model to empirical data (in which the hubs are not necessary the oldest nodes). The general argument in favour of the model~\cite{barabasi_network_2016} is that it focus on one feature (the scale free degree distribution), that additional features would need to be included in specific cases, and that the ubiquity of observations of the \index{scale-free networks} scale-free networks (statistical law) is generally explained by the fact that preferential attachment appears naturally in many contexts (e.g., the more citations one paper has, the easier it is to be found and cited again).  \index{power-law distribution}

\paragraph{Consequences} 

There are numerous statistical properties of networks that are critically affected by a power-law degree distribution~\cite{newman_networks_2018,barabasi_network_2016}. Possibly the most important is the effect on critical values for percolation and related transitions, that make (random) scale-free networks robust against random failures but susceptible to deliberate attacks and the spreading of diseases. Intuitively, this can be understood by the role played by the hubs in maintaining the connectivity of the network. The benefit of the \index{scale-free networks} scale-free-network law is that it allows for analytical calculations and estimations that would not be possible without a simple parametric function.

\subsection{Other power-law distributions}\label{ssec.otherpowerlaws}

Pareto, Auerbach, and especially Zipf, initiated the study of power-law distributions in a variety of settings.
Nowadays, there are an even larger number of settings in which power-law distributions~(\ref{eq.powerlaw}) have been proposed to describe observations, in the same spirit of the statistical laws revised here~\cite{mitzenmacher_brief_2004,newman_power_2005,simkin_re-inventing_2011}. Some of the early and more prominent examples include:  \index{power-law distribution}
 
\begin{itemize}
\item Yule's law of number of species in different genera~\cite{simon_class_1955,simkin_re-inventing_2011}.
 \item In geography, power-laws were proposed to describe the frequency of length of rivers~\cite{dodds_unified_1999} and area of lakes and islands (see Ref.~\cite{perline_strong_2005} and references therein).
\item Richardson law of war magnitudes~\cite{richardson_variation_1948}.
\item Bibliometric data \index{bibliometric} on scientific publications~\cite{price_general_1976,simkin_re-inventing_2011}, including Bradford's Law \index{Bradford's law} on the number of articles in scientific journals, Lotka's law \index{Lotka's law} of scientific productivity (number of authors with at least $x$ publications), and the aforementioned Price's Law for the number of citations by scientific papers.
\item Size of neuronal avalanches and critical \index{brain} brain  hypothesis~\cite{chialvo_emergent_2010,bak_how_2013,beggs_being_2012}.
\item Intensity of solar flares~\cite{bak_how_2013}. \index{solar flares}
\item Frequency of features of molecules \index{Chemistry} (data from databases in Chemistry)~\cite{benz_discovery_2008}.
    \item Frequency of gene expression \index{gene expression} in single-cell transcriptomic data~\cite{lazzardi_emergent_2023}.
\item Various economic data~\cite{gabaix_power_2009}.
\item Measures of popularity of Internet items, including the number of view of memes~\cite{weng_competition_2012} or videos~\cite{crane_probabilistic_2018,miotto_predictability_2014} and signatures of online petitions~\cite{yasseri_rapid_2017}.
\end{itemize}

 Refs.~\cite{mitzenmacher_brief_2004,newman_power_2005,gabaix_power_2009,simkin_re-inventing_2011} provide reviews specifically on power-law distributions with many additional examples, but the number of additional claims keeps growing in defiance of systematic reviews.

Recent works have questioned the ubiquity of power-law distributions~\cite{stumpf_critical_2012}, calling for improved statistical methods to evaluate their validity. This point will be discussed in further details in Sec.~\ref{ssec.ml-freq} and Chap.~\ref{chap.synthesis}.

\paragraph{Mechanistic Models}

As in the examples discussed above, different mechanistic models were proposed to explain these observations~\cite{newman_power_2005,gabaix_power_2009,simkin_re-inventing_2011}, typically variations and adaptations of the processes discussed at the start of Sec.~\ref{sec.powerlaw}. Important for our argument, these mechanistic models are developed and adapted to specific problems on a phenomenological level (i.e., trying to justify their assumptions based on what is known in each case) and their comparison to the data is essentially based on their ability to reproduce the statistical law (or, in some cases, comparing the numerical values of the parameters of the law estimated from data). There is no further data-model comparison in the sense of inference of model parameters from the data.

An example of a mechanistic model motivated by the fat-tailed distribution is the proposal of a stochastic growth process with linear growth to describe the evolution of the view of YouTube videos~\cite{miotto_stochastic_2017}. Interestingly, while the linear preferential growth element was observed in the data, the fluctuations around these values are themselves heavy tailed (and modeled by a L\'evy-distributed stochastic variable). This shows that the heavy-tailed distributions is not only due to preferential growth.  \index{fat-tailed distribution}

Another example of mechanistic model is the model of how scientific papers gather citations~\cite{wang_quantifying_2013}, which includes not only the preferential-attachment mechanism discussed in Sec.~\ref{ssec.scalefree} but also the fitness and (temporal dependent) novelty of the work. The incorporation of these additional mechanisms follows many of the characteristics of the mechanistic models proposed to explain statistical laws, such as the claim {\it "that all papers tend to follow the same universal temporal pattern".}~\cite{wang_quantifying_2013}.

\paragraph{Consequences}

The combination of the statistical laws and the models they motivated lead to improved methods of forecasts and analysis, for instances of the asymptotic number of citations a paper will receive~\cite{wang_quantifying_2013} or of the probability of an online video to become viral~\cite{miotto_stochastic_2017}.

\section{Scaling laws}\label{ssec.scaling}

Scaling laws are statistical laws that specify a power relationship between two or more variables observed in a population of $i=1,\ldots, N$. In its simplest and most common form, one variable $y_i$ of interest is set to depend or scale with a size variable $x_i$, for any $i$, as
\begin{equation}\label{eq.scaling}
y \sim x^\beta, 
\end{equation}
where $\beta \in \mathbf{R}$ is a parameter. The linear $\beta=1$ scaling is often expected while the non-linear $\beta \neq 1$ scaling is divided into the sub- ($\beta<1$) and super- ($\beta>1$) linear cases. By growing a system from size $x$ to size $\lambda x$, the observable $y$ changes or scales by a factor $\lambda^\beta$. In particular, two ($\lambda=2$) systems of size $x$ are different from a system of size $2x$ as $2x^\beta \neq (2x)^\beta$, for $\beta \neq 1$. A simple geometrical example of non-linear scaling is the scaling of the area of objects with their volume or mass ($\beta = 2/3$). Often, exponents $\beta$ given by simple fractions are proposed to explain the relationship between different variables.

Statistical laws propose scalings and associated mechanistic explanations that go beyond geometrical relationships. This tradition goes back at least to the birth of \index{social physics} Social Physics -- discussed in Sec.~\ref{ssec.socialphysics} -- with Quetelet's proposal of $\beta=5/2$ to describe the scaling between the weight $y$ and height $x$ of humans. Nowadays, $\beta=2$ is used in the computation of the Body Mass Index, also known as Quetelet's index, widely used to determine whether individuals are under- or over-weighted. More generally, the study of the scaling of different animal properties $y$ with body size $x$ is known in biology as \index{allometric laws} allometry, giving rise to many interesting statistical laws from the early 20th century on (to be discussed in Sec.~\ref{ssec.allometry}). Going back to the same time, and following this socio-physics tradition, the scaling of the area and population of cities was studied~\cite{stewart_suggested_1947}, a tradition expanded through new proposals of \index{Urban scaling laws} urban scaling laws in the 21st century. 

Scaling laws as in Eq.~(\ref{eq.scaling}) are particularly significant when the values of $x$ in the population vary over many orders of magnitude. This is so because the scaling analysis is typically intended to determine the leading dependence between the variables and a (non-linear) scaling becomes relevant (visible) when large variations in $x$ exist. This provides a connection with the statistical laws described in Sec.~\ref{sec.powerlaw} because the fat tails of the power-law distribution ensure that different magnitudes of the quantity of interest  are available. For instance, one of the consequences of the Auerbach-Lotka-Zipf\index{Auerbach-Lotka-Zipf's law}'s law of city sizes discussed in Sec.~\ref{ssec.alz} is that the population of cities ranges over at least $5$ orders of magnitudes, from small villages ($10^2$) to huge metropolis ($10^7$). This motivates us to first consider the case of scaling laws associated to cities, before going to other examples of scaling laws.  \index{power-law distribution}

\subsection{Cities (\index{Urban scaling laws}urban scaling law)}\label{ssec.urbanscaling}

Statistical laws of scaling type have been long proposed to describe observations of cities. The most traditional analyses use the population $P$  of cities as a measure of their size $x$. Another measure of the size of city is their area $A$. In the sociophysics tradition~\cite{stewart_suggested_1947}, the scaling law~(\ref{eq.scaling}) was proposed to describe how $y=A$ scales with $x=P$, with a sub-linear scaling $\beta=\beta_A<1$.

\paragraph{Empirical Evidence}

The 21st century brought renewed interest on scaling laws in urban systems~\cite{bettencourt_growth_2007,rybski_urban_2019}, with the proposal that many different socio-economic observables $y$ of cities  show non-linear scalings with their population $x$. The proposal that such relationships are observed in cities of different countries (with similar exponents) is known as \index{Urban scaling laws} urban scaling laws. In the stronger version of urban scaling laws~\cite{bettencourt_growth_2007}, the same type of scaling or the same universal exponents are proposed to describe a large class of observables $y$ as follows:
\begin{itemize}
    \item observables related to economic (e.g. GDP\index{Gross Domestic Product, GDP}), scientific (e.g. patents), and artistic (e.g., books, plays) innovation show super-linear scaling (sometimes claimed to show the same exponent $\beta \approx 1.15$); 
    
    \item observables related to infrastructure (e.g., road sizes) show a sub-linear scaling~(sometimes claimed to show the same exponent $\beta\approx 0.85$). 
\end{itemize}
Modern studies also use both area $A$ and population~$P$ (and their combination) to obtain improved descriptions of how observables~$y$ scale with city sizes~\cite{ribeiro_effects_2019}.

\begin{figure*}[!t]
  \includegraphics[width=0.5\linewidth]{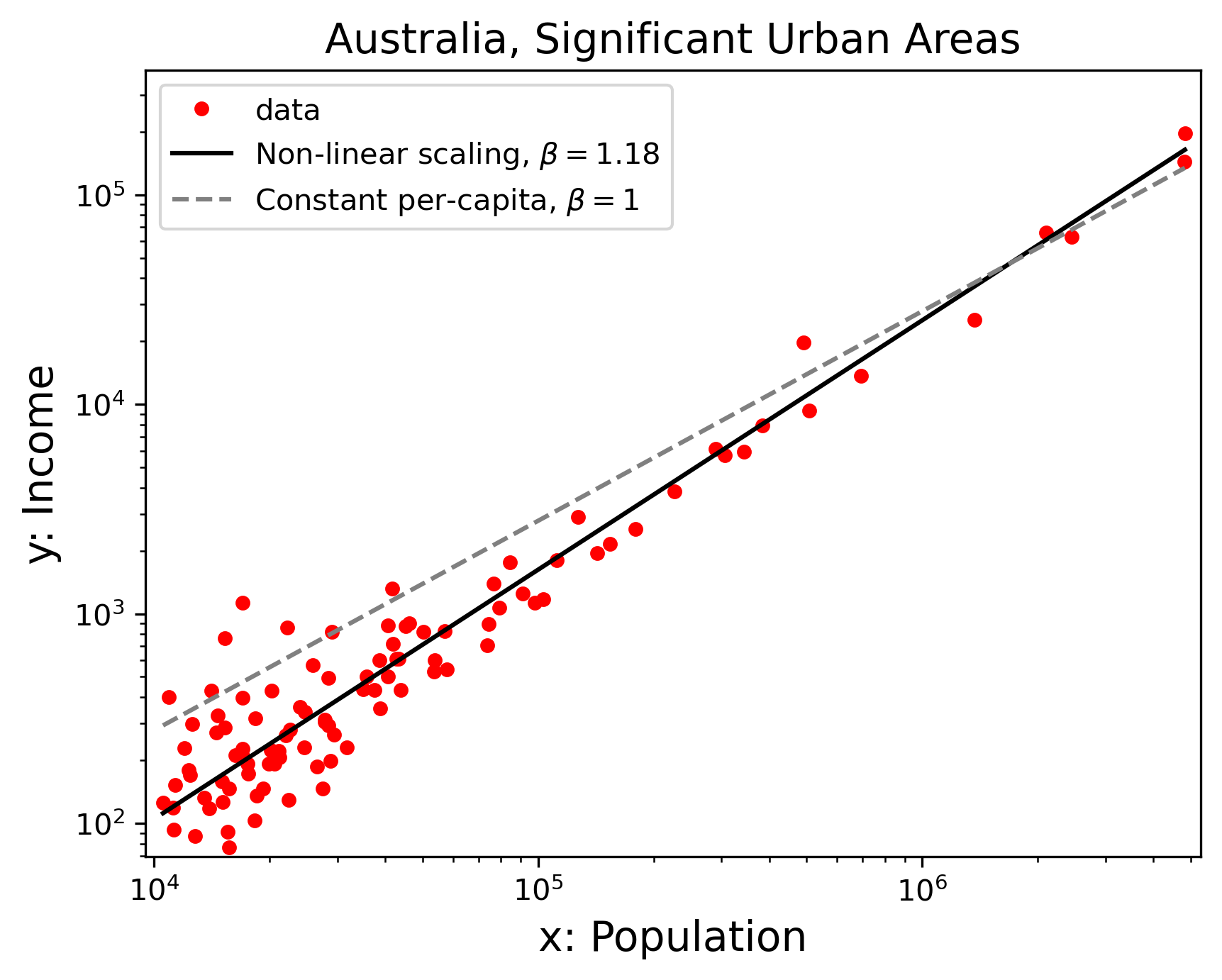}   \includegraphics[width=0.5\linewidth]{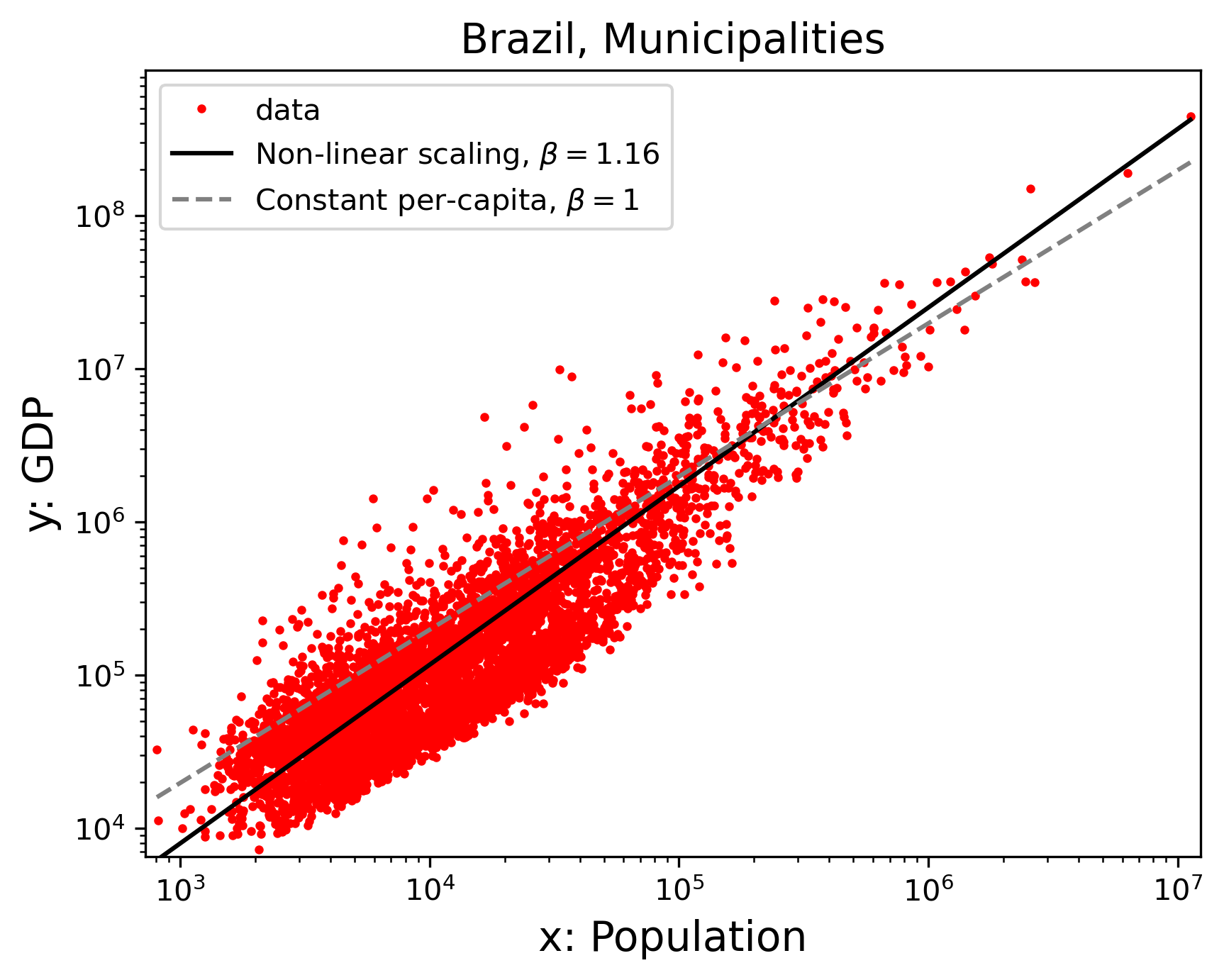}
  \includegraphics[width=0.5\linewidth]{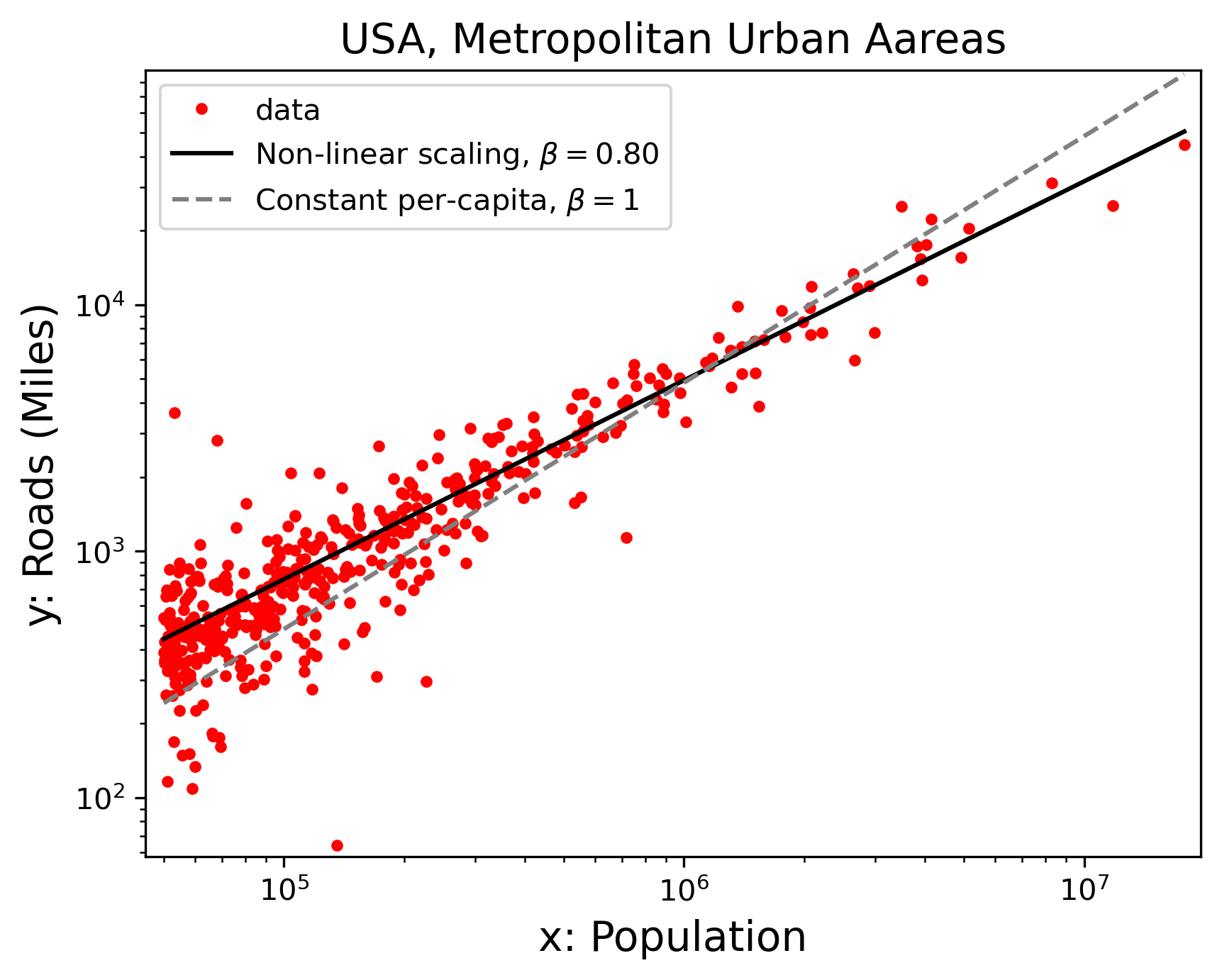}   \includegraphics[width=0.5\linewidth]{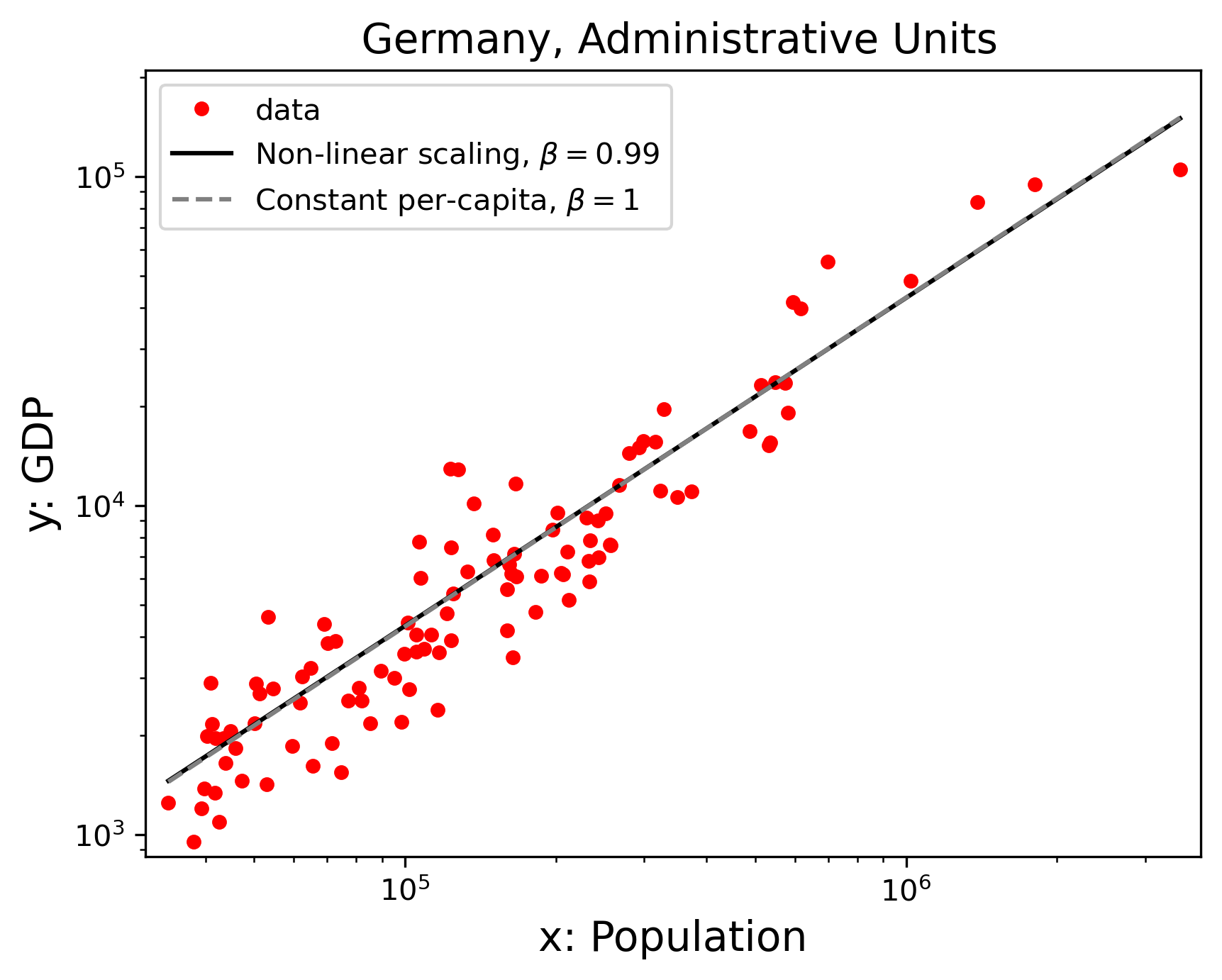}
  \caption{\index{Urban scaling laws} urban scaling law in four different countries.  Different observables $y$ scale with the population $x$ of urban areas with exponent $\beta$ as in Eq.~(\ref{eq.scaling}). The dashed line corresponds to a constant per-capita division. The solid line is a non-linear scaling with $\beta$ estimated using a model of attributing tokens to individuals, as described in Fig.~\ref{fig.illustration-tokenUrban} and Sec.~\ref{ssec.inference} below. Top left: number of individuals at the top bracket in income in Australian largest urban areas (2021). Top right: gross domestic product (GDP \index{Gross Domestic Product, GDP}) of Brazilian municipalities (2010). Bottom Left: extension of roads in metropolitan areas in the USA (2013). Bottom Right: GDP of German administrative units (2012). See Appendix~\ref{chap.appendices} for information on data and code used in this figure.}
\label{fig.urbanscaling}%
\end{figure*}

Figure~\ref{fig.urbanscaling} shows a sample of four different datasets and countries. It confirms a superlinear scaling $\beta>1$ for the income of Australian cities and for the GDP \index{Gross Domestic Product, GDP} of Brazilian municipalities, in agreement with the general expectation mentioned above and the results shown for the USA in the introduction (Fig.~\ref{fig:urban}). Interestingly, the scaling analysis of the GDP of German administrative units seems compatible with a linear scaling $\beta=1$, i.e., the same GDP per-capita in all cities. The case of the length of roads in Metropolitan Areas of the USA provides an example of sub-linear scaling $\beta<1$.

A critical point in the study of \index{Urban scaling laws} urban scaling laws, and in the quantitative investigations of urban systems more generally~\cite{rozenfeld_area_2011}, is the definition of what a city is (i.e., the urban area appropriate for the analysis). What are the boundaries of cities? Is there a minimum population size to an urban area to be counted as a city for scaling analysis? Importantly, estimations of $\beta$ and even conclusions about their non-linearity depend on how these questions are answered~\cite{arcaute_constructing_2015,louf_scaling_2014,leitao_is_2016}. We will discuss this point in further detail, and provide a statistical explanation for these observations, in Secs.~\ref{ssec.caveatsLinear} and \ref{ssec.likelihoodscaling} below.

\paragraph{Mechanistic models}

The observation of non-linear \index{Urban scaling laws} urban scaling laws has motivated the proposal of mechanistic explanations~\cite{bettencourt_origins_2013,ribeiro_mathematical_2023}. In line with the explanations of scaling laws in physical objects and of \index{allometric laws} allometry in animals, many of the explanations related how spatial variables scale with each other and with population. For instance, assuming that (large) cities grow vertically, and therefore the population is distributed in three dimensions (a volume), the scaling exponent $\beta_A$ for the area vs. population of cities is derived as $\beta_A = 2/3$~\cite{batty_new_2017,ribeiro_mathematical_2023}. 

\index{social physics}
The simple scaling theories for area, traditional in socio-physics~\cite{stewart_empirical_1947}, do not provide much insight about the underlying urban system, arguably failing our definition of statistical laws in Sec.~\ref{ssec.definition}. Urban scaling laws become thus {\it bona fide} examples of statistical laws when their claims extend to other observables $y$.
Mechanistic explanations for other observables $y\neq A$ typically rely on the idea that they depend on the opportunities that exist for people to interact~\cite{bettencourt_origins_2013,ribeiro_mathematical_2023}. Cities with larger population (density) provide more opportunities to their citizens to interact, reducing the per-capita need for infrastructure (e.g., length of roads, $\beta<1$) and increasing their individual productivity (e.g., GDP $\beta>1$). The recent review~\cite{ribeiro_mathematical_2023} lists dozens of mechanistic models proposed to explain urban scaling laws, classifying them between those that focus in intra- and inter-urban processes. 

\paragraph{Consequences}
One of the applications of \index{Urban scaling laws} urban scaling law is the proposal of indicators of city performances that go beyond per-capita reports~\cite{bettencourt_urban_2010,gudipudi_efficient_2019}. In fact, a (strong) non-linear scaling $\beta\neq 1$ implies that ranking cities according to the per-capita $y/x$ observations will be strongly correlated with the population $x$ of cities themselves and thus of limited interest. Instead, if a urban scaling law is valid, the re-scaled variable $y_i/x_i^\beta$ would provide a better estimation of the deviation of the values of city $i$ from the expectation (based on their population).

Urban scaling laws suggest also that they can be used to predict how observables $y$ of cities will change as the cities grow or shrink. This should be done carefully as the results over an ensemble of cities may differ from what is observed in a single city. In fact, Ref.~\cite{depersin_global_2018} reports significant changes in the scaling observed when analyzing how congestion-induced delays in different cities scale with city size and in time. The connections between urban scaling laws and ALZ's law of city sizes was discussed in Ref.~\cite{gomez-lievano_statistics_2012}.

\subsection{Words (Herdan-Heaps' law)}\label{ssec.heaps}

Herdan's and Heaps' laws can be \index{Herdan-Heaps' law} viewed as scaling-laws between the vocabulary size $y$ (number of word types or unique words) and the corpus size $x$ (number of word tokens or length of text)~\cite{egghe_untangling_2007}. Herdan's proposal \index{Herdan-Heaps' law} is part of his seminal work on "Quantitative Linguistics"~\cite{herdan_quantitative_1964} that looked for statistical laws and other invariant properties in texts and proposed \index{linguistic laws}
$$\beta = \frac{\ln y}{\ln x},$$
with $0 \le \beta \le 1$. Heaps' work focused on information retrieval and considered $y$ to be the new information (key words) obtained by increasing the sample of new documents, with the typical case of $\beta<1$ representing a law of diminishing returns. More generally, such type-token relationship can be viewed as the scaling between unique elements in a population~\cite{egghe_untangling_2007}. 

\paragraph{Empirical evidence}

\begin{figure*}[!t]
\begin{center}
  \includegraphics[width=0.8\linewidth]{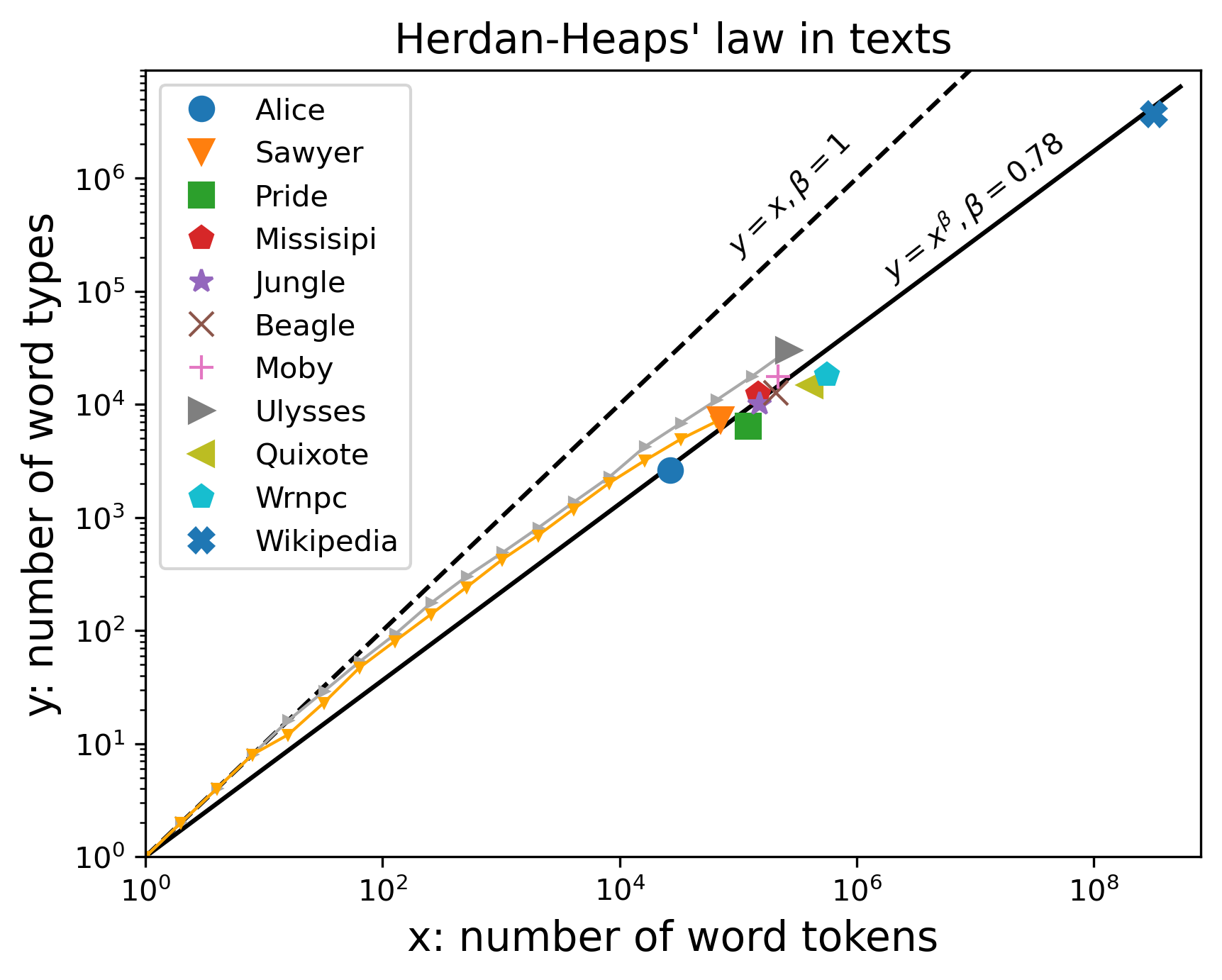}
\end{center}
  \caption{Herdan-Heaps' law in \index{Herdan-Heaps' law} different texts in English. The number of unique words ($y$ axis) is plotted as a function of the text size, measured in number of word tokens ($x$ axis). The symbols correspond to ten different novels (see legend and Tab.~\ref{tab.exponentzipfword} for details) and the complete Wikipedia. For two novels ("The Adventures of Tom Sawyer" by Twain and "Pride and Prejudice" by Austen) values of $(x,y)$ are plotted along the text, i.e., for the first $x$ word tokens of the novel . The straight lines correspond to Eq.~(\ref{eq.scaling}) with $\beta=0.78$ and $\beta=1$ (and prefactor one). See Appendix~\ref{chap.appendices} for the data and code used in this figure.}
\label{fig.heaps}%
\end{figure*}

In the usual linguistic \index{linguistic laws} analysis, Herdan-Heaps' law \index{Herdan-Heaps' law} can be viewed both within a document --  counting how many unique words $y_i$ are there in the first $x_i=i$ words of the text -- or over an ensemble of $N$ documents -- computing the size $x_i$ and vocabulary $y_i$ of the $i-th$ document with $i=1,\ldots N$. These two representations are shown in Fig.~\ref{fig.heaps}, where Herdan-Heaps' law corresponds to a straight line relationship. The first representation -- growing vocabulary within a text -- shows initially a linear growth ($\beta=1$) before slowing down to a sublinear scaling (see Ref.~\cite{gerlach_stochastic_2013} for a characterization of this transition), while the data of different texts (symbols) show the sub-linear scaling $\beta<1$ to provide a better description over $4$ orders of magnitude, from short novels to the complete English Wikipedia. Similar observations have been reported in a variety of cases~\cite{baayen_word_2001,egghe_untangling_2007,font-clos_scaling_2013}, different languages~\cite{petersen_languages_2012,gerlach_stochastic_2013,font-clos_scaling_2013}, and even in key-words used in Internet-based datasets~\cite{tria_dynamics_2014}.

Herdan-Heaps' law is nowadays widely interpreted to be valid for large text sizes $x \gg 1$ and variations are expected for short $x$ (e.g., $y=1$ for $x=1$ in any text, leading to a trivial $\beta=1$). For $x\rightarrow \infty$ it predicts $y \rightarrow \infty$ (for any $\beta>0$), i.e., an infinitely large vocabulary size. This contradicts the common assumption (e.g., in information theory) of finite vocabulary and also the bound imposed by the finite number of (finite-length) words composed from the finite number of existing phoneme (or letters). However, the analysis~\cite{gerlach_stochastic_2013} of extremely large corpora ($x> 10^{11}$), involving millions of books (Google n-gram corpus) and articles (complete English Wikipedia), show no indication of a convergence of the $y(x)$ to a constant and in fact suggest the practical validity of Herdan-Heaps' law and an effectively infinite vocabulary size (for practical purposes).
\index{Herdan-Heaps' law}

\paragraph{Mechanistic models}
    
The intimate connection between Herdan-Heaps' law and Zipf's law\index{Zipf's law} of word frequencies has been noted at least since the Simon-Mandelbrot's debates~\cite{mandelbrot_informational_1953,simon_class_1955,simon_size_1958,mandelbrot_note_1959}. \index{Mandelbrot, Benoit} Mandelbrot argues that Simon's \index{Mandelbrot, Benoit} \index{Simon, Herbert} explanation for a Zipfian distribution with $\alpha>1$ requires a probability of adding a new word ($p_{\text{new}}$ in Sec.~\ref{ssec.zipf}) to decay with text length (time) as $x^\beta$, with $\beta = 1/\alpha$~\cite{mandelbrot_note_1959,zanette_dynamics_2005}. This shows how Herdan-Heaps' law can lead to Zipf's law via \index{Simon, Herbert} Simon's model, a result that has been extended also to urn models~\cite{simkin_re-inventing_2011,tria_dynamics_2014}. Reversely, assuming $x$ word tokens are sampled from a Zipfian distribution of word-type frequencies  -- i.e., Zipf's law $F_r$ in Eq.~(\ref{eq.powerlaw}) for an arbitrarily large vocabulary $r=1, 2, \ldots R \rightarrow \infty$ -- Herdan-Heaps' law is obtained for large $x$ and $\beta = 1/\alpha$~\cite{eliazar_growth_2011}.
\index{Herdan-Heaps' law}

The connection to Zipf's law has been extended~\cite{gerlach_scaling_2014} to the double power-law (dp) extension of Zipf's law introduced in Eq.~(\ref{eq.modeldp}), which leads to a corresponding two-regime extension of the Herdan-Heaps law  \index{power-law distribution}
\begin{equation}\label{eq.heapsdp}
y_{dp}(x;\beta,b) = C_n 
\begin{cases}
x & x < b\\
b^{\alpha-1}x^{-\beta} & r\geq b,
\end{cases}
\end{equation}
where $\beta=1/\alpha$, $\alpha$ (Zipfian exponent) and $b$ (core vocabulary size) are the parameters of the double-power law distribution~(\ref{eq.modeldp}),  $C_n = C/n$ is a constant with $C\approx F_1$ (frequency of the most frequent word) and $n\gg 1$ (threshold applied to the word count of a word to include it in the count of $y$).

\paragraph{Consequences}
\index{Herdan-Heaps' law}
Applications of Herdan-Heaps' law include the prediction of size of unique words (e.g., for memory allocation when mining data) or for normalization of quantities as a function of data size (e.g., complexity of vocabulary measures depend on corpus size~\cite{gerlach_scaling_2014}). 

\subsection{Metabolism (Kleiber's law) and allometric scaling}\label{ssec.allometry}

A remarkable property of the diversity of life is that species of the same class can vary dramatically in size $x$. For instance, there is a difference of $3$ orders of magnitude between the body size of the smallest and largest mammal -- from the $3$cm small bumble-bee bat to the $30$m long blue whale -- and a remarkable $21$ orders of magnitude in weight between all organisms~\cite{west_general_1997,da_silva_allometric_2006}. It is thus possible to evaluate in which extent different properties $y$ across different species scale with their size $x$, potentially revealing non-linear scalings. \index{allometric laws} Allometric scalings exist for different $y$ (e.g., heart rate, bone sizes), the most famous being the metabolic rate which is known as \index{Kleiber's law} Kleiber's law. The importance of this observable is that it is directly related to the efficiency of different species in processing energy, a key physical quantity affecting their evolution and that can thus be expected to be highly optimized. 

The key element of \index{Kleiber's law} Kleiber's law is the value of the exponent~$\beta$ -- in particular Kleiber's claim of $\beta=3/4$ -- and not necessarily its non-linearity. This is so because the null model in this case is already nonlinear, $\beta=2/3$, obtained considering the simple geometrical and thermodynamical argument that the loss of heat depends on the surface area~\cite{da_silva_allometric_2006}. In fact, one of Kleiber's contribution from 1932~\cite{kleiber_body_1932} was to propose the study of metabolism rate as a function of the mass instead of the "surface law" traced back to 1839 (almost a century earlier). A further significance of the $\beta=3/4$ exponent of Kleiber's law is that it underlies a series of other "quarter-laws" related to it through other \index{allometric laws} allometric scalings~\cite{west_general_1997}.

\paragraph{Empirical evidence}

\begin{figure}
    \centering
    \includegraphics[width=0.95\columnwidth]{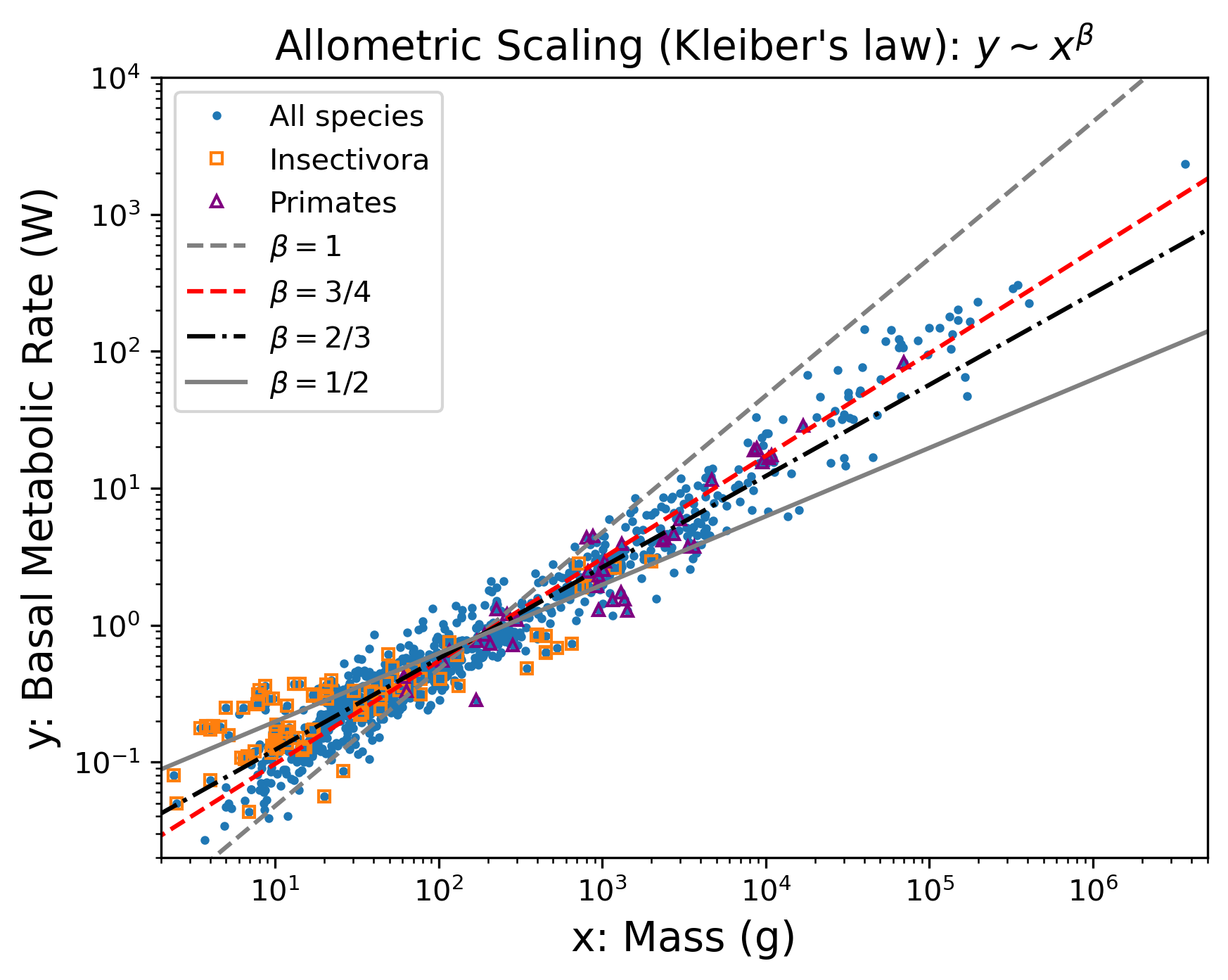}
    \caption{\index{Kleiber's law} Kleiber's law \index{Kleiber's law} for the metabolic rate of mammals. The data corresponds to the basal metabolic rate ($y$, measured in Watts) and the mass ($x$, measured in grams) of $1,006$ mammals. Results for species in $2$ distinct orders are highlighted with different symbols (see legend): {\it Primates} (39 species) and {\it Insectivora} (86 species). \index{species} The straight lines correspond to the scaling law~(\ref{eq.scaling}) with different $\beta$ values (see legend) -- $\beta=2/3$ (area law) and $\beta=3/4$ (Kleiber's law)-- with a prefactor chosen in such a way that they intersect at the same point $(10^{\langle \log x \rangle},10^{\langle \log y \rangle})$. The data corresponds to measurements reported in several publications and compiled in Appendix 1 of Ref.~\cite{savage_predominance_2004}, see Appendix~\ref{chap.appendices} for the data and code used in this figure.}
    \label{fig.allometric}
\end{figure}

\begin{figure}
    \centering
    \includegraphics[width=1\columnwidth]{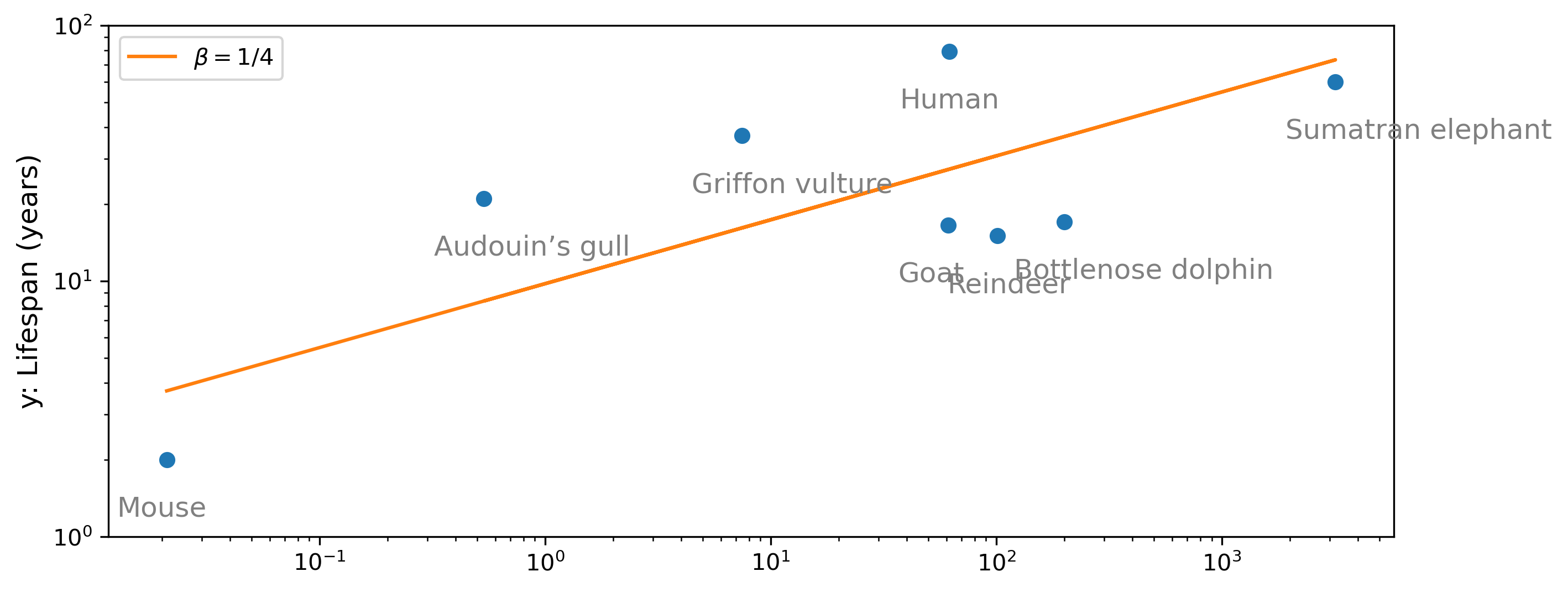}
    \includegraphics[width=1\columnwidth]{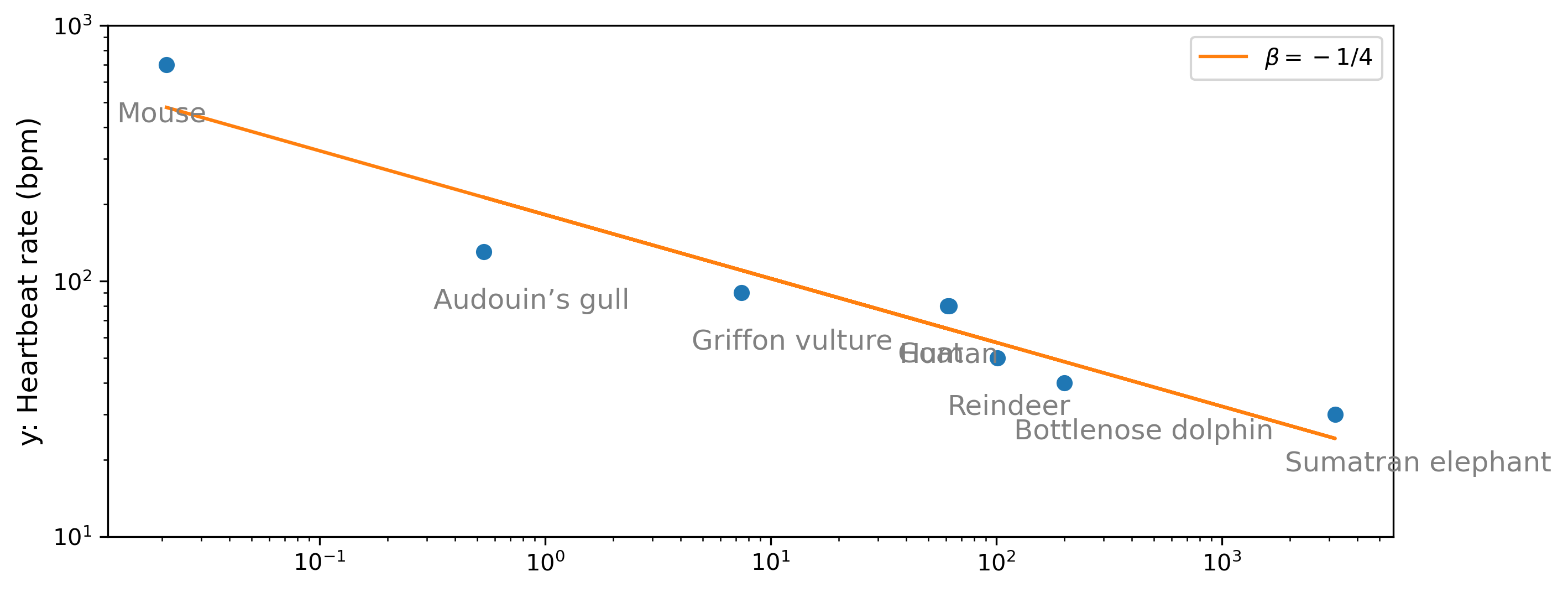}
    \includegraphics[width=1\columnwidth]{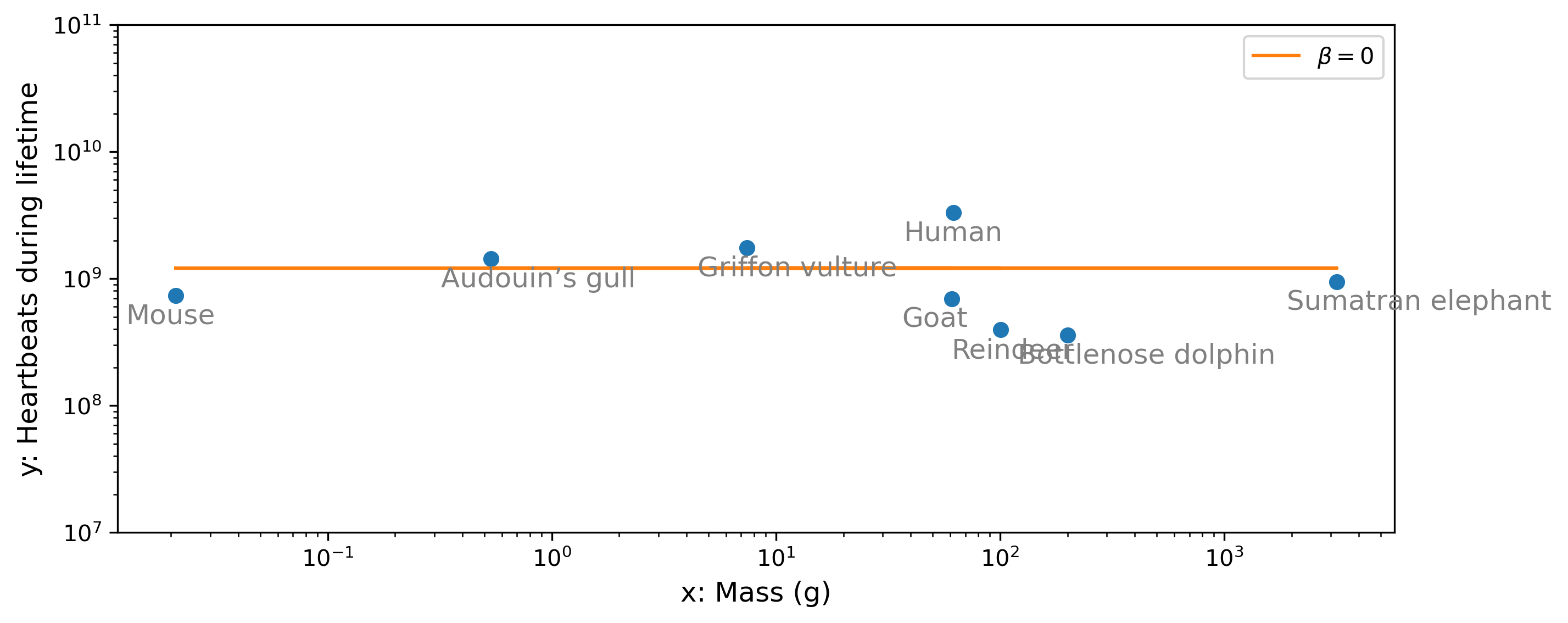}
    \caption{Allometry in mammals. The plots show the scaling of different quantities $y$ (top: average lifespan in years; middle: heart rate in beats per minute; bottom: average beats in lifetime) with weight $x$ (measured in grams) for $8$ different species of mammals. \index{species} The solid lines correspond to the scaling law~(\ref{eq.scaling}) with the quarter exponents $\beta$ (see legends) in line with Kleiber's law~\cite{west_scale_2018} and with a proportionality factor chosen so that the lines pass through ($(10^{\langle \log x \rangle},10^{\langle \log y \rangle})$). Data retrieved from Table S1 of Ref.~\cite{whittemore_telomere_2019}, see Appendix~\ref{chap.appendices} for the data and code used in this figure.}
    \label{fig.allometric2}
\end{figure}

Figure~\ref{fig.allometric} shows a compilation of modern data for $1,006$ mammalian species. It shows a strong dispersion of points around the two proposed scaling relationships. \index{Kleiber's law} Kleiber's proposal of $\beta=3/4$ appears to have a better agreement for large masses $x$, while specific genera (Insectivora) show a slower scale closer to the alternative $\beta=2/3$ proposal. 

Kleiber's data analysis~\cite{kleiber_body_1932} seems to be the first to strongly favours the non-trivial exponent $3/4$ and was viewed for a long time as the key departure over the earlier $2/3$ prediction from the area law.  By rounding the estimated result $\beta=0.74$ to a simple fraction, Kleiber implicitly suggests the existence of a simple and universal explanation similar to the one behind the $2/3$ surface expectation and other scaling relationships, the starting point of later theoretical attempts to explain it. 

The debate between the validity of $\beta=3/4$ and $\beta=2/3$ resurfaced in the end of the 20th century and was again particularly lively at the start of the 21st century \cite{dodds_re-examination_2001,savage_predominance_2004,white_allometric_2005,white_allometric_2007}. We refer to these publications for further historical accounts and references on the rich history of this dispute, which involves choices of the type of metabolism and measurement (e.g., standard vs basal metabolic rates~\cite{da_silva_allometric_2006}), the set of species used \index{species} in the analysis, different fitting methods~\cite{dodds_re-examination_2001,savage_predominance_2004,da_silva_allometric_2006} (more on this in Sec.~\ref{sec:traditional-methods} below), the interval in $x$ in which the analysis is performed, dependence on habitat regions (e.g., geography, diet, temperature), and the analysis of the models proposed to explain the different cases. Some of the challenging and contentious issues on this dispute are seen in Fig.~\ref{fig.allometric}: a larger $\beta$ for large masses $x$~\cite{dodds_re-examination_2001}, the uneven distribution of species along the $x$ axis that bias fits towards low values of $x$~\cite{savage_predominance_2004}, the dependency of the fit on different groups of species~\cite{white_allometric_2007}, and the correlation in the data of philogenetically close species~\cite{savage_predominance_2004} (e.g., about half of the species are from the order {\it Rodentia}).

Besides the defendants of the $\beta=2/3$ and $\beta=3/4$, a third position that emerged is that of lack of universality of the relationship between metabolic rate and mass (or of the exponent $\beta$). The review paper~\cite{da_silva_allometric_2006} indicates that the maetabolic rates of mammals yield values of $\beta$ between $2/3$ and $3/4$ {\it "depending on the selected data and on the statistical procedure chosen to examine the data"}. A review of $24$ different birds and mammals datasets from $12$ different references shows $\beta \in [0.65,0.96]$ (Ref.~\cite{da_silva_allometric_2006}, Table 2). The meta-analysis in Ref.~\cite{white_allometric_2007} finishes by saying that 

\begin{quote}
{\it "Our analysis of 127 exponents suggests that there is no single true \index{allometric laws} allometric exponent relating metabolic rate to body mass and no universal metabolic allometry."}\cite{white_allometric_2007}
\end{quote}

\paragraph{Mechanistic models}

The evidence in favour of \index{Kleiber's law} Kleiber's law ($\beta=3/4$) has been the main driver behind the search for theoretical (mechanistic) explanations of this unexpected exponent. Many different models have been proposed, more recently relating the {\it "quarter exponents"} to {\it "fractal-like networks"} which {\it "effectively endow life with an additional fourth spatial dimension"}~\cite{west_general_1997,west_fourth_1999,west_scale_2018}. This argument is based on the optimization of branching biological distribution networks (e.g., circulatory, respiratory, vascular system) and the similarity of the components and challenges faced by all mammals \index{species} \index{mammals} or species in the same group. \index{species}

\paragraph{Consequences}

One of the main consequences of Kleiber's law and the associated models to explain it is that they simultaneously explain other allometric relationships~\cite{west_general_1999,west_scale_2018}. This is done either using traditional scaling arguments or as part of the mechanistic models. Examples are shown in Fig.~\ref{fig.allometric2}, which plots the predicted scaling of the life expectancy ($\beta=1/4$), heartbeat rate (with $\beta=-1/4$), and heartbeats during lifetime ($\beta=0$) with the mass of $8$ of different mammals. The quarter scaling in this case explains the remarkable constancy (in the last panel) of the expected number of heartbeats during the total lifetime of species, with a small relative variation over several orders of magnitude in mass~\cite{levine_rest_1997,west_scale_2018}. More practically, these different scaling laws can be used to scale the amount of food or medicine needed by species of different mass. The success of \index{allometric laws} allometric scaling in describing different observations and combining theory to data has motivated studies of allometry in urban data (discussed in Sec.~\ref{ssec.urbanscaling}) and different areas~\cite{west_scale_2018}, a recent example being the metabolic scaling in human cancer cells~\cite{perez-garcia_universal_2020}.

\subsection{Other scaling laws}

Numerous other type vs token relationship have been proposed to follow a scaling law:
\begin{itemize}
\item The scaling of the number of unique chemical elements~\cite{benz_discovery_2008}.
  \item The scaling of the number of expressed genes in single-cell transcriptomic data~\cite{lazzardi_emergent_2023}.
\item The species-area relationship in \index{species} \index{Ecology} ecology~\cite{gleason_henry_allan_relation_1922,brainerd_relation_1982}.
\item The onset of novelties on the Internet and in social media~\cite{tria_dynamics_2014}.
\end{itemize}
These examples are motivated by, and analogous to, Herdan-Heaps' law discussed in Sec.~\ref{ssec.heaps}. Accordingly, they have been directly related to an associated  power-law frequency distribution (similar to Zipf's law) and a suitable sampling processes.  \index{power-law distribution}

\section{Inter-event times}\label{sec.interevent}

Temporal regularities in the occurrence of events are the source of several statistical laws, the simplest ones focusing on the times $\tau$ between successive events. The proposal of statistical laws to describe the inter-event time distribution $P(\tau)$ has a long tradition in the study of the distribution of words in texts~\cite{zipf_human_2012} and has been more recently proposed to describe the time between large \index{earthquakes} earthquakes~\cite{bak_unified_2002,christensen_unified_2002,corral_long-term_2004}, extreme events more generally~\cite{bunde_long-term_2005}, and  bursty human dynamics~\cite{karsai_bursty_2018}. Following the  pattern of other statistical laws, as discussed in Sec.~\ref{ssec.reasoning}, each specific law has motivated the proposal of mechanistic models to explain them. Before discussing in detail each of the cases, we introduce a common notation and discuss the general properties of inter-event times.

\begin{figure}
    \centering
    \includegraphics[width=1.0\columnwidth]{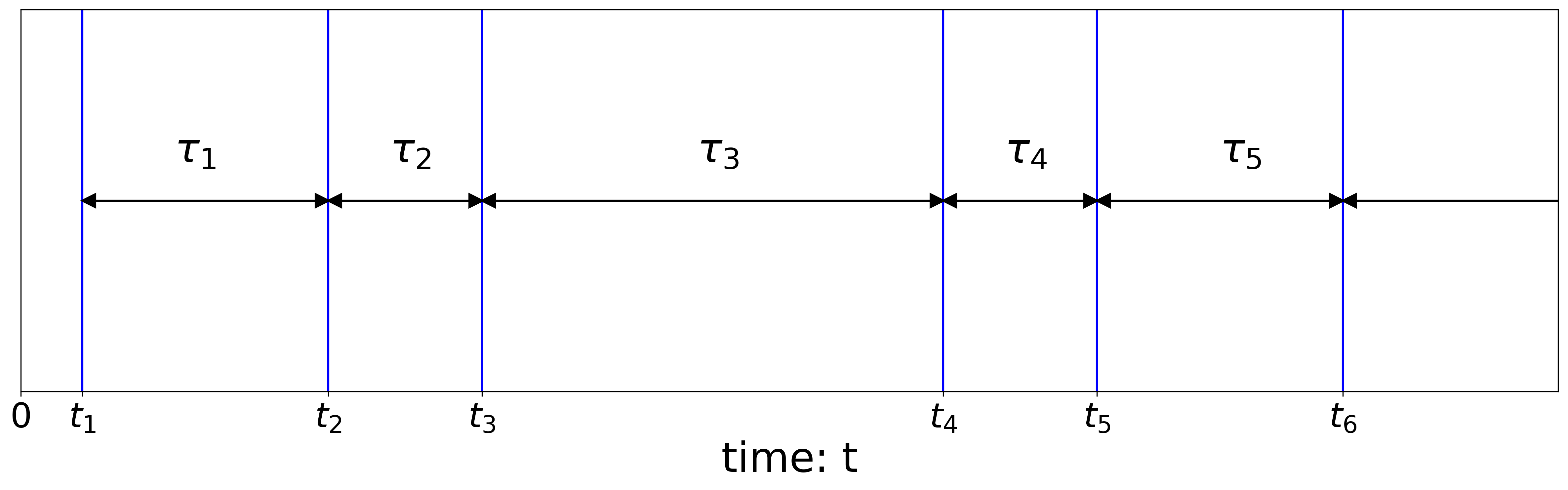}
    \caption{Sequence of events and inter-event times. The inter-event (or recurrrence) \index{recurrence} times~$\tau$ are  computed as defined in Eq.~(\ref{eq.interevent}).}
    \label{fig.interevent1}
\end{figure}

The inter-event time -- often denoted recurrence time or first return time -- $\tau$ is defined as the time between two {\it successive} occurrences of the event of interest.  More formally, consider that the sequence ${\bf t} = \{t_1, t_2, \ldots, t_{N}\}$  indicates the time of occurrence of $N$ events in a time series of length $T\ge T_N$ (total time of observation). The $i-th$ inter-event time (return interval) of the event is then defined as
\begin{equation}\label{eq.interevent}
    \tau_i \equiv t_{i+1}-t_{i}, \text{ for } i\in [1,N],
\end{equation}
where, for mathematical convenience, we define $t_{N+1} = T+t_1$ (periodic boundary conditions). Figure~\ref{fig.interevent1} illustrates the computation of inter-event times~$\tau_i$ form the event times $t_i$.

The time appearance of the events is completely represented by  the position they appear, given by the sequences ${\bf t}$ or, equivalently, by the return sequences ${\bf \tau}$ (and the first occurrence of the event). The study of inter-event times focuses on the statistical analysis of the sequences ${\bf \tau}$, the premise being that the statistics of simple properties of ${\bf \tau}$ provides universal or useful information about the dynamics leading to the appearance of the event. Examples of simple properties include the distribution (histogram) $P(\tau)$ of $\tau \in {\bf \tau}$ (i.e., ignoring the ordering) and moments of $P(\tau)$ such as the average $\langle \tau \rangle$ or standard deviation. 

When computing statistical properties of the sequences $\{\tau_i\}$ it is important to determine how they depend on the frequency of the event and how they compare to a null models (e.g., random appearance or Poisson process). \index{Poisson distribution} The average value of $\{\tau_i\}$ does not depend on the ordering of the sequence and it is simply given by the inverse of the (normalized) frequency $f=N/T$ of the event. This can be seen from this simple calculation
\begin{equation}\label{eq.kac}
    \langle \tau \rangle \equiv \dfrac{\sum_{i=1}^{N} \tau_i}{N}  = \dfrac{T}{N} = \dfrac{1}{f},
\end{equation}
where the periodic boundary conditions defined after Eq.~\ref{eq.interevent} is used in the second equality (sum of the return intervals equal to length, $T = \sum \tau$). This simple result can be seen also as a particular case of Kac's lemma~\cite{kac_probability_1959,altmann_recurrence_2004}.
More information about the temporal patterns of events is obtained  counting the number of times that each interval $\tau$ appears in $\{\tau_i\}$. Statistical laws typically focus on the distribution $P(\tau)$ of inter-event times (or recurrence-time distribution), which  describes the fraction of intervals in $\{\tau_i\}$ that are of type $\tau$.

The random expectation of $P(\tau)$ (e.g., obtained shuffling the sequence of observations or time series) can be computed considering a Poisson model in which a constant probability $\mu$ (with $\mu=f =1/\langle \tau \rangle$) of the event occurring at time $t$. Assuming, for simplicity, observations at discrete times $t$ we can compute the probability of an appearance (probability $\mu$) for the first time at time $\tau$ (i.e. after $\tau-1$ non-appearance with probability $1-\mu$) as 
\begin{equation}\label{eq.altmannexp}
    P(\tau) = \mu (1-\mu)^{\tau-1} \approx \mu e^{-\mu \tau} = \frac{e^{-\tau/\langle \tau \rangle}}{\langle \tau \rangle},
\end{equation}
where the approximation holds for $\mu=f=1/\langle \tau\rangle \ll 1$. The (complementary) cumulative distribution is given as
\begin{equation}\label{eq.altmann.cumulative}
    P(\tau^*>\tau) \equiv \sum_{\tau*=\tau}^\infty P(\tau^*) = \sum_{\tau*=\tau}^\infty \mu(1-\mu)^{\tau^*-1} \approx e^{-\tau/\langle\tau\rangle}.
\end{equation}
Distribution of inter-event times decaying more slowly than~(\ref{eq.altmann.cumulative}) are considered a signature of {\it burstiness} \index{burstiness} or a {\it bursty} dynamics~\cite{goh_burstiness_2008,karsai_bursty_2018}. Since
the average $\langle \tau \rangle$ is fixed by Eq.~(\ref{eq.kac}), such distributions -- in comparison to the Poisson assumption -- show not only larger than expected long $\tau$'s but also small $\tau$'s.

\begin{figure}[bt]
    \centering
    \includegraphics[width=0.48\columnwidth]{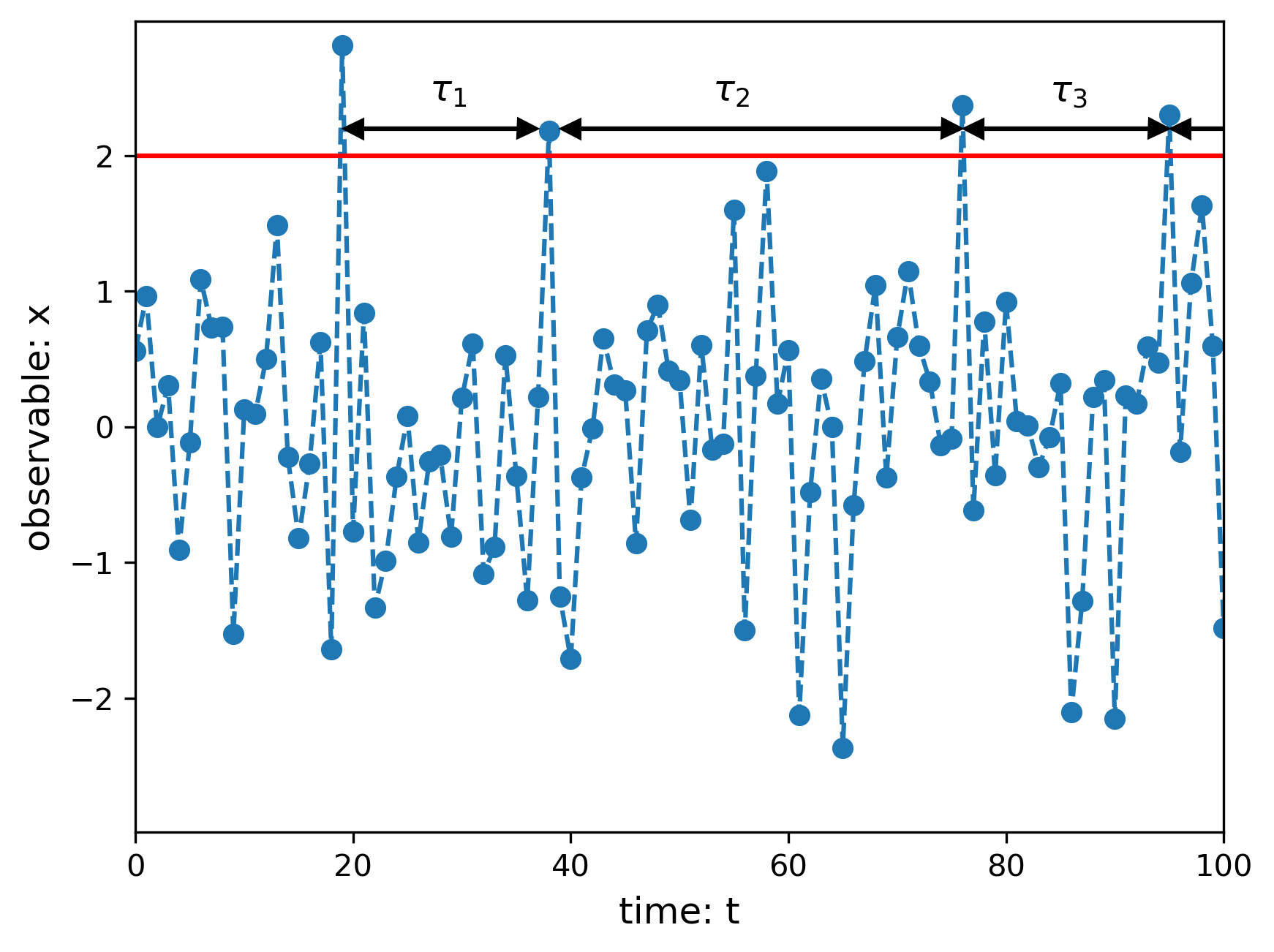}   \includegraphics[width=0.48\columnwidth]{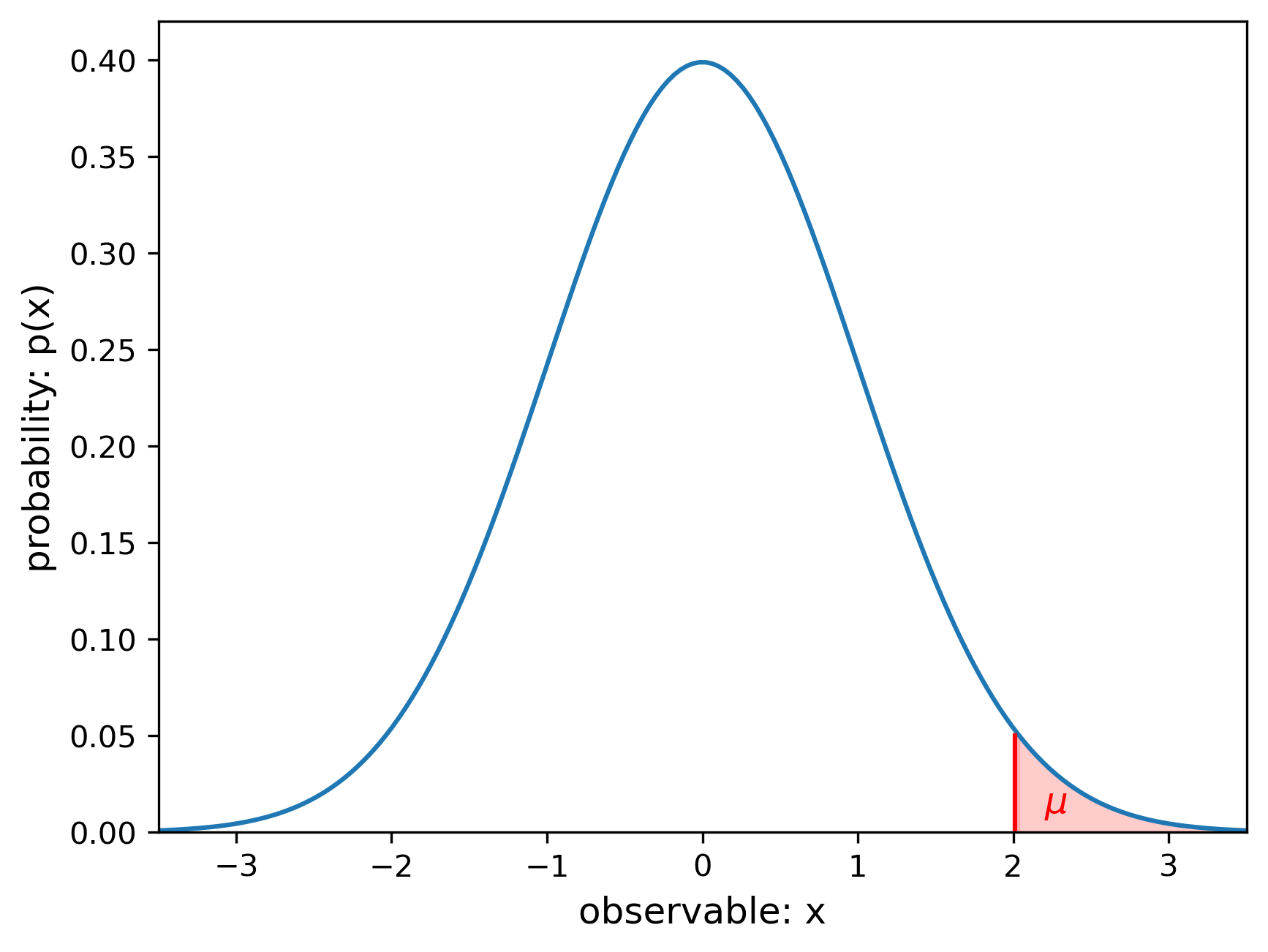}
    \caption{Sequence of inter-event times~$\tau$ between extreme events of a time series $x(t)$ . The inter-event (or recurrrence) times~$\tau$ are  computed as the time intervals between successive values of $x$ larger than a threshold $x^*$. In the figure, a sequence of $100$ Gaussian  \index{Gaussian distribution} distributed values $x$ is shown and the inter-event times are computed using $x^*=2$.}
    \label{fig.interevent2}
\end{figure}

So far we considered the time of occurrences of events ${\bf t} = \{t_1,t_2,\ldots,t_N\}$ to be known. In several studies of inter-event times, this happens only after choosing the definition of the events of interest. For instance, the event can be the extreme value of a time series (i.e., $x>x^*$ for an arbitrary threshold $x^*$) or the appearance of a specific word type in a text (in which case word tokens play the role of time). Figure~\ref{fig.interevent2} illustrates the procedure for a real-valued time series $x(t)$.  In the same dataset, one is typically interested in the inter-event times of different events, such as \index{earthquakes} earthquakes of different magnitudes or different word types. Eq.~(\ref{eq.kac}) connects the interevent times to the frequency of events through the probability of occurrence of events $\mu=f$ in Fig.~\ref{fig.interevent2} and in Eq.~(\ref{eq.kac}) and thus to the statistical laws that govern the distribution of frequencies~$p(f)$, such as the power-law distributions discussed in Sec.~\ref{sec:power-law-distr}. Statistical laws for inter-event are thus connected and complimentary to the statistical laws of the distribution of frequencies $p(f)$. As discussed in the examples below, inter-event times and frequency distribution laws are often proposed to describe the same system: Gutenberg-Richter law for earthquake magnitudes and statistical laws of inter-earthquakes times; and Zipf's law\index{Zipf's law} of word frequency and Weibull \index{Weibull distribution} law for inter-word intervals.  \index{power-law distribution}

\subsection{Words}\label{ssec.burstywords}

One of the first proposals of statistical laws in the inter-event time considers the distance between successive appearances of the same word $w$ in a text (also known as word returns).  For instance,  the text

\begin{quote}
{\it All human beings are born free and equal in dignity and rights. They are endowed with reason and conscience and should act towards one another in a spirit of brotherhood.}
\end{quote}
has length $T=30$ word tokens, the word $w="and"$ appears $4$ times in the locations ${\bf t}^{and}=\{7,11,18,20\}$ and therefore its inter-event times are ${\bf \tau}^{\text{and}} = \{4,7,2,17\}$ and its frequency is $f^{and}= 4/30$. Analogously, the word $w= in$ has $f^{in}=2/30, {\bf t}^{in}=\{9,26\},$ and ${\bf \tau}^{in}=\{17,13\}$. 

Different statistical laws for the distribution $p(\tau)$ were proposed to describe empirical observations. Looking at words listed in the index of books, Zipf suggested \cite{zipf_human_2012}
\begin{equation}\label{eq.altmann.zipf}
p(\tau) = a \tau ^{-\bar{\gamma}},
\end{equation}
where $a$ can be seen as a normalization constant and $\bar{\gamma}$ is the scaling exponent of interest\footnote{We use a bar notation on the exponents used to describe inter-event statistical laws (e.g., $\bar{\gamma}$ instead of $\gamma$). This is done to avoid confusion with the variables used in the two previous sections but to retain the original notation used when these laws were proposed. Proportionality constants and normalization factors are often denoted by the same variables (e.g., $a,b,$ and $c$) and should be interpreted in context.}.
More recent studies in Quantitative Linguistics~\cite{kohler_quantitative_2005} proposed a generalization of Zipf's proposal in form of a so-called Zipf-Alekseev distribution
\begin{equation}\label{eq.altmann.zipfalekseev}
    p(\tau) = a \tau^{-\bar{\gamma} +\bar{\gamma}' \ln(\tau)},
\end{equation}
which includes a faster decay for long $\tau$'s when compared to Eq.~(\ref{eq.altmann.zipf}).

Recent studies focused on longer texts, performed a more systematic studies of different words $w$, and suggested that $p(\tau)$ of all words can be better described by a Weibull \index{Weibull distribution} distribution~\cite{altmann_beyond_2009,corral_universal_2009,tanaka-ishii_long-range_2016}
\begin{equation}\label{eq.altmann.weibull}
    p(\tau) = a \bar{\beta} \tau^{\bar{\beta}-1} e^{-b \tau^{\bar{\beta}}}.
\end{equation}
Assuming the distribution~(\ref{eq.altmann.weibull}) to be valid for all $\tau$,  the parameters $a,b$ can be computed by imposing normalization $\sum_{\tau=1}^\infty P(\tau) = 1$ and its average through Eq.~(\ref{eq.kac}) as 
$a=b=\left(f^w\Gamma(\bar{\beta}+1)/\bar{\beta}\right)^{\bar{\beta}}$
(where $\Gamma$ is the Gamma function, see \cite{altmann_beyond_2009}). The distribution~\ref{eq.altmann.weibull} is then dependent only on $\bar{\beta}$ (and on the frequency $f^w$ of the word) and the cumulative distribution is given by
\begin{equation}\label{eq.altmann.weibull.cumulative}
    P(\tau^*>\tau) = e^{-a\tau^{\bar{\beta}}}.
\end{equation}
For $\bar{\beta}=1$, it recovers the random expectation computed in Eqs.~(\ref{eq.altmannexp})-(\ref{eq.altmann.cumulative}), while for $\bar{\beta} \rightarrow 0$ it approaches Zipf's proposal in Eq.~(\ref{eq.altmann.zipf}) (with $\bar{\gamma} \rightarrow 1$).

This example shows how a much simpler description (with two less parameters) is obtained under the assumption that the same statistical law describes the same $p(\tau)$ for all $\tau$. However, in practice this is often not satisfied because for short inter-event times $\tau$, syntactic rules will typically have a strong effect on the distribution of $\tau$'s (e.g., forbidding repeated words implies $P(\tau=1)=0)$). This short $\tau$ deviations leads to strong deviations from all proposed $P(\tau)$ -- which are monotonically decaying functions with a maximum at $\tau=1$ -- and can strongly impact the computation of the normalization factor and mean. In fact, statistical laws are often intended to describe the long $\tau$ behaviour of $P(\tau)$ (tail of the distribution, see Fig.~\ref{fig.representation-weibull} for an example). Therefore, often the additional parameters of the proposed statistical laws are independently fitted to the data (i.e., without imposing the normalization and mean as constraints).

\begin{figure}
 \begin{center}
    \includegraphics[width=0.75\columnwidth]{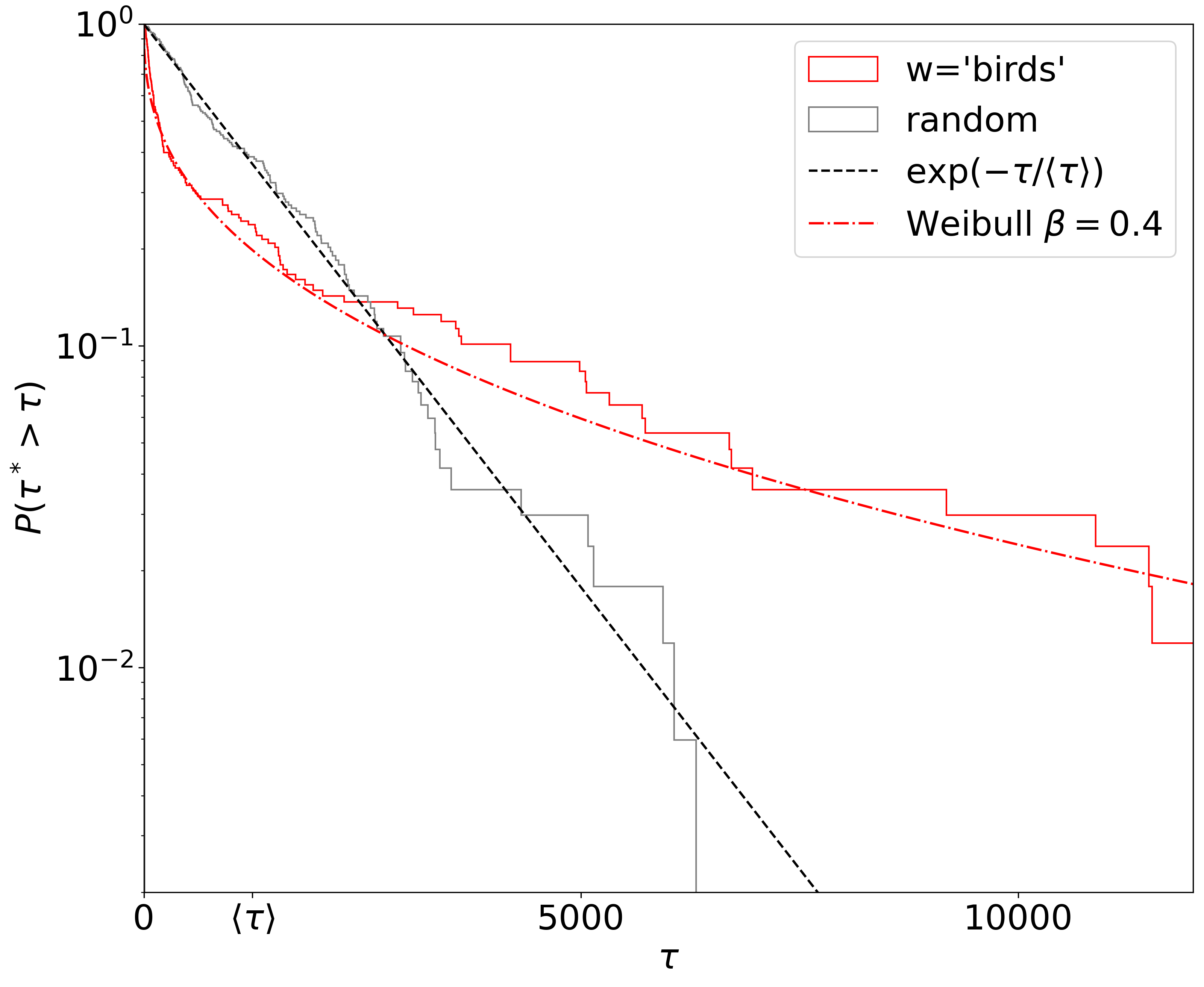}
        \includegraphics[width=0.75\columnwidth]{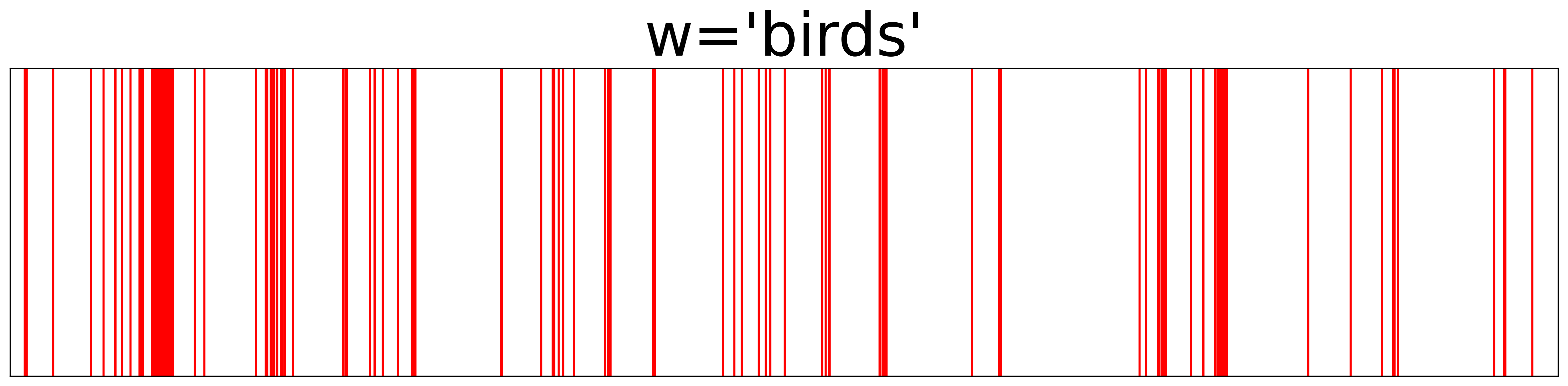}
         \includegraphics[width=0.75\columnwidth]{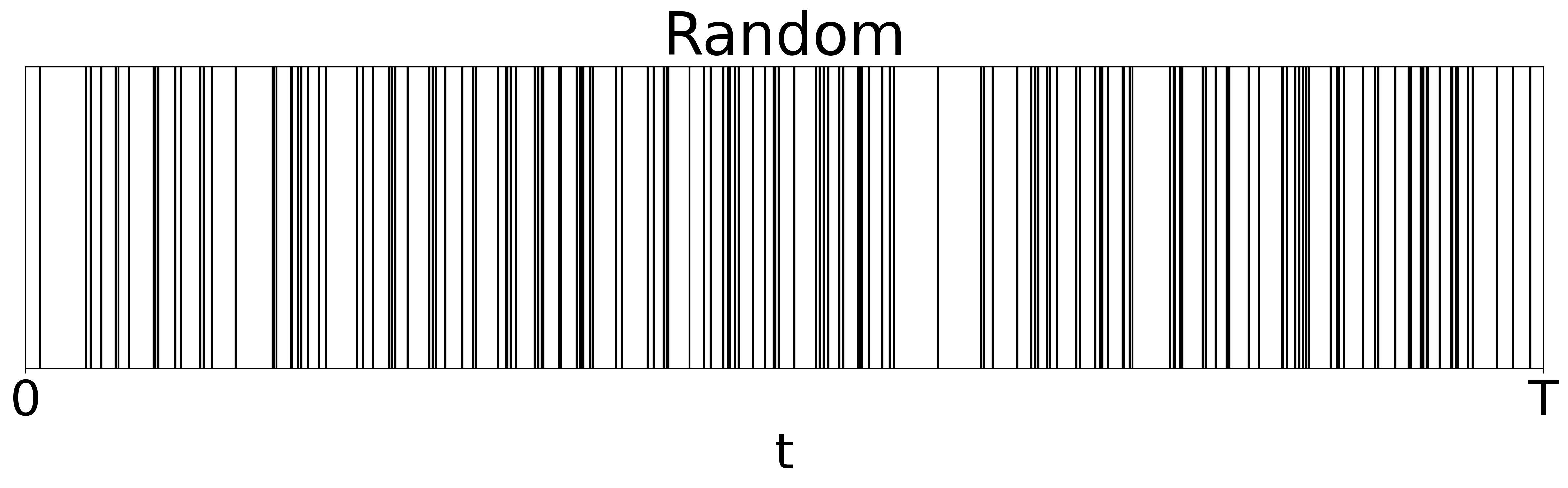}
    \caption{The bursty appearance of words in a text described by a stretched exponential distribution. \index{stretched exponential distribution} The inter-event times $\tau$ of the word $w=$'birds' (red) in 
    the book "The Voyage of the Beagle", by Charles Darwin, is compared to the random expectation (black) and to the predictions of the cumulative Weibull distribution~(\ref{eq.altmann.weibull.cumulative})  \index{stretched exponential distribution} with $\bar{\beta}=0.4$.}\label{fig.burstyWords}
 \end{center}
\end{figure}

\paragraph{Empirical evidence}

Figure~\ref{fig.burstyWords} illustrates the bursty appearance of words in text by comparing the location of a word in a long novel to the random expectation. The bursty (intermittent) distribution is clearly visible and quantified by the cumulative distribution $P(\tau'>\tau)$ of inter-event times. It deviates considerably from the random expectation in Eq.~(\ref{eq.altmann.cumulative}), with a more slowly (sub-exponential) tail. Comparison to the one-parameter Weibull \index{Weibull distribution}distribution~(\ref{eq.altmann.weibull.cumulative}) suggests that this simple one-parameter function accounts for the main deviations.

\paragraph{Mechanistic models}

The inter-event time distribution $p(\tau)$ can be obtained from stochastic processes proposed to model the appearance of words in texts, beyond the simple Poisson process used in Eq.~(\ref{eq.altmannexp}). A simple stochastic process that leads to the Weibull's law of word returns considers that the time-dependent probability $\mu(t)$ of the word appearing at location $t$ depends only on the time since last occurrence of the word. This corresponds to a renewal process and the Weibull \index{Weibull distribution} distribution~(\ref{eq.altmann.weibull}) is obtained with the choice of the (hazard) function~\cite{santhanam_return_2008,altmann_beyond_2009}
\begin{equation}\label{eq.renewal}
    \mu(t) = a \bar{\beta} t^{-(1-\bar{\beta)}},
\end{equation}
which corresponds to a power-law decay of the probability of use since last occurrence. 
The simple renewal model in Eq.~(\ref{eq.renewal}) does not explain the appearance \index{long-range correlations} long-range correlation in texts~\cite{altmann_beyond_2009,altmann_origin_2012}, showing also that the distributions of returns~$P(\tau)$ for each word contains only part of the information contained in the sequences of returns $\{\tau\}$ (which are themselves long-range correlated). Long-range correlations are known to exist in texts~\cite{schenkel_long_1993,altmann_origin_2012,tanaka-ishii_statistical_2021}, and both text characteristics -- long-range correlation and Weibull return distributions -- have been connected in Ref.~\cite{tanaka-ishii_long-range_2016}.  \index{power-law distribution}

\index{Weibull distribution}

\paragraph{Consequences}

The study of word returns connects to other quantitative studies of words in texts. The connection to  Zipf's law\index{Zipf's law} of word frequencies (see Sec.~\ref{ssec.zipf}) is established through Eq.~(\ref{eq.kac}). Beyond word frequencies, the model of inter-event times in Eq.~(\ref{eq.renewal}) was used in Ref.~\cite{lijffijt_significance_2016} to design significance tests for the distribution of the appearance of words in corpora.


\subsection{Earthquakes}\label{ssec.burstyearthquakes}

The distribution $P(\tau)$ of inter-event times $\tau$ between successive large \index{earthquakes} earthquakes is used to understand the variable probability of earthquakes over time~\cite{bak_unified_2002,christensen_unified_2002,corral_long-term_2004},  beyond the traditional distinction between main events and aftershocks (a central point of the 19th century \index{Omori's law} Omori's law of aftershocks, discussed in Sec.~\ref{ssec.omori} below).  The universality of $P(\tau)$ appears as part of the same research program which connected the \index{Gutenberg-Richter's law} Gutenberg-Richter law to critical phenomena, see Sec.~\ref{ssec.gutenberg} above.

Translating the complex spatial-temporal data of earthquakes to a simple inter-event distribution \index{threshold} $P(\tau)$ requires additional assumptions and data processing steps. Not only the threshold at a given magnitude is required to focus on the desired large events, as shown in Fig.~\ref{fig.threshold} above, one needs also to make choices about the spatial location of interest. The search for a generic description of inter-event times led to a focus on identifying suitable re-scalings of data that map different thresholds (magnitude and spatial size) to universal curves~\cite{bak_unified_2002,christensen_unified_2002,corral_local_2003}. The universality of statistical laws is then reported in form of the collapse of the data from different regions and magnitude thresholds after suitable re-scalings. These re-scalings allow for the unified treatment of data with a variable rate of large \index{earthquakes} earthquakes, connecting the average recurrence time $\langle \tau \rangle$ to Gutenberg-Richter law~(\ref{eq.gutenberg}) via Eq.~(\ref{eq.kac}).

The reports of universal curve collapse for different data have initially focused on the appearance of power-law scaling regimes and the thresholds between them~\cite{bak_unified_2002,corral_local_2003,davidsen_are_2004}.
A single explicit parametric function -- in the tradition of statistical laws reviewed here -- was proposed by Corral as~\cite{corral_long-term_2004}  \index{power-law distribution}
\begin{equation}\label{eq.corral}
    P(\tau) = C \frac{1}{\tau^{1-\bar{\gamma}}} e^{-\tau^{\tilde{\delta}/B}},
\end{equation}
where $B,C,\tilde{\bar{\gamma}},\tilde{\delta}$ are parameters. This stretched exponential distribution  \index{stretched exponential distribution}reduces to the Weibull \index{Weibull distribution} distribution~(\ref{eq.altmann.weibull}) by taking $\bar{\gamma}=2-\bar{\delta}$. It describes a decay that is slower than the Poisson \index{Poisson distribution} prediction~\ref{eq.altmannexp}  (for $\delta<1$) but faster than a power-law decay (which is observed for small $\tau$ or large $B$).

\paragraph{Empirical evidence}
The empirical evidence in support of Eq.~(\ref{eq.corral}) is provided in Ref.~\cite{corral_long-term_2004}, for different datasets, as an overlay of the curve collapse and a fit. Further empirical evidence of universal properties of $P(\tau)$ are described in Refs.~\cite{bak_unified_2002,davidsen_are_2004,de_arcangelis_statistical_2016}.

\paragraph{Mechanistic models}
An explanation and generalization of the waiting-time distribution between large earthquakes in Eq.~(\ref{eq.corral}) was provided in Ref.~\cite{saichev_universal_2006}. It builds on a theoretical model called epidemic-type after shock sequence, which incorporates the Gutenberg-Richter law (Sec.~\ref{ssec.gutenberg} above) and \index{Omori's law} Omori's law (Sec.~\ref{ssec.omori} below). This can thus be sen as a mechanistic explanation that connects different statistical laws in an unified framework. An alternative approach proposed in Ref.~\cite{davidsen_earthquake_2006} views large earthquakes as record-breaking events in a continuous spatio-temporal process. From the more general perspective of the complicated spatio-temporal dynamics of \index{Omori's law} earthquakes~\cite{kawamura_statistical_2012,de_arcangelis_statistical_2016}, the inter-event time distribution between large earthquakes is viewed as one emergent statistical regularity among many others, possibly emerging from the superposition of multiple processes (such as aftershocks and main events). Several models have included aftershocks in self-organized critical models, see Ref.~\cite{de_arcangelis_statistical_2016} for a review. \index{earthquakes}

\paragraph{Consequences}

The main interest in the study of statistical signatures in \index{earthquakes} earthquakes data is to make probabilistic forecasts about future events~\cite{kawamura_statistical_2012,de_arcangelis_statistical_2016}. The interest in the inter-event time distribution is that it can be connected to the expected time until the next earthquake~\cite{sornette_paradox_1997} (e.g., considering a renewal process as in Sec.~\ref{ssec.burstywords}).

\subsection{Extreme events}\label{ssec.bunde}

One of the main motivations to the study of the inter-event times $\tau$ between extreme events $x>x^*$ in time series $x(t)$, as defined in Fig.~\ref{fig.interevent2}, is the cluster of natural disasters in time (e.g., floods, draughts). As in the case of words -- Eq.~(\ref{eq.altmann.weibull}) -- and \index{earthquakes} earthquakes -- Eq.~(\ref{eq.corral}) --, the main statistical laws proposed to describe $P(\tau)$ between extreme events are in form of stretched exponential distributions ~\cite{bunde_effect_2003,bunde_long-term_2005}  \index{stretched exponential distribution}
\begin{equation}\label{eq.stretched}
    P(\tau) \sim  e^{(\tau/\langle \tau \rangle)^{\bar{\beta}}},
\end{equation}
with $0 < \bar{\beta} < 1$. In comparison to a Poissonian null-model~(\ref{eq.altmannexp}), the distribution~(\ref{eq.stretched}) with the same average $\langle \tau \rangle$ predicts a larger number of short $\tau \ll \langle \tau \rangle$ and long $\tau \gg \langle \tau \rangle$, i.e., a clustering of extreme events.

The Weibull distribution~(\ref{eq.altmann.weibull}) is a particular case of the stretched exponential distribution that has also been proposed to describe extreme events~\cite{santhanam_return_2008}. An alternative -- non-stretched exponential-- proposal is the gamma distribution~\cite{bunde_precipitation_2012} \index{gamma distribution}
\begin{equation} \label{eq.gamma}
    P(\tau) = C \tau^{\bar{\alpha}-1}e^{-\bar{\lambda} \tau}.
\end{equation}
All these distributions can be seen as a special case of the distribution~(\ref{eq.corral}) proposed to describe inter-event between \index{earthquakes} earthquakes. These different distributions have in common the fact that they describe a decay that is slower than exponential (at least for a large interval of intermediate $\tau$'s) but, asymptotically, decay faster than a simple power-law decay~$P(\tau) \sim \tau^{-\bar{\alpha}}$. While there is no unique distribution or precise definition of the data for which such distributions apply, there are multiple aspects of the study of the inter-event time distributions between extreme events that resemble the use of statistical laws proposed more generally (as defined in Sec.~\ref{ssec.definition}): its focus on simple parametric functions, the claim of universality in different observations, and the connection to theoretical aspects of the underlying dynamical system.   \index{power-law distribution}

\paragraph{Empirical evidence}

Stretched-exponential distributions were proposed to describe the inter-event distributions in different time-series long-range correlations~\cite{bunde_effect_2003,bunde_long-term_2005,santhanam_long-range_2005,altmann_recurrence_2005,eichner_statistics_2007}, i.e., with an auto-correlation function that decays as  \index{stretched exponential distribution}
\begin{equation}\label{eq.longrangecorrelation}
    C(\delta t) \equiv \langle x(t) x(t+\delta t) \rangle \sim (\delta t)^{-\bar{\alpha}}.
\end{equation} 
Bunde and co-workers~\cite{bunde_effect_2003} identified the decay in correlation $\bar{\alpha}$ with the exponent $\bar{\beta}$ in Eq.~(\ref{eq.stretched}) for extreme events, i.e., $\bar{\beta}=\bar{\alpha}$. For non-extreme events in the centre of the  distribution, $\bar{\beta}<\bar{\alpha}$ was proposed to hold in Ref.~\cite{altmann_recurrence_2005}.
These studies were often based on synthetic time series constructed to have the desired correlation properties, such as a Gaussian process  \index{Gaussian distribution} with a specified exponent $\bar{\alpha}$ in Eq.~(\ref{eq.longrangecorrelation}). The  analyses of empirical data that includes extremes in temperature~\cite{bunde_return_2004} and wind gusts~\cite{santhanam_long-range_2005} -- well described by~(\ref{eq.stretched})--, as well as precipitation and river flow rivers~\cite{bunde_precipitation_2012} -- better described by Eq.~(\ref{eq.gamma}).

\paragraph{Explanation}

The proposed explanation for the stretched exponential distribution of interevent time is the presence of long-range correlations~(\ref{eq.longrangecorrelation}) or 1/f noise in time series. The ubiquity of such characteristics in different time series has long been reported (see Refs.~\cite{montroll_1f_1982,bunde_return_2004} and references therein) and the claim of its widespread appearance shares characteristics with the use of statistical laws. The explanation for Eq.~(\ref{eq.stretched}) is thus not a mechanistic model for each of the observations, as common in other statistical laws, but instead a connection to other widely observed statistical features of data.   \index{stretched exponential distribution}

\paragraph{Consequences}

As in the case of \index{earthquakes} earthquakes, the main interest in the study of extreme events is to obtain probabilistic forecasts of natural disasters. The inter-event time distribution $P(\tau)$ can be directly connected to a hazard function under the assumption of independent sampling of $\tau$'s (renewal process, as discussed in Sec.~\ref{ssec.burstywords}). However, this asumption is often violated in empirical time series as they show correlations in the sequence of $\tau_i$'s. This violation is particularly important in the case of long-range correlated series $x(t)$~\cite{bunde_effect_2003,bunde_long-term_2005,altmann_origin_2012}.

\subsection{Burstiness of social activities}\label{ssec.burstysocial}

The recent availability of large records of human activities has motivated the quantitative study of the inter-event time~$\tau$ between successive individual activities (e.g., sending messages, accessing webpages)~\cite{barabasi_origin_2005,vazquez_modeling_2006,goh_burstiness_2008}. The proposal of universal distributions is in line with the statistical-law traditions. The main proposed functional form is a power-law distribution for the inter-event times \index{burstiness}  \index{power-law distribution}
\begin{equation}\label{eq.taupowerlaw}
    P(\tau) \sim \tau^{-\bar{\gamma}},
\end{equation}
with $\bar{\gamma} \approx 2$. This proposal was further generalized to consider that Eq.~(\ref{eq.taupowerlaw}) describes a wide range of $\tau$'s, but that for very large $\tau \gg \langle \tau \rangle$ an asymptotic cut-off in form of an exponential decay is observed. This corresponds to the Gamma distribution in Eq.~(\ref{eq.gamma}) with small $\bar{\lambda}$. More generally, cut-offs and truncations appear in other statistical laws and can have a strong influence on the data analysis~\cite{perline_strong_2005}.  \index{gamma distribution}

\paragraph{Empirical evidence}
Power-law distributions~(\ref{eq.taupowerlaw}) with and without cut-offs were used to describe the inter-event time distribution of a variety of human activities, both online (sending e-mails~\cite{barabasi_origin_2005,vazquez_modeling_2006}, visiting a web-portals~\cite{vazquez_modeling_2006}, being on a phone call~\cite{karsai_universal_2012}) and offline (e.g., sending letters~\cite{oliveira_darwin_2005}, library loans~\cite{vazquez_modeling_2006}).  \index{power-law distribution}

\paragraph{Mechanistic models}

The main mechanistic model proposed to explain the appearance of Eq.~(\ref{eq.taupowerlaw}) in human activities, introduced simultaneously to the claims of universal validity of this statistical law, considered a queuing system in which humans attribute different prioritizations to tasks~\cite{barabasi_origin_2005}. At each time step a task is performed and a new task is added to the queue as follows:
\begin{itemize}
    \item[(i)]~the priority of tasks are drawn from a uniform distribution;
    \item[(ii)]~with probability $p$ the highest-priority task is solved and with probability $1-p$ a random task is solved. 
\end{itemize}
The waiting time~$\tau$ for tasks in this model was shown in Ref.~\cite{vazquez_exact_2005} to follow the power-law distribution~(\ref{eq.taupowerlaw}) for $p \rightarrow 1$, a Poisson \index{Poisson distribution} distribution~(\ref{eq.altmannexp}) for $p=0$, and a power-law with exponential cut-off for intermediate $p$.  \index{power-law distribution}

An alternative explanation for the non-Poissonian behaviour was introduced in \index{Poisson process} Ref.~\cite{malmgren_poissonian_2008}. It considers a non-homogeneous Poisson process which incorporates periodic (circadian) patterns which are known to affect the Poissonian rate of event generation.  For instance, the probability to perform an activity (e.g., send an E-mail) depends directly on day-night and weekly cycles. It was argued that this simpler model can reproduce the observed waiting time distribution, which for a range of $\langle \tau \rangle$ resembles and can be-confused with a fat-tailed distribution.   \index{fat-tailed distribution}

While there are different stochastic processes used to capture the non-Poissonian behaviour of many human activities~\cite{karsai_bursty_2018}, the claims of universal validity of the power-law inter-event time distribution played an important role in the study of burstiness, following the same characteristics observed in the study of other statistical laws. \index{burstiness}  \index{power-law distribution}

\section{Other statistical laws}\label{sec.otherlaws}

Here we list statistical laws that do not directly fall in one of the three main classes used in the previous section (i.e., power-law frequency distributions, scaling laws, and inter-event times). We stick to the definition of statistical law proposed in Sec.~\ref{ssec.definition}, to maintain our focus on the cases of interest. It is not always clear whether a certain observation meets all the points of our definition and often there is room for debate whether some empirical observations should be treated as a statistical law in our sense.  For instance, observations of {\bf \index{long-range correlations} long-range correlations} and {\bf $1/f$ noise} is widespread~\cite{montroll_1f_1982,bunde_return_2004}, but only in some cases (e.g., in text analysis) it leads to the proposal of mechanistic models. The shape of the adoption of innovations over time as an S-curve \index{S-curves} can also be seen as a statistical law, and has been used to distinguish between mechanistic models in the case of the adoption of vocabulary~\cite{ghanbarnejad_extracting_2014}. The statistical laws discussed here share also similarities with the statistical laws observed in fluid dynamics, turbulence, weather, and climate~\cite{lovejoy_weather_2018}.   \index{power-law distribution}

A famous borderline case is the famous {\bf Benford's law}, \index{Benford's law} which in its simplest form states that the frequency of the {\it significant} digit $d=1,2, \ldots, 9$ of numbers that appear in texts or databases is given by 
\begin{equation}
    p(d) = \log_{10}(1+1/d).
\end{equation}
It is proposed to be applicable in different settings (tables, corpora, etc.) and large datasets, in line with the "universality" conditions (i) in the definition in Sec.~\ref{ssec.definition}. Still, it  ultimately describes only $9$ points, not the  "large number of data points" mentioned  in conditions (i). More importantly, its theoretical explanation is predominantly of a statistical-mathematical nature~\cite{hill_base-invariance_1995,hill_statistical_1995} (i.e., not directly connected to mechanistic models as specified in condition (iii) of Sec.~\ref{ssec.definition}).

\subsection{Earthquake aftershocks (Omori's law)}\label{ssec.omori}

The \index{Gutenberg-Richter's law} Gutenberg-Richter law discussed in Sec.~\ref{ssec.gutenberg} is only one of the many statistical laws proposed to describe empirical observation of \index{earthquakes} earthquake data~\cite{kawamura_statistical_2012,davidsen_generalized_2015,de_arcangelis_statistical_2016}. This tradition goes back to \index{Omori's law} Omori's law proposed in the late 19th century~\cite{omori_after-shocks_1895,guglielmi_omoris_2017}. It states that the frequency $n$ of aftershocks after a main shock decays as function of time $t\gtrapprox 0$ since the main shock as 

\begin{equation}
    n(t) = \frac{C}{(K+t)^p},
\end{equation}
with $C,K,p\approx 1$ constants.

Omori's law plays an important role in the debates on the existence of a universal inter-event time distribution between large \index{earthquakes} earthquakes, discussed in Sec.~\ref{ssec.burstyearthquakes}. More generally, the different statistical laws of earthquakes led to proposals of a unified description (statistical law)~\cite{bak_unified_2002,christensen_unified_2002,davidsen_generalized_2015} and proposals of mechanistic models~\cite{saichev_universal_2006} and experiments~\cite{lherminier_et_al_continuously_2019} to simultaneously explain them.

\subsection{Linguistic laws}\label{ssec.linguistic-laws}

\index{Herdan-Heaps' law}

Zipf's law\index{Zipf's law} of word frequencies -- Sec.~\ref{ssec.zipf} -- and Herdan-Heaps' law \index{Herdan-Heaps' law} vocabulary size -- Sec.~\ref{ssec.heaps} -- are just two of the most famous statistical laws in quantitative linguistics~\cite{herdan_quantitative_1964,kohler_quantitative_2005,altmann_statistical_2016,tanaka-ishii_statistical_2021}. In fact, The substantial knowledge on this subject in the filed of quantitative linguistics is useful and often overseen in the analysis of statistical laws more generally. \index{linguistic laws} \index{Herdan-Heaps' law}

Further examples of statistical laws in linguistics  include both laws specific to language and applications of existing laws to linguistic data:

\begin{itemize}
    \item The Menzerath-Altmann \index{Menzerath-Altmann law} law which provides a parametric function that describes Menzerath's principle that "the greater the whole, the smaller the parts"~\cite{altmann_gabriel_prolegomena_1980,kohler_quantitative_2005,tanaka-ishii_statistical_2021}.
    \item Information-theoretic analysis of texts show non-trival scalings of the information of words~\cite{zipf_human_2012, piantadosi_word_2011} and texts~\cite{ebeling_entropy_1994,debowski_hilbergs_2006} with their size. Mechanistic explanations for these findings typically involve an optimization process.
    \item Taylor's scaling law \index{Taylor's law} between  fluctuation and mean of signals was applied to the size of vocabularies~\cite{eisler_fluctuation_2008,gerlach_scaling_2014,tria_zipfs_2018,tanaka-ishii_taylors_2018}.
    \item Attempts to quantify and model the "S-curve" of language change~\cite{blythe_s-curves_2012,ghanbarnejad_extracting_2014,amato_dynamics_2018}
\item The observation of \index{long-range correlations} long-range correlations in texts~\cite{schenkel_long_1993,tanaka-ishii_long-range_2016}, with a mechanistic explanation related to the cascade of information over different scales~\cite{altmann_origin_2012}.
\end{itemize}

The parametric functions proposed in these laws are reported in Tab.~\ref{tab.linguisticlaws}. 
These laws have recently been investigated also for corpora of oral language~\cite{hernandez-fernandez_linguistic_2019}, acoustic signals~\cite{torre_emergence_2017}, and automated "machine-generated" texts~\cite{takahashi_evaluating_2019,lippi_natural_2019}.

\begin{table*}
  \centering
  \footnotesize
  \begin{tabular}{|l|c|c| c|}
\hline
    {Linguistic law}
&{Observables}
&{Functional form}
&{References}\\
\hline
{Zipf}
&{$f$: freq. of word $w$; $r$: rank of $w$ in $f$}
&{$ f(r) \sim r^{-\alpha}$}
&{\cite{zipf_human_2012,piantadosi_zipfs_2014}}\\
{Heaps}
&{$V:$ number of words; $N:$ database size}
&{$V\sim N^{\beta}$}
&{\cite{herdan_quantitative_1964,egghe_untangling_2007,baayen_word_2001}}\\
{Recurrence}
&{$\tau:$ distance between words }
&{$P(\tau)\sim \exp{(a \tau)}^{\bar{\beta}}$}
&{\cite{altmann_beyond_2009,corral_universal_2009}}\\
\hline
{Menzerath-Altmann}
&{$x:$ length of the whole; $y:$ size of the parts}
&{$y= \alpha_M x^{\beta_M} e^{-\gamma_M x}$}
&{\cite{altmann_gabriel_prolegomena_1980,cramer_parameters_2005}}\\
{Long-range correlation}
&{$C(\tau)$: autocorrelation at lag $\tau$} \index{autocorrelation function}
&{$C(\tau) \sim \tau^{-\lambda}$}
&{\cite{schenkel_long_1993,altmann_origin_2012,tanaka-ishii_statistical_2021}}\\
{Entropy Scaling}
&{$H:$ Entropy of text with blocks of size $n$}
&{$H \sim \alpha^\dagger n^{\beta^\dagger} + \gamma^\dagger n$}
&{\cite{ebeling_entropy_1994,debowski_hilbergs_2006}}\\
{Information content}
&{$I(l):$ Information of word with length $l$}
&{$I(l) = A + B l$}
&{\cite{zipf_human_2012,piantadosi_word_2011}}\\ 
{Taylor's law}  \index{Taylor's law}
&{$\sigma$: standard deviation around the mean $\mu$}
&{$\sigma \sim \mu^\delta$}
&{\cite{gerlach_scaling_2014,tria_zipfs_2018}}\\
{S-curves} \index{S-curves}
&{$\rho(t)$ frequency of linguistic variant}
&{$\rho(t) \sim(1-e^{at})^{-1}$}
&{\cite{blythe_s-curves_2012,ghanbarnejad_extracting_2014,amato_dynamics_2018}}\\
\hline
  \end{tabular}
  \caption{Parametric function of linguistic laws. The three examples above the line were reviewed in Secs.~\ref{ssec.zipf},~\ref{ssec.heaps}, and~\ref{ssec.burstywords}, respectively. Table adapted from Ref.~\cite{altmann_statistical_2016}. }\label{tab.linguisticlaws}
\end{table*}

\subsection{Gravitational laws in urban systems}

The proposal that the strength of the interaction between populations can be described using expression similar to Newton's gravitation law has a long and very active tradition. It goes back to the birth of socio-physics \index{social physics} and social sciences in the early 18th century~\cite{carrothers_historical_1956} and is still used, for instance, in studies of human mobility~\cite{barbosa_human_2018,schlapfer_universal_2021}. A simple formulation considers that the flow of population between two cities $i$ and $j$ is described by
\begin{equation}
    w_{ij} = c \frac{P_i P_j}{d_{i,j}^2},
\end{equation}
where $P_i$ ($P_j$) is the population of city $i$ ($j$), $d_{i,j}$ is the distance between the cities, and $c$ is a constant. Generalizations consider powers different from $2$ in the denominator, different types of distances $d_{i,j}$, and different functional forms for the effect of the populations. 

Gravity-type mobility models are used as null-models against which more sophisticated models are compared to, for instance, for migration patterns~\cite{prieto-curiel_diaspora_2024}. Gravity model have been considered also as part of explanations for \index{Urban scaling laws} urban scaling laws discussed in Sec.~\ref{ssec.urbanscaling}, as discussed in Ref.~\cite{altmann_spatial_2020,ribeiro_effects_2019}.

\chapter{From data to laws}\label{chap.data}

This chapter introduces and critically discusses the quantitative (statistical) methods used to study statistical laws. So far, we avoided details on the methods used to analyze data, assess the validity, and estimate parameters of statistical laws, focusing mostly on plots and remarks about specific cases. This was deliberately done in order to provide -- in this chapter --the methodological and statistical discussion in an unified and comparative way. This unified treatment is in line with the similarity of the use of different statistical laws across different disciplines, as summarized in Sec.~\ref{ssec.reasoning} and emphasized throughout the last chapter. This unified approach to statistical laws -- including interpretation and methodology -- has a mixed legacy: on the one hand, it builds on a tradition that dates back hundreds of years, that led to the creation of new knowledge and paradigms, and that continues to be a source of inspiration; on the other hand, it shows controversies and disputes that are not only widespread across disciplines but also persistent and difficult to be resolved over time. To better understand these controversies and limitations of different methods, in this Chapter we introduce the different quantitative and statistical approaches in increasing order of sophistication, which roughly correlates with their chronological introduction.

\paragraph{Controversies}

Six examples of controversies we encountered in the last chapter, all of them taking place in the 21st century, illustrate the difficulty in finding consensus on the assessment of statistical laws:  \index{power-law distribution}

\begin{itemize}
   \item Debates over \index{allometric laws} allometric scaling exponents and the validity of Kleiber's law from 1932 -- a work building on the area law from 1830s -- persists into the 21st century. Kleiber's $\beta=3/4$ exponent (and other quarter exponents in other allometric scaling laws) has been disputed, in favour of the geometrically-expected case $\beta=2/3$~\cite{dodds_re-examination_2001,da_silva_allometric_2006} (see also Sec.~\ref{ssec.allometry}).
    \item The validity of \index{Auerbach, Felix} Auerbach-Lotka-Zipf\index{Auerbach-Lotka-Zipf's law}'s law\index{Zipf's law} of city sizes has been questioned after decades of multiple studies on this law, with authors arguing for the more natural log-normal distribution~\cite{eeckhout_gibrats_2004,levy_gibrats_2009,eeckhout_gibrats_2009,malevergne_testing_2011} (see also Sec.~\ref{ssec.alz}).
       \item The ubiquity of \index{scale-free networks} scale-free networks (i.e., power-law degree distribution) was reported and celebrated in numerous papers in the first decade of the 21st century, to later be directly questioned~\cite{amaral_classes_2000,khanin_how_2006,stumpf_critical_2012,broido_scale-free_2019,klarreich_clara_scant_2018} (see also Sec.~\ref{ssec.scalefree} and Ref.~\cite{serafino_true_2021}).
    \item The significance and explanation for the origin of the Zipf's law\index{Zipf's law} of word frequencies remains open after a century of intensive work~\cite{piantadosi_zipfs_2014} (see also Sec.~\ref{ssec.zipf}).
    \item The ubiquity of \index{Urban scaling laws} urban scaling laws has been questioned~\cite{louf_scaling_2014,shalizi_scaling_2011,arcaute_constructing_2015} after a large number of observations and confirmations of the general proposal (see also Sec.~\ref{ssec.urbanscaling}).
    \item The observation of power-law distributed avalanches of neuron activities, and its connection to a mechanistic explanation based on critical phenomena, are the basis of the soc-called ``critical brain hypothesis''. \index{brain} The extent of the validity of this statistical law, and of evidence of a critical state, remains controversial~\cite{chialvo_emergent_2010,beggs_being_2012}.
\end{itemize}

As we will see in this Chapter, quantitative data-analysis methods play a crucial role in all these crises and disputes, with similar issues arising independently in communities working on different statistical laws. 
The lack of agreement on the validity, ubiquity, and significance -- even after decades of study --, indicates also that their solution is not simply a matter of obtaining larger datasets or using the "right" statistical method, but that it involves a connection between the application of different methods and the interpretations (or intended use) of statistical laws. One of the main goals of this monograph is to show that the persistence of such controversies is due to a mismatch between the interpretation of statistical laws and the quantitative methods used to study them, a point we will discuss below and come back in Chap.~\ref{chap.synthesis}.


\section{Graphical methods}

The visual comparison between points and curves is a powerful method to evaluate the agreement between data and the functional form proposed in a statistical law. Such graphical methods (visual-inspection techniques) have been the main source of evidence in support of statistical laws, as presented in our case studies in Chap.~\ref{chap.examples} and in most (if not all) historical works proposing new laws. For instance,  Persky's retrospective ~\cite{persky_retrospectives_1992} on \index{Pareto's law} Pareto's law in the late 20 century mentions: 

\begin{quote}
    {\it "Pareto used no quantitative measure of goodness of fit, visual inspection suggested that these linear equations worked quite well ... Pareto emphasized ... the fundamental difference from a normal curve"}.
\end{quote}  
 Pareto reports also estimations of the parameters~$A$ and $\tilde{\gamma}$ in~(\ref{eq.pareto}) for different datasets~\cite{pareto_cours_1897}.
Altogether, this shows how graphical methods were used to evaluate the validity of laws, to compare them to alternatives (model comparison), and to estimate parameters.

\subsection{Linear representations}\label{ssec.linearRepresentations}

Plots of data in logarithmic paper or scale can be attributed to the discovery of many of the statistical laws, including the work by Pareto~\cite{pareto_cours_1897}, \index{Auerbach, Felix} Auerbach~\cite{auerbach_felix_gesetz_1913}, and Kleiber~\cite{kleiber_body_1932}. \index{Kleiber's law}
Underlying this approach there is a choice of representation of the data and of the proposed law that highlights the regularity in the data, typically following a straight line.

In the case of power-law relationships   \index{power-law distribution}
-- including both power-law distributions 
$P(x) \sim x^{-\gamma}$ discussed in Sec.~\ref{sec.powerlaw}  and scaling laws $y \sim x^\beta$ discussed in Sec.~\ref{ssec.scaling} -- a linear relationship is achieved simply using a log-transformation of variables (or, equivalently, logarithmic paper or scale) as: 
\begin{equation}
    y = a x^\lambda \Rightarrow \log y = \log a + \lambda \log y \Rightarrow Y = A + \lambda X, 
\end{equation}
with $Y=\log y, X=\log x$, and  $A =\log a$. Another interesting property of power-law relationships $y = a x^\lambda$, such as those in Eqs.~(\ref{eq.powerlaw}) and (\ref{eq.scaling}), is that their functional form remains the same (i.e., apart from multiplicative constants) after re-scaling the independent variable: $x \mapsto b x \Rightarrow y = a'x^{\lambda}$, where $a'=a b^\lambda$. This means that all scales are equally appropriate or, alternatively, that there is no characteristic scale of the data.

The example of power-law relationships can be seen as an example of the more general approach of finding a transformation of variables that maps the data to a plot in which the functional form of the proposed law is a straight line. This has been used in the analysis of statistical laws with functional forms beyond a power-law, such as the stretched exponential distribution~\cite{bunde_effect_2003,altmann_beyond_2009}  \index{power-law distribution}  \index{stretched exponential distribution}
\begin{equation}\label{eq.transformationStreteched}
    y = y_0 \exp(\alpha x^{\beta}) \Rightarrow \log y/ y_0 = \alpha x^\beta \Rightarrow \log \log y/y_0 = \log \alpha +\beta \log x \Rightarrow Y = A + \beta X,
\end{equation}
with $Y=\log \log y/y_0, X= \log x,$ and $A=\log \alpha$.

\begin{figure}[!h]
\begin{center}
\includegraphics[width=0.48\textwidth]{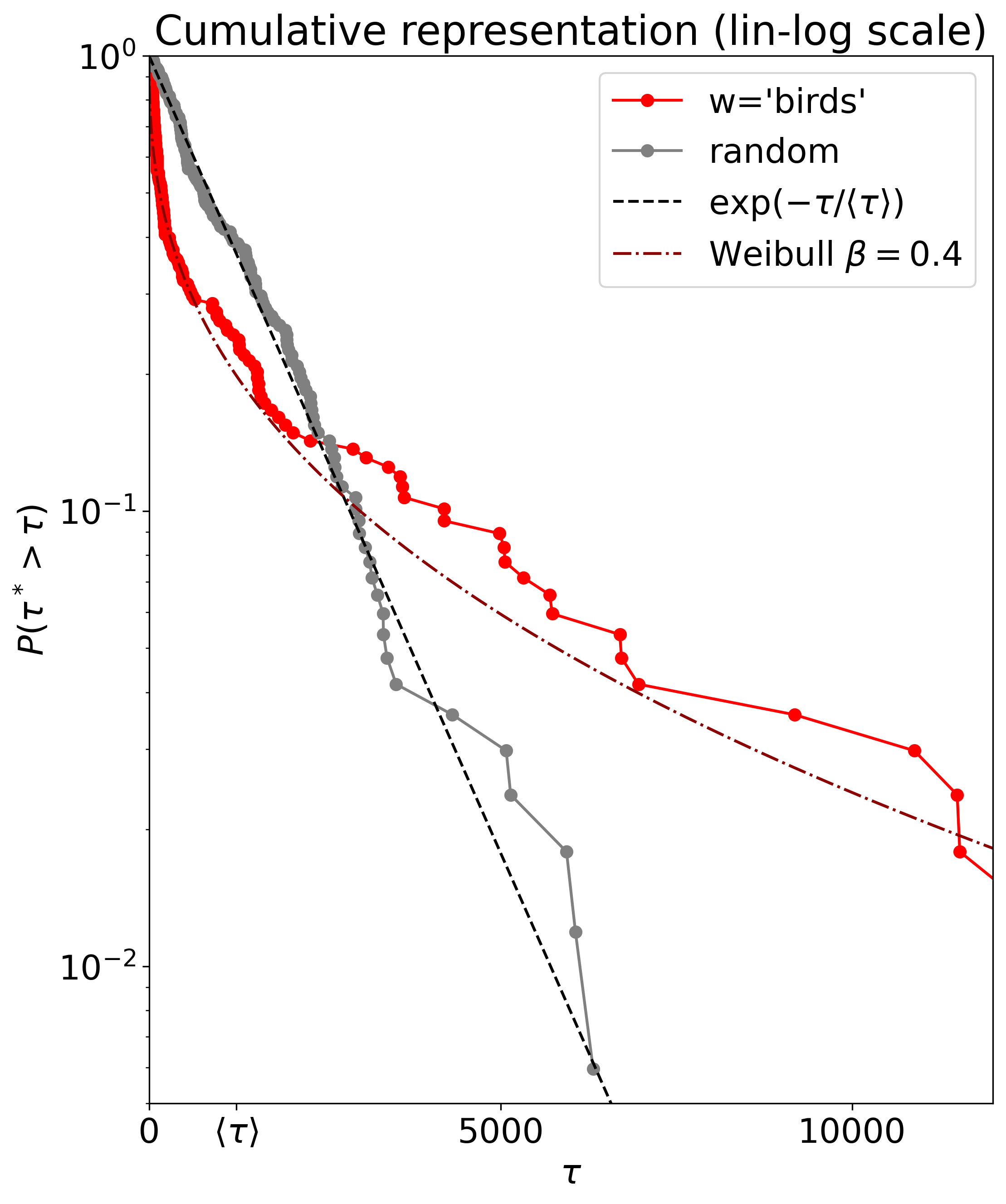}
\includegraphics[width=0.48\textwidth]{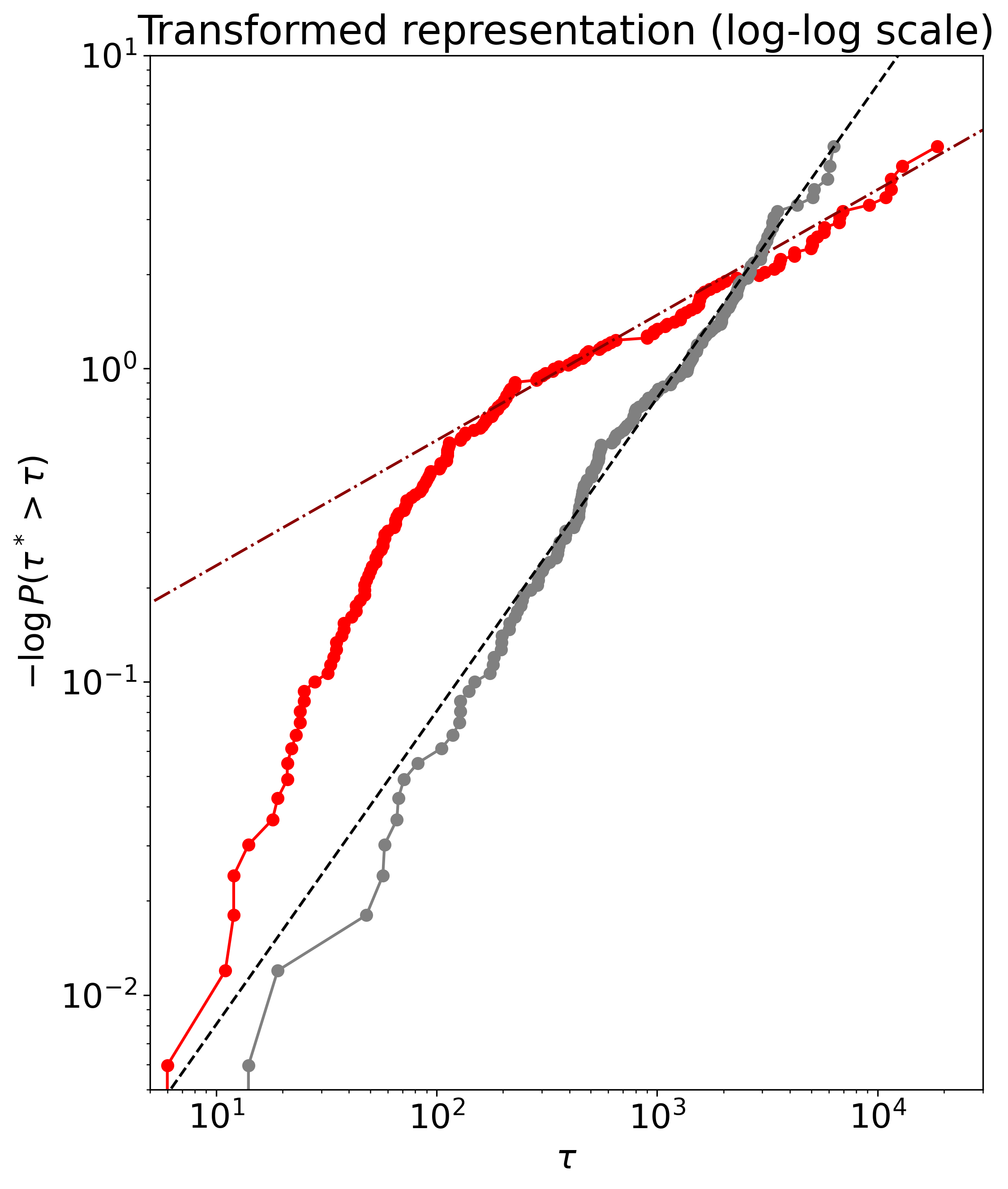}
\caption{Two different representations of the Weibull \index{Weibull distribution} distribution and its comparison to the inter-event times $\tau$ between words. The results for one word (``w=bird'') and its random expectations are shown together with the corresponding theoretical curves. The left plot corresponds to the representation depicted in Fig.~\ref{fig.burstyWords}, which contains further information about the data and proposed statistical law. The right plot corresponds to the same data and functions as the left plot, obtained after the application of the transformation~(\ref{eq.transformationStreteched}). }\label{fig.representation-weibull}
\end{center}
\end{figure}

In Fig.~\ref{fig.representation-weibull} we show a data-law comparison for the inter-event time $\tau$ of a word in a book. The two plots correspond are obtained before and after the application of the transformation~(\ref{eq.transformationStreteched}) that linearizes it. It is clear that the visual comparison between the data and the curves is strongly affected by these changes of variable: regions of small recurrence times $\tau$ (x-axis) are highlighted in the transformed representation (see Sec.~\ref{ssec.burstywords} for a discussion). The overall agreement suggested in the original representation manifests itself only in a range of large $\tau$ values, suggesting that the stretched exponential describes only the tail of the distribution. While the law itself is uniquely mapped through the transformation, the evaluation of its agreement between to the data is strongly affected by it. As we argue below, this is not only a property of graphical methods: it affects also other statistical methods used to compare the laws to data.  

\subsection{Rank frequency and frequency distribution}\label{ssec.rankrepresenation}

In the case of power-law distributions, there are two representations leading to a straight line in double-logarithmic plots: the rank frequency $F_r \sim r^{-\alpha}$ and the frequency distribution $p(x) \sim x^{-\gamma}$. These distributions were introduced in Eq.~(\ref{eq.powerlaw}) as functional forms representing a variety of statistical laws, as reviewed in Sec.~\ref{sec.powerlaw}. Analytically, the one-to-one connection between these representation can be seen as follows~\cite{adamic_zipf_2000,mitzenmacher_brief_2004,cristelli_there_2012}. Assuming the rank representation, the expected $x$ value of the $r-$th largest value scales as $\mathbb{E}(x_r) \sim  r^{-\alpha}$, i.e., we expect to find $r$ other entries with $x \ge C_1 r^{-\alpha}$ (for a constant $C_1$) and thus   \index{power-law distribution}
$$P(x \ge C_1 r^{-\alpha}) \sim r.$$
Changing variables to $y=C_1 r^{-\alpha} \Rightarrow r \sim  y^{-1/\alpha}$ we obtain
$$P(x\ge y) \sim y^{-1/\alpha} = y^{-\tilde{\gamma}}$$
which corresponds to the cumulative distribution. The probability distribution $p(y)=dP(y)/dy$ is thus
$$p(y) \sim y^{-(1+1/\alpha)} = y^{-\gamma},$$
with 
\begin{equation}\label{eq.relationexponents}
    \tilde{\gamma} = \gamma-1 = 1/\alpha,
\end{equation}
as enunciated in the beginning of Sec.~\ref{sec.powerlaw}.

\begin{figure}[!h]
\begin{center}
\includegraphics[width=0.4\textwidth]{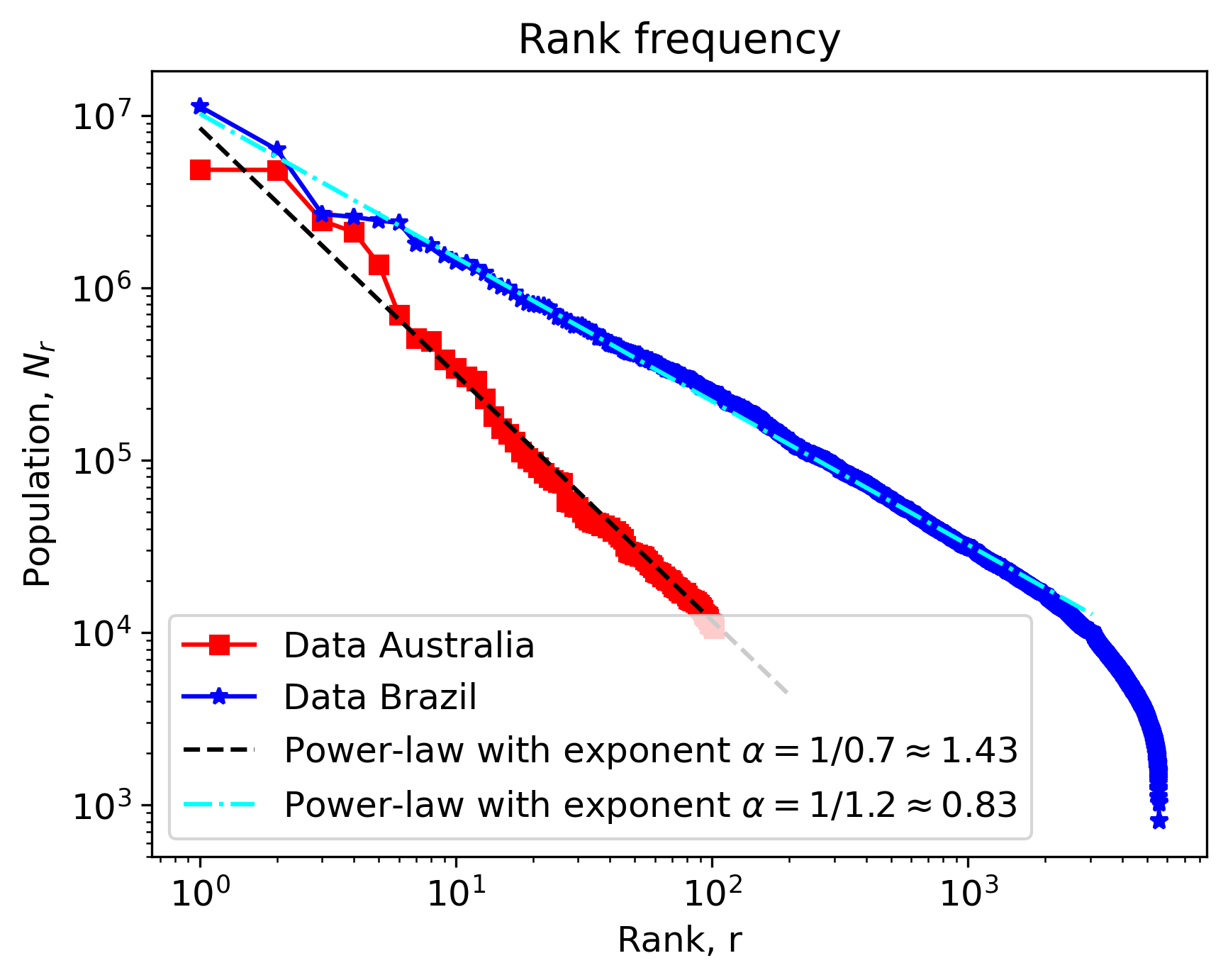}\\
\includegraphics[width=0.4\textwidth]{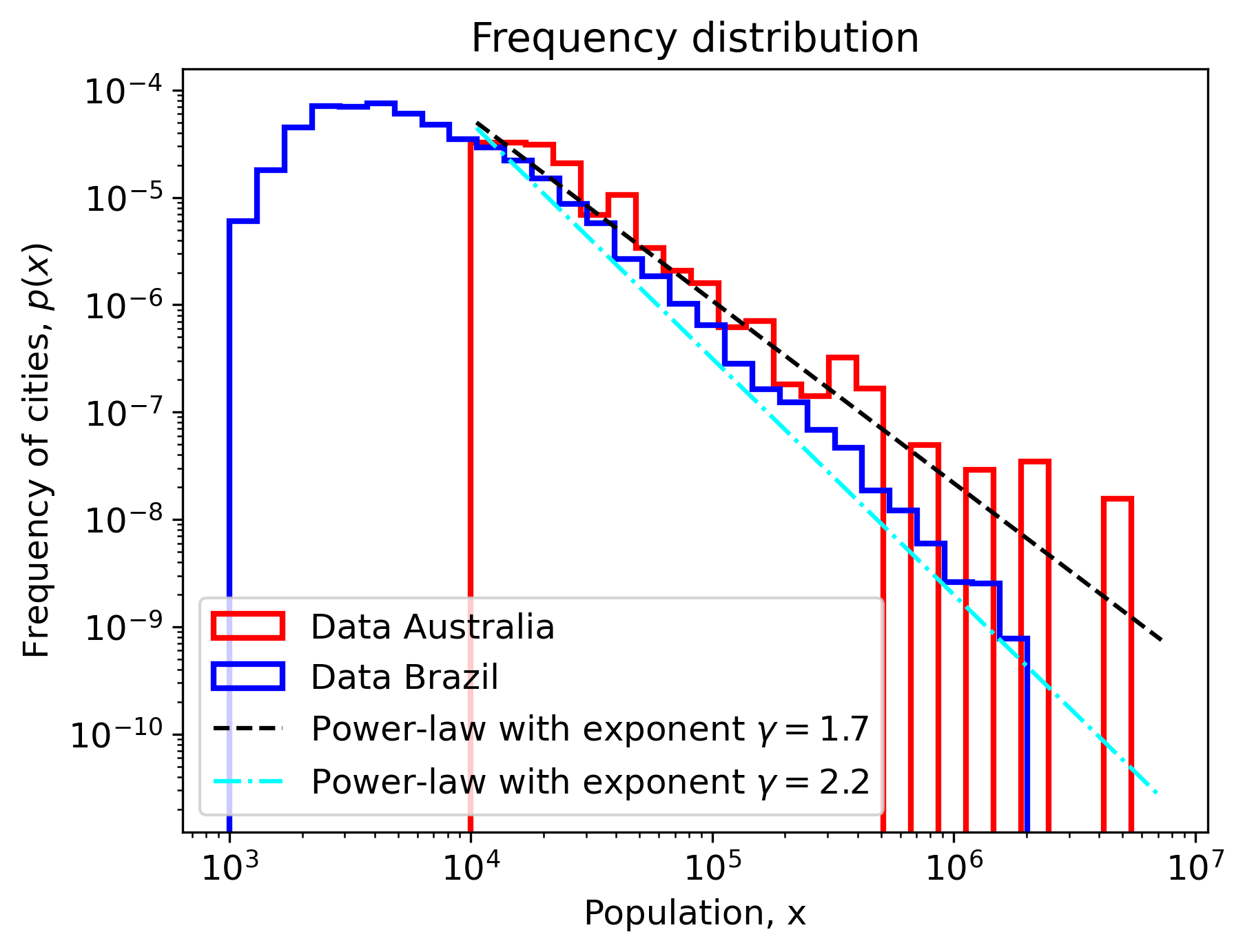}\\
\includegraphics[width=0.4\textwidth]{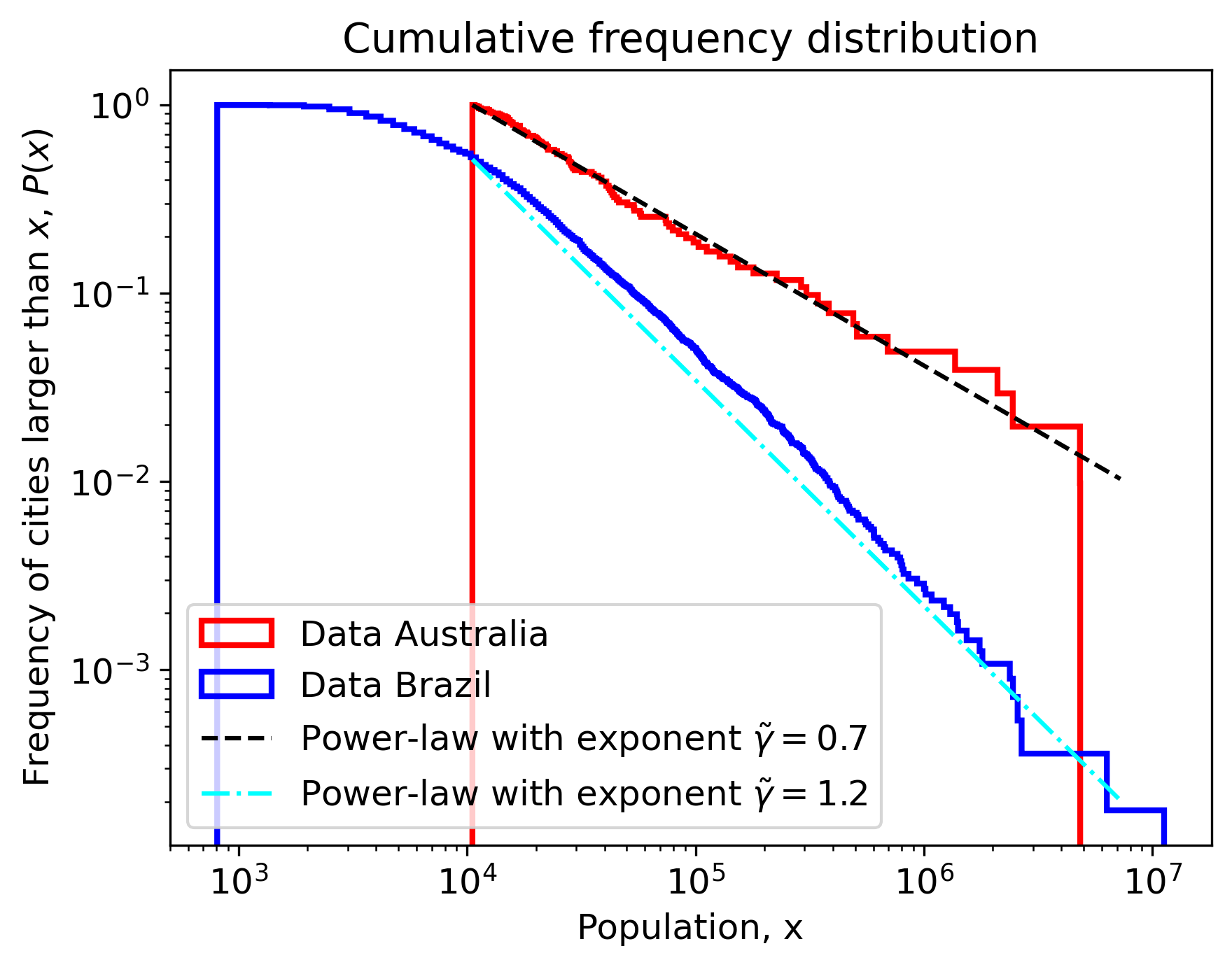}
\end{center}
\caption{
Different representations of scaling laws in city-size distributions. The different representations of Auerbach-Lotka-Zipf\index{Auerbach-Lotka-Zipf's law}'s law (see Sec.~\ref{ssec.alz}) in three different representations (see Secs.~\ref{sec.powerlaw}  and~\ref{ssec.rankrepresenation}): (Top) Rank frequency representation; (Middle)  Distribution (frequency) $p$ of cities  with population $x$. (Bottom) Cumulative distribution $P(x)$ of cities with population at least $x$. Cities from Australia and Brazil are shown, see legend and Fig.~\ref{fig:zipfcities}. The straight lines correspond to power-laws (as predicted by the Auerbach-Lotka-Zipf\index{Auerbach-Lotka-Zipf's law}'s law) with exponents --  $\gamma=1.7$ for Australia and $\gamma=2.2$ for Brazil -- chosen based on visual inspection. The straight lines in the different plots were obtained mapping the exponents according to Eq.~(\ref{eq.relationexponents}). In the case of Australia, all cities were used (natural threshold is $10^4$). In the case of Brazil, only cities with $x>10^4$ were used to compare the curve and data (i.e., in the choice of $\gamma$ and in the computation of the normalization in the last plot). See Appendix~\ref{chap.appendices} for information on the data and code used in this figure.}\label{fig.representation-rank}
\end{figure}

The calculation above shows the one-to-one relationship between power-law distributions in $p(x)$, its complementary cumulative version $P(x)$, and the power-law rank-frequency distribution $F_r$. This general relationship is illustrated in Fig.~\ref{fig.representation-rank} for data of city-size distributions (ALZ law discussed in Sec.~\ref{ssec.alz}). This point was clear already for Zipf, who used both representations and referred to the cumulative distribution $P(x)$ as the Paretian school. In the analysis of the degree distribution of networks, Ref.~\cite{huberman_growth_1999} states that their fat-tailed \index{fat-tailed distribution} distribution were not in the traditional Zipfian sense, but the authors later~\cite{adamic_zipf_2000} recognize the unity of representations. For distributions different from power law, a similar unique relationship between the rank frequency distribution and the (complementary) cumulative distribution exists, but in general their functional form changes and no simple relationship between parameters can be expected. The remarkably convenient aspect of power-law distributions is that they remain power-laws in all the three representations, with exponents related by Eq.~(\ref{eq.relationexponents}).  \index{power-law distribution}

The analytical equivalence between distributions does not imply that the representations are equivalent from a data-analysis perspective. In particular, deviations from the power-law distribution will manifest themselves very differently in each representation~\cite{gunther_zipf_1996,cristelli_there_2012}. This point is well illustrated in the two datasets shown in Fig.~\ref{fig.representation-rank}: the strong deviation from the ALZ\index{Auerbach-Lotka-Zipf's law} law -- discussed in Sec.~(\ref{ssec.alz}) -- due to the similar size of the two largest Australian cities is clearly seen in the top (rank-frequency) representation but less prominent in the other representation  (frequency distribution). Reversely, the deviation of Brazilian data from the ALZ law in the range of small cities is seen in the right-tail of the top (rank-frequency) representation and at the start of the other plots (frequency distribution). 

\subsection{Representation matters}\label{ssec.representationmatters}

Plots, representations, transformations of variables, and analytical manipulations of a statistical law change the extent into which the agreement between the function and the data is perceived. The choice between analytically-equivalent representations of statistical laws affects the conclusion drawn from the data analysis. This point is clear in the case of qualitative evaluations based on graphical methods, but it appears directly or indirectly in all quantitative analysis. The transformation of the functional forms of the laws can be seen as a change of variables or different choice of observable. These choices affect statistical evaluations, including the estimation of parameters and the agreement between the parametric functions and data. Analytical equivalence of representations does not imply statistical equivalence. 

An important caveat is that some representations tend to suggest more strongly the existence of regularities than others, often leading to a misleading sense of agreement between the data and the statistical law. For instance, (complementary) cumulative distributions and rank-frequency distributions are monotonic functions so that the fact that data follows such pattern is not an indication of any regularity but a feature of the representation. This point, combined with the fact that any continuous and smooth function can be locally approximated by a straight line, has often led to (over)interpretations of the agreement of data to power-laws. This has motivated the rule of thumb that the validity of power laws requires a linearity (in log-log scale) over several (or at least more than two) orders of magnitudes. 

The analysis of different representations of statistical laws is behind numerous controversies found in statistical laws and has been long recognized as such. This is evident from Persky's review\cite{persky_retrospectives_1992} of the debates around \index{Pareto's law} Pareto's law: it quotes Warren Persons 1909 as {\it ``an error in a logartihm gives a much larger error in the natural number''} and concludes that the accuracy of Pareto's law was {\it ``apparent and not real''}; it also cites Pigou 1920 as {\it ``Even if the statistical bases of the ``law'' were much securer than it is, the law would but rarely enable us to assert that any contemplated change {\it must} leave the form of the income distribution unaltered''... ``as things are, in view of the weakness of its statistical basis, it can never enable us to do this'.''}   Still, Persky concludes that {\it ``despite all the nitpicking, those double logarithmic curves still looked good''}, making Pareto's finding difficult to leave aside.  Quoting Norris Johnson, it states that {\it ``Pareto developed a fundamental yardstick. He found a useful simple description of the scheme of income distribution"}. 
Still, important questions remain: In which extent is the law valid? How come that it can be used in some cases (representations) but not in others? If the conclusions depend on the representations, can we trust consequences derived from the law? The situation is clearly not comfortable and difficult to interpret. We will return to this point in Chap.~\ref{chap.synthesis}, which includes also discussions on the consequences of the choice of representation to the formulation of (testable or falsifiable) statistical laws.

The lesson we learn is that the representation of the statistical laws matter: two representations that are equivalent from the functional form point of view are not equivalent from the statistical analysis point of view. The choice of representation is often associated to an implicit or explicit preference or focus on parts of the distribution. By focusing on the tails of the distribution of city sizes the focus is given to large cities while the tails of the rank-frequency distribution corresponds to small cities. A functional form that describes extremely well almost all cities may still be a very poor description for most of the population (if the exceptional cities are the largest ones). While so far we have illustrated this point based on visual inspection of the graph only, in the next sections of this chapter we will see how the effect of the representation of the statistical law strongly affects other quantitative methods used to study statistical laws.


\section{Regression}\label{sec:traditional-methods}

\subsection{Motivation}

The main motivations for the use of quantitative methods in the analysis of statistical laws is the qualitative nature of graphical methods, the difficulty of distinguishing between different distributions that are seemingly linear (in logarithmic scales)~\cite{perline_strong_2005}, the need to estimate the free parameters ${\bf \theta}$, and the desire to automatically test their validity and universality. Starting from graphical methods, the natural quantitative step is to mimic the visual inspection and consider the estimation of parameters based on the minimization of a suitably-defined distance between the data points and the parametric family of curves predicted by the statistical laws. For $i=1, \ldots, N$ data points $({\bf x}_i,{\bf y}_i)$, and an analytical expression of the predicted curve ${\bf y} = {\bf f}( {\bf x}|{\bf \theta})$, this distance can be written as
\begin{equation}\label{eq.norm}
    S = \sum_{i=1}^N ||{\bf y}_i-{\bf f}({\bf x}_i;{\bf \theta})||,
\end{equation} 
where $|| \ldots ||$ corresponds to the chosen norm. For instance, if ${\bf y} \in \mathbb{R}^d$ and ${\bf f}: \mathbb{R}^k \mapsto \mathbb{R}^d$, a popular choice is the $L^2$ norm $||{\bf y}|| = \sqrt{y_1^2+y_2^2+\ldots +y_d^2}$. The parameters ${\bf \theta}$ are then chosen as as the values ${\bf \hat{\theta}}$ which minimize $S=S({\bf \theta)}$. If the statistical law is formulated in form of a distribution (or a probability density function), the distances between the data and law can be computed also using a distance (or divergence) measure between the distribution and the histogram (or other estimator) based on the data (e.g., using an information-theoretic measure such as the Jensen-Shannon divergence).

\subsection{Linear regression}\label{ssec.linearregression}

As argued in Sec.~\ref{ssec.linearRepresentations}, graphical methods were typically employed in combination with a transformation of variables that resulted the proposed statistical law to be linear. This is not only convenient for visual inspection but also to the application of linear regression methods.  The ordinary least-squared fitting of a straight line in this representation provides thus a simple (closed-form) approach that has been early and widely used, e.g., alreday in the first half of the 20th century \index{Gutenberg-Richter's law} Gutenberg and Richter~\cite{gutenberg_earthquake_1942,gutenberg_frequency_1944} used linear regression to estimate the exponent of the law associated to their name.

Least-squared fitting considers the linearized representation of the statistical law and data, obtained after the suitable application of transformations as described in Sec.~\ref{ssec.linearRepresentations} (e.g., taking the logarithm of the observations or rank). Typically, the simplicity of statistical laws is such that there is only one independent variable and one dependent variable, so that after the suitable transformation the statistical law is given by $y = m x +c$, with parameters ${\bf \theta}=(m,c)$ and the transformed data points by $(x_i,y_i), \; i = 1, \ldots N$. 
The inferred parameters $\hat{m},\hat{c}$ are determined by 
\begin{equation}\label{eq.ols}
    (\hat{m},\hat{c}) = \arg\min(S(m,c)),
\end{equation}
with $S$ the sum of the squared difference between the points and the $(m,c)$ line 
$$S(m,c) = \sum_{i=1}^N (y_i-mx_i+c)^2,$$
in line with the choice of an $L^2$ norm in Eq.~(\ref{eq.norm}). The parameters of the statistical law in its original formulation are obtained from $(\hat{m},\hat{c})$, inverting the transformation of variables used to linearize the data and law. Contrary to other optimization procedures, that became feasible only after the recent expansion of computational power, the minimization in Eq.~(\ref{eq.ols}) has a  simple closed-form solution 
$$\hat{m}= \dfrac{\sum_{i=1}^N(x_i-\langle x \rangle)(y_i-\langle y\rangle)}{\sum_{i=1}^N (x_i-\langle x\rangle)^2},$$
$$\hat{c} = \langle y \rangle - \hat{m} \langle x \rangle,$$
where $\langle \ldots \rangle \equiv \frac{1}{N} \sum_{i=1}^N \ldots $ denotes the average. 

The coefficient of determination 
\begin{equation}\label{eq.R2}
R^2 = 1 - \dfrac{\sum_{i=1}^N(y_i-mx_i+c)^2}{\sum_{i=1}^N (y_i-\langle y \rangle)^2},
\end{equation}
is such that $R^2=1$ is obtained for a perfect linear alignment of $(x_i,y_i)$ and $R^2=0$ for uncorrelated $(x_i,y_i)$. This has motivated the use of $R^2$ as a "goodness-of-fit" measure, often viewed not only as a quantification of the extent into which the points are close to the fitted line but also as the agreement between the statistical law and the data. 

The linear regression approach to analyze statistical laws can be summarized as follows:

\begin{itemize}
\item[1.] Data transformations are performed so that the statistical law appears as a straight line, as discussed in Sec.~\ref{ssec.linearRepresentations}. For instance, for scaling laws in Eq.~(\ref{eq.scaling}), log-transformed variables $\ln y, \ln x$  are used. 
\item[2.] The parameters of the statistical law $\hat{\theta}$ are estimated based on the least-squared regression in Eq.~(\ref{eq.ols}). For instance, for scaling laws as in Eq.~(\ref{eq.scaling}), $\alpha,\beta$ are chosen such that $\sum_{i=1}^N (\ln \alpha x_i^\beta - \ln y_i)^2$ is minimized.
\item[3.] The quality of the fitting is quantified by the coefficient of determination $R^2$ in Eq.~(\ref{eq.R2}). $R^2$ close to $1$ is taken as evidence of the agreement between the fit and the data.
\item[3.] The 95\% confidence intervals $[\theta_{\text{min}}, \theta_{\text{max}}]$ on parameters are computed from the uncertainty of the linear fit (sum of residuals). Values inside (outside) the confidence interval are taken as evidence that the parameters of the law agree (disagree) with the possible value. For instance, for scaling laws as in Eq.~(\ref{eq.scaling}),  $1 \not \in [\beta_{\text{min}},\beta_{\text{max}}]$ is taken as an evidence that $\beta \neq 1$ (non-linear scaling law).
\end{itemize}
Examples of the use of this approach can be found in Refs.~\cite{bettencourt_growth_2007,bettencourt_urban_2010,um_scaling_2009,arbesman_scaling_2011,bettencourt_origins_2013,louf_scaling_2014,nomaler_scaling_2014} (\index{Urban scaling laws}urban scaling laws) and Ref.~\cite{savage_predominance_2004} (Kleiber's law and allometric scalings).

\subsection{Caveats and limitations of linear regression}\label{ssec.caveatsLinear}

The line obtained by the least squared regression passes as close as possible -- in the sense of an $L^2$ norm -- to the points in the transformed space and is thus often pleasing when evaluating the agreement through visual inspection. While statistical justifications for this approach are important in the discussion of statistical laws, and will be discussed in further detail in Sec.~\ref{sec.likelihood}, the use of this methodology in the study of statistical law is more intimately associated to the graphical methods and linear-transformation traditions underlying many of their discoveries. Still, there are two elements that can contribute to a discrepancy between the statistical law with parameters estimated from linear regression and the assessment of linearity performed looking at the graphical representation of the data and curve:

\begin{itemize}

\item {\bf Representation.} As discussed in Sec.~\ref{ssec.rankrepresenation}, statistical laws can be formulated in different representations and more than one representation may yield a linear relationship. As in the case of graphical methods, the representation chosen to apply the least square fitting matters. In particular, the transformation of variables that yields the statistical laws linear are typically non-linear (e.g., log-transformation) and therefore the estimation and minimization of the distance between data and point is {\it not} invariant under the transformation (representation).    \index{power-law distribution}

    \item {\bf Distribution of points.} Often the data points are not uniformly distributed in the $x$ or $\log x$ scale used in the plots. For instance, there are many more cities with small population $x$ in the ALZ analysis (Figs.~\ref{fig.urbanscaling} and~\ref{fig.representation-rank}) and many more words with low-frequency and thus high ranks $r$ in rank-frequency plots (Figs.~\ref{fig.zipfdata},~\ref{fig:zipf}, and \ref{fig.zipfcomparison}). Similarly, in scaling laws based on counting \index{Herdan-Heaps' law} (such as Herdan-Heaps' law in Fig.~\ref{fig.heaps}), the data points appear for all integers so that there are many more points at large portions of the x-axis; and in \index{allometric laws} allometric scaling laws (Fig.~\ref{fig.allometric2}) there are often more species concentrated (or sampled) around some intermediate masses.  Least squared fitting aims to reduce the sum of the distances over all points and therefore the estimated parameters will be mostly influenced by the regions (in $x$ or $r$) with higher density of points and not uniformly in the (logarithmic) scale of the plot (as often expected from visual inspection, in particular when a wide range of $r$ and $x$ values exist).
To address this problem, Ref.~\cite{savage_predominance_2004} introduced a modified binning procedure of log-transformed variable to analyze scaling laws (Kleiber's law) to give equal weight to all sizes intervals. This happens because there are many more data points on rodents (small mass) than on large mammals.  While this point is more clear in examples in which an exhaustive selection is either not possible or not obvious (such as the \index{allometric laws} allometric cases discussed in Sec.~\ref{ssec.allometry}), the consequences are effectively the same when the points are unevenly distributed (such as the \index{Urban scaling laws} urban scaling laws discussed in Sec.~\ref{ssec.urbanscaling}, in which the majority of cities are small). Log-distributed binning is also a pragmatic option to deal with this uneven distribution of points, affecting the estimation of parameters in the linear fit.

\end{itemize}

A direct consequence of the two points above is that the choice of thresholds and cut-offs often have a strong effect on the outcome of the analysis~\cite{perline_strong_2005,font-clos_perils_2015}. This is exemplified by the case of urban data, where a threshold in population $x_{min}$ determines which urban regions (those with $x>x_{min}$) are counted as cities, an explicit or implicit choice behind any urban data. As there are many more small cities than large cities, the fits will be optimized to pass close to the points immediately next to the chosen threshold $(x\gtrapprox x_{min})$. If small cities show a different behaviour than large cities, there will be a strong dependence on the choice of threshold $x_{min}$. This happens despite the fact that these large number of cities describe a relatively small fraction of the total population so that the estimation of the exponents is not dominated by where most people live. Consider the case of Brazil, whose data has a large number of municipalities ($5,565$) and a clear deviation of scaling (ALZ\index{Auerbach-Lotka-Zipf's law} law) for small cities, as shown in Fig.~\ref{fig.representation-rank}. Only 8\% of the population lives in the smallest half of all municipalities ($2,782$ cases), while half of the country's population lives in the largest $202$ cities. While the logarithmic scale distributes the data through their different scales (and guides visual inspections of graphical methods), the estimation based on regression will be dominated by smallest cities.

\begin{figure}[!h]
\begin{center}
\includegraphics[width=0.8\textwidth]{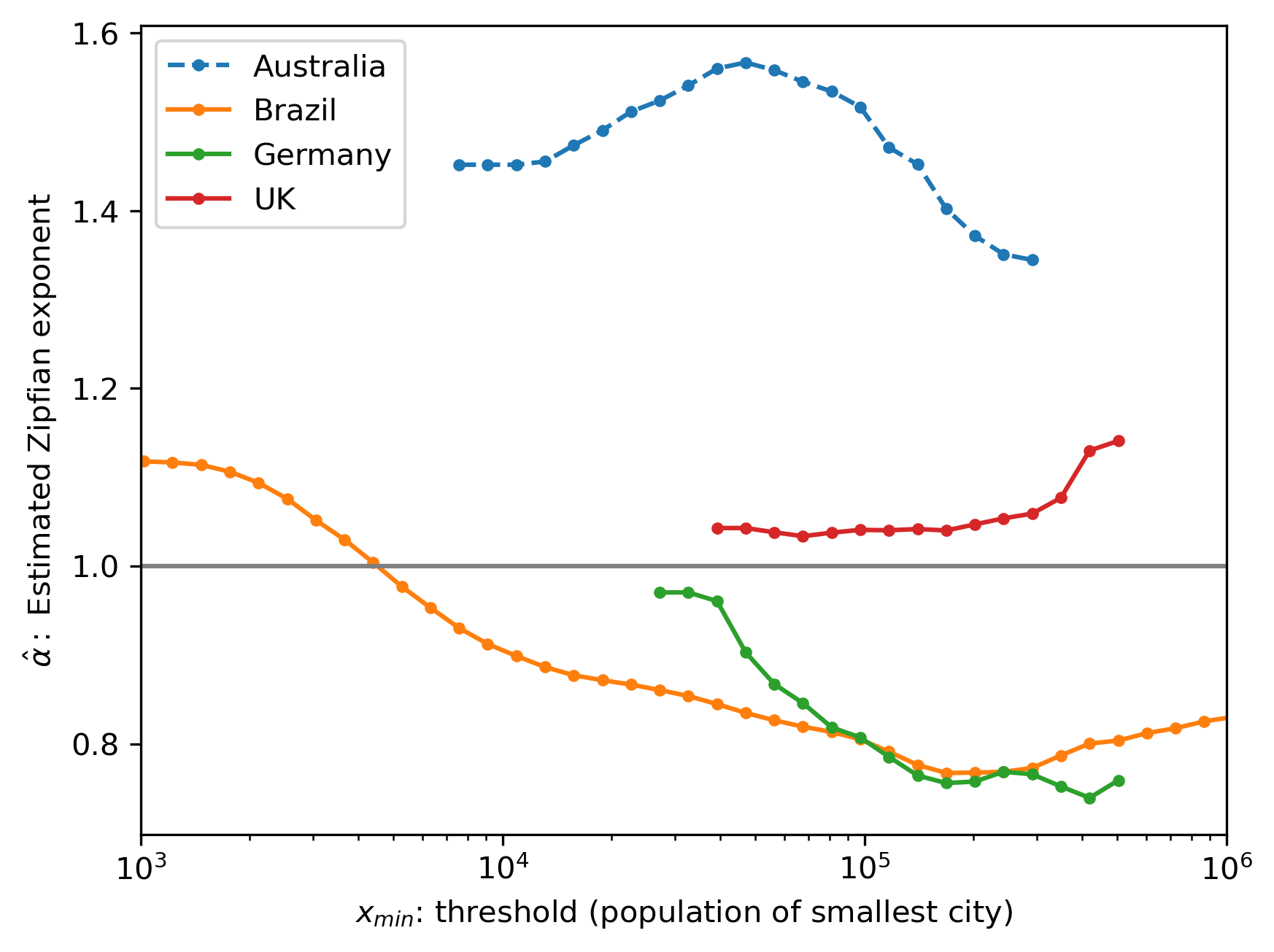}\\\includegraphics[width=0.8\textwidth]{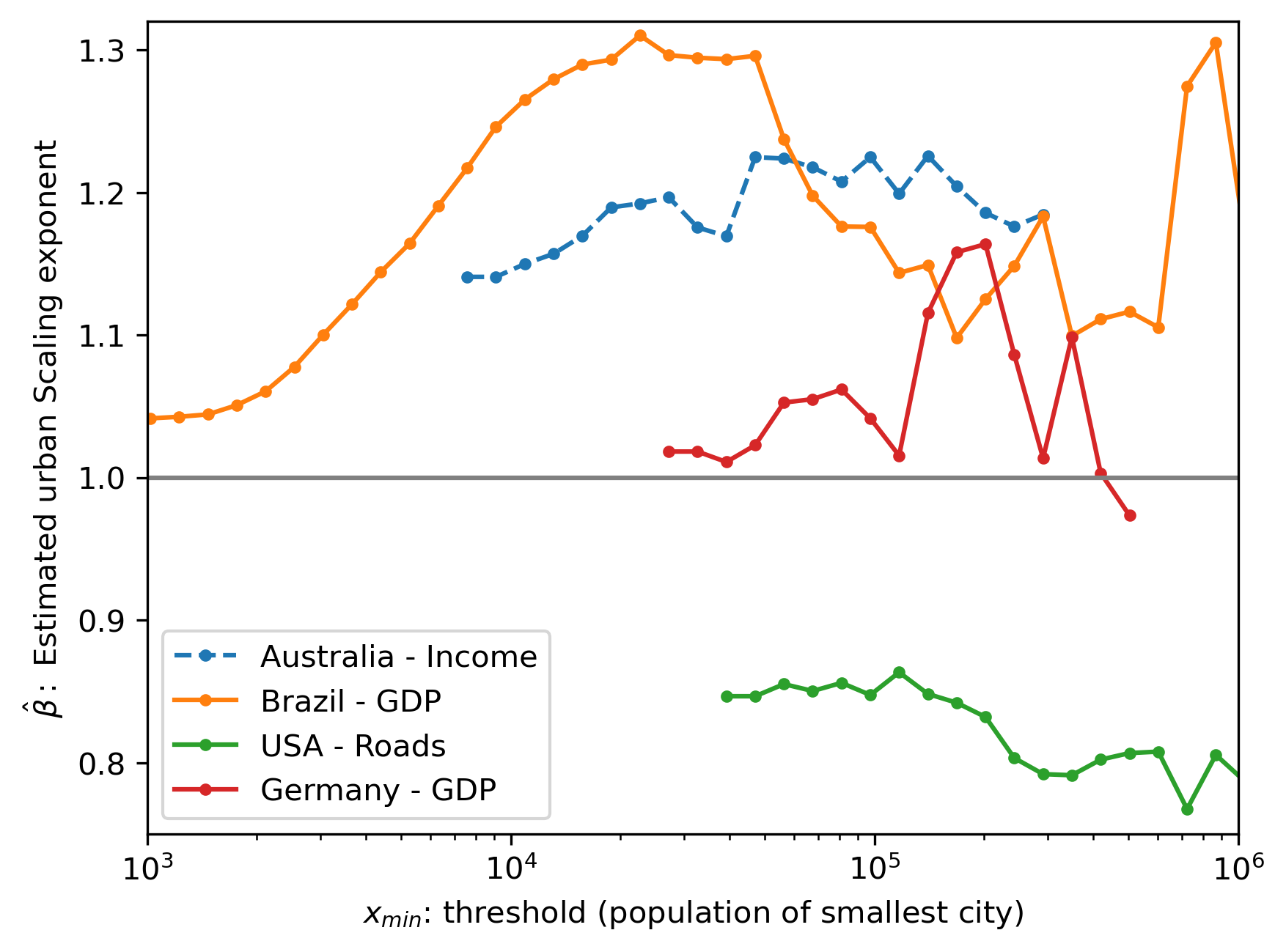}
\end{center}
\caption{Effect of thresholding (x-axis) \index{threshold} on the estimation of exponents (y-axis). Top: Zipfian exponent $\hat{\alpha}$ in~(\ref{eq.zipfcities}), estimated fitting the rank-frequency plots (data as in Fig.~\ref{fig.zipfcities}). Bottom: Urban scaling exponent $\beta$ in Eq.~(\ref{eq.urbanscaling}) (data as in Fig.~\ref{fig.urbanscaling}). In both cases the estimation was obtained using a linear regression (least-squared-fitting) of log-transformed variables using all cities with population $x>x_{min}$. The curves start at the threshold \index{threshold} in which all cities are used and end when less than $10$ cities were available. See Appendix~\ref{chap.appendices} for information on the code used in this figure.}\label{fig.threshold}
\end{figure}

In Tab.~\ref{tab.zipfexponents} and Fig.~\ref{fig.threshold} we show the practical effect of the general points mentioned above for the estimation of scaling exponents in urban data (using least-squared regression). Tab.~\ref{tab.zipfexponents} reports a variation of the estimation of the Zipfian exponent $\alpha$ in the ALZ law depending not only on the estimation methods (e.g., linear regression vs. maximum likelihood) but also on the representation of the statistical law. Figure~\ref{fig.threshold} focuses on the effect of the cut-off on the parameter estimation, not only in the estimation of the Zipfian exponent $\alpha$ but also on the exponent of the \index{Urban scaling laws} urban scaling law~$\beta$. It happens also on the data of Australian cities, which visually does not have strong deviation on small cities. The observed variations of the exponent $\approx 0.2$ are much larger than the standard error of the linear regression and the goodness of fit~$R^2$ is typically very high.

\begin{table}[h]
  \centering
  \small
\begin{tabular}{|c|c | c| c |}
\hline
 & Australia &  Brazil  & UK \\ \hline
 \multicolumn{4}{|c|}{ Data information }   \\ \hline
 Number of cities: $N$ & $102$ & $3,052$ & $100$\\ 
 Threshold: $x_{min}$ &  No &  Yes ($x_{min}=10,000$) & No \\ 
 Smallest city & $10,545$ &  $10,004$ & $50,030$\\ \hline
  \multicolumn{4}{|c|}{ Estimation of Zipfian exponent $\hat{\alpha} =1/(\hat{\gamma} -1)$}  \\ \hline
 Visual Inspection (rank)& 1.43 &  0.83 & 1.05 \\ 
  Linear fit (frequency, cumulative) & $1.468 \pm 0.016$ & $0.912\pm 0.001$ & $1.047 \pm 0.007$ \\ 
 Linear fit (rank) & $1.452 \pm 0.015$  & $0.908 \pm 0.001$ & $1.043\pm 0.007$\\
  Max. Likelihood (frequency) & 1.42 & 1.00 & 1.12 \\ 
 Max. Likelihood (rank, $r_{max}=N$) & 1.34 & 1.08 & 1.08  \\
 Max. Likelihood (rank, $r_{max} \rightarrow \infty$) & 1.55 & 1.19 & 1.42  \\
 \hline
\end{tabular}
\caption{Different estimations of the power-law exponent $\alpha$ in Eq.~(\ref{eq.powerlaw}) for data of city sizes (ALZ law discussed in Sec.~\ref{ssec.alz}). The linear regression method is discussed in Sec.~\ref{ssec.linearregression} and the different maximum-likelihood estimations in Sec.~\ref{ssec.ml-freq}. Estimations were performed in the indicated representations (parenthesis in the first column), with the exponents mapped to $\alpha$ through Eq.~(\ref{eq.relationexponents}) if needed. The uncertainty ($\pm$) in the linear fit cases was computed from the least-squared regression and propagated to $\alpha$. All linear-regression fits have a goodness-of-fit measure $R^2>0.99$, see Eq.~(\ref{eq.R2}). The data for Australia and Brazil is shown in Fig.~\ref{fig.representation-rank} together with the estimations based on visual inspection.  Graphical representations of the three datasets are shown in Figs.~\ref{fig.zipfcities},~\ref{fig.representation-rank}, and~\ref{fig.fitALZ} with some of the reported fits. See Appendix~\ref{chap.appendices} for the code and data used in this analysis. \index{power-law distribution}}
\label{tab.zipfexponents}
\end{table}

More generally, while the statistical approach centred around linear regression is appealing due to its simplicity and connection to graphical methods, it is important to remember that it contains limiting assumptions~\cite{leitao_is_2016}:

\begin{itemize}

\item[1.] $R^2$ does not quantify the statistical significance of the model, it quantifies the correlation between data and model (i.e., the amount of the total variance in the original $(x_i,y_i)$ observations that is explained by the linear fit). The use of a high $R^2 \lessapprox 1$ to justify the validity of a statistical law is problematic also because large values of $R^2$ are often observed for large $N$ even if the functional form is visually non-linear (provided $y$ varies substantially with $x$).  In particular, $R^2$ close to one is not an evidence that the data is a likely outcome of the model. Below we obtain that  datasets are typically not consistent with the model underlying the linear-regression approach.

\item[2.] The confidence interval around the estimated parameters $[\theta_{\text{min}}, \theta_{\text{max}}]$ is a range in which the true value of $\theta$ is expected to be found only if the model holds (i.e., if the data is generated by the model). In particular, for scaling laws in which the data is not compatible with the model, one cannot conclude a non-linear scaling $\beta \neq 1$ based on the observation that $1 \not \in [\beta_{\text{min}},\beta_{\text{max}}]$. In this case, both $\beta=1$ and $\beta \neq 1$ may be incompatible with the data.

    \item[3.] Log-transformations used to map the statistical law to a straight line (as discussed in Sec.~\ref{ssec.linearRepresentations}) imply that they cannot deal with "zero" observations, e.g., $y=0$ at a value of $x$ in scaling analysis or zero counts (frequencies) in distribution cases.  Typically these values are ignored, a pragmatic choice that bypasses the problem without addressing it. 
\item[4.] Distributions and probability density functions with the estimated parameters are not normalized, even if the data is. In particular, by fitting power-law distributions such that $\sum p(x) = 1$ or $\int p(x) dx=1$, the estimated parameters obtained fitting the distribution will not satisfy the same constraint.  \index{power-law distribution}

\end{itemize}

In addition to the considerations of these points, a simple and recommended test of the suitability of the linear regression is to inspect for trends in the residuals $\ln \alpha x_i^\beta -
\ln y_i$ which could characterize a deviation from the homogenous (Gaussian)  \index{Gaussian distribution} distribution predicted by the model underlying the linear regression (see Sec.~\ref{ssec.likelihoodscaling} below).

While the ordinary least-squared (OLS) regression has been and remains by far the most used technique, in particular in the case of scaling laws, alternative linear regression approaches have been considered as well. In the case of Kleiber's law (see Sec.~\ref{ssec.allometry}), this was done in~Refs.~\cite{zar_calculation_1968,dodds_re-examination_2001,warton_bivariate_2006}.  In particular, Ref.~\cite{dodds_re-examination_2001} considers the Kendall's non-parametric robust line fit method and the reduced major axis regression, finding that the estimations of the scaling exponent $\beta$ obtained with this alternative methods were within the confidence interval obtained using the least squared regression (applied to 3 different datasets and multiple cut-offs). This suggests that the choice of methods did not have a strong impact on the conclusions in that case. In the case of \index{urban scaling laws} urban scaling laws,  the validity of the hypothesis underlying the least-squared regression and alternative methods were discussed in Refs.~\cite{samaniego_cities_2008,bettencourt_urban_2010,gomez-lievano_statistics_2012,alves_distance_2013,nomaler_scaling_2014,gudipudi_efficient_2019}. For instance, Ref.~\cite{gudipudi_efficient_2019} proposes the Reduced Major Axis as an improved method to study the relations among scaling exponents.


\section{Likelihood-based methods}\label{sec.likelihood}

\subsection{Probabilistic approach}\label{ssec.probabilistic}

\paragraph{Probabilistic interpretation}

Linear fits and other regression models are motivated by graphical methods, visual inspection of data, and other heuristics. Their advantage is that they have a simple implementation and interpretation, directly linked to the representations that typically motivated the introduction of statistical laws (as discussed in Chap.~\ref{chap.examples}). Their disadvantage is that, alone, they do not allow for precise statistical statements about the validity of statistical laws, their agreement with data, and the estimation of parameters. Symptoms of these limitations discussed above include  the lack of invariance of estimations under transformations and the observation that different curves -- obtained using different parameters or functional forms -- can be significantly different from each other but still all show a high "goodness of fit", as measured by $R^2$ in Eq.~(\ref{eq.R2}).  

The limitations of linear regression and graphical methods motivate us to search for approaches that allow for more rigorous statistical analysis of data and more reliable conclusions on the agreement between data and proposed laws. This is typically achieved only after 
 a re-formulation of the problem of comparing a proposed statistical laws to data. This typically involves the following two inter-related steps:
 \begin{itemize}
     \item[(i)]~a reinterpretation of the observations ${\bf x}_i, i=1, \ldots N$, typically seen as realizations of random variables; 
     \item[(ii)]~a reformulation of the statistical law as a probabilistic statement  -- i.e., the probability $P({\bf x}_i | f, {\bf \theta})$ of the data ${\bf x}_i$, given the law $f$ and parameters ${\bf \theta}$.
 \end{itemize}  
 
 The re-interpretation of statistical laws under this framework allows for the computation of the probability of the data given the statistical law (and parameters)
\begin{equation}\label{eq.likelihod}
\mathcal{L}(\theta) = P({\bf x}_{i=0}, {\bf x}_{i=1}, \dots {\bf x}_{i=N} | f_{\bf \theta},{\bf \theta}),
\end{equation}
which corresponds to the {\it likelihood} function $\mathcal{L}(\theta)$. As further discussed later in this section, the application of likelihood-based methods to analyze statistical laws is based on an explicitly or implicitly reformulation of the law that is interpreted as the probability of observations or as their expected (or most likely) value.

\paragraph{Statistical analysis}

\begin{figure}[!h]
\begin{center}
\includegraphics[width=1.0\textwidth]{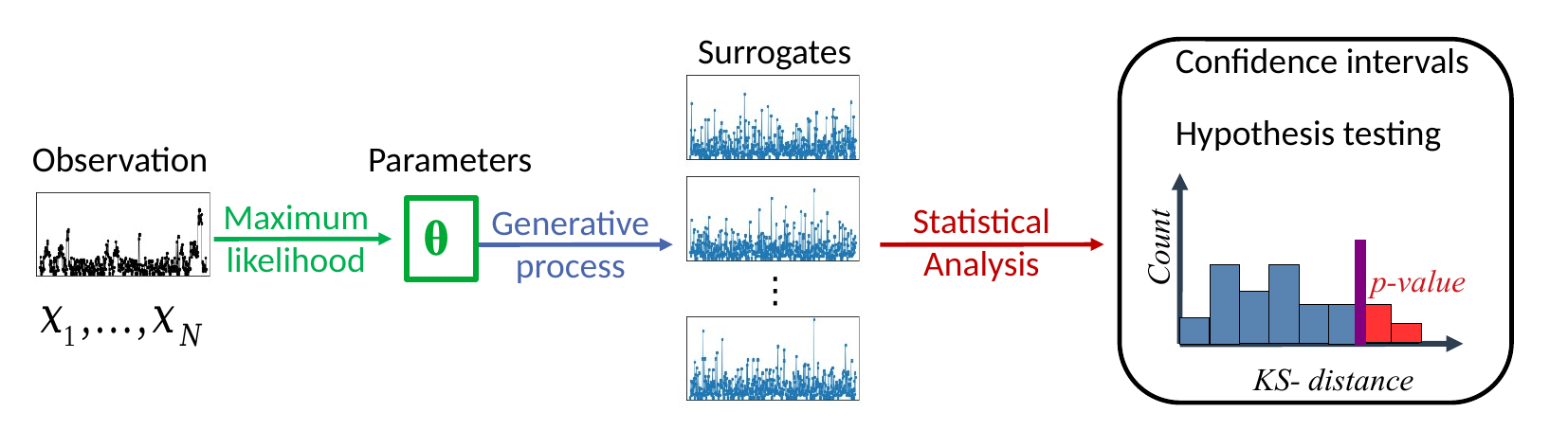}
\caption{Illustration of the steps employed in the analysis of statistical laws using likelihood-based methods. }\label{fig.illustration-ks}
\end{center}
\end{figure}

From the computation of the likelihood function~(\ref{eq.likelihod}), standard statistical approaches can be used to evaluate the statistical law~\cite{vuong_likelihood_1989,kass_bayes_1995,hastie_elements_2001,burnham_model_2002,sheskin_handbook_2003,clauset_power-law_2009}:
\begin{itemize}
    \item Fit: the parameters ${\bf \theta}$ are estimated considering the values that maximize $\mathcal{L}(\theta)$ (maximum-likelihood estimator) and their uncertainties based on the width of the likelihood function around the maximum.
    \item Model comparison: considers the evidence in favour of one model (curve) in comparison to another model (curve). For instance, the comparison between different model classes $M_1,M_2$ -- which correspond to different functional forms with parameters $\theta_1, \theta_2$, respectively--  can be done using the likelihood ratio~\cite{vuong_likelihood_1989} 
\begin{equation}\label{eq.likelihoodratio}
 \text{Likelihood Ratio =}   \frac{P({\bf x}_1, \ldots, {\bf x}_N)|M_1)}{P(({\bf x}_1, \ldots {\bf x}_N)|M_2)},
\end{equation}
with the likelihood of a model class $M_k$ obtained integrating over their free parameters ${\bf \theta}_k$
\begin{equation}\label{eq.marginalize}
    P({\bf x}_1, \ldots, {\bf x}_N |M_k) = \int P({\bf x}_1, \ldots, {\bf x}_N  | M_k,{\bf \theta}_k) d{\bf \theta}_k.
\end{equation}
Likelihood ratios larger (smaller) than one indicate a preference for model $M_1$ ($M_2$).
There is a variety of statistical methods to perform model comparison~\cite{kass_bayes_1995,burnham_model_2002,nakamura_comparative_2006,grunwald_minimum_2007}, including simplifications of Eq.~(\ref{eq.marginalize}), methods to account for different complexity (Bayesian Information Criteria, Akaike, etc.), and particular cases when the models are nested.
\item Hypothesis testing:  \index{hypothesis testing} a decision on whether the data can refute the law can be done computing the probability that the model leads to the observed deviation between the data and law (p-value). This is achieved defining a suitable measure (test statistic) that quantifies the data-law deviation, comparing the observed deviation to the deviation expected under a null model (compatible with the law), and setting a rejection threshold for the p-value (typically $5\%$ or $10\%$). This is often achieved by generating samples from the model with maximum-likelihood estimated parameters~$\theta$, which act as surrogates in time-series analysis, as depicted in Fig.~\ref{fig.illustration-ks}. \index{surrogate}
\end{itemize}

In Bayesian approaches~\cite{kass_bayes_1995,van_de_schoot_bayesian_2021}, the likelihood function~(\ref{eq.likelihod}) is combined with prior information on the proposed statistical law $M_k$ (and their parameters $\theta_k$), expressed in form of a prior probability $P(M,{\bf \theta})$, to obtain the posterior probability through Bayes' relationship as 
\begin{equation}\label{eq.bayes}
    P(M,{\bf \theta}| {\bf x}_{i=0}, {\bf x}_{i=1}, \dots, {\bf x}_{i=N} ) = P({\bf x}_{i=0}, {\bf x}_{i=1}, \dots, {\bf x}_{i=N} | f_{\bf \theta},{\bf \theta})  \frac{P(M, {\bf \theta})}{P({\bf x}_{i=0}, {\bf x}_{i=1}, \dots, {\bf x}_{i=N} )},
\end{equation}
 where the {\it evidence} $P({\bf x}_{i=0}, {\bf x}_{i=1}, \dots {\bf x}_{i=N})$ is a constant and does not affect the estimation of parameters and model comparison.

 \paragraph{Interpretation matters}
 
The reformulation of statistical laws to enable their probabilistic interpretation is a necessary step for their quantitative and statistical study. It forces us to be explicit about assumptions and expectations. In turn, this reveals ambiguities and weaknesses on the original formulations of statistical laws, which are incomplete and cannot be probabilistically evaluated on their own. At the same time, it is worth emphasizing that the process of reformulating statistical laws in probabilistic terms involves additional assumptions that are not unique and that are not present in the original (historical) formulation of the statistical law as reviewed in Chap.~\ref{chap.examples}. For instance, while \index{Gutenberg-Richter's law} Gutenberg-Richter's law can be interpreted as a power-law distributed probability of the energy released by an earthquake (see Sec.~\ref{ssec.gutenberg}), the original formulation of their law was done in terms of magnitudes and the logarithm of frequencies. Similarly, power laws can be formulated in the rank-frequency and frequency distribution representations. Graphically and analytically, both statements are uniquely connected, as shown in Sec.~\ref{ssec.rankrepresenation} above. Statistically and probabilistically, as we will show in Sec.~\ref{ssec.ml-freq} below, they suggest different sampling processes, the analysis is affected by these choices, and typically lead to different results. In the next sections we show how the main types of statistical laws can be re-interpreted probabilistically, that there is more than one way of doing so, and that the choice of the interpretation matters.  \index{power-law distribution}

\subsection{Scaling analysis}\label{ssec.likelihoodscaling}

Here we focus on statistical laws in which the parametric function $f_\theta:\mathbb{R} \mapsto \mathbb{R}$ prescribes the relationship between pair of observations ${\bf x}_i = (x,y)_i$ as $y_i = f_\theta (x_i),$ where $i=1, \ldots, N$ indicate different observations. This case includes the scaling laws discussed in Sec.~\ref{ssec.scaling}: in \index{Urban scaling laws} urban scaling laws $x$ is the population of cities, $y$ is an observable associated to the city (e.g., its GDP \index{Gross Domestic Product, GDP}), and $i$ is an index that goes through all $N$ cities in a country (or dataset); in Herdan-Heaps' law, $y$ is the number of unique words, $x$ is the size of the texts (in word tokens), and $i$ is either an index over different texts (books) or runs from the first $i=1$ to the last $i=N$ word token of one text; in Kleiber's law, $x$ is the mass, $y$ is the metabolism, and $i$ is an index over different species (in the dataset).
\index{Herdan-Heaps' law}

The simplest probabilistic formulation of these statistical laws interprets each of the $i=1, \ldots, N$ as an independent observation, $x$ as the independent variable, and $y$ as the dependent variable explained by $x$ and a model based on the statistical law $f_\theta$ as $y=f_\theta(x)$. A suitable probabilistic model should thus specify the probability of $y$ given $x$ and the laws with parameters $\theta$, represented as $P(y|x,\theta)$. We consider this probabilistic model compatible with a statistical law $f_\theta(x)$ -- as defined in Sec.~\ref{ssec.definition} -- if the expected value of $y$ according to $P(y|x)$ matches $y=f_\theta(x)$:
\begin{equation}
    \mathbb{E}(y|x) \equiv  \int P(y|x,{\bf \theta}) dy = f_\theta(x).
\end{equation}
$P(y|x,\theta)$ cannot be uniquely computed from $f_\theta(x)$ -- as the problem is under-determined -- and different $P(y|x,{\bf \theta})$ -- all compatible with the statistical law $f_\theta(x)$ -- can be  proposed based on different additional hypothesis.

The assumption of independent observations allow us to write the likelihood~(\ref{eq.likelihod}) as the product over observations (and the log-likelihood as the sum):
\begin{equation}\label{eq.likelihood-iid}
    \mathcal{L}(\theta) = \prod_{i=1}^N P(y_i|x_i) \Leftrightarrow \log \mathcal{L} (\theta) = \sum_{i=1}^N \log P(y_i|x_i).
\end{equation}
The monotonicity of the logarithmic function ensures that the maximum of the likelihood and log-likelihood coincide.

\paragraph{Connection to scaling laws}

In the case of scaling laws $y=f_\beta (x) \sim x^\beta$ -- as defined in Eq.~(\ref{eq.scaling}) and discussed in Sec.~\ref{ssec.scaling} -- we are looking for a probabilistic model $P(y|x,{\bf \theta})$ such that
\begin{equation}\label{eq.expectedscaling}
    \mathbb{E}(y|x) =  \int P(y|x,{\bf \beta}) dy \sim x^\beta.
\end{equation}

A natural way in which this is achieved is to consider 
\begin{equation}\label{eq.yfxe}
    y = Ax^\beta + \varepsilon,
\end{equation}
with parameters $\theta=\{A,\beta\}$ and $\varepsilon_i$ an independent and identically distributed random variable with zero mean. The observations $(x_i,y_i)$ are thus interpreted considering that $x_i$ is given and $y_i$ is obtained from $x_i$ and a random component $\varepsilon_i$ (noise) according to Eq.~(\ref{eq.yfxe})  (i.e., $\varepsilon_i = y_i - f(x_i)$).

\paragraph{Connection to linear fit}

The linear regression method described in Sec.~\ref{ssec.linearregression} can be connected to the probabilistic framework above. This is done based on the equivalence between the least-squared estimation of parameters of a linear model and the maximum-likelihood estimator, which is obtained assuming that the probability (uncertainty) of the independent variable is distributed around the expected value with a uniform width across all points (i.e., homoscedastic fluctuations \index{homoscedasticity} such as a Gaussian  \index{Gaussian distribution} with zero mean and constant standard deviation). In the case of scaling laws, the linear fit is obtained in the log-transformed variables $(\log x, \log y)$. Therefore, the equivalence to this case is obtained either considering that the observables $y$ and $x$ in Eq.~(\ref{eq.yfxe}) are the logarithmic of the original observations or, equivalently, that $P(y|x,\beta)$ is given by a log-normal as~\cite{leitao_is_2016}

\index{log-normal distribution}

\begin{equation}
\label{eq.P.lognormal}
P(y \mid x) = \frac{1}{\sqrt{2\pi}\sigma_{\LN}}\frac{1}{y} e^{- \frac{\left(\ln y-\mu_{\LN}(x) \right)^2}{2\sigma^2_{\N}}},
\end{equation}
with a fixed $\sigma_\LN$ and
\begin{equation} \label{eq.ln_parameters}
\mu_{\LN}(x) \sim \beta \ln x.
\end{equation}
The log-likelihood~(\ref{eq.likelihod}) is computed from Eq.~(\ref{eq.P.lognormal}) as
\begin{equation}\label{eq.LLN}
\ln \mathcal{L}(\beta) =  \sum_{i=1}^N - \ln (\sigma_{\LN} \sqrt{2\pi}) - \ln y_i - \frac{\left( \ln(y_i)-\mu_{\LN}(x_i)\right)^2}{2\sigma^2_{\LN}},
\end{equation}
This function is maximized when the squared difference $\sum_i (\ln(y_i) - \ln(Ax_i^\beta))^2$ is minimized, which is equivalent to the least-squared estimator in Eq.~(\ref{eq.ols}) once the log-transformation is applied. This shows the equivalence between the maximum-likelihood and the linear-regression estimators of $\beta$, obtained assuming Eq.~(\ref{eq.P.lognormal}).

\paragraph{Alternative approaches}

Through the discussion in this section we naturally encountered two different probabilistic models $P(y|x)$ compatible with scaling laws: Gaussian fluctuations  \index{Gaussian distribution} -- assuming Gaussian noise $\varepsilon$ in Eq.~(\ref{eq.yfxe}) -- and Log-normal fluctuations -- equivalent to a log-transformed observations with Gaussian fluctuations. Ref.~\cite{leitao_is_2016} considered two other models:

\begin{itemize}
    \item[(i)] (Taylor's law) The idea is to consider a conditional probabilities $P(y|x)$ that, in addition to the expected value satisfying the scaling law as in Eq.~(\ref{eq.expectedscaling}), have a variance satisfying the scaling~\cite{eisler_fluctuation_2008}
    \begin{equation}\label{eq.scaling3}
    \V(y|x) = \gamma \mathbb{E}(y|x)^{\delta},
    \end{equation}
where $\delta$ is a free parameter (typically $1\le \delta \le 2$). Scaling~(\ref{eq.scaling3}) corresponds to Taylor's law, observed in different datasets. It retrieves previous cases (Gaussian fluctuations for $\delta=1$, log-normal for $\delta =2$) and allows for a more flexible model of variable fluctuations (heteroscedasticity).
 \index{Taylor's law}

    \item[(ii)]$\;$ (Sampling tokens) The idea is to consider that $Y=\sum_{i=1}^N y_i$ tokens are sampled and attributed randomly to one of the $i=1, \ldots, N$ possible classes each with known $x=x_i$ (e.g., tokens of GDP \index{Gross Domestic Product, GDP} attributed to cities of populations $x_i$). Notice that $Y$ is fixed in this approach while it varies from realization to realization in the case in which $P(y|x)$ is defined. Under the assumption of independent sample of the tokens, the likelihood can be computed as shown in Sec.~\ref{ssec.inference} below.
\end{itemize}

\begin{figure}[!ht]
    \centering
    \includegraphics[width=0.75\textwidth]{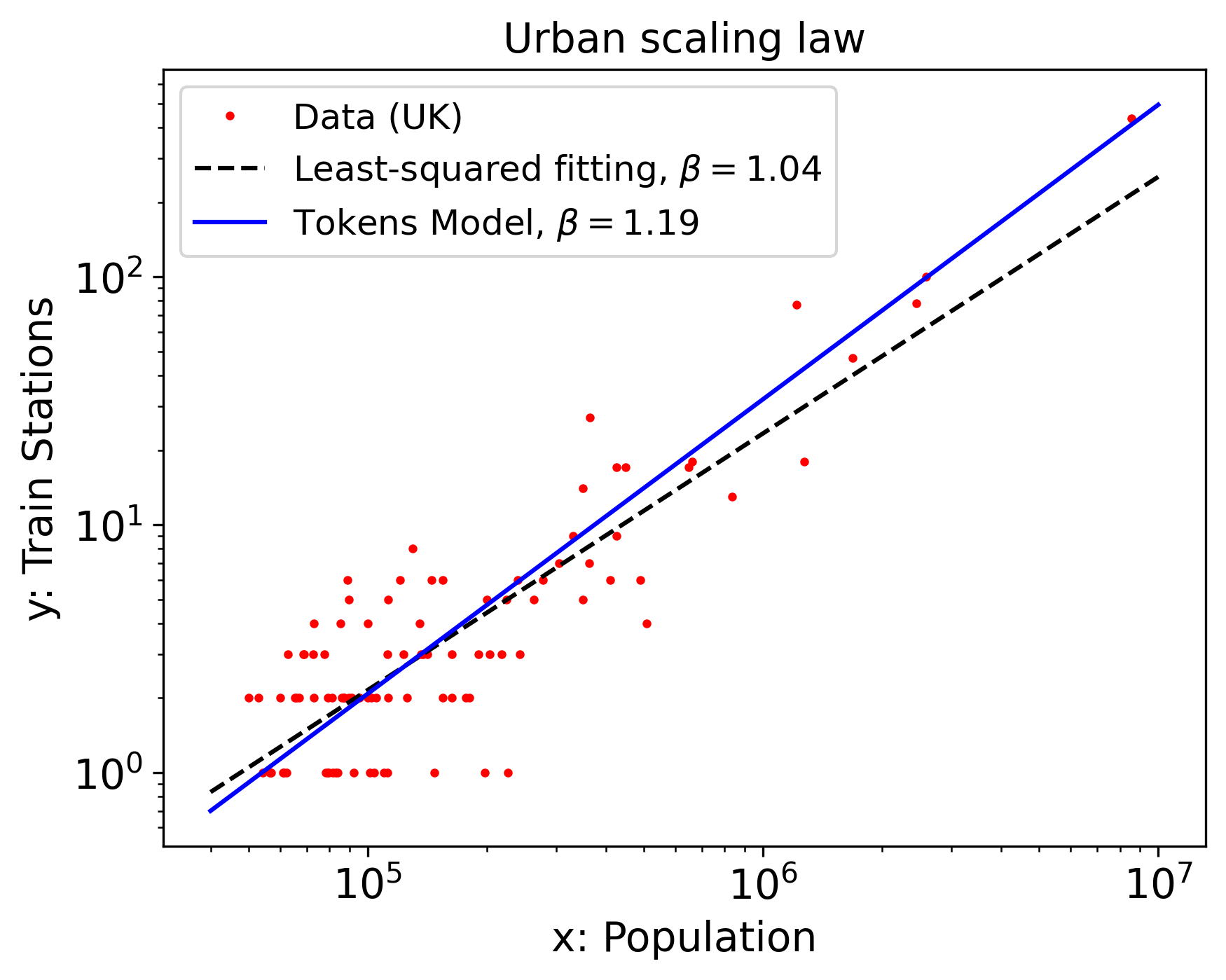}
    \caption{The estimated \index{Urban scaling laws} urban scaling law depends on the data-analysis method and underlying probabilistic model. The data corresponds to the number of train stations at cities in the United Kingdon. The linear regression method yields a line (dashed, $\hat{\beta}=1.04$) that describes better the small cities but considerably under-estimates the value observed in the large cities. The token model (solid line, $\hat{\beta}=1.19$) fits better the large cities.}
    \label{fig.fitScalingUK}
\end{figure}

These two approaches address also one of the characteristics of traditional linear regression identified in Sec.~(\ref{ssec.linearregression}) and Fig.~(\ref{fig.threshold}) as potential drawbacks: the fact that the estimation of the scaling parameter is dominated by the regions (in $x$) with a high-density of points (e.g., the large number of small cities in \index{Urban scaling laws} urban scaling laws, which account for very little of the total population, or the highly abundant species with small mass in Kleiber's law), not necessary the regions one is most interested in (for instance, the full range of $x$ values over many decades or the regions in which most population live). In approach (i) listed above (Taylor's law), $\delta < 2$ implies that the deviations between the line and the observations are more highly penalized at deviations around points with larger $x$; in approach (ii) (sampling tokens), the observations of a value $y$ are sampled $y$ times so that large $x$ are naturally more sampled (since $y\sim x^\beta, \beta>0$) and thus points with large $x$ exert a larger influence in the fit and estimation of $\beta$. An illustration of this point is shown in Fig.~\ref{fig.fitScalingUK} for the case of a urban scaling law in a very noisy datasets (the number of train stations in cities in the United Kingdom). The least-squared fitting better approximates the data for small cities but severely underestimating the number of train stations in London and other large cities in the UK. The maximum-likelihood estimation of $\beta$ in the token model has the opposite effect. While the stark contrast in the estimation of $\beta$ in this case ($1.04$ vs $1.19$) is due to the strong and population dependent fluctuations in the data, it is important to notice that variations across $x$ get amplified due to the fat-tailed distribution of cities sizes (ALZ\index{Auerbach-Lotka-Zipf's law} law) so that such variations can be expected in general.  \index{Taylor's law}  \index{fat-tailed distribution}

The main conclusion we take from the example discussed above is that there are different models $P(y|x,\theta)$ compatible with the same statistical law and that they can lead to different conclusions based on the data analysis. This applies not only to the estimation of the best parameters $\theta$ but also the evaluation of the extent into which a given dataset agrees with a law. While approaches (i) and (ii) were introduced in Ref.~\cite{leitao_is_2016} in the case of \index{Urban scaling laws} urban scaling laws, these ideas apply more generally to other scaling statistical laws. The choice between these (and other) models to evaluate the scaling law will depend on assumptions underlying each case -- Are $y=0$ observations possible? Are heteroscedastic \index{heteroscedasticity} fluctuations expected? -- and the decision about the most suitable model should ideally be performed using model comparison techniques that take into account the extent into which they describe the data well (likelihood of models) and also the complexity of the models. The results in Ref.~\cite{leitao_is_2016} indicate that different models are preferred on different datasets. 

\subsection{Frequency Distributions}\label{ssec.ml-freq}

A simpler probabilistic interpretation of statistical laws -- in line with the probabilistic approach proposed in Sec.~\ref{ssec.probabilistic} -- exists for the laws formulated as frequency distributions. The most notable cases are power-law distributions -- reviewed in Sec.~\ref{sec.powerlaw} --, but our discussion here applies to other types of parametric distributions and probability densities (e.g., log-normal, stretched-exponential), including their application to inter-event times -- reviewed in Sec.~\ref{sec.interevent}.  \index{power-law distribution}  \index{stretched exponential distribution}

The idea is to interpret the statistical law $f_\theta({\bf x})$ directly as the probability of the given observations. The most natural approach is to normalize the counts underlying the distributions to compute relative frequencies (e.g., of word types, of earthquake magnitudes, of city sizes, of people with a given income) which are interpreted as estimators of probabilities. The statistical law is then interpreted as a probability function proposed to describe the observations 
\begin{equation}\label{eq.normalization}
p_\theta (x) = \frac{f_\theta(x)}{\int f_\theta(x) dx} \Rightarrow \int p_\theta(x) dx =1,
\end{equation}
i.e., as a statement about the probability of a randomly-selected observations to have value in in the interval $[x,x+dx]$ with $x\in\mathcal{R}$ (or, similarly, $p_\theta(x) = f_\theta(x)/\sum f_\theta(x)$ for $x\in \mathcal{N}$).  In this interpretation, the ALZ\index{Auerbach-Lotka-Zipf's law} law for city sizes -- Sec.~\ref{ssec.alz} -- describes the probability of observing a randomly selected city with population $x$, Zipf's law\index{Zipf's law} of word frequency is a statement about the probability of observing a word type with a frequency $x$ in the text, \index{Pareto's law} Pareto's law of income describes the probability that a person has a certain income (above a threshold), \index{threshold} etc. Equivalently, the proposed parametric distributions $P(\tau)$ of waiting times $\tau$ describe the probability of observing a randomly selected inter-event time. 

The use of likelihood-based methods based on this probabilistic interpretation has been applied and advocated to study (power law) statistical laws in different publications at the start of the 21st century~\cite{goldstein_problems_2004,perline_strong_2005,bauke_parameter_2007,clauset_power-law_2009,deluca_fitting_2013,hanel_fitting_2017} and are increasingly used. In particular, Ref.~\cite{clauset_power-law_2009} was extremely influential because of its didactic review of methods and detailed connection to different power-law distributions. The adoption of likelihood-based methods in the study of (power law) frequency distributions is also due to the increased limitations of linear-regression models in this case. In addition to the limitations listed in Sec.~\ref{ssec.caveatsLinear}, simple linear fits of log-transformed variables leads to parameter estimations that do not respect natural normalizations of the data\footnote{Many statistical laws, such as the ALZ\index{Auerbach-Lotka-Zipf's law} law or Zipf's law\index{Zipf's law} were not formulated as normalized probability distributions. Still, it is natural and convenient to have parametric functions that share properties with the data, such as the total population of cities or words in the text. This is equivalent to the normalization imposed in likelihood-based approaches and is absent when linear regression is applied.} and are not maximum-likelihood estimators under reasonable assumptions (i.e., contrary to the case of simple scaling laws, there is no simple scenario in which the fluctuations around frequency distributions are uniform in the log-transformed variables). 

The usual approach is to again consider the $i=1, \ldots, N$ observations to be independent and identically distributed --according to $p(x_i|\theta) \equiv p_\theta$ in Eq.~(\ref{eq.normalization}) -- and thus write  the likelihood $\mathcal{L}$ as
\begin{equation}\label{eq.Liid}
\mathcal{L}(\theta)  = \prod_{i=1}^N p(x_i|\theta) \Leftrightarrow \log \mathcal{L} (\theta) = \sum_{i=1}^N \log p(x_i|\theta).
\end{equation}
The maximum-likelihood estimation $\hat{\theta}$ of the parameters $\theta$ are obtained maximizing~(\ref{eq.Liid}). For the (continuous) power-law distribution~(\ref{eq.powerlaw}), $p(x) \sim x^{-\gamma}$ for $x>x_{min}$, we have $\theta=\{\gamma, x_{min}\}$, and an explicit expression can be obtained for the maximum likelihood estimator of $\gamma$ at fixed $x_{min}$ as
\begin{equation}\label{eq.MLEgamma}
\hat{\gamma}=1+N\left(\sum_{i=1}^N \ln \frac{x_i}{x_{min}}\right).
\end{equation}
The derivation of this result and for the corresponding estimators for the case of discrete $x$ can be found, for instance, in Ref.~\cite{clauset_power-law_2009}.
The other parameter, $x_{min}$, is usually chosen in such a way to increase the range of validity of the power-law distribution~\cite{clauset_power-law_2009,deluca_fitting_2013}. The choice of $x_{min}$ is not only a choice of parameter, it effectively sets a truncation or threshold \index{threshold} that changes the number $N$ of points and is known to have important consequences to the evaluation and interpretation of statistical laws~\cite{perline_strong_2005,font-clos_perils_2015}. 

Standard statistical methods can also be applied to test the hypothesis that the data is sampled from a power-law distribution -- e.g., through the computation of the probability (p-value) that the observed distance between the histogram of the data and the proposed distribution is due to the finite-size observations, as reviewed in Ref.~\cite{clauset_power-law_2009} and illustrated in Fig.~\ref{fig.illustration-ks}-- and to compare the power law to alternative distributions -- e.g., using likelihood ratio in Eq.~(\ref{eq.likelihoodratio})~\cite{vuong_likelihood_1989} or accounting for model complexity~\cite{burnham_model_2002,grunwald_minimum_2007}.

 \paragraph{Rank-frequency representation}

Many power-law statistical laws admit both the rank-frequency $F(r)$ and the frequency-distribution $p(x)$ representations, as indicated in Eq.~(\ref{eq.powerlaw}) and discussed in Sec.~\ref{ssec.rankrepresenation}. The probabilistic interpretation described above -- adopted in Refs.~\cite{perline_strong_2005,clauset_power-law_2009} and in most likelihood-based analysis -- is based on the $p(x)$ representation. As noted in Ref.~\cite{gerlach_stochastic_2013}, the rank-frequency $F(r)$ representation can also be formulated probabilistically and be used for likelihood-based inference. Below we show how this approach is based on a different interpretation of the statistical law and leads to different estimations of parameters.

Similarly to the approach in Eq.~(\ref{eq.normalization}), the idea is to consider the normalized version of a rank-frequency statistical law $f_\theta(r)$ as
\begin{equation}\label{eq.normalization-rank}
F_\theta(r) = \frac{f_\theta(r)}{\sum_{r=1}^{r_{max}} f_\theta(r)} \Rightarrow \sum_{r=1}^{r_{max}} F_\theta(r)  =1,
\end{equation}
where $r_{max}\rightarrow \infty$ is taken if the law is assumed to be valid for an arbitrary number of cases (or, alternatively, $r_{max}$ can be kept equal to the number of observed items)\footnote{This is an important modeling choice that affects the quality of the fitting (as it affects the normalization and parameters $\theta$), and often changes model-comparison decisions. It is closely related to the issue of thresholding \index{threshold} data discussed in Refs.~\cite{perline_strong_2005,font-clos_perils_2015}. Choosing $r_{max}$ as the largest rank (i.e., number of word types, cities, etc.), corresponds to the assumption that the distribution applies only to the observed data and that $r_{max}$ is known. This is a stronger assumption that uses more information from the data and yields better agreement between the curve and the points (higher likelihood of the model). Choosing $r_{max} \rightarrow \infty$ corresponds to the assumption that the proposed law is valid for arbitrary large $r$'s (arbitrary large vocabulary or number of cities) and that current observations corresponds to those in which $F_r>0$ (the lack of observations for $r$ larger than the maximum observed ranking is considered in the normalization, thus affecting the whole fitted curve). This is a weaker assumption (stronger statement about the validity of the law) and yields worst agreements between the curve and the points (lower likelihood). Estimators reported in Tab.~\ref{tab.zipfexponents} shows that this choice strongly affects the estimation of the exponent in the ALZ law.
}. The statistical estimation and methods described above, in particular the likelihood in Eq.~(\ref{eq.likelihood-iid}) and the estimator~(\ref{eq.MLEgamma}), can then be directly applied to the rank representation in Eq.~(\ref{eq.normalization-rank}) considering the mapping $x \mapsto r$, $\hat{\gamma} \mapsto \hat{\alpha}$, $N \mapsto M=\sum_{i=1}^N x_i$, and $x_{min} \mapsto r_{min}$\footnote{While the estimators apply identifying the minimum $x$ with a minimum rank $r$, conceptually a minimum rank $r$ corresponds to a large $x$ while a small $x$ used as $x_{min}$ correspond effectively to cut-off at large $r=r_{max}$. See Refs~\cite{bauke_parameter_2007} for the inclusion of such cut-offs in maximum-likelihood estimators of power-law distributions.}.

The probabilistic interpretation of $F_\theta(r)$ in Eq.~(\ref{eq.normalization-rank}) is that it describes the probability of a randomly selected item to be of the type described by rank $r$, in contrast to the interpretation of $p(x)$ in Eq.~(\ref{eq.normalization}) which focuses on the probability of a randomly selected type. It is worth exemplifying this subtle yet crucial difference in some of the statistical laws discussed in Sec.~(\ref{sec.powerlaw}):
\begin{itemize}
    \item ALZ\index{Auerbach-Lotka-Zipf's law} law of city sizes: $p(x)$ describes the probability that a randomly selected {\it city} is of size $x$; $F(r)$ describes the probability that a randomly selected {\it person} lives in the r-th largest city.
    \item Zipf's law\index{Zipf's law} of word frequency: $p(x)$ describes the probability that a randomly select {\it word type} appears $x$ times in the text (or has frequency $x$); $F(r)$ describes the probability that a randomly selected {\it word token} is of type $r$ (i.e., of the $r$-th most frequent word type).
    \item Scale-free networks: $p(x)$ describes the probability that a randomly selected {\it node} has degree $x$; $F(r)$ describes the probability that a randomly selected (semi-)edge belongs to the $r$-th most central (highest degree) node. 
\end{itemize}
Despite the one-to-one correspondence of the power-law representations -- discussed in Sec.~(\ref{ssec.rankrepresenation}) -- their probabilistic interpretations are radically different. They correspond to different definitions of observation and sampling processes: the number of observations in the $p(x)$ case is the number of unique types (e.g., distinct words, different cities), which is much smaller than in the $F(r)$ case which focuses on the attribution of tokens (e.g., length of the text in number of word tokens, population of all cities).  
This distinction can be applied to any distribution describing the frequency of categorical types  (when $x \in \mathcal{N}$), such as the sales or preference of different products.  \index{power-law distribution}

\begin{table}[h!]
  \centering
  \small
\begin{tabular}{|l|c|c|c|}
\hline
\textbf{Corpus / Book} & Linear regression & Freq. dist. $\hat{\alpha}$ & Rank freq. $\hat{\alpha}$ \\
\hline
Alice's Adventures in Wonderland (L. Carroll) & 1.21 & 1.46 & 1.22 \\
The Voyage Of The Beagle (C. Darwin) & 1.29 &1.59 & 1.20 \\
The Jungle (U. Sinclair) & 1.22 & 1.45 & 1.21 \\
Life On The Mississippi (M. Twain) & 1.16 & 1.38 & 1.20 \\
Moby Dick; or The Whale (H. Melville) & 1.15 & 1.38 & 1.19 \\
Pride and Prejudice (J. Austen) & 1.35 & 1.66 & 1.21 \\
Don Quixote (M. Cervantes) & 1.12 & 1.29 & 1.21 \\
The Adventures of Tom Sawyer (M. Twain) & 1.12 & 1.29 & 1.21 \\
Ulysses (J. Joyce) & 1.03 &  1.15 & 1.18 \\
War and Peace (L. Tolstoy) & 1.44 & 1.84 & 1.20 \\
\hline
English Wikipedia & 1.58 & 1.60 & 1.17 \\
\hline
\end{tabular}
\caption{Different estimations of the power-law exponent $\alpha$ in Zipf's law\index{Zipf's law} of word frequencies discussed in Sec.~\ref{ssec.zipf}. Graphical representation of some of these datasets appears in Fig.~\ref{fig:zipf}. The second column reports results obtained using the linear regression of $\log F_r$ vs. $\log r$ as described in Sec.~\ref{ssec.linearregression} (the $R^2$ goodness-of-fit measure computed from Eq.~(\ref{eq.R2}) is larger than 0.97 in all cases). The third and fourth columns correspond to the maximum-likelihood estimators, as described in Sec.~\ref{ssec.ml-freq}. The results in the third column were obtained using the frequency distribution, with the estimated $\gamma$ in Eq.~(\ref{eq.MLEgamma}) mapped to $\alpha$ using Eq.~(\ref{eq.relationexponents}). The results in the fourth column were obtained using the rank-frequency representation in Eq.~(\ref{eq.normalization-rank}).  For the two maximum-likelihood estimators, the p-value computed as described in Fig.~\ref{fig.illustration-ks}, is smaller than $10^{-4}$ in all cases. Results from Ref.~\cite{altmann_statistical_2016}, the English translation of the books was used.}\label{tab.exponentzipfword}
\end{table}

\paragraph{Effect on estimation}

The choice of representation of statistical laws affects the estimations and conclusions obtained from likelihood-based analysis. This happens because one is typically analyzing large datasets with substantial fluctuations, which are not compatible with simple samples of any of the representations. A signature of this general point is the maximum-likelihood estimation of exponents using different representations. In Tab.~\ref{tab.zipfexponents}, discussed above, the estimation of the Zipfian exponent~$\alpha$ in the ALZ law (city sizes) was shown for different methods and countries. In Tab.~\ref{tab.exponentzipfword} we show the results for the Zipfian exponent $\alpha$ in Zipf's law (word frequencies) for different methods and books. The maximum-likelihood estimation based on the frequency distribution $p(x)$ yields larger values of $\hat{\alpha}$ than the maximum-likelihood estimation based on the rank frequency distribution $F_r$. This is compatible with our interpretation that the rank representation gives more weight to high-frequency words and the observation of faster decay of the rank-frequency plot for large $r$ (i.e., for small frequency words which affect more strongly the frequency distribution). 

\begin{table*}[!bt]
\centering
\small
\setlength\tabcolsep{2pt}
\begin{tabular}{ |c|c|c|c| } 
\hline
Model & $F_r \equiv F(r| \; \theta)$ & Parameter Estimates & $-\log \mathcal{L}/N$ \\ 
\hline
\hline
Simple       & $Cr^{-\alpha}$ & $\alpha$ = 1.19 & 7.515\\ 
\hline
Shifted Power Law  & $C(r+a)^{-\alpha}$ &$\alpha$ = 1.29, $a$ = 4.76  & 7.391  \\

Exponential cut off  & $C \text{exp}(-ar)r^{-\alpha}$ & $\alpha$ = 1.05, $a = 7.19 \; 10^{-6}$ & 7.351\\ 

Naranan    & $C \text{exp}(-a/r)r^{-\alpha}$  & $\alpha$ = 1.26, $a$ = 2.02 &  7.406\\ 

Weibull    & $C \text{exp}(-ar^{-\alpha})r^{\alpha - 1}$    & $\alpha$ = -0.344, $a$ = -2.85 &  8.369 \\ 
Log-normal & $C r^{-1} \text{exp}(-\frac{1}{2} (\text{ln}(r) - m)^{2}/s^{2})$   & $m$ = 1.02, $s$ =  1.80 &  7.339 \\ 

Double Power Law & 
$
C 
\begin{cases} 
    r^{-1} &  r \leq a\\
    a^{\alpha-1}r^{-\alpha} & r > a 
\end{cases}
$
& $\alpha$ = 1.77, $a$ = 8189 & {\bf 7.336}  \\
\hline
Double Gamma & 
$
C 
\begin{cases} 
    r^{-\alpha_1} &  r \leq a\\
    a^{\alpha_2-\alpha_1}r^{-\alpha_2} & r > a
\end{cases}
$
& $\alpha_1$ = 1.02, $\alpha_2$ = 1.80, $a$ = 10317.1 &  7.335\\
\hline
\end{tabular}
\caption{Model comparison of generalized Zipf's laws\index{Zipf's law}. The data is the frequency of words in Spanish books (Google n-gram database, $N=32,632,629,877$ tokens and $1,385,248$ types), shown in Fig.~\ref{fig.zipfcomparison}. Different models for the rank-frequency distribution $F_r\equiv F(r|\theta)$ were fitted to the empirical distribution $F_r$ using the maximum likelihood method in the rank-frequency representation~\cite{gerlach_stochastic_2013}. The parameters $\theta$ that maximize the likelihood $\mathcal{L}$ are reported together with the negative log-likelihood per token $-\log \mathcal{L}/M$ (at the given parameters). The preferred 2-parameter model (minimum $-\log \mathcal{L}$) -- based on the likelihood ratio test in Eq.~(\ref{eq.likelihoodratio}), evaluated at the maximum likelihood parameters $\hat{\theta}$ -- is the log-normal model and is highlighted in boldface. See Appendix~\ref{chap.appendices}for the data and code used in this analysis.}\label{tab.zipfspanish}
\end{table*}

\begin{figure}[!h]
\begin{center}
\includegraphics[width=0.85\textwidth]{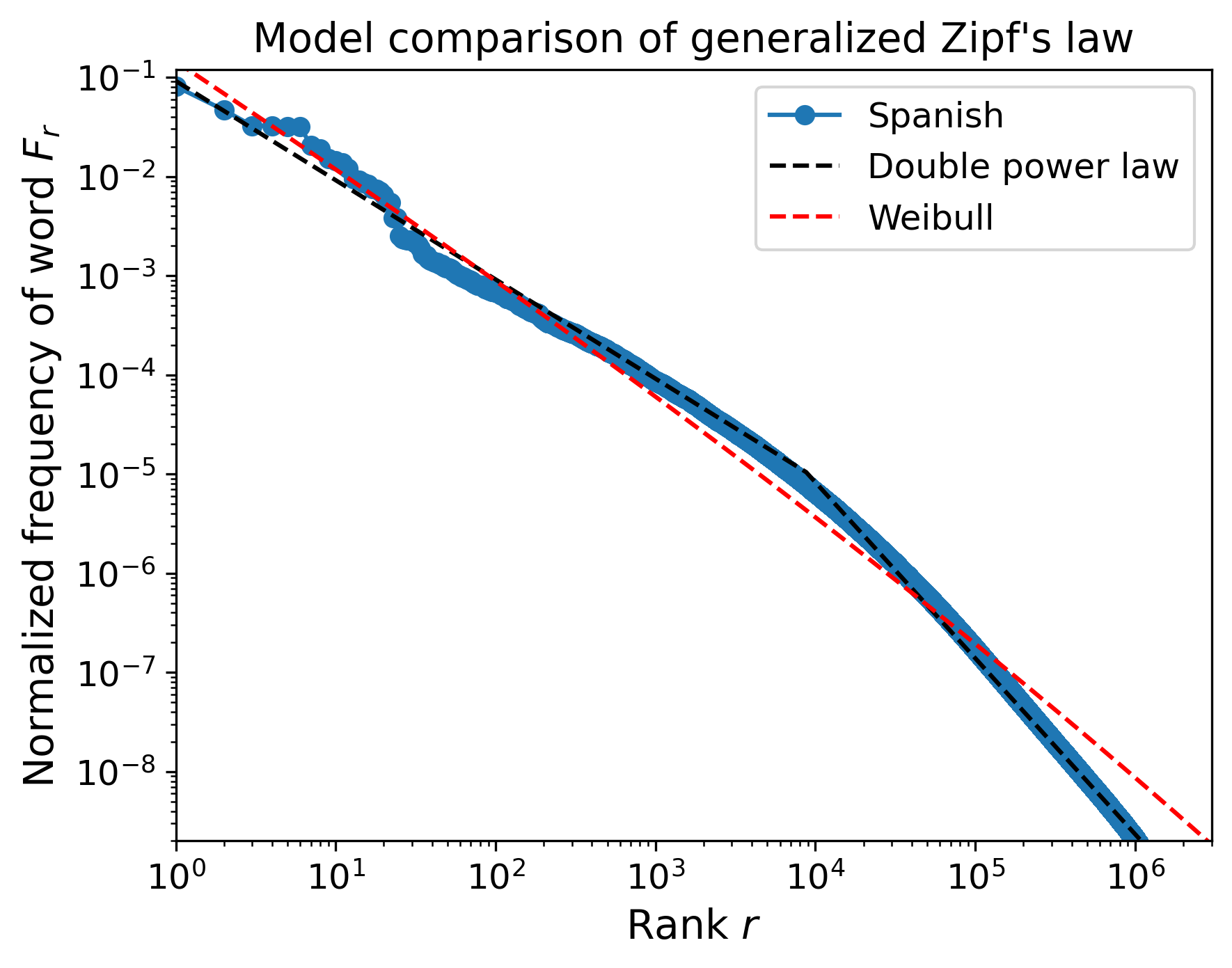}
\caption{Model comparison of different generalizations of Zipf's law\index{Zipf's law}. The data corresponds to the word frequency distribution obtained combining millions of Spanish books, as provided in the Google n-gram database and as used in Ref.~\cite{gerlach_stochastic_2013}. The two curves correspond to two of the $2$-parameter generalizations of Zipf's law described in Tab.~\ref{tab.zipfspanish}, with parameters estimated using the maximum-likelihood method in the rank representation and the reported $-\log \mathcal{L}/N$ values are evaluated at the maximum likelihood parameters $\hat{\theta}$. See Appendix~\ref{chap.appendices}for the data and code used in this analysis.}\label{fig.zipfcomparison}
\end{center}
\end{figure}

\paragraph{Model comparison}

We now show how model-comparison \index{model comparison} methods in the rank representation can be used to analyze generalizations of Zipf's law\index{Zipf's law}, reproducing in new datasets the findings first reported in Ref.~\cite{gerlach_stochastic_2013}. We consider $6$ two-parameter functions that have been previously proposed as a generalization of the simple power-law in Zipf's law of word frquencies. These functional forms are provided in Tab.~\ref{tab.zipfspanish} together with the maximum-likelihood parameter estimates for one dataset. A graphical comparison in this case is given in Fig.~\ref{fig.zipfcomparison}, including the best and the worst model. We see that the graphical analysis agrees with the model comparison based on the likelihood ratio test, but that the distinctions are relatively small even considering the comparison of the best and words model (the distinctions become difficulty to discern by eye when considering some of the other models).  \index{power-law distribution}

\begin{table*}[!bt]
\centering
\small
\setlength\tabcolsep{2pt}
\begin{tabular}{ |c||c|c|c|c||c|c|c|c| } 
\hline
     & \multicolumn{8}{c|}{$-\log \mathcal{L}/N$}\\
     
 Model     & \multicolumn{4}{c||}{Google n-gram data} & \multicolumn{4}{c|}{Books (in English)} \\
 &  English & French & German & Russian & War\&Peace & Beagle & Sawyer & Alice \\ 
\hline
\hline
N (word tokens) & $222\; 10^9$ & $28\;10^9$   & $25\;10^9$ & $21\;10^9$& $565 \; 10^3$  & $208\; 10^3$&$71\; 10^3$ &$27\; 10^3$ \\
Word types & $4 \; 10^6$ &$1 \; 10^6$ & $3 \; 10^6$& $2 \; 10^6$& $18\; 10^3$& $13\; 10^3$& $7\; 10^3$& $3\; 10^3$\\
\hline
\hline
Simple        & 7.794 & 7.376 & 8.614& 9.206& 6.976& 7.024 & 6.901 & 6.444\\ 
\hline
Shifted Power Law  & 7.689 & 7.300 & 8.459& 9.078& 6.771& 6.874& 6.659& 6.140\\
Exponential cut off  & 7.619 & 7.224& 8.410& {\bf 8.901}& {\bf 6.679}& {\bf 6.715}& {\bf 6.564}& {\bf 6.048}\\ 
Naranan    & 7.710 & 7.307& 8.488& 9.114& 6.823& 6.917& 6.721& 6.238\\ 
Weibull    & 8.647 & 8.277& 9.423& 10.008& 7.771& 7.849& 7.677& 7.205\\ 
Log-normal & 7.594 & 7.241& {\bf 8.384}& 8.928& 6.699& 6.774& 6.591& 6.083\\ 
Double Power Law & {\bf 7.570} & {\bf 7.223}& 8.396& 8.907& 6.697& 6.724& 6.570& 6.082\\
\hline
Double Gamma & 7.569 & 7.218 & 8.393& 8.892& 6.695& 6.723& 6.569& 6.072\\
\hline
\end{tabular}
\caption{Model comparison \index{model comparison} of generalized Zipf's law\index{Zipf's law} for different datasets. The functional form of the models is given in Tab.~\ref{tab.zipfspanish}, together with the results for the Spanish Google n-gram data. The preferred 2-parameter model (minimum $-\log \mathcal{L}$) is highlighted in boldface. The estimations used the maximum-likelihood fit in the rank-frequency representation (with no upper cut-off $r_{max}$)}\label{tab.zipfcomparison}
\end{table*}

The claim of universal validity underlying statistical laws suggest that the same functional form should describe also different datasets. To test this claim, we consider $8$ other corpora of different sizes -- books in English of  various lengths-- and in different languages -- Google n-gram corpora in 5 languages. The results for the model comparison \index{model comparison} of these additional corpora is given in Tab.~\ref{tab.zipfcomparison}. It suggests that the model of Zipf's law with an exponential cut-off is better for small corpora but that the double-power law discussed in Sec.~\ref{ssec.zipf} -- Eq.~(\ref{eq.modeldp}) -- is the best model for large datasets, remaining reasonably competitive also for books. 

The new results reported here corroborate the Zipfian view that simple parametric functions can describe a variety of word-frequency distributions for different datasets and languages. This is remarkable as our analysis contains datasets involving millions of books, beyond the possibilities of analysis in the early 20th century. Our findings corroborate also Ref.~\cite{gerlach_stochastic_2013}'s preference for the double-power-law generalization of Zipf's law. However, our results show that a nuanced interpretation on the universal validity of statistical laws is needed. Overall, we see that there is no single best functional form describing the observations in all cases and that different functional forms do reasonably well. The essential ingredient behind Zipf's law, and the success of its generalizations, is that functional forms with a broad distribution are needed to characterize the observations, with a roughly $1/r$ decay for small $r$'s and a faster decay for large $r$. \index{power-law distribution}

\paragraph{Advantages and disadvanatages of the rank representation}

From the (functional form of the) statistical law in one representation we can compute  the law in other representations, as shown in Sec.~\ref{ssec.rankrepresenation} for the case of power-law distributions. This analytical relationship between the functional forms does not mean that the probabilistic formulation of the law in both representations is equivalent. In particular, the statistical analysis and tests of the different representations of the law can lead to very different results and estimations of the exponents~\cite{altmann_statistical_2016,corral_truncated_2020}, shown in Tabs.~\ref{tab.zipfexponents} and~\ref{tab.exponentzipfword} above. The choice of the representation reflects different views about the observations of interest underlying the law and the generative process. The choice comes also with different advantages and disadvantages.

An advantage of the rank-representation $F(r)$ of power-law distributions  is that sampling types is often not realistic~\cite{cristelli_there_2012}: could we imagine countries in which their capital or more populous cities are not sampled? (e.g., a France without Paris?) Or texts in which some of the most frequent word types do not appear? (a text in English without "of"). If we interpret the sampling process as individual realizations of arbitrary sample size, in the spirit of Fig.~\ref{fig.illustration-ks}, this would be likely outcomes. Instead, in the $F(r)$ sampling, the large number of token samples ensure that the probability of having a sample in which they do not appear is negligibly small. Sampling word tokens is also more natural if one identifies this process with the order of word tokens in a text, i.e., a book written from start to finish by sampling each word token randomly with a fixed probability of attributing it to different word types (no similarly natural interpretation exists for the sampling process underlying the $p(x)$ representation). 

Another advantage of the $F(r)$ formulation is that the most frequent types (cities or words) play a more important role on the computation of the likelihood and thus on the estimation of the parameters. Likelihood-based methods based on the $p(x)$ representation suffer from the same problems identified in Sec.~\ref{ssec.caveatsLinear} for the linear regression: they are mostly influenced by the large number of types with small frequencies which often compose only a small fraction of the total system (i.e., a small fraction of the text or of the population of the country). This happens because in the $p(x)$ representation the observation is defined to be a type, there are many more small-frequency types in power-law distributed data, and each type contributes to one term in the likelihood function. Instead, a likelihood based method based on the $F(r)$ representation considers tokens to be observations and therefore the types with more tokens (e.g., large cities, frequent words) naturally contribute more. The crucial issue of performing model comparison \index{model comparison} in ALZ\index{Auerbach-Lotka-Zipf's law}'s law is the difference between small and large cities, as encapsulated in the choice of $x_{min}$. As discussed in Sec.~\ref{ssec.alz}, depending on the analysis, the large number of small cities or the few large cities (with most of the population) will dominate (leading to different conclusions about whether log-normal or power-law distributions provide a better fit). Similarly, in the analysis of fat-tailed data  \index{fat-tailed distribution} the choice of representation and statistical methods will often be dominated either by the many types with small frequency or by the few types with large frequency.  \index{power-law distribution}

\begin{figure}[!ht]
    \centering
    \includegraphics[width=0.75\textwidth]{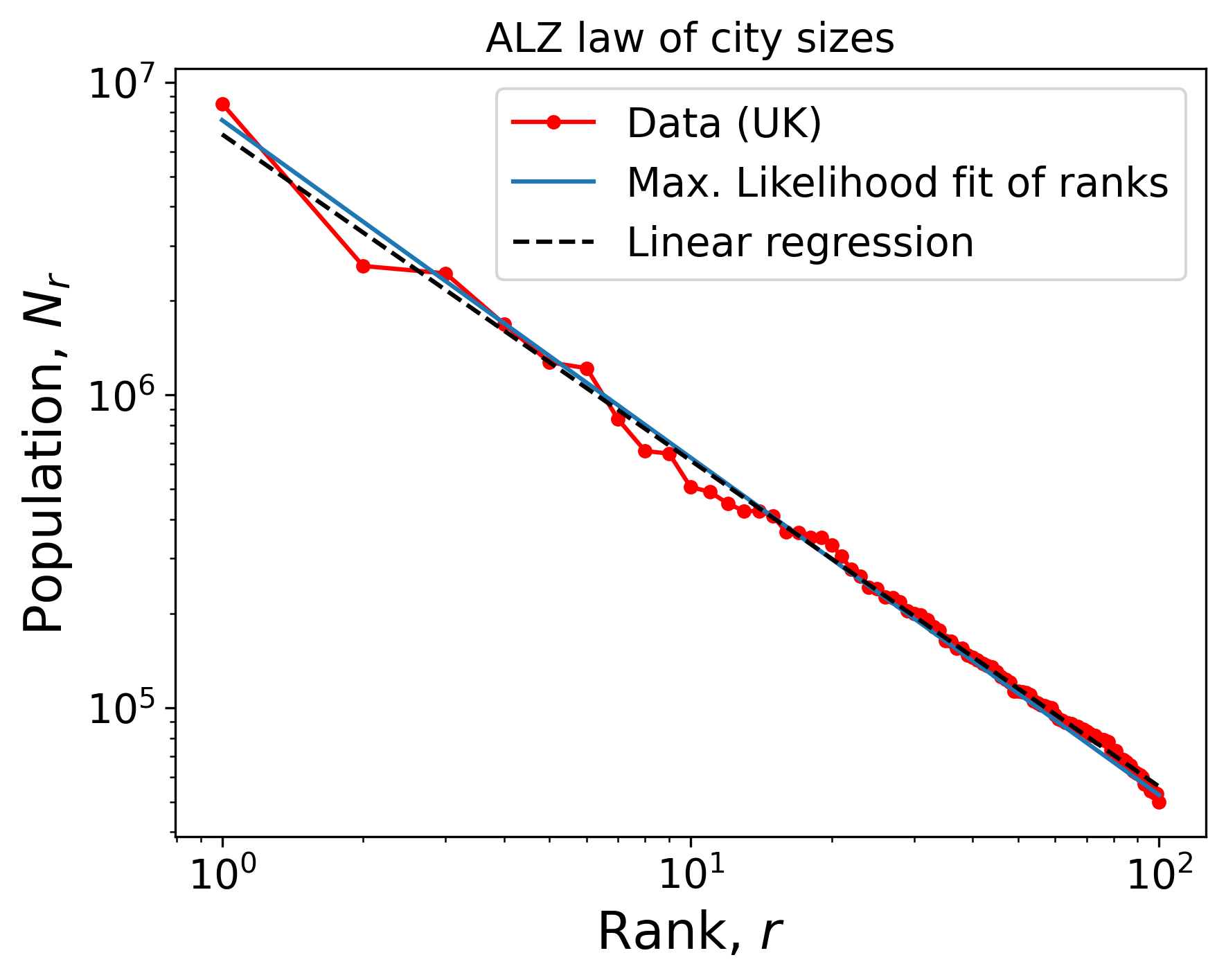}
    \caption{Dependence of estimated power laws on the data-analysis method. The data corresponds to the $100$ largest cities in the UK and the different estimations were reported in Tab.~\ref{tab.zipfexponents}. Linear regression (dashed line $\hat{\alpha}=1.04$) yields a curve that describes better the majority of cities (large rank), but underestimates severely the estimation for the largest city ($r=1$, London). The maximum-likelihood estimation using the rank-representation (in $r\in[1,r_{max}=100]$) yields a distribution (solid line, $\hat{\alpha} =1.08$), that is more influenced by the largest cities and describes better the largest city. }
    \label{fig.fitALZ}
\end{figure}

An illustration of the points above for the case of the ALZ\index{Auerbach-Lotka-Zipf's law} law of city sizes is provided in Fig.~\ref{fig.fitALZ} (see also the previously presented results in Tab.~\ref{tab.zipfexponents}). The maximum likelihood estimation using the rank representation preserves the total population of the largest cities and is more strongly influenced by the larger-than-expected size of London (the largest city). In contrast, regression methods are dominated by the smaller cities. 

A disadvantage of the $F(r)$ interpretation is that it works directly with ranked variables~\cite{gunther_zipf_1996}. In its direct implementation, it assumes that the rank of the types is known a priori and can be used as labels for the types in the computation of the likelihood. This is not the case as the ranks are attributed based on the data. If we interepret the data as a realization of an underlying process with (asymptotic) ranks used as node labels, any finite-size realization will lead to empirical ranks that differ from the true ranks and thus to mis-attribution of their probability by the model $F(r)$. This problem of rank mis-attribution is particularly important for small data sizes and large ranks (when the number of samples is small and $F(r)$ cannot be estimated accurately). 
Refs. \cite{gunther_zipf_1996,cristelli_there_2012} investigate the effect of ranking finite-samples of a Zipfian  distribution, showing how strong deviations appear and are connected to observations of Zipf's law\index{Zipf's law}. This is a crucial issue when considering what is the region in which Zipf's law will be tested (e.g., in a state, country, or continent) or whether all cities have been included~\cite{gunther_zipf_1996}. 

\subsection{Caveats and limitations of likelihood-based methods}\label{ssec.limitationslikelihood}

The likelihood function is the essential element in a data-model comparison and thus in a probabilistic evaluation of statistical laws. 
The limitation of likelihood-based methods -- in particular for "curve fitting" and hypothesis test -- is that they often rely on simplistic assumptions and interpretations that are not part of the statistical law and that are often not explicitly discussed. 
In fact, statistically-focused publications~\cite{perline_strong_2005,clauset_power-law_2009} reduce the question about the validity of power-law statistical laws to the evaluation of the goodness-of-fit between parametric distributions and the data. This simplistic view ignores both the simplifying assumptions underlying the statistical tests and central points in the study of statistical laws (as defined in Sec.~\ref{ssec.definition}): the fact that the same statistical law admits different (probabilistic) interpretations (formulations), the ambiguity in the choice of representation and definition of observed quantities, the claims of universal validity in different datasets and settings, and the role statistical laws play in mechanistic models and theories.

\paragraph{Independence hypothesis}

The conflict between the study of statistical laws and the naive statistical-test approach becomes clear noting that one of the assumptions underlying the simple likelihood strategies discussed above (and also linear regression) is the assumption of independence of the observed data, i.e., that each of the observations (data points $x_i,y_i$) is the outcome of a process that is independent of the other observations and of other variables not explicitly considered in the model (such as time or location).  In the analysis of data coming from complex systems, this assumption is violated in virtually every case of interest. There are numerous examples of  statistical laws in which the lack of independence appears explicitly in the data:

\begin{itemize}
    \item Kleiber's law and \index{allometric laws} allometric scaling (Sec.~\ref{ssec.allometry}): data from philogenetically close species will be naturally correlated~\cite{savage_predominance_2004}.
    \item \index{Gutenberg-Richter's law} Gutenberg-Richter law (Sec.~\ref{ssec.gutenberg}): sequence of magnitudes of earthquakes are (spatially and temporally) correlated,  affecting the estimation and tests of power-law distributions~\cite{gerlach_testing_2019,moore_nonparametric_2022}. 
    \item The words in a text or corpus are not randomly distributed, a point that affects the study of statistical laws in \index{linguistic laws} linguistics~\cite{altmann_statistical_2016}.
    \item The sequence of inter-event times~$\tau_1,\tau_2, \ldots, \tau_N$ (Sec.~\ref{sec.interevent})  is typically correlated so that $P(\tau)$ is not a complete characterization of burstiness~\cite{bunde_long-term_2005}. In particular, \index{burstiness} the sequence of recurrence times between words discussed in Sec.~\ref{ssec.burstywords} is \index{long-range correlations} long-range correlated~\cite{altmann_beyond_2009,altmann_origin_2012}. 
    \item Urban data:  urban centres (cities) affect each other (e.g., through cultural or geographical proximity), therefore affecting the ALZ\index{Auerbach-Lotka-Zipf's law} law (Sec.~\ref{ssec.alz}) and \index{Urban scaling laws} urban scaling laws (Sec.~\ref{ssec.urbanscaling}) (as discussed in Sec.~\ref{ssec.inference} below).
\end{itemize}

More formally, the maximum-likelihood goodness-of-fit tests in frequency distribution -- discussed in Sec.~\ref{ssec.ml-freq} and reviewed in Ref.~\cite{clauset_power-law_2009}-- is based on the standard "independent and identically distributed" (iid) assumption, which corresponds to two hypotheses  \index{hypothesis testing} on the observations $x_i$, $i=1, \ldots, N$~\cite{gerlach_testing_2019}:

\begin{itemize}
\item[]H1: they are distributed as $p(x|\theta)$, e.g. for a power law $p(x|\gamma) = C x^{-\gamma}$;
\item[]H2: they are independent (e.g., of $i$ or $x_{i-1}$). 
\end{itemize}
While H1 is specified by the statistical law, H2 is a strong simplifying assumption not contained in the historical formulations of the statistical laws and that is known to be violated. When a statistical test leads to a rejection (small p-value), as used in the recent claims~\cite{khanin_how_2006,stumpf_critical_2012,broido_scale-free_2019} of violation of power laws, it rejects the compound hypothesis (H1+H2). It is not clear if it is due to a systematic deviation of the parametric-form of the law (H1), or, instead, due to the well-known fact that observations are not independent (H2).  

\begin{figure*}[!ht]
\begin{center}
  \includegraphics[width=0.53\linewidth]{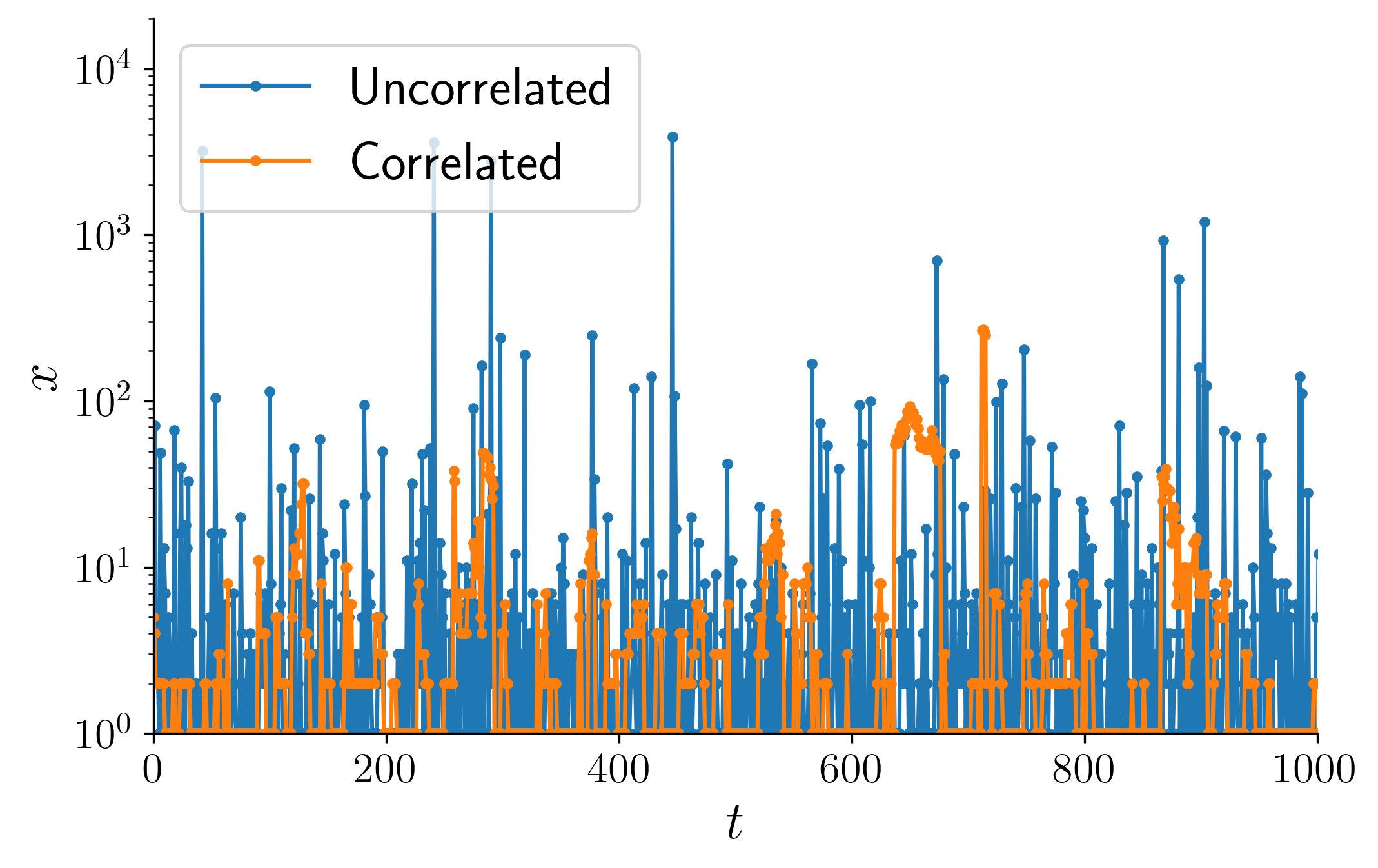}
  \includegraphics[width=0.53\linewidth]{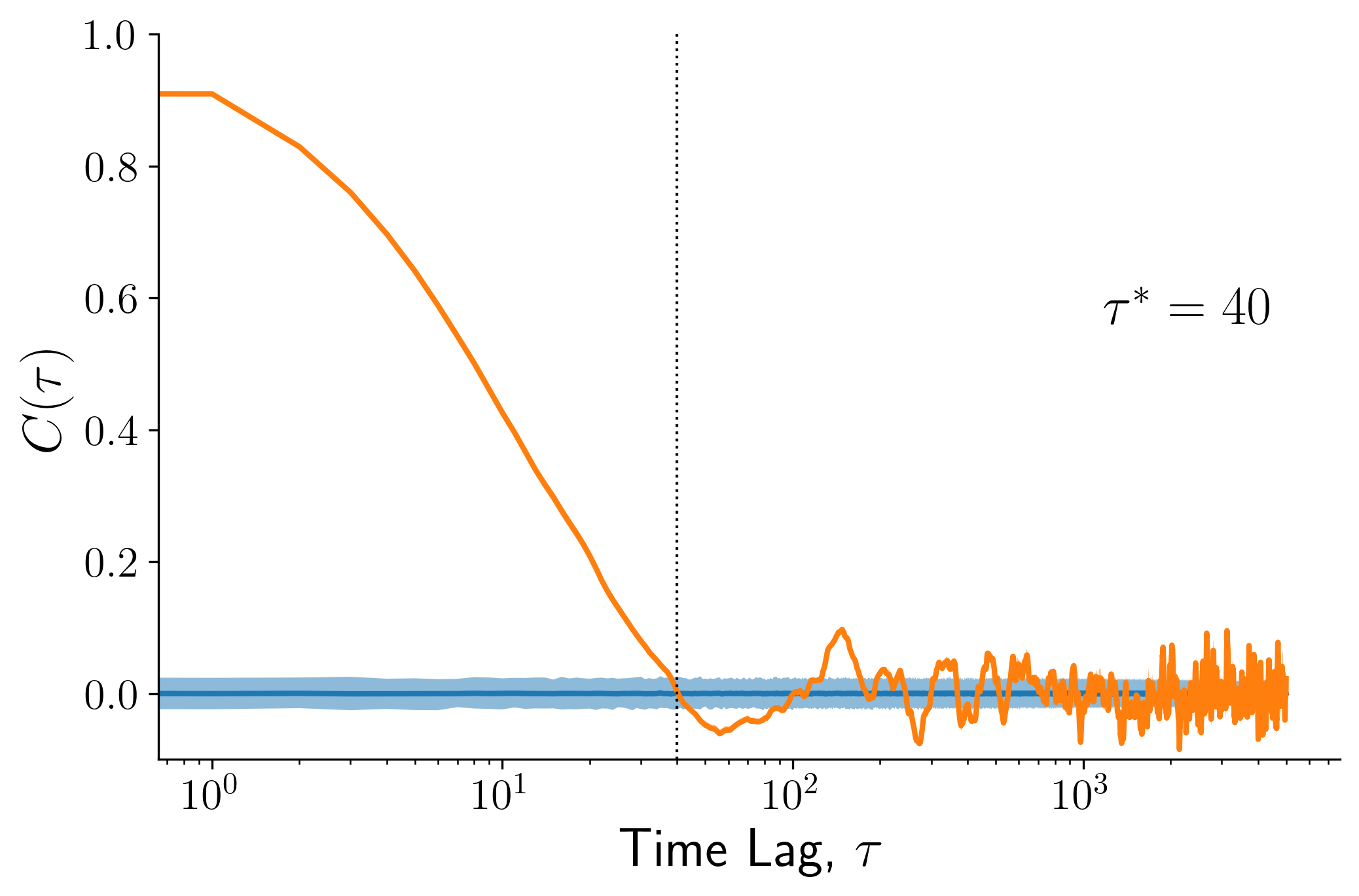}
  \includegraphics[width=0.53\linewidth]{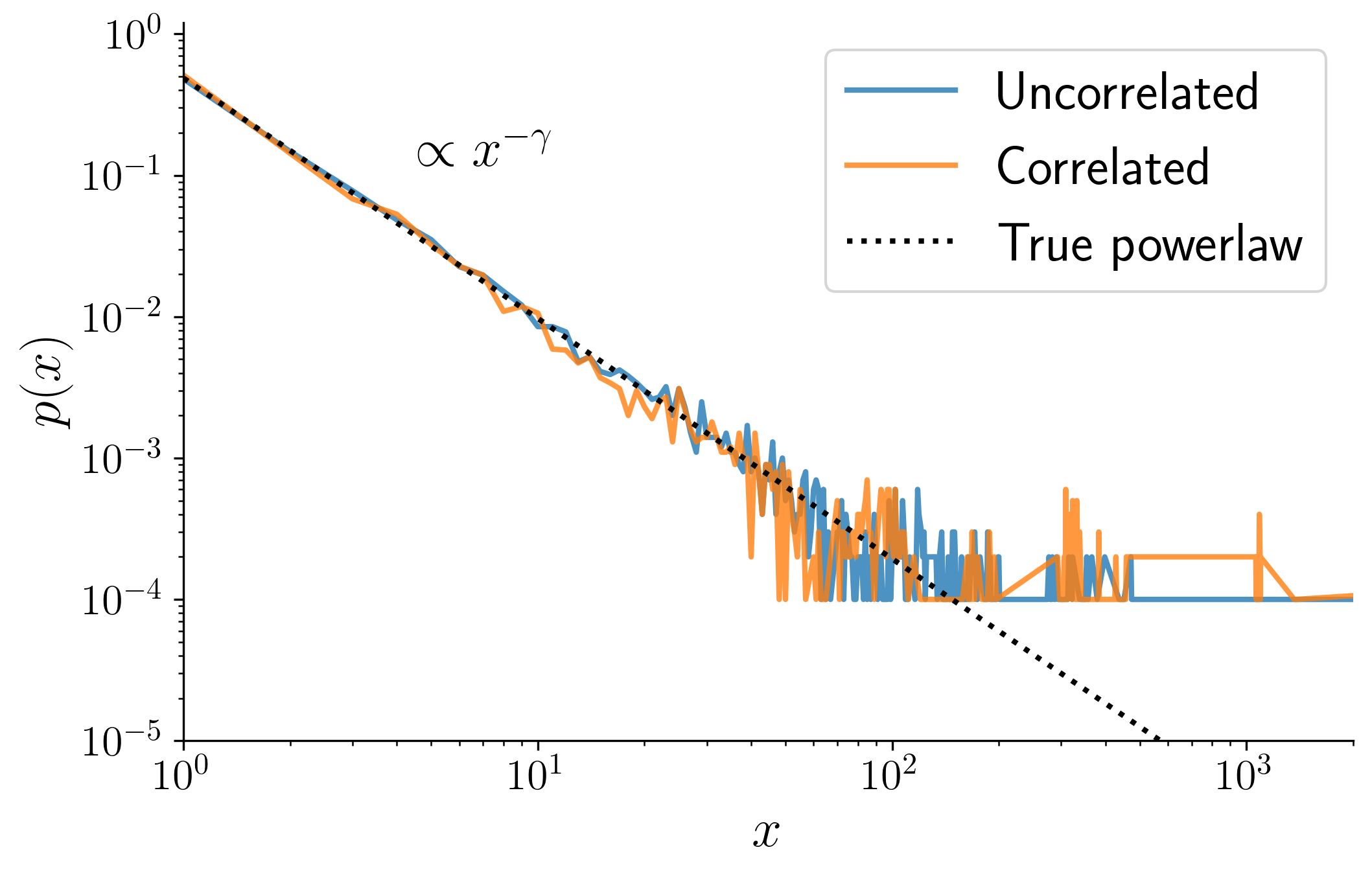}
\end{center}
  \caption{Comparison between correlated and uncorrelated data with a power-law distribution. Results are shown for two time series $x(t)$ with a power-law distribution: $p(x) = C x^{\gamma}$, $\gamma=1.7$, $x$ an integer value $x\in[1,2\;10^5]$,  and $t\in [1,N]$: one in which $x(t)$ are independently sampled (uncorrelated, in blue) thus satisfying both H1 and H2 mentioned in the text; and one in which $x(t)$ is a Markov process of order one so that $x(t+1)$ depends on $x(t)$ (correlated, in orange) thus satisfying H1 but not H2. Top: time series $x(t)$. Middle: autocorrelation \index{autocorrelation function} function $C(\tau)$ as a function of the delay time $\tau$. Bottom: histogram $p(x)$ obtained for $N=10^4$ with the theoretical ($N\rightarrow \infty$) distribution as a dashed line. See Ref.~\cite{gerlach_testing_2019} for  details and Appendix~\ref{chap.appendices} for information on the code used in this figure. }\label{fig.synthetic-correlated}
\end{figure*}

\begin{figure*}[!ht]
\begin{center}
  \includegraphics[width=0.53\linewidth]{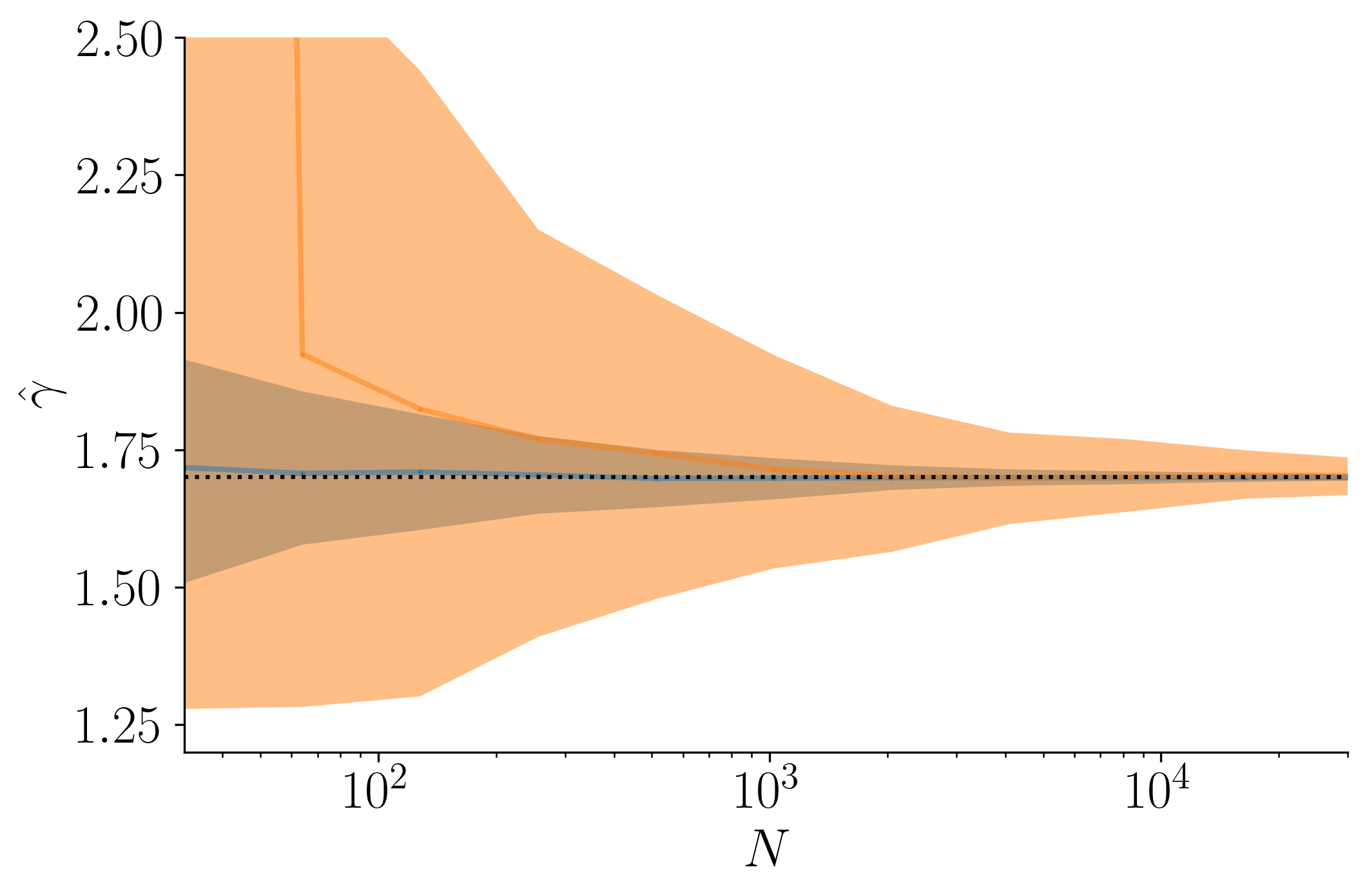}
    \includegraphics[width=0.53\linewidth]{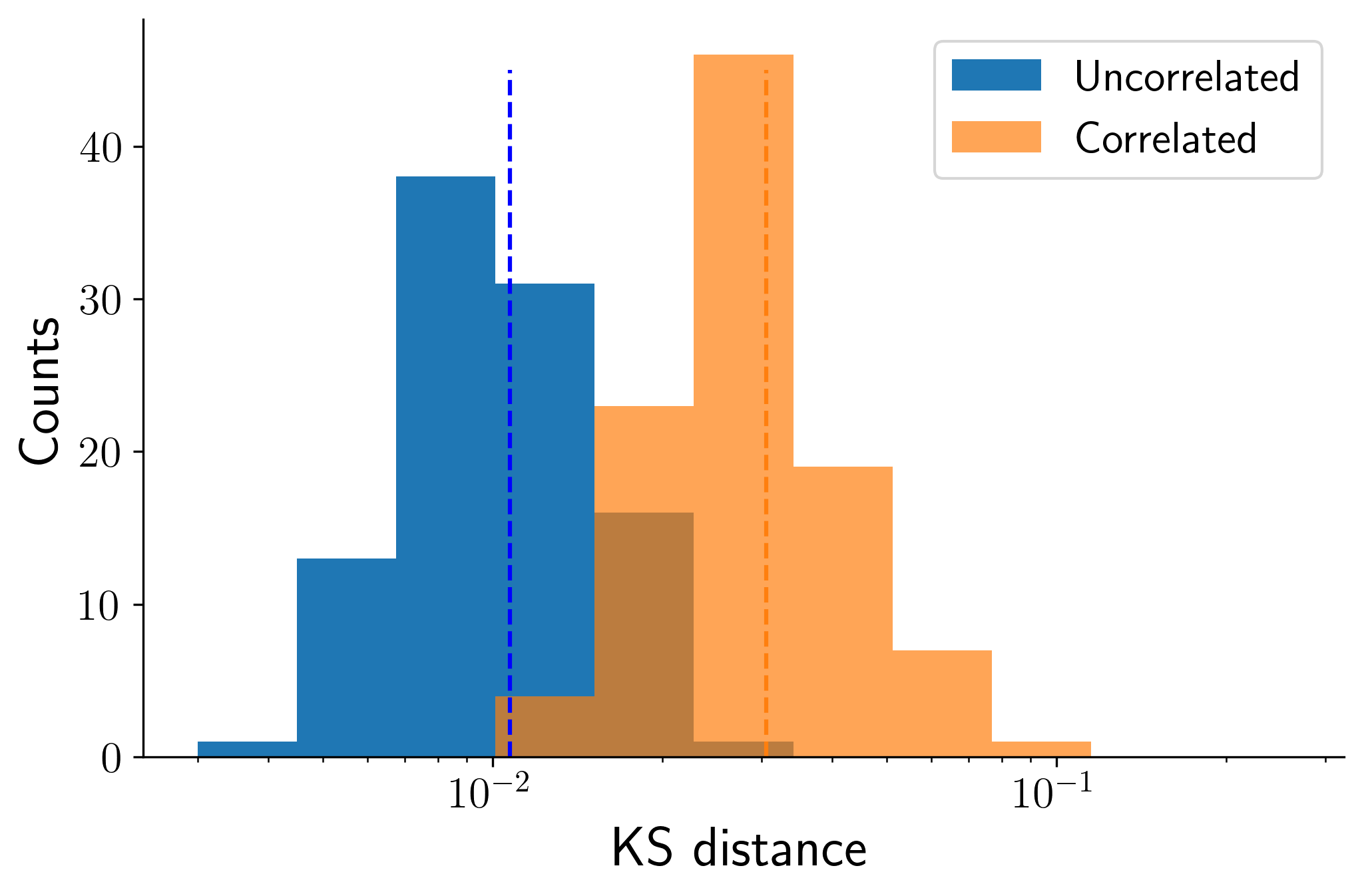}
    \includegraphics[width=0.53\linewidth]{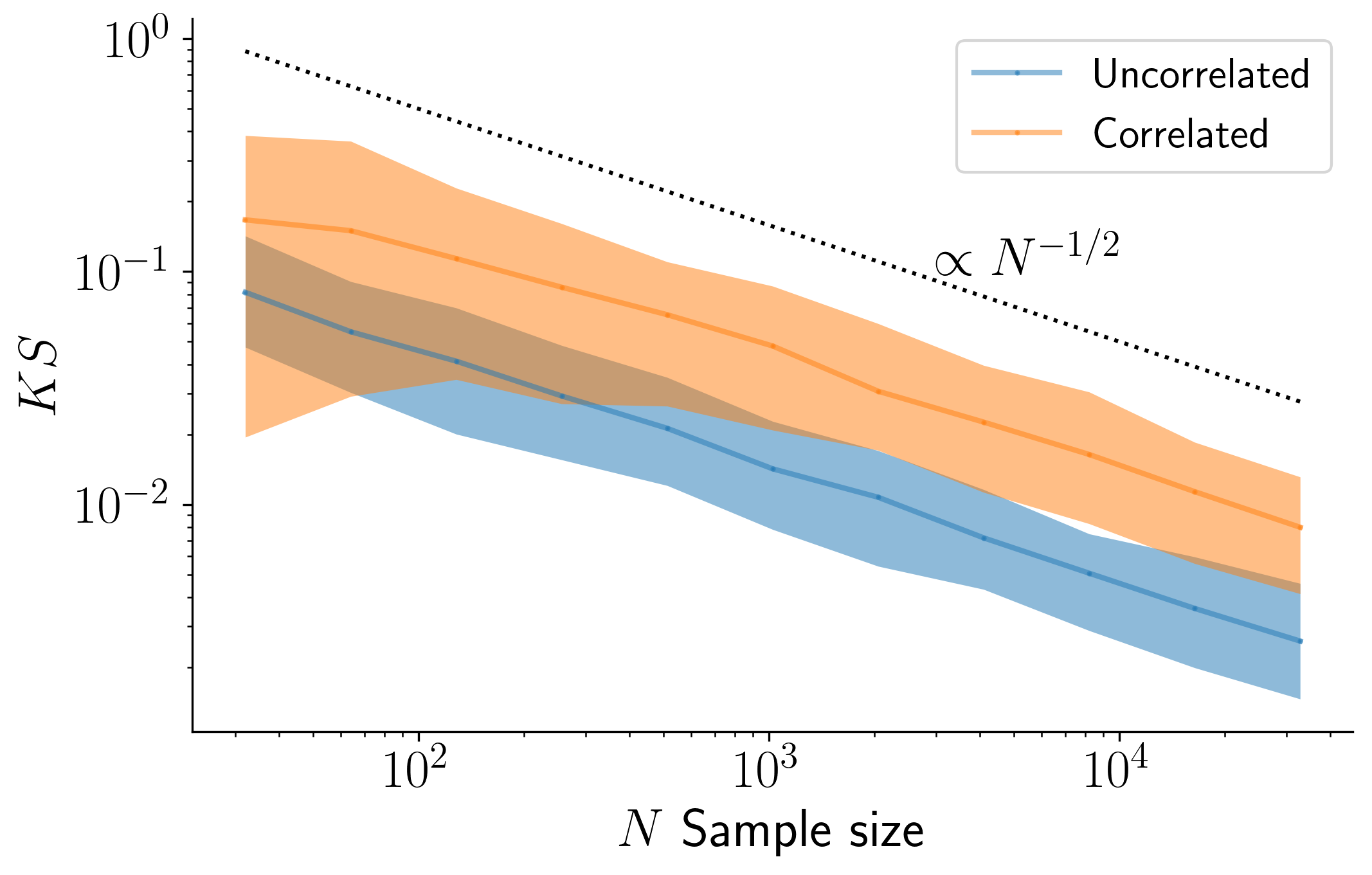}
\end{center}
  \caption{Correlations affect the analysis of statistical laws. Results are shown for the two time series $x(t)$ with a power-law distribution: $p(x) = C x^{\gamma}$, $\gamma=1.7$ shown in Fig.~\ref{fig.synthetic-correlated}. Top: estimation of the power-law exponent $\hat{\gamma}$ for increasing data size $N$. Middle: histogram \index{Kolmogorov-Smirnov distance} of the Kolmogorov-Smirnov distance (KS)~\cite{clauset_power-law_2009} between data and power-law distribution (with $\gamma=\hat{\gamma}$) obtained over $100$ independent time-series of length $N=1,000$. Bottom: dependence of the expected and $95\%$-percentiles of the KS distance (computed over independent realizations) as a function of the sample size $N$. Applying the hypothesis-testing  \index{hypothesis testing} method suggested in Fig.~\ref{fig.illustration-ks} and Ref.~\cite{clauset_power-law_2009} leads to a rejection (e.g., p-value<0.05) for the correlated case for all $N>100$. See Ref.~\cite{gerlach_testing_2019} for  details and Appendix~\ref{chap.appendices} for information on the code used in this figure.}\label{fig.synthetic-correlated2}
\end{figure*}

To investigate this point, following our approach in  Ref.~\cite{gerlach_testing_2019}, we compare two time series that satisfy H1: one that satisfies H2 (independent samples) and on that violates H2 (Markov process of order 1). These two time series are shown in Figure~\ref{fig.synthetic-correlated}, together with their auto-correlation and histogram (distribution). The analysis of these time series in Figure~\ref{fig.synthetic-correlated2} shows that violations of H2 lead to much larger fluctuations of the data around the statistical law than when H2 is satisfied. These fluctuations lead to biased and more uncertain estimations of $\gamma$ and to a rejection of the joint hypothesis. A naive application of statistical tests based on the iid hypothesis, as illustrated in Fig.~\ref{fig.illustration-ks} and proposed in Ref.~\cite{clauset_power-law_2009}, would consider this to be a rejection of the power-law and thus of the statistical law, even though the time series, by construction, follows a power-law distribution exactly (for $N\rightarrow \infty$).

More generally,  \index{hypothesis testing} hypothesis testing of goodness-of-fit are only significant if they lead to a rejection of the tested hypothesis because a non-rejection is {\it not} a confirmation of the hypothesis. The strength of this approach depends on how general the hypothesis being tested is: the more general the hypothesis is (weaker assumptions), the more surprising (significant) a rejection is. By including a very strong assumption that is known to be violated (such as H2 above), the outcome of the hypothesis test is invariably weak (if not meaningless).

\paragraph{Family of distributions~$\theta$ vs. maximum-likelihood distribution $\theta=\theta^*$}

The use of the maximum-likelihood estimator $\hat{\theta}$ in the generation of the surrogate sequences in Fig.~\ref{fig.illustration-ks} \index{surrogate} restricts the analysis that can be performed for the family of distributions in the statistical law (i.e., for all parameters $\theta$). In particular, the model-data comparison applies to the full family of distribution only if the choice of the test statistic (used to quantify the distance between the data and the curve) is so that it remain invariant under different choices of $\theta$ (i.e., a pivotal test statistics. Otherwise, the analysis is restricted to the maximum-likelihood estimated parameters $\hat{\theta}$ and not the complete $\theta$-family of distributions $P(x|\theta)$.

\paragraph{Sample size}

Another characteristic of the hypothesis-testing approach  \index{hypothesis testing} is that it critically depends on the number of observations~$N$. For small $N$, virtually no distribution is rejected -- correctly reflecting the lack of evidence available -- but for large $N$ any small deviation of the proposed law becomes statistically significant, regardless of the size of the effect (for $N\rightarrow \infty$). 

Contrary to controlled experiments or specific observations, in the study of statistical laws the value of $N$ is often not strictly specified. Based on the universality assumption underlying statistical laws, there is a choice of the (size of the) dataset in which they will be tested and it is often possible to increase $N$ by adding more data (larger texts in the analysis of \index{linguistic laws} linguistic laws, more species in the analysis of Kleiber's law, longer time series, etc.). In addition to that, the different representations of statistical laws change the definition of observation with dramatic effects on $N$ (there are many more word tokens than types, many more citizens than cities, etc.). One is often faced with the contradictory situation that the modern availability of larger datasets confirm the observations that motivated the proposition of the laws (using graphical methods), but leads to a rejection of the statistical laws based on statistical tests.

Underlying this point is the idea that the proposed statistical laws are not intended to describe the system in detail, but to capture one non-trivial effect. This is a widespread idea in complex-systems and mathematical modeling, which is difficult to formalise probabilistically and test statistically. It suggests that instead of using hypothesis testing  \index{hypothesis testing} methods based on goodness of fit, one should favour model  comparison \index{model comparison} between simple models (e.g., Kleiber's 3/4 law or the 2/3 scaling). The choice of which models to use in the comparison often includes the question about their theoretical underpinning so that the evaluation of the statistical laws goes beyond standard discussions involving statistical tests. 

\paragraph{Networks}

The limitations mentioned above acquire special characteristics when the data is in form of a network. In particular, the controversial case of ubiquity of the \index{scale-free networks} "scale-free-networks"~\cite{amaral_classes_2000,broido_scale-free_2019,klarreich_clara_scant_2018,serafino_true_2021}- see Sec.~\ref{ssec.scalefree} -- is another example of the limitations of naive maximum-likelihood tests. The iid assumption corresponds to sampling independently node degrees from a power-law distribution. This not only does not correspond to the process in which most networks are sampled, it often leads to unrealistic realizations. In fact, typical realizations of an iid process will lead to non-graphical degree sequences, for instance when the sum is an odd number and therefore no network can be created~\cite{gerlach_testing_2019}.  \index{power-law distribution}

The sampling of networked data plays a key role in the analysis of networks~\cite{crane_probabilistic_2018}, as often an exhaustive sample is impossible (e.g., of the www). In particular, the degree distribution of networks is strongly affected by undersampling the complete network and the \index{scale-free networks} scale-free property discussed in Sec.~\ref{ssec.scalefree} is not invariant under typical undersampling cases~\cite{stumpf_subnets_2005,stumpf_sampling_2005,lee_statistical_2006}.   The question of how the network data was obtained plays a key role in the extent in their comparison to random network models~\cite{crane_probabilistic_2018}. The role of effective sample size in network modeling has also been considered in Ref.~\cite{kolaczyk_question_2015}.

\paragraph{Limitations of statistical tests}
The incompatibility between statistical tests based on the independent  \index{hypothesis testing} hypothesis~\cite{clauset_power-law_2009} with the points raised above should be taken into account when interpreting the implications of statistical tests to the evaluation of the compatibility of the data with statistical laws: a rejection of the hypothesis may be a consequence of the correlations and not necessarily of the deviations of power-law distribution. Testing the validity of a statistical law involves not only the shape of the distribution but also on the generative models because the measured deviations have to be confronted with the expected fluctuations of the generative model. Ultimately, this shows that the question of the validity of a statistical law is interconnected with the question of the generative process that gave origin to it, in violation of the usual statistical-laws approach summarized in Sec.~\ref{ssec.reasoning}.  \index{power-law distribution}

A more pragmatic approach is to consider that the correlation between observations is beyond the scope of the analysis of the distribution alone and accept that the statistical law cannot be easily tested in full generality (even in one given dataset). Often the best we can do is to compare alternative parametric functions (assuming that the assumption of independence will affect all of them similarly) and report whether the proposed law or another distribution provides the best description. Fortunately, many of the relevant questions underlying statistical-laws studies can be addressed based on such model-comparison approach, bypassing the enticing (yet ill-defined) question of absolute validity of a law.


\section{Statistical methods for complex data}\label{sec:stat-meth-compl}

The previous sections review the three traditional data-analysis approaches used to evaluate statistical laws: graphical methods, linear regression, and likelihood-based approaches. This section will briefly discuss other approaches that have been proposed either more recently or in specific contexts. A series of methods -- presented in Secs.~\ref{ssec.undersampling}-\ref{ssec.inference} -- can be seen as addressing the limitations of na\"ive maximum-likelihood methods discussed in Sec.~\ref{ssec.limitationslikelihood}. Other approaches -- presented in Sec.~\ref{ssec.otherDataMethods} -- go back to the sociophysics roots of statistical laws and apply more general scaling and statistical-mechanics arguments. 

\subsection{Undersampling}\label{ssec.undersampling}

\begin{figure*}[!ht]
\begin{center}
  \includegraphics[width=0.75\linewidth]{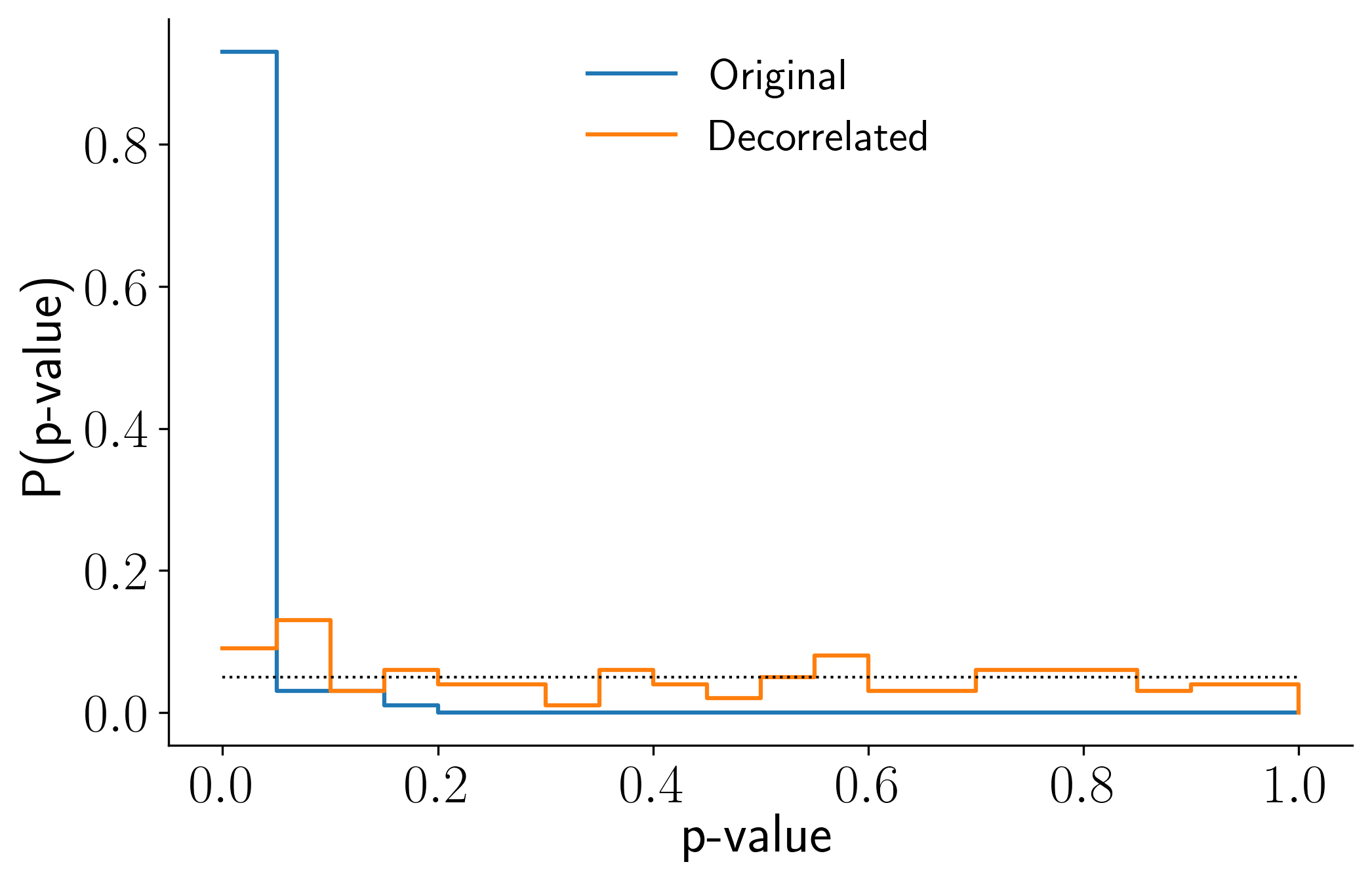}
\end{center}
  \caption{Undersampling correlated data can change the outcome of statistical tests. The goodness-of-fit test described in Fig.~\ref{fig.illustration-ks} was applied for the synthetic data shown in Figs.~\ref{fig.synthetic-correlated} and~\ref{fig.synthetic-correlated2}. The results show the $p-value$ obtained over $100$ realizations of the process. When the test is applied to the original data, the p-value is concentrated at small values and leads to a rejection of the  \index{hypothesis testing} hypothesis. When the data is undersampled, the p-value is uniformly distributed and leads to a rejection in a fraction of cases similar to the chosen threshold value (e.g., $0.05$). See Appendix~\ref{chap.appendices} for the code used in this figure.}
\label{fig.synthetic-correlated3}
\end{figure*}

The limitations of likelihood-based methods discussed in Sec.~\ref{ssec.limitationslikelihood} are typical and well-known in Statistics, which has a broad literature and many methods that address each of the specific raised points. For instance, statistical tests that go beyond the assumption of independent data are discussed in Refs.~\cite{gasser_theo_goodness--fit_1975,weiss_marc_s_modification_1978,chicheportiche_goodness--fit_2011} and possibilities to account for composite hypothesis in Ref.~\cite{shaffer_multiple_1995}.

One of the key ideas how to address the issue of correlated samples is to estimate an "effective sample size" $N^*$ that can be treated as independent observations. This can be achieved, for instance, by quantifying a correlation time $\tau^*$ in the data, computing the effective sample size as $N^* = N/\tau^*$, and undersampling the original sequence to size $N^*$.  This approach was proposed and tested in the case of (frequency-distribution) statistical laws in Ref.~\cite{gerlach_testing_2019}. Starting from a time series $x_t$ for $t=1, \dots, N$, the first step is to compute the autocorrelation \index{autocorrelation function} function $C(\tau)$  at lag time $\tau$ as~\cite{kantz_nonlinear_2004}
\begin{equation}\label{eq.autocorrelation}
    C(\tau) = \frac{1}{\sigma_x^2}\langle (x_t-\langle x \rangle)(x_{t-\tau}-\langle x\rangle)\rangle = \frac{\langle x_tx_t-\tau\rangle -\langle x \rangle^2}{\sigma_x^2},
\end{equation}
and compute the the correlation time $\tau^*$ as the time $C(\tau=\tau^*) \approx 0$ or the characteristic scale of decay of $C(\tau)$.

The key point obtained from the application of the undersampling method is that tests that lead to a rejection at sample size $N$ often do not reject at $N^*<N$. This reflects that there is a reduced evidence against the law due to the correlation in the data, an effect that is increasingly important as the correlation $\tau^*$ increases. For instance, in Fig.~\ref{fig.synthetic-correlated3} we applied this approach to the synthetic time-series analyzed previously -- in Figs.~\ref{fig.synthetic-correlated}-\ref{fig.synthetic-correlated2}-- and find that it succeeds in preventing the false rejection of power-law distribution due to correlation. In Ref.~\cite{gerlach_testing_2019}, the auto-correlation time for a sequence of earthquakes was estimated to be of more than 2 years, dramatically reducing the sample size and thus changing the associated p-value of the analysis of the \index{Gutenberg-Richter's law} Gutenberg-Richter law to be not significant. This, alone, is not an evidence of the validity of \index{Gutenberg-Richter's law} Gutenberg-Richter law, it reflects simply the lack of evidence to reject it in the considered data, contrary to the conclusion drawn if correlations are ignored.   \index{power-law distribution}

\subsection{Constrained Surrogates}\label{ssec.constrained}

\index{surrogate}

\begin{figure}[!h]
\begin{center}
\includegraphics[width=1.1\textwidth]{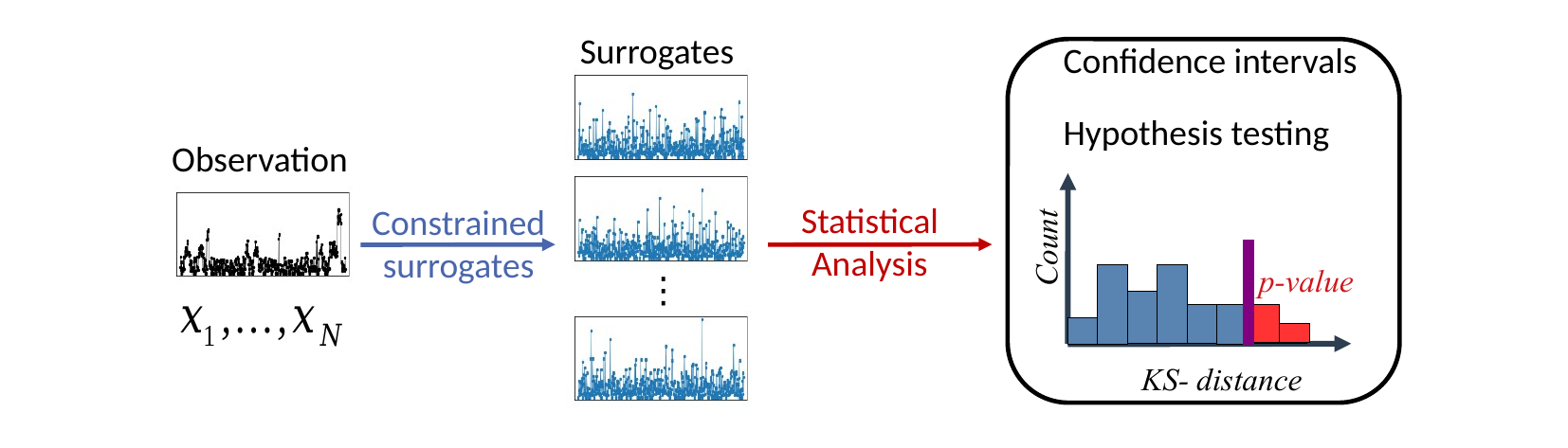}
\caption{Illustration of the use of constrained surrogates for the analysis of statistical laws. In contrast to the standard case shown in Fig.~\ref{fig.illustration-ks}, constrained surrogates are generated directly from the original data (time series) and are not restricted to the maximum likelihood exponent $\hat{\theta}$. The statistical analysis accounts therefore for the full family of distribution in addition to other constraints imposed (e.g., temporal correlations). The case of surrogates constrained to power-law distribution has been introduced and applied to statistical laws in Ref.~\cite{moore_nonparametric_2022}.}\label{fig.illustration-surrogates}
\end{center}
\end{figure}

Another approach to improve over standard likelihood-based methods is to use surrogate methods~\cite{kantz_nonlinear_2004,theiler_using_1991,theiler_testing_1992,small_applying_2002}. Starting from a sequence of observations (typically a time series $x_t$), the idea is to generate surrogate sequences $\tilde{x}_t$ that can then be directly used in the statistical analysis as illustrated in Fig.~\ref{fig.illustration-surrogates}. The difference to the traditional approach -- Fig.~\ref{fig.illustration-ks} -- is that the generation of the surrogates can be based on more general null hypotheses and do not necessarily need to involve the maximum-likelihood estimation of parameters $\hat{\theta}$ and the generation of data based on an iid sample. 

A simple example of the approach underlying surrogates is to test whether $x_t$ is correlated. A suitable surrogate in this case is obtained shuffling the original time series. In order to perform the  \index{hypothesis testing} hypothesis testing step, a test statistic that quantifies the temporal dependence of the data should be chosen, such as the value of the autocorrelation function~(\ref{eq.autocorrelation}) at a suitable lag time $\tau$, e.g., $C(\tau=1)$. Comparing the value in the original time series $x_t$ and in a sequence of shuffled surrogates, we can estimate the probability that the observation in the original time series is compatible with observations in an uncorrelated (finite-size) sequence. The surrogate obtained shuffling the original sequence is based on the null-hypothesis that the data is independently sampled and can thus be used to test this null hypothesis in the data. Shuffling does not generate suitable surrogates to test for frequency-distribution statistical laws because the estimated distribution (histogram) remains unchanged. More generally, methods of constrained surrogates~\cite{theiler_using_1991,theiler_testing_1992,small_applying_2002} consider surrogates that fix (constrain) properties of time series compatible with a chosen null-hypothesis, at the same time allowing all other aspects to vary randomly. 

In Ref.~\cite{moore_nonparametric_2022} we applied constrained surrogates to the study of statistical laws in form of power laws. 
The idea key idea is to generate surrogate sequences for which the likelihood function~(\ref{eq.Liid}) for the proposed frequency distribution (under the independence assumption) is the same for {\it all} power-law exponents $\gamma$. This implies that any likelihood-based inference applied to any of the surrogate sequences would lead to the same outcome as their application to the original sequence $x_t$. As such, comparing test statistics between the surrogates and the sequence allows us to test for power-law distribution in the data but it is not restricted to a single exponent. This addresses one of the limitation discussed in Sec.~\ref{ssec.limitationslikelihood}\footnote{In particular, this method gives the freedom to use more general test statistics, that focus on properties of interest, while the traditional approach is restricted to pivotal test statistics.}. The surrogate method proposed in Ref.~\cite{moore_nonparametric_2022} is valid for time series $x_t$ in which $x\in \mathbb{N}$ and samples uniformly sequences of $N$ values which preserve the product $\Pi_{t=1}^N x_i$ (notice that the likelihood function~(\ref{eq.likelihod}) for the power law distribution in this case depends only on this product). 

\begin{figure*}[!ht]
\begin{center}
  \includegraphics[width=0.42\linewidth]{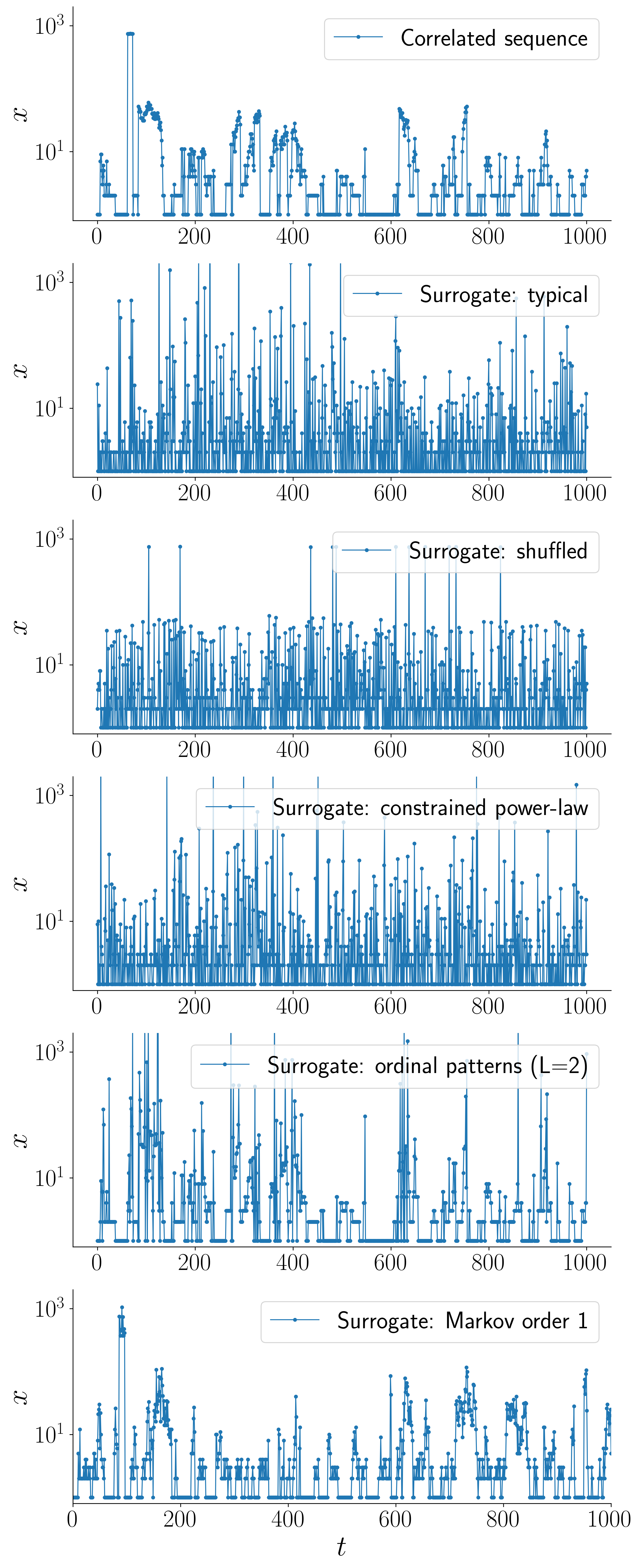}\includegraphics[width=0.42\linewidth]{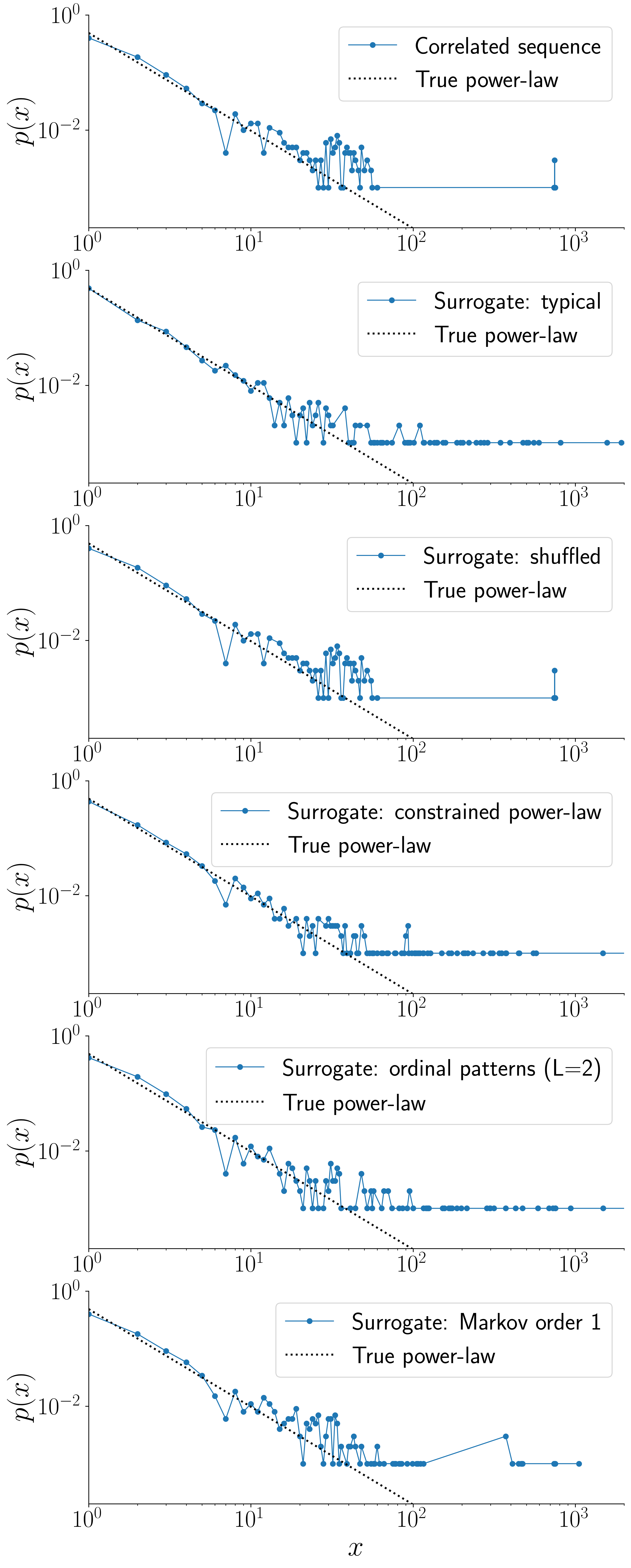}
\end{center}
  \caption{Different power-law surrogates. The first row corresponds to the synthetic time series $x_t$ constructed as a Markov process that leads to correlations and have $p(x) \sim x^{-\gamma}, \gamma=1.7$, as used in Figs.~(\ref{fig.synthetic-correlated})-(\ref{fig.synthetic-correlated3}). The other five rows show different surrogate sequences computed applying four different methods to the series in the first row: typical (independent sampling from $p(x)$ with maximum likelihood exponent $\hat{\gamma}$, as used in Ref.~\cite{clauset_power-law_2009} and illustrated in Fig.~\ref{fig.illustration-ks}), shuffling, constrained power-law, ordinal patterns with $L=2$, and Markov of order 1 (see Sec.~\ref{ssec.constrained} and Ref.~\cite{moore_nonparametric_2022} for details). The left column shows the complete time series $x(t)$, with $t \in [1,\ldots, N]$. The right column shows the normalized histogram of $x$ in the corresponding time series, together with the theoretical power-law as a dotted line. See Appendix~\ref{chap.appendices}for the data and code used in this analysis.}
\label{fig.constrained}%
\end{figure*}

In addition to the constraint in the likelihood function, constraints on the temporal order of appearance of $x_t$ can be imposed. In Ref.~\cite{moore_nonparametric_2022}, this is done for the discrete power-law case by imposing Markov transition probabilities or the rank order  of events (ordinal patterns). This can be imposed up to an arbitrary order (window size)  allowing for a tuning on the strictness of the constraints on the correlation. Figure~\ref{fig.constrained} shows different types of surrogates obtained from the synthetic-correlated time series discussed in Sec.~\ref{ssec.limitationslikelihood}. The typical surrogate -- generated from independent sampling as represented in Fig.~\ref{fig.illustration-ks} and suggested in Ref.~\cite{clauset_power-law_2009} -- shows much smaller deviations from the true power-law than the other surrogates, leading to a rejection of the hypothesis if goodness-of-fit tests are applied. In contrast, constraints which include temporal correlations of the original time series lead to surrogate series that more closely resemble the input sequence and show similar fluctuations in $p(x)$.  \index{power-law distribution}

Constrained surrogates overcome also the limitation of simply shuffling the data as it allows for the generation of previously unobserved values of $x$ (in particular, in the tail).  Ref.~\cite{moore_nonparametric_2022} found (using synthetic series) that the statistical tests based on constrained surrogates are particularly useful for small $N$ and when one is interested in more general test statistics.

While constrained surrogates overcome the most simplistic assumptions of likelihood-based methods, it is important to note that some limitations of  \index{hypothesis testing} hypothesis-testing methods remain. Useful information is obtained from hypothesis-testing methods when they lead to a rejection of the null hypothesis. Accordingly, traditional uses of surrogates are designed based on {\it null} models that do not include key properties of the time series that we wish to test and highlight. For instance, a traditional application is to show the presence of non-linearities in the dynamics by constructing surrogates based on the null hypothesis of linearity~\cite{theiler_using_1991}. In contrast, surrogates for hypothesis testing of statistical laws -- not only the surrogates based on the independence hypothesis in Fig.~\ref{fig.illustration-ks} but also in the case of constrained surrogates in Fig.~\ref{fig.illustration-surrogates} -- arise from the distributions proposed by the functional form of the law and do not omit the properties we wish to test and highlight. \index{hypothesis testing} \index{surrogate}

\subsection{Statistical inference of mechanistic models}\label{sec:stat-meth-comp}\label{ssec.inference}

One of the motivations for the use of constrained surrogates is the possibility to incorporate additional properties of the data (e.g., temporal correlations) into the (null) models underlying the surrogates. A natural extension of this idea is to formulate generative models that contain essential features of the process generating the data and perform statistical inference using standard (likelihood-based) techniques. In the case of statistical laws, this involves breaking the traditional division between statistical law (as an empirical law) and the mechanistic models (as a theoretical explanation), summarized in Sec.~\ref{ssec.reasoning} and illustrated throughout Chap.~\ref{chap.examples}. The underlying models can be probabilistic versions of the traditional mechanistic models used to explain the law or they can incorporate the statistical law of interest explicitly. 

In line with the tradition of simple models to explain the statistical laws, the models suitable for such statistical inference will typically contain severe simplifications of the underlying generative process. It is thus, again, expected that the deviations -- between the outputs of these models and the real data -- will be statistical significant (for sufficiently large number of observations $N$). Instead of looking for a statistical test of the validity of the model, the focus is thus on performing model comparison \index{model comparison} between different such models, ideally including examples in which the statistical law is present and examples in which it is absent. 

While a tendency towards inferential approaches to study statistical laws is common in studies of different laws, the models and methods are often specific to each case as they try to capture case-specific characteristics. Below we discuss examples of this approach in two prominent cases of statistical laws: \index{scale-free networks} scale-free networks and \index{Urban scaling laws} urban scaling laws. 

\paragraph{Scale free networks}

In the analysis of networks, statistical inference allows for a rigorous connection between data and random-graph models. The importance of such inferential approaches has a long tradition in Statistics and social-network analysis~\cite{crane_probabilistic_2018}, and is increasingly being recognized in the study of "complex networks" and "network science"~\cite{peel_statistical_2022}. 

The main statistical law proposed to describe complex networks is the power-law degree distribution leading to \index{scale-free networks} scale-free networks, reviewed in Sec.~\ref{ssec.scalefree}. The inferential approach to study this case aims to go beyond the simple maximum-likelihood analysis of degree distribution (see Sec.~\ref{ssec.ml-freq} and Ref.~\cite{clauset_power-law_2009,broido_scale-free_2019}). The starting point of the analysis are the mechanistic models proposed to explain the statistical law, in particular the preferential attachment model proposed in Ref.~\cite{barabasi_emergence_1999}. Here it is important to note that networks generated from preferential attachment model are very special in the space of random-graph models with a scale-free degree sequence~\cite{judd_what_2013,zhang_exactly_2015,small_growing_2015,chakraborty_searching_2022}. This emphasizes once more the difference between claiming (and testing) the ubiquity of (i) scale-free networks (regardless of the generative process) vs. (ii) the preferential-attachment process.

In Refs.~\cite{pham_pafit_2015,falkenberg_identifying_2020}, different approaches are proposed to estimate parameters for preferential-attachment type models from data of (temporal) networks. The key point from our point of view is that this involves a direct comparison between data and network model which is not mediated by the evaluation of whether the degree-distribution is power-law or not, i.e., in contrast to the traditional approach to study statistical laws (see Sec.~\ref{ssec.reasoning}).

\paragraph{Urban scaling laws}

Another example in which an inferential approach to study statistical laws has been recently applied is the case of \index{Urban scaling laws} urban scaling laws, reviewed in Sec.~\ref{ssec.urbanscaling} and which illustrated the general methodological discussions of Sec.~\ref{ssec.likelihoodscaling}. The idea we advance here is to propose a probabilistic model for the generation of the observable $y_i$ in city $i$ with population $x_i$ which allows for an explicit computation of a likelihood function~(\ref{eq.likelihod}) that can be used for the statistical analysis of (any) observed data. From approximately 20 models of urban scaling laws reviewed in Ref.~\cite{ribeiro_mathematical_2023}, only our models-- from Refs.~\cite{leitao_is_2016,altmann_spatial_2020}, which  we review below  -- follows this approach. 

\begin{figure}[!ht]
    \centering
    \includegraphics[width=1\textwidth]{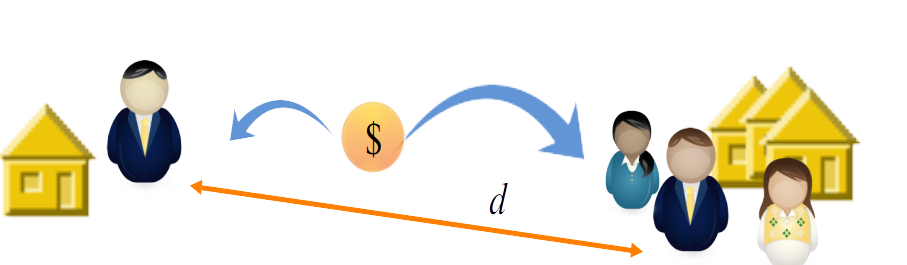}
    \caption{Illustration of the generative model used in the inferential approach to \index{Urban scaling laws} urban scaling laws. Instead of directly modeling how the values $y_i$ are attributed to each city $i$ with population $x_i$, the model specifies how the $Y=\sum_i y_i$ tokens (\$) are distributed to the $X=\sum_i x_i$ inhabitants of different cities. The probability of attribution of a token to an inhabitant depends on the size of the city in which she lives and on the distance $d$ between cities.}
    \label{fig.illustration-tokenUrban}
\end{figure}

A common element of many of the models is the explanation of the non-linear scaling $y \sim x^\beta, \beta \neq 1$ in \index{Urban scaling laws} urban scaling laws based on the increased possibilities of interactions to citizens of larger cities. The argument is that these interactions make their per-capita production more efficient and reduce the per-capita need of infrastructure. Accordingly, the starting point for the generative model is to consider the probability $p(j)$ that a token is attributed to individual $j$ who lies in a city $c(j)$, as illustrated in Fig.~\ref{fig.illustration-tokenUrban}. Depending on the data $y$, a token can be, for instance, a dollar of GDP \index{Gross Domestic Product, GDP} or an unit of CO2 emission.

More formally, consider $j=1,...,M$ individuals living in  $i=1,...,N$ cities. A total of $Y \equiv \sum_i y_i$ tokens are (randomly) assigned to the individuals with probability 
\begin{equation}\label{eq.pj}
p(j) = \frac{A_j^{\beta-1}}{Z(\beta)},
\end{equation}
where $A_{j}$ is the total attractiveness due to all interactions of $j$ and $Z({\beta})$ is the normalization constant so that  $\sum_j^M p(j)=1$.  The attractiveness $A_j$ is computed as the sum of pairwise interactions $a_{j,j'}$ between individuals $j$ and $j'$ separated by a distance $d=d_{j,j'}$. We obtain $A_j$ as the total interaction of $j$ and all other individuals $j'$ by summing over all $j'$
\begin{equation}\label{eq.Aj}
A_j = \sum_{j'\neq j} a_{j,j'}(d_{j,j'}).
\end{equation}
Different interaction functions $a(d)$ define different models of interaction of individuals, ranging from simple cases -- such as a  constant per-capita ratio ($\beta=0$, arbitrary $a(d)$) or interactions restricted only within cities --  to models incorporating the spatial location of the cities -- such as the ones specifying a gravitational law or exponential decay in $a(d)$.

Assuming that the tokens are attributed at random to each individual and that all individuals in the same city are indistinguishable, we computed the log-likelihood of these models as~\cite{altmann_spatial_2020}
\begin{equation}\label{eq.likelihoodUrban}
\ln \mathcal{L}(\theta) = \ln Y! - \sum_{i=1}^N \ln(y_i!) + \sum_{i=1}^N y_i \ln
 \left(\frac{x_i A_i^{\beta-1}}{Z(\beta)}\right),
\end{equation}
where
\begin{equation}\label{eq.Ai}
A_i = \sum_{j',c(j)=i} a(d_{j,j'})= \sum_{i'} x_{i'} a(d_{i\equiv c(j),i'\equiv c(j')}).
\end{equation}
is the attractiveness of the city $i$.

This allows for models of increasing complexity to be considered, leaving for the data analysis not only to estimate the scaling parameter $\beta$ but also the comparison \index{model comparison} between the different models. In Tab.~\ref{tab.urbanscaling} the results for four different models are shown for one dataset. They show that the estimated value of $\beta$ differ from the one estimated through linear regression (similar to what we reported in Fig.~\ref{fig.fitScalingUK}) and that the more complex model, which account for the spatial interactions of close by cities, provides a better description of the data (smaller description length).

\begin{table}[!bt]
  \begin{center}
    \small
  \begin{tabular}{|c c | c | c|| c | c | c|}
    \hline
    \multicolumn{4}{|c|}{Model}  & \multicolumn{3}{|c|}{Results in Australian Data}\\
    \hline
    \multicolumn{2}{|c|}{} & Interaction $a(d)$ & Parameters $\theta$ & $\hat{\alpha}$& $\hat{\beta}$ & Description length $\mathcal{D}$  \\
    \hline
    Per capita & P & - & - & - & 1 & 2852512 \\
     City & C & $\delta(d)$ & $\beta$ & - & $1.19 \pm 0.04$ & 2830289 \\
    Gravitational & G & $1/(1+(d/\alpha)^2)$ & $\alpha,\beta$ & $8.3$km & $1.20 \pm 0.05$ & {\bf 2830210} \\
    Exponential & E & $e^{-d\ln 2/\alpha}$ & $\alpha,\beta$  & $9.5$km &$1.20 \pm 0.04$&2830271\\
    \hline
  \end{tabular}\label{tab.urbanscaling}
    \caption{Four probabilistic models of \index{Urban scaling laws} urban scaling law applied to data of the income of Australia's cities. The left columns of the table specify the models discussed in Sec.~\ref{ssec.inference} and in Ref.~\cite{altmann_spatial_2020}, including the $a(d)$ introduced in Eq.~(\ref{eq.Aj}).  The parameter $\alpha$ can be interpreted as a characteristic distance of interaction between individuals (measured in $km$). The results reported in the right columns of the table were obtained using the likelihood~(\ref{eq.likelihoodUrban}) with the data of Australia's 102 significant urban areas. The last column corresponds to the description length~\cite{grunwald_minimum_2007,altmann_spatial_2020} of each model, with the lowest valued (best model) obtained for the gravitational model. For comparison, the linear regression of $\log y$ vs. $\log x$ yields the scaling exponent $\hat{\beta}=1.15$. See Ref.~\cite{altmann_spatial_2020} for details on the models and Appendix~\ref{chap.appendices}for the data and code used in this analysis.}
\end{center}
\end{table}

\subsection{Other methods}\label{ssec.otherDataMethods}

The methods proposed in this section so far-- i.e., in Secs.~\ref{ssec.undersampling},~\ref{ssec.constrained}, and~\ref{ssec.inference} -- embrace the probabilistic formulation of statistical law used in likelihood-based methods and try to overcome the limitations of the more simplistic maximum-likelihood recipes discussed in Sec.~\ref{ssec.limitationslikelihood}. This tendency inevitably leads to a data analysis of statistical laws that evaluates not only the law itself but that is intrinsically connected to the underlying model of the generative process. This is an important realization that shows that the upport for the validity of a law estimated from a given data is not independent of the generative model proposed to explain the law. At the same time, this approach goes against the historical and proposed use of statistical laws reviewed in Sec.~\ref{ssec.reasoning} and reported in the examples of Chap.~\ref{chap.data}, where statistical laws are viewed as empirical laws that motivate and justify the introduction of simple mechanistic models to explain them. This motivates us to consider also data-analysis methods that follow a different approach, as the ones listed below.

\paragraph{Characterizing fluctuations}

One of the key observations in the data analyses of this section (e.g., in Figs.~\ref{fig.synthetic-correlated2} and~\ref{fig.constrained}) is the discrepancy between the fluctuations and deviations expected from the simplistic models and those observed in data, including data that is closely described by the proposed statistical law. Instead of ignoring these observations, considering them as a sign that the laws are violated, or trying to model them in detail, an alternative approach is to recognize the existence of these fluctuations,  try to characterize them, and understand their origin. Very often, these fluctuations show statistical regularities that can be described through simple models, a process that is similar to the traditional approach used in statistical laws themselves. 

Consider the case of statistical laws proposed to describe different patterns related to the appearance of words in texts, as discussed in  Secs.~\ref{ssec.zipf},~\ref{ssec.heaps},~\ref{ssec.burstywords}, and~\ref{ssec.linguistic-laws}. A typical simplifying assumption included in models and estimations is to consider that words appear randomly distributed in a text or, similarly, that they are used with a fixed probability (independent of time) equal to its overall frequency. This random (or "bag-of-words") assumptions are obviously violated in texts, but they allow for analytical calculations that reveal insightful connections, such as the relationship between Zipf's and Herdan-Heaps' laws discussed in Secs.~\ref{ssec.zipf} and~\ref{ssec.heaps}. Naturally, the limitations of this simplifying assumption appears when looking quantitatively at data-model comparisons or when considering different properties of the text. Typically, the observation is that the fluctuations and variations in the data are much broader than the expectation under the random assumption. Two examples in which this was observed and led to further proposals how to characterize these laws are:

\index{Herdan-Heaps' law}
\begin{itemize}
    \item The typical vocabulary size $y$ (number of word types) for corpora (books) of a given length $x$ (number of word tokens) shows a very similar scaling -- Herdan-Heaps' law -- as the one predicted assuming Zipf's law\index{Zipf's law}  and the bag-of-word assumption (see, for instance, Fig.~\ref{fig.heaps} or Ref.~\cite{gerlach_stochastic_2013}). However, when looking at the variation of $y$ over different texts (books, articles) with similar $x$, they show much larger variation than the predicition based on the random assumption (in particular, for large $x$). We showed in Ref.~\cite{gerlach_scaling_2014} that these fluctuations scale with text size $x$ with a different exponent than the one based on random prediction, and characterized this scaling using Taylor's law~\cite{eisler_fluctuation_2008}. In turn, a mechanistic explanation based on quenched disorder or topics is able to explain these observations.  \index{Taylor's law}
    \item The assumption of random appearance of words in texts predicts a narrow (Poisson) distribution \index{Poisson distribution} of the space between two consecutive appearances of the same word. As discussed in Sec.~\ref{ssec.burstywords}, observations in texts lead to a broader variation that is better characterized by a stretched-exponential distribution~\cite{altmann_beyond_2009,corral_universal_2009}. The proposal of such new statistical law has motivated the proposal of a renewal process for the appearance of words in a text, i.e., with independent inter-event times. Again, this simplifying assumption allows for analytical derivation of the new law but it is not only violated at closer inspection it is incompatible~\cite{altmann_beyond_2009,altmann_origin_2012} with the observations and proposal that texts show \index{long-range correlations} long-range correlations~\cite{schenkel_long_1993,tanaka-ishii_long-range_2016}.  \index{stretched exponential distribution}
\end{itemize}

These examples show the recursively application of the standard statistical law approach to overcome the limitations of simplistic assumptions. Another similar case was reported in Ref.~\cite{miotto_stochastic_2017}, when models to explain the power-law distribution in the distribution of view of online videos have revealed non-Gaussian fluctuations around the typical growth process. In turn, the fluctuations were characterized by L\'evy distributions and led to a modification of generative model from a Gaussian to  \index{Gaussian distribution} a Levy-noise stochastic differential equations.  \index{power-law distribution}

\paragraph{Scaling and critical phenomena}

Methods based on critical phenomena have a long tradition in the study of statistical laws~\cite{bak_how_2013}. They are particularly useful when the scale of analysis can be varied or arbitrarily chosen, in which case scale invariance and the appearance of scaling laws can be investigated quantitatively by comparing (graphically) the collapse of curves obtained for different scales (possibly after re-scaling). 
These re-scaling techniques have been applied to investigate different statistical laws:

\begin{itemize}
    \item in urban systems, the analysis of (cluster) population distributions~\cite{rozenfeld_area_2011,fluschnik_size_2016} and flows of population~\cite{schlapfer_universal_2021} was performed  varying the spatial scale (or area). 
    \item in \index{earthquakes} earthquake data, Refs.~\cite{bak_unified_2002,christensen_unified_2002,corral_local_2003,corral_long-term_2004} considered the analysis of earthquake magnitudes and inter-event times.
    \item in networks, Ref.~\cite{serafino_true_2021} applied these techniques varying data sizes to analyze the controversy of ubiquity of \index{scale-free networks} scale-free distributions, finding scale-invariance compatible with expectations of scale-free networks. 
\end{itemize}

\chapter{Synthesis: statistical laws in context}\label{chap.synthesis}

In the previous Chapters we described how statistical laws have been used in complex-systems research: Chap.~\ref{chap.introduction} provided a brief historical overview and a working definition of statistical laws; Chap.~\ref{chap.examples} listed examples from different disciplines, focusing on the similar role played by statistical laws in connecting data and models; and Chap.~\ref{chap.data} introduced increasingly sophisticated quantitative methods that have been employed to test, fit, and explore statistical laws. In this last chapter, we will move away from the description of how statistical laws have been used and, instead, focus on the role they can and should play in the study of complex systems. We will propose different ways in which methods and interpretations can be coherently employed, highlight possible pitfalls, suggest good practices, and speculate about the future of statistical laws in data-driven research.

\section{An unified view on statistical laws}

The main motivation and crucial point of this monograph is to argue for an unified treatment of statistical laws (in complex-system research). This has long been done for the case of power-law distributions~\cite{zipf_human_2012,simon_class_1955,mitzenmacher_brief_2004,newman_power_2005,simkin_re-inventing_2011} and scaling laws~\cite{west_scale_2018}, but it is further expanded here not only to combine these two types but also to include statistical laws more generally (as defined in Sec.~\ref{ssec.definition}). The justification for this unified approach is not that the same functional forms or generative models apply for different laws, as has been the motivation for the unified treatment of power-law distributions (e.g., underlying rich-get richer mechanisms) and scaling laws (e.g., connections to fractal geometry and critical phenomena). Instead, the more abstract commonality we explore in this monograph is based on the conceptual use of statistical laws in different settings and by various research communities. This involves both the traditional uses of statistical laws -- as summarized in Sec.~\ref{ssec.reasoning}, highlighted throughout the examples in Chap.~\ref{chap.examples}, and further explored in Sec.~\ref{ssec.traditional} below-- and the methodological debates around the validity and role of statistical laws -- discussed at the start of Chap.~\ref{chap.data} and reviewed in Sec.~\ref{ssec.persistent} below.

\subsection{Traditional approach: potential and limitations}\label{ssec.traditional}

The most common use of statistical laws in complex systems can be summarized as follows.

\begin{tcolorbox}
\noindent Traditional approach to statistical laws in complex systems: \\ an initial stage of {\bf 1. formulation and empirical validation of the law} is followed by the proposal of {\bf 2. generative models} that explain the origin of the law and by explorations of {\bf 3. consequences of the law}. 
\end{tcolorbox}

\noindent This schematic description mimics the (re-constructed and idealized) chronological steps in which the investigation of a statistical law typically happened, even if (more recently) more than one of these steps are done already in the same paper.

{\it A posteriori}, the causal description becomes
\begin{tcolorbox}
\noindent {\bf generative model} $\mapsto$ {\bf data satisfies empirical law} $\mapsto$ {\bf consequences of the law}.
\end{tcolorbox}
\noindent Often, the statistical law is considered as a prediction of the generative model so that new observations of the law are not only taken as a corroboration of its validity but also as evidence that the mechanistic generative model is in action in the system underlying the new observation. A recent example in which this (often problematic) simplification has been applied is the case of considering observations of \index{scale-free networks} scale-free networks as evidence of the preferential-attachment model (as discussed in Sec.~\ref{ssec.scalefree} and Ref.~\cite{small_growing_2015}).

\paragraph{Benefits of the traditional approach}

This traditional approach is both convenient and useful, as it splits the problem in parts and allows each of them to be investigated separately. For instance, when discussing the mechanistic models that generate the law one can focus on simplified settings (e.g., \index{Simon, Herbert} Simon's model discussed in Sec.~\ref{ssec.pareto}) or general scaling arguments (e.g., in the role of city areas in urban scaling laws discussed in Sec.~\ref{ssec.urbanscaling}), leaving the cumbersome analysis of data aside. Similarly, the consequences of the statistical laws (e.g., fat-tailed  \index{fat-tailed distribution} distributions) can be explored using simple distributions (e.g., power-laws) instead of working directly with the data or with the generative process. In fact, a major motivation and benefit of having simple parametric functions in the formulation statistical laws is to allow for analytical calculations and estimations that allow for an exploration of the consequences of the data feature the law aims to capture. For instance,  power-law distributions allow for the estimation of the probability of unobserved extreme events and to establish an analytical relationship between the exponents of Zipf's and Herdan-Heaps' laws.   \index{power-law distribution}
\index{Herdan-Heaps' law}

Despite the recurring controversies around the validity of statistical laws, recent reviews on historical laws recognize their overall contribution to their fields. For instance, reflecting on \index{Pareto's law} Pareto's law almost a century after its proposal, Persky indicates~\cite{persky_retrospectives_1992}

\begin{quote}
  {\it`For all the excesses of the Paretian camp followers, there remains the significant insight that the history of all hitherto existing society is a history of social hierarchies. There is the feel of structure behind income distributions. Almost all income distributions are continuous, unimodal, and highly skewed. We have no examples of uniform distributions or egalitarian distributions or strikingly trimodal distributions. Something is going on here.''}
\end{quote}

Similarly, the significance of the connection between Gibrat's process and Zipf's law\index{Zipf's law} is recognized by Gabaix as~\cite{gabaix_power_2009}

\begin{quote}
  {\it ``regardless of particulars driving the growth of cities (e.g., their economic role), as soon as cities satisfy Gibrat's law with very small frictions, their population distribution converges to Zipf's law\index{Zipf's law}. Power laws give the hope of robust, detail-independent economic laws''}
  \end{quote}
Common to this two quotes is the indication that the contribution is on capturing general tendencies that exist in the data or system, and not in the ``particulars'' or in a strict obedience of the system to the law, interpretations that often lead to ``excesses'' in the expectations around what statistical laws can deliver.

\paragraph{Limitations of the traditional approach}

The traditional approach has also important limitations that affect the applicability of statistical laws. A clear example is that, alone, it is unable to decide between different generative models that explain the same statistical law. Another important limitation, discussed in detail in Sec.~\ref{ssec.limitationslikelihood}, is the difficulty to reach a decision about the validity or falsification of statistical laws based alone on the application of standard statistical tests to data. In a na\"ive application of the traditional approach one would seek to have a definite yes/no answer on the validity of the law, in order to safely move on towards the next two stages (explanation of the law and exploration of its consequences). Ideally, such a decision would be  based only on the application of statistical methods to compare the agreement between the proposed curve and data (i.e., independently from the proposed generative model). One limitation of this approach, discussed in Secs.~\ref{ssec.representationmatters}, is that different conclusions can be reached depending on the representation of the law chosen for the analysis -- even if they are analytically equivalent -- and depending on the statistical methods used in the analysis. Unavoidably, methods that ignore detailed generative models will contain simplifications that can lead statistical methods to reject the law if sufficient data is available. 
Attempts to go beyond the traditional approach have led to similar methodological developments in different cases, in particular the tendency towards more sophisticated quantitative methods that incorporate aspects of the data and mechanistic models (as discussed in Sec.~\ref{sec:stat-meth-compl}). Ultimately, this tendency breaks down the separation between data analysis and mechanistic models present in the traditional approach. 

The limitations discussed above were also highlighted in Piantadosi's review of Zipf's law\index{Zipf's law} of word frequencies~\cite{piantadosi_zipfs_2014} 

\begin{quote}
  {\it  ``Word frequencies are not actually so simple. They show a statistically reliable structure beyond Zipf's law that likely will not be captured with any simple model. At the same time, the large-scale structure is robustly Zipfian.''} While models focus {\it ``very narrowly on deriving the frequency/frequency rank power law, while ignoring these types of broader features of word frequencies.''}
    In conclusion, {\it ``we have a profusion of theories to explain an empirical phenomenon, yet very little attempt to distinguish those theories using scientific methods. This is problematic precisely because there are so many ways to derive Zipf's law that the ability to do so is extremely weak evidence for any theory.'' }
\end{quote}
While these conclusions were drawn for the case of Zipf's law\index{Zipf's law} of word frequencies, they apply more broadly as they are a consequence of treating mechanistic explanations separated from the data analysis of statistical laws, and provide a warning to research in statistical laws more broadly.

\paragraph{Probabilistic interpretations of statistical laws}

While complex-systems researchers have (mostly) liberated themselves from the "physicalism" of early sociophysics (i.e., the reduction of social aspects to physics concepts such as force and energy~\cite{mainzer_berechnung_2014}), the steps and logic of the traditional approach sketched above still resemble the sociophysics \index{social physics} program (see Sec.~\ref{ssec.socialphysics}) of repeating to other areas the idealized view of the development of classical mechanics: different empirical laws (Kepler's laws) \label{Kepler's law} are eventually explained by a model/theory (Newton) \index{Newton's law} and explored for further consequences/generalizations. This analogy between "statistical laws" and "empirical laws" exposes a fundamental misconception: the traditional approach ignores the statistical and probabilistic nature of these laws. This nature appears in at least two related facets:

\begin{itemize}
    \item[i)] Probabilistic formulation. Testable (falsifiable) formulations of statistical laws are only possible when they can be effectively interpreted as statements about the probability or expected value of observations (as discussed in Sec.~\ref{sec.likelihood}). This applies, in particular, to frequency-distribution laws -- e.g.,  \index{Pareto's law} Pareto law of inequality in Sec.~\ref{ssec.pareto} is viewed as a statement of the probability of a random person to have a given income  -- but also to other statistical laws such as scaling laws -- urban scaling laws in Sec.~\ref{ssec.urbanscaling} determine the expected output of cities of a given size.

    \item[ii)] The typicality of the setting in which observations are made. The condition of ``universal'' validity of a statistical law is often not explicitly formulated so that it is not clear what conditions a system needs to fulfill to show the law. In contrast, within a consistent Physics \index{Physics} theory, these conditions are specified and it is not possible to design experiments for which the theory does not apply. For instance, Newton's gravitation theory predicts that it is not possible to have bodies with gravitational mass that do not attract each other. All bodies are subject to the same laws and fluctuations around empirical laws (e.g., non-elliptic orbit of Uranus) are due to non-accounted effects that can in principle be incorporated  (e.g., the gravitational effect of Neptune). No similar impossibility exists in the case of statistical laws in complex systems. For instance, it is perfectly possible to write a perfectly understandable text that violates Zipf's law\index{Zipf's law} or to conceive  economical systems that violate \index{Pareto's law} Pareto distribution.  As discussed in Sec.~\ref{ssec.alz}, \index{Auerbach, Felix} Auerbach-Lotka-Zipf\index{Auerbach-Lotka-Zipf's law} law of city sizes has a clear counter-example in Australia data (two largest cities with approximately the same population), but that is often viewed as a particularity of the country and not as a violation (falsification) of the law as a whole. This shows that there is an implicit assumption that the law holds in typical (most probable) cases, an assumption similar (but not identical) to the the "ceteris paribus" assumption, typical in economics.

\end{itemize}

These two points are ignored in na\"ive interpretations of the traditional approach to statistical laws. There is no explicit statement about what are the settings in which the law should apply, what are the conditions that need to be satisfied for the law to be observed, or what are the expected fluctuations (around the most-probable values) determined by these laws. Without explicit statements about these points, it is unclear how the proposed laws can be falsified. This point was made precise in Secs.~\ref{ssec.limitationslikelihood}, which shows that using a na\"ive \index{hypothesis testing} hypothesis testing method (based on the assumption of independent data) leads to the wrong rejection of statistical laws in synthetic data compatible with the law (but correlated). In the traditional approach, the falsiability of statistical laws is essential because it lies at the foundation of the mechanistic models and applications. Instead, as we argue here, statistical laws are not {\it per se} falsifiable and their evaluation needs to be taken more generally within their role in the proposed model, research program, or application. 

More generally, the traditional formulations of statistical laws are ambiguous and admit different probabilistic interpretations. For instance, one can conceive different probabilistic models that have the same asymptotic or expected behaviour (compatible with the statistical laws) but different fluctuations around it (or different joint probabilities). Examples of this general property are obtained when considering probabilistic formulations of the different representations of the same law (e.g., rank-frequency vs. frequency distribution representations of power laws). Despite the one-to-one (analytical) relationship between the representations, they correspond to different probabilistic formulations compatible with the law. Estimations and conclusions drawn from data based on one of the formulations may not apply to the other.

\subsection{Persistent controversies}\label{ssec.persistent}

Another reason for an unified treatment of statistical laws in Complex Systems is the fact that debates about their validity are persistent and share similar characteristics. At the start of Chap.~\ref{chap.data} we listed 6 cases of controversies involving the validity of statistical laws, most of them continue over many decades and are periodically revived. The review of data-analysis methods in that chapter reveals that these controversies are intimately connected to different choices of quantitative methods to study statistical laws:

\begin{itemize}
    \item The disputes around the validity of \index{Kleiber's law} Kleiber's law reported in Refs.~\cite{dodds_re-examination_2001,da_silva_allometric_2006} show also the interesting interplay between simplicity, data analysis, and theoretical model for the acceptance and challenge of statistical laws. As mentioned in ~\cite{dodds_re-examination_2001}, the  reason behind the $\beta=3/4$ law being {\it "accepted and used as a general rule for decades"} ({\it "...often been taken as a fact"}) relied heavily on a consensus among practitioners {\it "that simple fractions would be a more convenient standard"}, emphasizing the simplicity and analytical tractability as a major reason for the choice of statistical laws.  The reversal of the conclusion by~\cite{dodds_re-examination_2001} after decades of consensus relied on the use of new statistical techniques  \index{hypothesis testing} (hypothesis-testing, fitting methods) and on the comparison to more sophisticated models (in particular, the one with different cut-offs or upper bounds for fitting, used to partition the data in different fitting intervals).

\item The debate around the validity of ALZ\index{Auerbach-Lotka-Zipf's law} law of city sizes involved the comparison of the proposed power-law to the alternative log-normal distribution, as discussed in Sec.~\ref{ssec.alz} and Refs.~\cite{eeckhout_gibrats_2004,levy_gibrats_2009,eeckhout_gibrats_2009,malevergne_gibrats_2009}. The evidence for log-normal in Ref.~\cite{eeckhout_gibrats_2004} uses the frequency of cities of a given size while \index{Auerbach, Felix} Auerbach's traditional observation (and the analysis in Ref.~\cite{levy_gibrats_2009}) focuses on the rank frequency representation. As discussed in Sec.~\ref{ssec.representationmatters} and~\ref{ssec.ml-freq}, the choice of representation is part of the data-analysis method and affects the conclusions obtained from their application.

\item The reversal of claims about the validity of power-law distributions in the early 21st century~\cite{stumpf_critical_2012,broido_scale-free_2019} are closely associated to the introduction and popularization (through Ref.~\cite{clauset_power-law_2009}) of maximum-likelihood methods. This led to the application of statistical tests to empirical data that ignited debates of the validity of these laws, in particular about the ubiquity of \index{scale-free networks} scale-free networks discussed in Sec.~\ref{ssec.scalefree}.  \index{power-law distribution}

\item The sensitivity of urban-scaling laws to definitions of Urban areas -- see Sec.~\ref{ssec.urbanscaling} and Ref.~\cite{arcaute_constructing_2015} -- are related to the lack of a generative model of the expected fluctuations around the scaling behaviour and also to the sensitivity of linear regression methods to the regions in which most cities are present -- see Sec.~\ref{ssec.caveatsLinear}.
\end{itemize}

A point often ignored in the choices of data-analysis methods is that they include commitments to different interpretations of the statistical laws and also assumptions regarding the process generating the data. 
The methodological developments underlying the controversies listed above tend to advocate for the use of more rigorous statistical arguments that better explore the modern availability of computational power. Naturally, they are generally interpreted as better than the traditional graphical and linear-regression methods. A subtler yet critical point is that these methods often require a re-interpretation of the statistical laws in ways that allow for the methodology to be applied. This is particularly clear in the case of naive applications of  \index{hypothesis testing} hypothesis testing methods for the analysis of power-law distributions, as discussed in Sec.~\ref{ssec.limitationslikelihood}, which includes the assumption of independent observations that is obviously violated in data. It is important to recognize that the historical interpretation of statistical laws were not committed to any probabilistic interpretation and that graphical methods are potentially suitable for exploratory, qualitative, or semi-quantitative interpretations (e.g., that a scaling-like behaviour is observed over different orders of magnitude or that the tails of a distribution decays slower than exponential). 

In this monograph we argue that the crises and controversies in the use of statistical laws in complex-systems research stem from the failure to recognize the limitations of the traditional approach and from a na\"ive interpretation of statistical laws. More precisely, the difficulty in reaching consensus is, in a great extent, due to the failure to acknowledge how different (legitimate) interpretations of statistical laws affect the methodological choices and lead to different conclusions. This is explicit also in the log-normal vs. power-law debate on city sizes: while in a na\"ive interpretation the alternative representation of ALZ\index{Auerbach-Lotka-Zipf's law} law (rank-frequency vs. population distribution) are equivalent, in practice they affect the data analysis and correspond to different interpretations of the same law.  The difficulty in evaluating the validity of statistical laws is also intrinsically connected to the impossibility of decomposing complex systems into simple parts. For instance, the idealized situations in which data can be hypothesized to come from independent observations would typically also destroy the very same interactions in the underlying system leading to the non-trivial laws.  One of the main points of this monograph is thus to emphasize the importance of fully assimilating the statistical nature of these laws (e.g., focusing on the fluctuations of the data around the predicted curves) and choosing data-analysis methods that are consistent with the interpretation, conclusions, and intended use of statistical laws.
 \index{power-law distribution}

\section{Statistical laws well done}

The critical focus we have employed so far can lead to a skeptical or cynical view on statistical laws, such as the conclusion that they are not a scientific concept because they are not falsifiable. Despite problematic interpretations and uses of statistical laws, it is important to recognize the many achievements obtained through their use. The goal of this section is thus to formulate recommended practices that avoid problematic uses and allow for a more balanced evaluation of the potential and limitations of this concept.  When formulating recommendations below, we have in mind someone who is interested in evaluating the use of statistical laws to a specific application or a particular dataset. We will not formulate recipes, but instead will recommend practices for the analysis of data and the study of statistical laws, discussing different alternatives and focusing on the consistency between the interpretation, application, and choice of data-analysis method.

In new and exploratory data analysis, it is natural and convenient to retain some level of division between empirical analysis of the statistical law from the question about a mechanistic model or application. This is the key element of the traditional approach (see Secs.~\ref{ssec.reasoning} and~\ref{ssec.traditional}), but we should now proceed with care to avoid the dangers and misuses that happen when considering statistical laws to be valid in absolute terms and deriving conclusions uncritically (i.e., ignoring its statistical-probabilistic nature). The controversies and limitations of methods discussed above show not only the importance of abandoning na\"ive views on the validity of statistical laws but act also as a warning against the blind application of statistical recipes. There is no single "right" way of studying statistical law, alternatives with increasing levels of sophistication were introduced in Chap~\ref{chap.data}. The traditional maximum likelihood fitting is one of them, but we showed how often one needs to go beyond its own limitations by including additional feature of the system (e.g., temporal correlations) in the data analysis --Sec.~\ref{sec:stat-meth-compl} -- and distinguishing between models based on inference and \index{model comparison} model comparison~\ref{ssec.inference}.

The traditional statements of statistical laws -- as evident from the examples reviewed in Chap.~\ref{chap.examples} -- are typically based on very simple data-analysis methods and formulated in analytical/absolute terms. As we learned throughout this monograph, such formulations, alone, are incomplete, not falsifiable, and open to different interpretations. Such interpretations will typically make additional assumptions that were not contained in the formulation of the law, but that are essential for evaluating, testing, and using the statistical laws. The application and test of validity of statistical laws can only be performed in their expanded setting and it is thus paramount to have clarity and consistency about the intended use and interpretation of statistical laws. Accordingly, we start our discussions with questions about the desired interpretation of statistical laws, before moving to more practical questions about the choice of data-analysis methods.

\subsection{Setting the interpretation}\label{ssec.interpretation}

Before considering the comparison of data to a functional form proposed as a statistical law, an important question to be considered is the motivation or goal of the analysis. In increasing order of sophistication or ambition, common reasons to use statistical laws include (more than one may apply to the same analysis):

\begin{tcolorbox}
\begin{itemize}
    \item[1] Use as a summary statistics of the data. For instance, the parameters of the fitted statistical law will be estimated and their values will be compared (in different cases).
    \item[2] Comparison between alternative models. This can be done in different degrees:
      \begin{itemize}
      \item[2a] showing that distributions or scalings are not simple or do not belong to simple classes. For instance, showing that the function is non-linear or that the distribution is non-Gaussian  or fat-tailed.  \index{fat-tailed distribution}
      \item[2b] showing that one proposed function is better than an alternative one. For instance,  comparing power-law,  lognormal, and stretched exponential distributions. \index{log-normal distribution}  \index{stretched exponential distribution}
      \end{itemize}
    \item[3] Perform analytical computations and estimations using the functional form of the law.  For instance, using the fitted curve to estimate the probability of unobserved events.
    \item[4] Obtain information about the generative process underlying the data. This can be done in different degrees:
    \begin{itemize}
        \item[4a] justify the inclusion of one process in a mechanistic model (e.g., a linear or non-linear term in the model).
        \item[4b] validate the connection between datasets and generative process. For instance, this could include the comparison of specific model parameters (e.g., exponents) to specific generative processes (e.g., types of phase transitions) or specific data classes.
    \end{itemize}
\end{itemize}
\end{tcolorbox}

Another key aspect of the interpretation of the statistical law is to be clear about which cases or data are potentially described by the statistical law. Common options include:

\begin{tcolorbox}
\begin{itemize}
    \item[A] Particular observations within a sample, such as the ones with the largest or smallest values (e.g., tails of distributions, values above or below some threshold).
    \item[B] Typical observations within a sample, such as the expected value or the majority of the cases (e.g., most cities or typical cities).
    \item[C] Typical observations in data that spans different orders of magnitudes (e.g., from small villages to large cities, texts of different sizes).
    \item[D] Samples obtained when specified conditions are met (e.g., texts in a specific language by one author, cities within the same country, earthquakes in the same region). 
    \item[E] All the data in all the samples (e.g., any text in any language).
\end{itemize}
\end{tcolorbox}

\subsection{Choosing the data-analysis methods}\label{ssec.choosemethod}

The choice of data-analysis method depends not only on the available data but also on the interpretation and motivation for the study specified in the previous section.  In most cases, a graphical visualization of the data is recommended as a visual tool to test whether patterns are visible and further analysis is justified. Here it is important to choose a representation that not only favours the detection of patterns but that is consistent  with the motivation and choice of data. A typical choice for the detection of patterns is to transform variables and plot axes in such a way that the statistical law appears as a linear curve, see Sec.~\ref{ssec.linearRepresentations}. If the goal is to compare different curves, cases 2 and 4a above, different data representations could be used to detect whether any of them shows the predicted pattern. If the focus is on describing a wide range of scales, case C above, the plot should use logarithmic scales and the data should be chosen so that it covers a wide range of values. If the focus is on describing only the tails of a distribution, one should consider applying a threshold or choosing variables that highlight these cases (e.g., a rank-frequency representation for power-law distributions). A key point is to ensure that threshold and cut-offs are chosen in such a way that a wide range of values (e.g, at least two decades for plots in logarithmic scale) rmeains accessible to test the data-model agreement (any smooth curve looks locally linear).   \index{power-law distribution}

Going beyond graphical analysis is needed if one is interested in drawing more ambitious conclusions from the statistical-law analysis, in particular motivations 2b, 3, and 4. In fact, if only motivations 1 and 2a listed above apply, one should consider whether the proposal of a statistical law is indeed needed and consider the possibility of, instead, using alternative summary statistics (instead of parameters estimated by fitting parametric functions) or statistical tests (e.g., about non-linearity or non-Gaussian behaviour). When choosing the statistical method it is important to consider the motivation for the analysis, the available data, and realistic formulations of the sampling/stochastic process (e.g., which captures how the data was measured). A consistent way of proceeding is to consider a probabilistic interpretation of the statistical law (e.g., distributions as probability of observations), the most suitable representations of the law, and the best formulation of a sampling process to write down the likelihood of observations. This process will involve simplifying assumptions, such as the uniformity of fluctuations (in log-transformed variables) and the independence of the observations. It is important to be aware of these assumptions, test them in the data if possible, and consider whether the assumptions that are clearly violated in the data have an implication to the conclusions. In the typical case in which not all assumptions are satisfied in the data, it is important to abandon the expectation of a full compatibility between the data and the statistical-law model. This implies, in particular, that one cannot rely on statistical computations that assume that the data is a realization of the model.

\paragraph{Choice of representation}

A critical choice is on the representation and interpretation of the law and data, which should be done consistently. For instance, when looking at the properties of word frequencies one can choose to focus on word types (unique words) or word tokens. The word-type choice leads to a frequency-distribution representation of Zipf's law\index{Zipf's law}, is consistent with a sampling of unique words (treating each of them as observations), and will have the statistical analysis dominated by the large number of low-frequency words that together compose only a small fraction of the whole text. The word-token choice leads to a rank-frequency representation of Zipf's law, is consistent with a sampling of word tokens such as the one obtained by going through a text, and will have the statistical analysis dominated by the small number of high-frequency words that compose a large fraction of the whole text. As discussed in Secs.~\ref{ssec.representationmatters},~\ref{ssec.linearRepresentations}, and~\ref{ssec.rankrepresenation}, this choice affects the application and outcome of data analysis methods, including the estimation of parameters and evaluateion of the data-law agreement. It is thus important to determine what are the observations considered of interest or typical in the data-application options A, B, and C. The choice of representation is expected to have a significant impact on the outcome of the analysis in all datasets with fat-tailed distributions.  \index{fat-tailed distribution} For instance, if urban data is used -- such as in the ALZ\index{Auerbach-Lotka-Zipf's law} law of city-size distribution discussed in Sec.~\ref{ssec.alz} or in the urban scaling laws discussed in Sec.~\ref{ssec.urbanscaling} -- the choice is between cities and inhabitants. If the focus is on cities, the data analysis is dominated by the large number of small cities where a small fraction of the population live. If the focus is on inhabitants, the data analysis is dominated by the few large cities where most of the population live. As shown in Secs.~\ref{ssec.likelihoodscaling} and~\ref{ssec.ml-freq}, this can strongly affect the data analysis. 

\paragraph{Model Comparison}

In most cases involving motivations 2b and higher, a statistical comparison \index{model comparison} between the data and different functional forms is the most indicated approach. The preference for such comparison, in opposition to a test of validity of the law, has been emphasized by Gabaix in the study of city-size distributions as

\begin{quote}
  {\it ``some of the debate on Zipfs law should be cast in terms of how well, or poorly, it fits rather than whether it can be rejected.''}~\cite{gabaix_power_2009}
\end{quote}

As formulated by Gabaix and Ioannides in their analysis of the \index{Auerbach, Felix} Auerbach-Lotka-Zipf\index{Auerbach-Lotka-Zipf's law}'s law of city sizes: 

\begin{quote} {\it "The main question of empirical work should be how well a theory fits, rather than whether or not it fits perfectly (i.e., within the standard errors). With an infinitely large data set, one can reject any non-tautological theory. Consistently with this suggestion, some of the debate on Zipf's law\index{Zipf's law} should be cast in terms of how well, or poorly, it fits, rather than whether it can be rejected or not."} ~\cite{gabaix_chapter_2004}
\end{quote}

Model  comparison \index{model comparison} should be performed using the most suitable representation of the statistical law and be based on similar assumptions for each of the functional forms. These are essential points to ensure that the simplifying assumptions or representation choices are not unintentionally affecting the decision about which of the curves best describes the data. For instance, when using likelihood methods to investigate statistical laws in form of frequency distributions (Sec.~\ref{ssec.ml-freq}), the assumption of independent observations can strongly affect  \index{hypothesis testing} hypothesis testing (Sec.~\ref{ssec.limitationslikelihood}) but still allow for a fair comparison between alternatives using likelihood-ratio tests. Model comparison is also important in more general inferential approaches (Sec.~\ref{ssec.inference}) in which statistical laws are not directly tested to alternatives but instead they are tested together with more realistic (mechanistic) models for the generation of the data. 

An essential consideration in model comparison is the complexity of the different models under consideration. The likelihood of the more complex model will never be smaller in nested models (i.e., when one is reduced to the other for particular parameter choices). Beyond nested model, it is important to consider whether the association between the number of parameters and the model complexity is justified (see Ref.~\cite{piantadosi_one_2018} for an example of a single-parameter function that is able to (over)-fit any number of points with arbitrary precision) and whether methods that penalize for parameters can be applied. While more sophisticated model-comparison models are recommended, they are not always easily applicable. Moreover, statistical laws are, by definition in Sec.~\ref{ssec.definition}, restricted to a small number of parameters and it is expected that in the presence of large datasets will be best described by more complicated functional forms (the log-likelihood term increases linearly with $N$ and the advantage of describing even small fluctuations become statistically significant). In practice, a pragmatic procedure is to restrict the comparison of different functional forms to alternatives that share properties with the proposed statistical law (same number of parameters, simple functions, etc.).

A particular case of interest for model comparison is when one is interested in specific parameters of the statistical laws. Examples include the debate on the exponent of \index{Kleiber's law} Kleiber's law, claims of universality in Urban scaling laws, and exponents of power-law distributions connected to specific explanations (critical phenomena, preferential attachment). In these cases, one should consider not only reporting the values of the estimated parameters but also a model comparison between alternative descriptions at fixed exponents (e.g., $\beta=3/4$ vs $\beta=2/3$) or fixed vs. arbitrary (e.g., $\beta=1$ vs. $\beta>1$).

\paragraph{Hypothesis testing and goodness-of-fit tests}

 \index{hypothesis testing}
The incompatibility between simplifying assumptions used in the computation of the likelihood function usually compromises statistical tests of the compatibility between data and statistical-law model (see Sec.~\ref{ssec.limitationslikelihood}). Still, in situations in which this is intended, an important point to emphasize is all the hypothesis that the test involves and to try (as much as possible) to clarify whether the reasons for the simplifying outcome can be associated to the functional form of the statistical law or to another  hypothesis.

\subsection{Formulating the conclusions}\label{ssec.formulateconclusions}

Statistical laws cannot be determined as valid in an absolute sense, independent of their use, their representations, and the proposed generative model. 
Conclusions about the applicability of statistical laws to a specific data or problem should consider the context in which they appear (e.g., past work on the topic), the motivation of the analysis (e.g., what will be done with it), and both the choice and outcome of the data-analysis methods. In virtually all cases, conclusions such as ``the law is true or valid'' or ``the data corroborates the validity of the law'' are, at best, imprecise and misleading.  Fortunately, such exaggerated claims on the validity of statistical laws are typically not needed in the evaluation of how useful a statistical law is for a scientific program or application. One should thus focus on whether ``the law provides a useful or reasonable description of the data'' and ensures that the reported data-analysis provides support for it based (e.g., based on comparisons to alternative descriptions). An important point is to emphasize that the insights obtained from the law that would not be possible based on data-analysis only because often the conclusions on the value of statistical laws depend on their use. For instance, a statistical law in form of a power-law distribution might be useful in order to distinguish between processes leading to short- and long-tailed distributions but it might fail to associate different "universal" exponents to different cases or to critical values.

When following the traditional approach to statistical law, a weaker sense in which they are considered to be valid is usually needed. For instance, statistical laws can be used as inspirations for the proposal of mechanistic models, possibly identifying how key simplifying asumptions need to be changed. Statistical laws in form of power laws may be useful as a distinction of uni-modal distributions or as capturing fat tails, but the exponents of power-law fits may not be universal or helpful. SImilarly, ubran-scaling laws may indicate an overall tendency for large cities, but they may not be predictive of how cities evolve over time and thus be of limited use for urban planning. In general, one needs to proceed with care when formulating conclusions derived from statistical laws and avoid assuming their absolute validity or to consider them as empirical laws in a traditional sense.

The stronger the formulation on the applicability of a statistical law, the stronger the statistical evidence needed to support it. Two possible types of claims are:

\begin{itemize}
  
\item[(i)] {\bf The statistical law (e.g., a power-law distribution) provides a much better explanation than other expected curves that act as null models (e.g., a Gaussian or exponential distribution)}.  \index{Gaussian distribution} In this case, the statistical analysis will be based on statistical model comparison\index{model comparison}. Mechanistic models reproducing the law should be contrasted to those reproducing the null models and should be interpreted as one possible mechanism explaining this feature. The goal here is to reveal plausible mechanisms, while the plausibility of the mechanism and the comparison to alternatives will depend also on the extent into which they are realistic and explain other observations.  \index{power-law distribution}

\item[(ii)]{\bf The statistical law is expected to be satisfied in some idealized limit (e.g., infinitely many observations, time to infinity, idealized setting).} In this case, in addition to the model-comparison analysis of the previous point, the statistical test could consist in measuring the distance between the data and statistical law as we approach the idealized case (e.g., as $N\rightarrow \infty$). Mechanistic models should ideally incorporate aspects that explain the deviation from the ideal case (e.g., finite size sample).
\end{itemize}

Conclusions about the support for the proposed mechanistic model to explain a statistical law should consider the existance of alternative models and need to combine quantitative methods and other theoretical considerations.  This has consequences also to the extent into which evidence for the law can be considered as evidence for the model. An example is the Poissonian  explanation \index{Poisson distribution} of burstiness proposed in Ref.~\cite{malmgren_poissonian_2008} and discussed in Sec.~\ref{ssec.burstysocial}). An alternative and more radical approach is to abandon the traditional approach to statistical laws, accept that they cannot be fully evaluated independently from the generative model, and proceed to an inferential approach to their study (Sec.~\ref{ssec.inference}), i.e., the formulation of probabilistic generative models that can be rigorously compared to each other through statistical methods.


\paragraph{\it My data shows a strong pattern, can I call it a law?}

Before falling into the temptation of having your very own law, a few cautionary steps are recommended:

\begin{itemize}
    \item Recall the definition of statistical laws in Sec.~\ref{ssec.definition} and check if all points are fulfilled, including the "theoretical connection" and "universality" requirements. 
    \item Check if there is a simple explanation for your observation, such as a general statistical arguments (e.g., the central limit theorem) or a simple connection to an existing law.
    \item Evaluate the evidence in support of the law, including the statistical analysis of the data-model agreement and the usefulness of the law in providing theoretical insights or applications.
    \item Consider the extent into which the law is "universally" applicable, including the family of cases in which it is expected to be valid and the amount of data in support of such claims. 
\end{itemize}

\paragraph{\it What is a valid explanation for a law? How should we choose between competing explanations?}

Typically, multiple theories (models) explain the same statistical law, a feature typical of any scientific theories. Positive aspects of an explanation include:

\begin{itemize}
    \item Simplicity of the proposed mechanism (Occam's razor).
    \item Realistic assumptions (rooted in theory, compatible with the data, and independently verifiable).
    \item Non-circular (i.e., not directly implied by the statistical law that it aims to explain).
    \item New predictions (testable and independent).
\end{itemize}

In most cases only a few of these aspects are met. Ultimately, the extent to which each of them is relevant, and the decision in favour of an explanation, depends on the law, the empirical evidence supporting it, its use, and the context in which it appeared.

\subsection{Summary of recommendations}

\begin{itemize}
\item[1.] Set the interpretation (Sec.~\ref{ssec.interpretation}), choose the data-analysis method (Sec.~\ref{ssec.choosemethod}), and formulate the conclusions (Sec.~\ref{ssec.formulateconclusions}) in such a way that they are mutually consistent and aligned with the role you intend the statistical law to play in your research and problem. 
\item[2.] Avoid thinking that the law is true or valid in an absolute sense. Instead, consider whether (i) it provides a useful description of the data and (ii) it brings new insights about the generative process. 
\item[3.] Instead of testing whether one can reject or not the law (one typically can, with enough data~\cite{gabaix_power_2009}), in most cases one should focus on comparing the proposed law to alternatives (not only null models, but also other similarly simple functions) and in quantifying in which extent (and conditions) a simple function describes well the data. Such model comparison \index{model comparison}  should consider also (as much as possible) the generative process of the data. For instance, if correlations are known to exist in the data, one should consider whether they affect each model differently or whether methods that account for them exist (such as the ones suggested in Sec.~\ref{chap.data}); if the data is from a network, models of networks should be preferred. If the law is identified as better than similar alternative, it does not mean that the law is precisely valid in the sense that in some limit the data will be exactly described by the curve or that the finite-size deviations should be comparable to samples of a naive null model based on the law.
\item[4.] Look beyond the law and quantify fluctuations around it (e.g. studying residuals) and other statistical features of the data (e.g., fluctuation scaling). Examples of studies using these techniques exist for \index{Kleiber's law} Kleiber's law~\cite{dodds_re-examination_2001}, Herdan-Heaps' law~\cite{gerlach_scaling_2014}, and urban scaling~\cite{bettencourt_urban_2010}. \index{Urban scaling laws} \index{Herdan-Heaps' law}
\item[5.] The law might be useful to make derivations, estimations, and analytical reasoning. Here it is important to consider that a good agreement with the law in your representation does not necessarily imply a similarly good agreement for the derived quantities (e.g., deviations that are irrelevant in double-logarithmic plots of data can become extremely relevant for other observables of interest). Uncertainties and fluctuations around the statistical law need to be quantified and propagated into the quantities of interest. Conclusions derived from the law need to be independently checked against the data.
\item[6.] Mechanistic models proposed to explain the law should be presented as one of the plausible explanations. The hypotheses of the model should be justified based on additional knowledge or data analysis. Independent predictions of the model should be formulated and, if possible, tested (in independent data). It is important to consider a simple alternative (null model) and remember that there are other models that explain the same law (data). One should consider carefully what components of the model are essential and how to compare the different alternatives. The comparison between mechanistic models that explain a statistical law will typically not rely only on the agreement to data, but also on their simplicity and their agreement with other known properties of the underlying system.
  \item[7.] Conclusions should be formulated consistently with the statistical evidence in support of the law and of the theoretical explanation. 
\end{itemize}

It is worth providing short answers to some of the recurring questions in the study of statistical laws:

\paragraph{\it Can a statistical law be falsified or proven wrong?}

Not in a simple \index{hypothesis testing} "hypothesis-testing" sense of falsification.  Statistical laws allow for multiple (probabilistic) interpretations because the specification (by the law) of the average tendency or marginal distribution is not enough to compute the likelihood of the complete observations (it requires additional assumptions, such as the hypothesis that the observations are independent of each other). A statistical test of the validity of a law is thus always a test contingent on these additional hypotheses (i.e., of a specific formulation or interpretation of the law that specifies the generative stochastic process or joint distribution of the observations). As there are infinitely many possible models compatible with the law, it is not possible to test (reject) all of them.

This does not mean that there are no good reasons to discard a proposed law or that there are no sensible ways of evaluating proposed laws. For instance, simple visual inspection of plots and comparisons to different simple curves \index{model comparison} can reveal strong disagreements with the proposed statistical laws that indicate that they are not helpful to understand that dataset. An example of a proposed statistical law that is abandoned through this method is Zipf's proposal of power-law distribution for the burstiness \index{burstiness} of words, discussed in Sec.~\ref{ssec.burstywords}. The use of regression and likelihood methods can also identify whether alternative proposals outperform the proposed law, in which case the statistical law should be discarded or re-interpreted. The point we want to make here is that evaluations of statistical laws should not blindly follow a single recipe, but instead they should emphasize the compatibility between the hypotheses underlying the methods and the interpretation of the statistical law. 

\paragraph{When can we say that a statistical law is valid?}

As any other scientific law, the validity of statistical laws is not only a data-analysis or empirical question: it needs to be considered together with its use and the theories that allow for its interpretation. The validity of statistical laws should consider (i) the more general theoretical and applied context in which they appear; and (ii) an interpretation and evaluation that takes into account their probabilistic nature. A statistical law that is contributing to a research program is expected to provide:

\begin{itemize}
\item[1.] a better description of the data than equally simple alternatives. 
\item[2.] insights on the mechanistic model underlying the data, ruling out other natural alternatives.
\item[3.] improved predictions or estimations for unobserved data or cases.
\end{itemize}

For instance, in the case of scaling laws -- such as the urban scaling laws discussed in Sec.~\ref{ssec.urbanscaling} or Herdan-Heaps' law for vocabulary sized discussed in Sec.~\ref{ssec.heaps} -- these points could be: 1. a comparison to a linear scaling or an exponential convergence to a constant; 2. comparison to a model of constant-per capita use or of finite vocabulary; and 3. useful metrics to compare cities or to estimate vocabulary size of unobserved datasets. Datasets in which these conditions apply can be said to ``follow'' the statistical law. If different datasets follow a statistical law, the law is effectively a useful (valid) tool within that research program.

\paragraph{Why is it so difficult to reach consensus?}

There are multiple factors that contribute to the difficulty in reaching a consensus around the validity and interpretation of statistical laws (such as the six controversies listed at the start of Chap.~\ref{chap.data}):

\begin{itemize}
\item the ambiguity that exists in the formulation of statistical laws which leads to different interpretations and representations;
\item the use of different data-analysis methodologies;
\item and the availability of better datasets.
\end{itemize}

In some cases -- such as the Zipf's proposal for the burstiness \index{burstiness} of words discussed in Sec.~\ref{ssec.burstywords}-- better data and computers contribute to a new view on the problem. More often, it is the use of new quantitative methods that leads to new conclusions. Different methods are associated to different applications and interpretations of the law, and also involve different assumptions on the generation of the data.  Underlying these controversies is the traditional division of the analysis of statistical laws into the validation of the empirical curve, the development of mechanistic models, and the interpretation of the law. While this separation used in the traditional approach to statistical laws is convenient and didactic, and has been proven useful in the study of many statistical laws, it has limitations (see Sec.~\ref{ssec.persistent}). Ultimately, a robust and stable understanding of a specific statistical law can only come if the mechanisms underlying it and the comparison to data are both established.

\section{The future of statistical laws}

\subsection{From stylized facts to inferential approaches}
 
\begin{figure}
    \centering
    \includegraphics[width=1.0\textwidth]{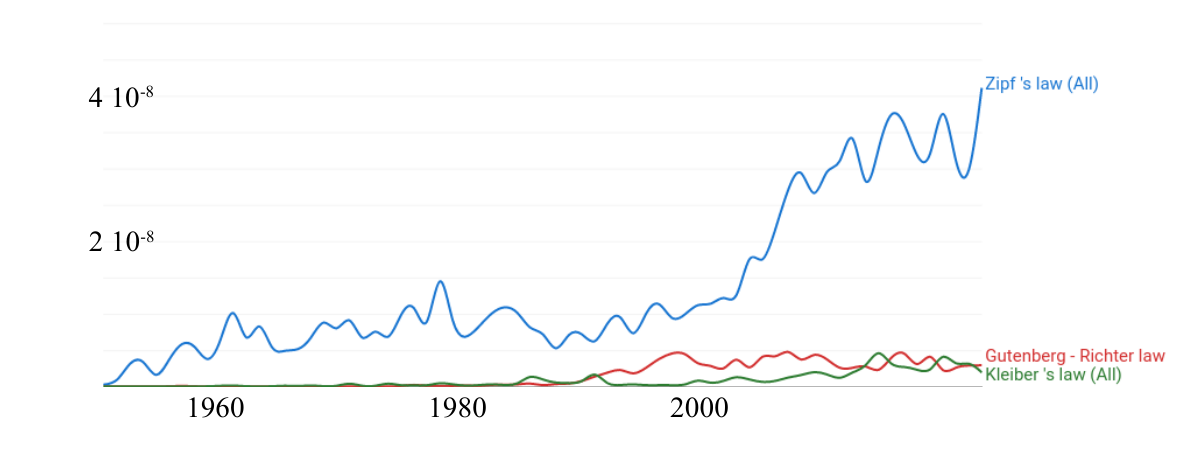}
    \caption{Frequency of mentions to statistical laws in books in English (Google n-gram database).}
    \label{fig.ngram}
\end{figure}

\begin{figure}
    \centering
    \includegraphics[width=0.8\textwidth]{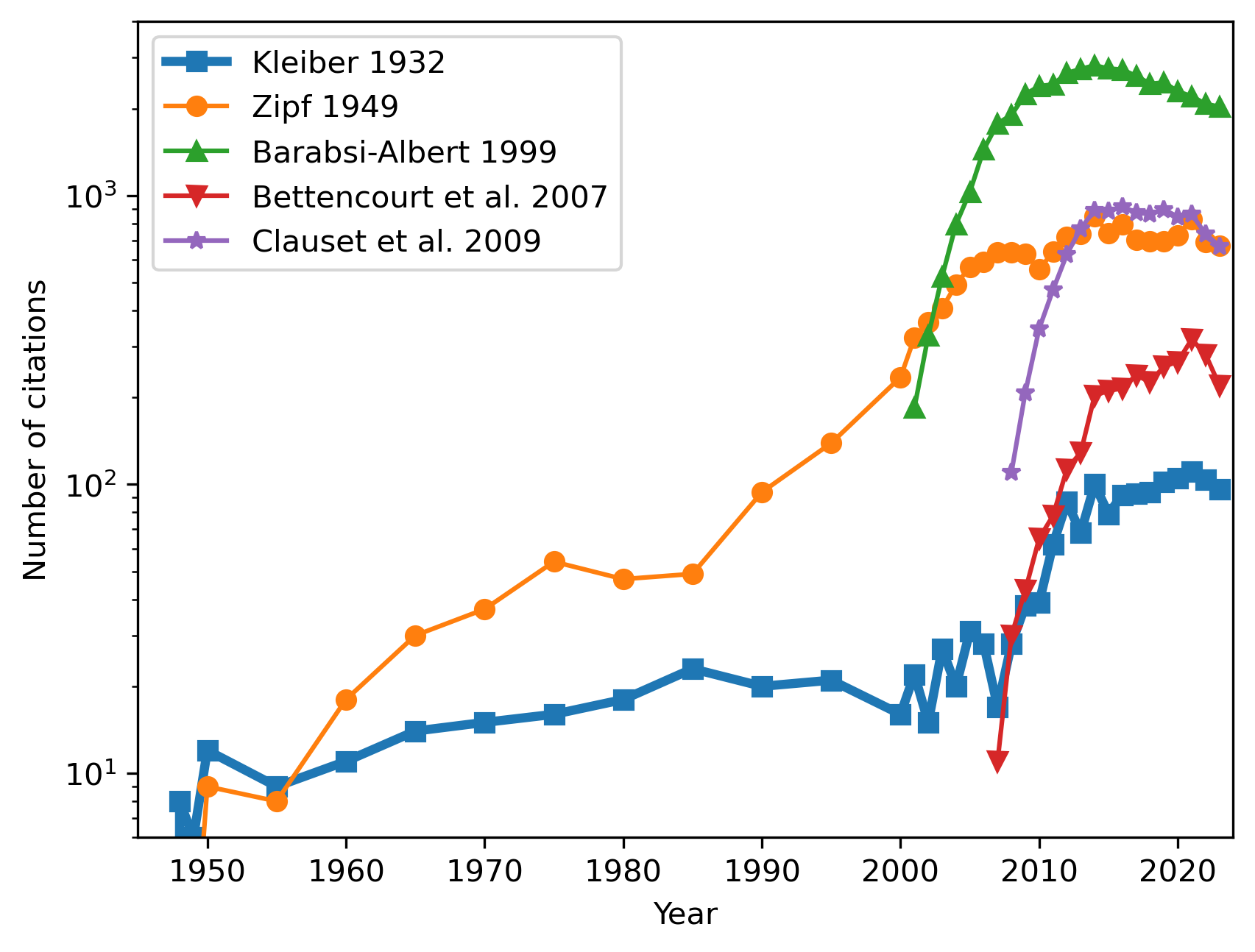}
    \caption{Number of citations to influential publications in statistical laws. Each point corresponds to the number of publications in the Google scholar database (retrieved June 7, 2024) that cited the publications indicated in the legend. These five publications and their total citations are \index{Kleiber's law} Kleiber~\cite{kleiber_body_1932} $2,774$; Zipf~\cite{zipf_human_2012} $18,173$; Barabasi and Albert~\cite{barabasi_emergence_1999} $46,152$; Bettencourt et al.~\cite{bettencourt_growth_2007} $3,048$; Clauset et al.~\cite{clauset_power-law_2009} $11,232$. See Appendix~\ref{chap.appendices}for the data and code used in this figure.
}\label{fig.bibliometric}
\end{figure}

What will be the role for statistical laws in the future? An informed speculation about this question needs to consider how statistical laws have been used throughout the recent years. Figure~\ref{fig.ngram} shows that mentions to statistical laws in published books have increased considerably since the 1990s and that there is no sign of decay of interest in recent years. In terms of scientific publications, Fig.~\ref{fig.bibliometric} shows that the number of citations to classical papers in the field of statistical laws increased in the 1990s and more clearly in the early 2000s, achieving very large numbers from the 2010s on, and possibly peaking in the recent years. 
This bibliometric data \index{Bibliometry} provides also a quantification of the amount of work and the overall interest in the subject. Zipf's seminal book~\cite{zipf_human_2012} alone has been cited more than $18,000$ times with $\approx 700$ new publications citing it every year\footnote{The magnitude of publications in the subject makes it evident that this monograph does not provide an exhaustive review of the literature in statistical laws. In particular, the selection of papers and problems published in the last two decades is unavoidably biased towards the work of the author.}. These observations, and the two-centuries tradition, strongly suggest that statistical laws is a healthy area of study that provides an useful approach in various disciplines and that will continue to flourish in the (near) future.

The traditional uses of statistical laws are to summarize data (stylized facts), allow for analytical reasoning, and motivate the proposal of mechanistic understandings of important (unexpected) features of the system underlying the data. The review in Chap.~\ref{chap.examples} shows numerous successful cases of these usages and we expect statistical laws will continue to serve this purpose into the future. Here it is important to consider that many of the successful uses have an exploratory nature, i.e., a weaker sense in which statistical laws are said to be valid. Statistical laws are also increasingly used in inferential approaches based on probabilistic generative models (Sec.~\ref{ssec.inference}). Here the laws are either introduced in models or the mechanistic models proposed to describe them are formulated probabilistically. The advantage here is that: (i) stronger (more rigorous) statements about the selection between alternative models can be made based on their performance in model-comparison tests; and (ii) probabilistic models can be used beyond the description of the law or observed data (for instance, for prediction of unobserved events). Statistical laws are important in the proposal and creation of these models. 

\subsection{Data science, machine learning, and artificial intelligence}

From mere curiosities to quantitative applications and theoretical models, statistical laws are used whenever large (observational) data is available. The increase in the availability and economical importance of such "big data" is a hallmark of the 21st century and it is partially responsible for the recent renewal of interest in statistical laws. At the same time, it is important to recognize that statistical laws, and the scientific fields traditionally related to them (such as sociophysics or complex systems), \index{social physics} play a limited role in both the applications and scientific studies triggered by big data. Instead, the emerging field of "data science" is currently dominated by areas of computer science, mostly \index{machine learning} machine learning which is also the dominant approach to achieve artificial intelligence.  \index{Data Science}
The combination of machine-learning methods, large datasets, and computational power have led to breakthroughs in scientific problems and applications, ranging from the success of deep learning techniques to predict protein structures~\cite{jumper_highly_2021,abramson_accurate_2024} to the remarkable ability of large language models to generate realistic text.

As the dominant data-driven paradigm, machine learning is taking roles and sharing aims that in the past have been attributed to statistical laws. For instance,  in a famous popular-science book on machine learning~\cite{domingos_pedro_master_2015}, the current state of the algorithms used in this field is compared to Kepler's law \index{Kepler's law} and as a preparation for the imminent arrival of general purposed artificial intelligence, that will play the role of Newton's theory for classical mechanics. This is precisely the role attributed to statistical laws in the traditional socio-physics tradition, as mentioned in \index{social physics} Sec.~\ref{ssec.socialphysics}. This leads to the question: what is the role of statistical laws in view of the increasingly important role played by machine learning in data-driven research?

The relevance of this question is accentuated by noting that the Machine Learning (ML) approach to data analysis is radically different from the statistical laws (SL) approach reviewed in this monograph. Table~\ref{tab.MLSL} highlights some of the most salient distinctions, which reflect not only the different goals of machine learning approaches but also their different relationship with scientific knowledge and theory. \index{machine learning} Machine learning methods typically do not intend to create or be based on realistic descriptions of how the data was generated, they instead focus on the improvement of generic and efficient algorithms that can be widely and flexibly applied~\cite{domingos_pedro_master_2015}. 

\begin{table}[!h]
  \centering
  \small
    \begin{tabular}{c|c|c}
         & Statistical Laws  &  Machine Learning \\
         \hline 
Parameters        & $<10$, typically $1$ or $2$ & unbounded, $>10^{12}$ in large language models \\
 & (restriction is a path to simplicity) & (growth is a path to improved methods) \\
 \hline
 Functional form & Simple parametric & generic representations  \\
 & (interpretability and tractability) & (capture arbitrary statistical patterns)\\
\hline
 Mechanistic Model &  Step to understand & Oblivious  \\
 of underlying system& & \\ 
\hline
 Scientific tradition & Natural Sciences & Engineering \\
  & (simple theories explain complex data) & (develop tools for problem solving) \\
 \hline
    \end{tabular}
    \caption{Schematic list of distinctions between statistical-laws and \index{machine learning} machine-learning approaches to data science.}
    \label{tab.MLSL}
\end{table}

An instructive example of the difference between the statistical-laws and \index{machine learning} machine-learning approaches is obtained looking at the analysis of large collections of written text. In the statistical-laws approach -- as discussed in Secs.~\ref{ssec.zipf},~\ref{ssec.heaps}, and~\ref{ssec.linguistic-laws} --  the acceptance of Zipf's law\index{Zipf's law} triggered the creation of simple generative models of text generation (such as \index{Simon, Herbert} Simon's model) that intended to capture how repeated and new words are used (connecting it to Herdan-Heaps' law). \index{Herdan-Heaps' law} Instead, in machine-learning approach that now culminated in large language models, generic methods (transformers, attention models, etc) were unsupervisedly trained in large datasets to create models with trillions of parameters. This approach does not deliberately include statistical laws, grammatical rules, or any other theoretical properties of language. The models ``learn'' from the data and their outputs satisfy (most of the time) the properties observed in the data, including statistical laws~\cite{takahashi_neural_2017,takahashi_evaluating_2019,lippi_natural_2019}. These laws and rules are not explicit coded nor mathematically derived from the model. Their empirical outputs reproduce the properties of real text, but the reasons or mechanisms remain unclear. There is no theory of language in these models, neither as an input nor as an output. There is no ambition to code or reproduce the mechanism humans use to generate language; this is not how large language models were conceived, programmed, or designed. 

Reflecting on the natural-science experiences of the past millennia is important to better set expectations and to understand the limitations of data-driven research, such as machine learning and the use of statistical laws. Firstly, as empirical science has long established, data and theory are entwined: the measurement and interpretation of data are contingent upon theoretical frameworks -- there exists no "theory-neutral" algorithm or data-analysis methodology.
Secondly, theoretical models, along with compatible computational methods, are essential not only to fulfill the scientific quest of a mechanistic understanding but also to explore scenarios, extrapolate predictions to unobserved settings, and consider interventions. Challenges of interpreting and manipulating \index{machine learning} machine-learning methodology frequently stem from a combination of these elements. 
In this context, the study of statistical laws assumes significance as it exemplifies a data-driven approach rooted in the natural sciences: they naturally benefit from the increasing large availability of data but at the same time they aim at a scientific (theoretical) understanding of the underlying systems.

The reasoning above indicates that statistical laws can contribute to an alternative, science-based approach to data driven research. This monograph has discussed in detail the subtleties and difficulties in matching statistical laws, data, and models, often portraying them as limitation of a na\"ive application of the traditional approach of statistical laws. More broadly, they reflect the difficult interplay between theory and data that exists in all scientific fields and that needs to be taken into account if theoretical (generalizable) understanding is set as the scientific goal. In particular, we have seen how the analysis of the data and a decision on the validity of a statistical law cannot be done independently from a theoretical framework. This is well-known in natural sciences, but is often ignored in \index{machine learning} machine learning approaches in which the methodology to study the data is allegedly theory-free. The lack of an explicit connection between machine-learning methods and theoretical models is a limitation of these approaches. 

Beyond the broad opposition between the two approaches, statistical laws can be used in combination with machine-learning methods to address data-science problems. 
An example of this approach is the role played by Zipf's law in the development of improved topic-modelling methods for unsupervised text analysis\index{Zipf's law}~\cite{sato_topic_2010,lim_nonparametric_2016}. Simple parametric functions -- such as the ones used in statistical laws -- are also employed in the "Bayesian machine scientist" approach of model discovery~\cite{guimera_bayesian_2020,fajardo-fontiveros_fundamental_2023}. With the growing importance of automated discovery and machine-learning methods, a critical test for the relevance of statistical laws is in which extent they will remain relevant in the development of such methods. As these methods are expected to be increasingly complex and relevant, both inside and outside science, the relevance of statistical laws becomes exemplary to the broader question of the relevance of theory (and simple models) to the creation of knowledge and the development of applications.

\appendix

\chapter{Appendix: Datasets and Codes}\label{chap.appendices}
\section{Repositories}

The repository: \\ \url{https://github.com/edugalt/StatisticalLaws} contains the data and codes used in this monograph. It builds on the codes and data developed previously for specific studies: 

\begin{itemize}
    \item Urban scaling laws: \\ Repository~\url{https://github.com/edugalt/TwitterHashtags} \\ Refs.~\cite{leitao_is_2016,altmann_spatial_2020}.
    \item Fitting fat-tailed distributions: \\ Repository~\url{https://github.com/edugalt/TwitterHashtags} \\ Ref.~\cite{gerlach_stochastic_2013}.  \index{fat-tailed distribution}
    \item Effect of correlations: \\ Repository~\url{https://github.com/martingerlach/testing-statistical-laws-in-complex-systems} \\ Ref.~\cite{gerlach_testing_2019}.
    \item Constrained surrogates: \\ Repository:\url{https://github.com/JackMurdochMoore/power-law/} \\ Ref.~\cite{moore_nonparametric_2022}. \index{surrogate}
\end{itemize}

\section{Source of figures}

All figures of this monograph that contains data analysis can be reproduced using the code and data of our repository. The list below contains the name of the Jupyter notebooks available in repository~\url{https://github.com/edugalt/StatisticalLaws}, together with the figure numbers of this monograph that they reproduce:

\begin{itemize}
    \item {\it allometric.ipynb} contains the analysis of Kleiber's law and allometric scaling laws -- Sec.~\ref{ssec.allometry} -- including Figs.~\ref{fig.allometric} and \ref{fig.allometric2}.
    \item {\it bibliometric-data.ipynb} contains the analysis of the bibliometric data shown in Fig.~\ref{fig.bibliometric}. \index{Bibliometry}
    \item {\it burstinessWords.ipynb} contains the analysis of the inter-event time between words -- \index{burstiness} Sec.~\ref{ssec.burstywords} -- including Figs.~\ref{fig.burstyWords} and ~\ref{fig.representation-weibull}.
    \item {\it cities.ipynb} contains the analysis of all urban data, including the ALZ law -- Sec.~\ref{ssec.alz} --, urban scaling laws -- Sec.~\ref{ssec.urbanscaling} --, \index{Auerbach-Lotka-Zipf's law}  \index{Urban scaling laws} Figs.~\ref{fig:urban},~\ref{fig.zipfcities},~\ref{fig.urbanscaling},~\ref{fig.representation-rank},~\ref{fig.threshold},~\ref{fig.fitScalingUK}, and~\ref{fig.fitALZ}, and Tab.~\ref{tab.urbanscaling}.
      \item {\it constrained-powerlaw.ipynb} contains the code to generate constrained surrogates -- Sec.~\ref{ssec.constrained} -- including Fig.~\ref{fig.constrained}.
    \item {\it heaps.ipynb} contains the analysis of Herdan-Heaps' law -- Sec.~\ref{ssec.heaps} -- including Fig.~\ref{fig.heaps}. \index{Herdan-Heaps' law}
    \item {\it pareto.ipynb} contains the analysis of Pareto's law of inequality \index{Pareto's law} -- Sec.~\ref{ssec.pareto} -- including Fig.~\ref{fig.pareto}
    \item {\it synthetic-powerlaw.ipynb} contains the generation and analysis of synthetic power-law datasets with correlation -- Sec.~\ref{ssec.limitationslikelihood} -- including Figs.~\ref{fig.synthetic-correlated} and~\ref{fig.synthetic-correlated2}.
    \item {\it zipf.ipynb} Contains the analysis of Zipf's law of word frequencies \index{Zipf's law} -- Sec.~\ref{ssec.zipf} -- including Figs.~\ref{fig:zipf}-\ref{fig.zipfcomparison} and Tab.~\ref{tab.zipfspanish}-\ref{tab.zipfcomparison}.
\end{itemize}

\newpage


\addcontentsline{toc}{chapter}{Bibliography}
\begingroup
\renewcommand{\addcontentsline}[3]{}

\newcommand{\etalchar}[1]{$^{#1}$}
\providecommand{\bysame}{\leavevmode\hbox to3em{\hrulefill}\thinspace}
\providecommand{\MR}{\relax\ifhmode\unskip\space\fi MR }
\providecommand{\MRhref}[2]{%
  \href{http://www.ams.org/mathscinet-getitem?mr=#1}{#2}
}
\providecommand{\href}[2]{#2}

\endgroup



\addcontentsline{toc}{chapter}{Index}

\begin{theindex}

  \item allometric laws, \hyperpage{9}, \hyperpage{19}, 
		\hyperpage{40, 41}, \hyperpage{45, 46}, 
		\hyperpage{48, 49}, \hyperpage{62}, \hyperpage{71}, 
		\hyperpage{90}
  \item Auerbach, Felix, \hyperpage{9, 10}, \hyperpage{27, 28}, 
		\hyperpage{30}, \hyperpage{62}, \hyperpage{64}, 
		\hyperpage{110, 111}, \hyperpage{117}
  \item Auerbach-Lotka-Zipf's law, \hyperpage{9, 10}, \hyperpage{16}, 
		\hyperpage{27--29}, \hyperpage{40}, \hyperpage{62}, 
		\hyperpage{66, 67}, \hyperpage{72}, \hyperpage{80}, 
		\hyperpage{82}, \hyperpage{84}, \hyperpage{89}, 
		\hyperpage{91}, \hyperpage{110--112}, 
		\hyperpage{116, 117}, \hyperpage{130}
  \item autocorrelation function, \hyperpage{61}, \hyperpage{92}, 
		\hyperpage{96}

  \indexspace

  \item Benford's law, \hyperpage{59}
  \item bibliometric, \hyperpage{38}
  \item Bibliometry, \hyperpage{123}, \hyperpage{130}
  \item Binomial distribution, \hyperpage{11}, \hyperpage{24}, 
		\hyperpage{28}
  \item Bradford's law, \hyperpage{36}, \hyperpage{38}
  \item brain, \hyperpage{38}, \hyperpage{63}
  \item burstiness, \hyperpage{51}, \hyperpage{58, 59}, \hyperpage{91}, 
		\hyperpage{122, 123}, \hyperpage{130}

  \indexspace

  \item Chapernowne, \hyperpage{26}
  \item Chemistry, \hyperpage{38}

  \indexspace

  \item Data Science, \hyperpage{6, 7}, \hyperpage{126}

  \indexspace

  \item earthquakes, \hyperpage{19}, \hyperpage{34--36}, \hyperpage{49}, 
		\hyperpage{52}, \hyperpage{55--57}, \hyperpage{59, 60}, 
		\hyperpage{105}
  \item Ecology, \hyperpage{49}

  \indexspace

  \item fat-tailed distribution, \hyperpage{6}, \hyperpage{27}, 
		\hyperpage{36}, \hyperpage{39}, \hyperpage{59}, 
		\hyperpage{66}, \hyperpage{80}, \hyperpage{89}, 
		\hyperpage{107}, \hyperpage{114}, \hyperpage{116}, 
		\hyperpage{129}

  \indexspace

  \item gamma distribution, \hyperpage{57, 58}
  \item Gaussian distribution, \hyperpage{11}, \hyperpage{16}, 
		\hyperpage{34}, \hyperpage{51}, \hyperpage{57}, 
		\hyperpage{75}, \hyperpage{79}, \hyperpage{105}, 
		\hyperpage{119}
  \item gene expression, \hyperpage{38}
  \item Gross Domestic Product, GDP, \hyperpage{10}, \hyperpage{41, 42}, 
		\hyperpage{78}, \hyperpage{80}, \hyperpage{101}
  \item Gutenberg-Richter's law, \hyperpage{6}, \hyperpage{19}, 
		\hyperpage{34, 35}, \hyperpage{55}, \hyperpage{59}, 
		\hyperpage{69}, \hyperpage{77}, \hyperpage{90}, 
		\hyperpage{97}

  \indexspace

  \item Herdan-Heaps' law, \hyperpage{19}, \hyperpage{31}, 
		\hyperpage{34}, \hyperpage{43--45}, \hyperpage{60}, 
		\hyperpage{71}, \hyperpage{78}, \hyperpage{104}, 
		\hyperpage{108}, \hyperpage{121}, \hyperpage{127}, 
		\hyperpage{130}
  \item heteroscedasticity, \hyperpage{81}
  \item homoscedasticity, \hyperpage{79}
  \item hypothesis testing, \hyperpage{15}, \hyperpage{77}, 
		\hyperpage{91}, \hyperpage{93--97}, \hyperpage{100}, 
		\hyperpage{110--112}, \hyperpage{117, 118}, 
		\hyperpage{122}

  \indexspace

  \item Kepler's law, \hyperpage{12}, \hyperpage{126}
  \item Kleiber's law, \hyperpage{13}, \hyperpage{19}, 
		\hyperpage{45, 46}, \hyperpage{48}, \hyperpage{64}, 
		\hyperpage{111}, \hyperpage{118}, \hyperpage{121}, 
		\hyperpage{125}
  \item Kolmogorov-Smirnov distance, \hyperpage{93}

  \indexspace

  \item linguistic laws, \hyperpage{43}, \hyperpage{60}, \hyperpage{91}, 
		\hyperpage{94}
  \item log-normal distribution, \hyperpage{28}, \hyperpage{79}, 
		\hyperpage{114}
  \item long-range correlations, \hyperpage{19}, \hyperpage{55}, 
		\hyperpage{59, 60}, \hyperpage{91}, \hyperpage{104}
  \item Lotka's law, \hyperpage{36}, \hyperpage{38}

  \indexspace

  \item machine learning, \hyperpage{6}, \hyperpage{15}, 
		\hyperpage{126, 127}
  \item mammals, \hyperpage{48}
  \item Mandelbrot, Benoit, \hyperpage{12}, \hyperpage{19}, 
		\hyperpage{23}, \hyperpage{31}, \hyperpage{33}, 
		\hyperpage{45}
  \item Menzerath-Altmann law, \hyperpage{60}
  \item model comparison, \hyperpage{85, 86}, \hyperpage{88, 89}, 
		\hyperpage{94}, \hyperpage{100}, \hyperpage{102}, 
		\hyperpage{113}, \hyperpage{116, 117}, \hyperpage{119}, 
		\hyperpage{121, 122}
  \item Monkey typist, \hyperpage{33}

  \indexspace

  \item Newton's law, \hyperpage{109}

  \indexspace

  \item Omori's law, \hyperpage{55, 56}, \hyperpage{59}

  \indexspace

  \item Pareto's law, \hyperpage{6}, \hyperpage{9}, \hyperpage{19}, 
		\hyperpage{21}, \hyperpage{24--26}, \hyperpage{63}, 
		\hyperpage{68}, \hyperpage{82}, \hyperpage{108--110}, 
		\hyperpage{130}
  \item Physics, \hyperpage{11}, \hyperpage{110}
  \item Poisson distribution, \hyperpage{11}, \hyperpage{24}, 
		\hyperpage{50}, \hyperpage{56}, \hyperpage{58}, 
		\hyperpage{104}, \hyperpage{119}
  \item Poisson process, \hyperpage{58}
  \item power-law distribution, \hyperpage{9}, \hyperpage{13}, 
		\hyperpage{19, 20}, \hyperpage{22}, \hyperpage{28, 29}, 
		\hyperpage{31}, \hyperpage{34}, \hyperpage{36}, 
		\hyperpage{38}, \hyperpage{40}, \hyperpage{45}, 
		\hyperpage{49}, \hyperpage{52}, \hyperpage{55}, 
		\hyperpage{57--59}, \hyperpage{62}, \hyperpage{64--66}, 
		\hyperpage{71}, \hyperpage{73}, \hyperpage{75}, 
		\hyperpage{78}, \hyperpage{81}, \hyperpage{84}, 
		\hyperpage{86, 87}, \hyperpage{89}, \hyperpage{94, 95}, 
		\hyperpage{97, 98}, \hyperpage{105}, \hyperpage{108}, 
		\hyperpage{111, 112}, \hyperpage{115}, \hyperpage{119}

  \indexspace

  \item recurrence, \hyperpage{50}

  \indexspace

  \item S-curves, \hyperpage{59}, \hyperpage{61}
  \item scale-free networks, \hyperpage{6}, \hyperpage{9}, 
		\hyperpage{36--38}, \hyperpage{63}, \hyperpage{94, 95}, 
		\hyperpage{100, 101}, \hyperpage{105}, \hyperpage{107}, 
		\hyperpage{111}
  \item Simon, Herbert, \hyperpage{12}, \hyperpage{19}, 
		\hyperpage{22, 23}, \hyperpage{26}, \hyperpage{29}, 
		\hyperpage{31}, \hyperpage{36}, \hyperpage{45}, 
		\hyperpage{107}, \hyperpage{127}
  \item social physics, \hyperpage{11, 12}, \hyperpage{40}, 
		\hyperpage{42}, \hyperpage{61}, \hyperpage{109}, 
		\hyperpage{126}
  \item solar flares, \hyperpage{38}
  \item species, \hyperpage{46--49}
  \item stretched exponential distribution, \hyperpage{6}, 
		\hyperpage{19}, \hyperpage{54}, \hyperpage{56, 57}, 
		\hyperpage{64}, \hyperpage{81}, \hyperpage{104}, 
		\hyperpage{114}
  \item surrogate, \hyperpage{77}, \hyperpage{93}, \hyperpage{97}, 
		\hyperpage{100}, \hyperpage{129}

  \indexspace

  \item Taylor's law, \hyperpage{60, 61}, \hyperpage{80}, 
		\hyperpage{104}
  \item threshold, \hyperpage{55}, \hyperpage{74}, \hyperpage{82, 83}

  \indexspace

  \item Urban scaling laws, \hyperpage{9, 10}, \hyperpage{13}, 
		\hyperpage{16}, \hyperpage{40--43}, \hyperpage{61}, 
		\hyperpage{63}, \hyperpage{71, 72}, \hyperpage{78}, 
		\hyperpage{80, 81}, \hyperpage{91}, 
		\hyperpage{100--103}, \hyperpage{121}, \hyperpage{130}
  \item urban scaling laws, \hyperpage{75}

  \indexspace

  \item Weibull distribution, \hyperpage{52, 53}, \hyperpage{55, 56}, 
		\hyperpage{65}

  \indexspace

  \item Zipf's law, \hyperpage{6}, \hyperpage{13}, \hyperpage{18}, 
		\hyperpage{30--33}, \hyperpage{45}, \hyperpage{52}, 
		\hyperpage{55}, \hyperpage{60}, \hyperpage{62, 63}, 
		\hyperpage{82}, \hyperpage{84--88}, \hyperpage{90}, 
		\hyperpage{104}, \hyperpage{108--110}, 
		\hyperpage{116, 117}, \hyperpage{127, 128}, 
		\hyperpage{130}

\end{theindex}

\end{document}